\documentclass[braunschweig, accepted, englishtitle]{thesis}		

\usepackage{graphicx}
\graphicspath{ {images/} } 
\usepackage{float}

\usepackage{subfigure} 

\usepackage{natbib}
\bibpunct{(}{)}{;}{a}{}{,}	
\usepackage{txfonts}

\usepackage[hang,splitrule]{footmisc}
\usepackage[figuresright]{rotating}

\usepackage[CaptionAfterwards]{fltpage}

\usepackage{lipsum}
\usepackage{ccaption}

\usepackage{microtype}

\usepackage{bm} 

\usepackage{epsf} 

\usepackage{xcolor} 

\usepackage{epigraph}

\usepackage[NoDate]{currvita}

\title{A search for signatures of Europa's atmosphere and plumes in Galileo charged particle data}
\author{Hans L.F. Huybrighs}
\town{Etterbeek, Belgium}  

\thesisadvisorycommitteea{Prof. Dr. Karl-Heinz Glassmeier}
\instthesisadvisorycommitteea{Technische Universit{\"a}t Braunschweig, Braunschweig, Germany}

\thesisadvisorycommitteeb{Dr. Norbert Krupp}
\instthesisadvisorycommitteeb{Max-Planck-Institut f\"ur Sonnensystemforschung, G{\"o}ttingen, Germany}

\thesisadvisorycommitteec{Dr. Elias Roussos}
\instthesisadvisorycommitteec{Max-Planck-Institut f\"ur Sonnensystemforschung, G{\"o}ttingen, Germany}

\thesisadvisorycommitteed{Dr. Yoshifumi Futaana}
\instthesisadvisorycommitteed{Swedish Institute of Space Physics, Kiruna, Sweden}

\refereea{Prof. Dr. Karl-Heinz Glassmeier}
\instrefereea{Technische Universit{\"a}t Braunschweig, Braunschweig, Germany}

\refereeb{Assoc. Prof. Dr. Yoshifumi Futaana}
\instrefereeb{Swedish Institute of Space Physics, Kiruna, Sweden}

\refereec{}
\instrefereec{}

\submitteddate{8.10.2018}
\submittedyear{2018}
\examinationdate{5.12.2018}
\publicationyear{2018}

\begin{document}

\maketitle

\chapter*{Publications}
\addcontentsline{toc}{chapter}{Publications}
\label{ch_publications}
Partial results of this work were published in advance in the following contributions:

\section*{Peer-reviewed publications}

\textbf{Huybrighs, H.L.F.}, Futaana, Y., Barabash, S., Wieser, M., Wurz, P., Krupp, N.,  Glassmeier, K.H., Vermeersen. B., 2017, On the in-situ detectability of Europa's water vapour plumes from a flyby mission. Icarus 289, 270-280. 
https://doi.org/10.1016/j.icarus.2016.10.026
\newline

\noindent
Carnielli, G., Galand, M., Leblanc, F.,  Modolo, R., Beth, A., \textbf{Huybrighs, H.L.F.}, Jia, X., First 3D test particle model of Ganymede's ionosphere. \textit{Submitted} to Icarus.

\section*{Poster and oral presentations}
\noindent
Carnielli, G., Galand, M., Modolo, R., Leblanc, F., Beth, A., \textbf{Huybrighs, H.L.F.}, Jia, X., 2018, Constraining Ganymede's exosphere through numerical simulations of its ionosphere and Galileo observations. European Planetary Science Congress 2018, Berlin.
\newline

\noindent
Carnielli, G., Galand, M., Modolo, R., Leblanc, F., Leclerq, L., Beth, A., \textbf{Huybrighs, H.L.F.}, 2018, Constraining Ganymede's neutral and plasma environment through numerical simulations of its ionosphere. Magnetospheres of Outer Planets 2018, Boulder. Poster presentation.
\newline

\noindent
\textbf{Huybrighs, H.L.F.}, Roussos, E., Krupp, N., Fraenz, M., Futaana, Y., Barabash, S. Glassmeier, K.-H., 2018, Signatures of Europa's atmosphere in Galileo EPD data during the E12 flyby. Magnetospheres of Outer Planets 2018, Boulder. Poster presentation.
\newline

\noindent
\textbf{Huybrighs, H.L.F.}, Roussos, E., Krupp, N., Fraenz, M., Futaana, Y., Barabash, S. Glassmeier, K.-H., 2017, The search for active Europa plumes in Galileo plasma particle detector data: the E12 flyby. AGU fall meeting 2017, New Orleans. Poster presentation.
\newline

\noindent
\textbf{Huybrighs, H.L.F.}, Roussos, E., Krupp, N., Fraenz, M., Futaana, Y., Barabash, S. Glassmeier, K.-H., 2017, Are there signatures of active Europa plumes in Galileo in-situ data? European Planetary Science Congress 2017, Riga. Oral presentation.
\newline

\noindent
\textbf{Huybrighs, H.L.F.}, Roussos, E., Krupp, N., Fraenz, M., Futaana, Y., Barabash, S. Glassmeier, K.-H., 2017, The search for Europa plume signatures in Galileo plasma particle data. Magnetospheres of Outer Planets 2017, Uppsala. Poster presentation.
\newline

\noindent
\textbf{Huybrighs, H.L.F.}, Roussos, E., Krupp, N., Fraenz, M., Futaana, Y., Barabash, S. Glassmeier, K.-H., 2017, The search for Europa plume signatures in Galileo in-situ data. DPG Fr{\"u}hjahrstagung 2017 (German Physical Society Spring Meeting), Bremen, March 14, 2017. Poster presentation. 
\newline

\noindent
\textbf{Huybrighs, H.L.F.}, Roussos, E., Krupp, N., Futaana, Y., Barabash, S. Glassmeier, K.-H., 2016, Feasibility study of the in-situ detectability of Europa's neutral and plasma plumes from a flyby mission. Europa-Enceladus Plumes Workshop, Caltech, October 15, 2016. Oral presentation.
\newline

\noindent
\textbf{Huybrighs, H.L.F.}, Roussos, E., Krupp, N., Futaana, Y., Barabash, S. Glassmeier, K.-H., 2016, Feasibility study of in-situ measurements of Europa's neutral and plasma plumes with JUICE/PEP. In: Geophysical Research Abstracts Vol. 18, EGU2016-13425, 2016 EGU General Assembly 2016, Vienna. Oral presentation.
\newline

\clearpage\null\newpage

\tableofcontents

\clearpage\null\newpage

\section*{Summary\markboth{Summary}{Summary}}
\addcontentsline{toc}{chapter}{Summary}

Europa, the fourth largest moon of Jupiter, 
is thought to harbour a potentially habitable ocean of water under its icy surface. 
Remote sensing observations indicate that Europa is surrounded by a tenuous atmosphere. Furthermore, recent observations and historic data from Galileo 
hint at the occurrence of water vapour eruptions originating from the interior that create 200 km high plumes. Taking samples of the atmosphere, and more directly, the plumes, could offer a "window" into Europa's interior, without even having to land.
Due to the lack of adequate measurements large uncertainties exist in the properties of Europa's atmosphere (e.g. loss rate and structure) and plumes (e.g. structure, source location, duration), of the latter one the existence is not even confirmed.
Unexploited opportunities to constrain properties of the atmosphere or plumes remain in the in-situ data collected by the Galileo mission, the only mission that has surveyed Europa from close by. 
As it was not equipped with a neutral particle detector, no direct detections of the particles of the atmosphere or the plumes could have been made. However, it carried two charged particle detectors, the energetic particle detector (EPD) and the plasma instrumentation (PLS), and a magnetometer (MAG) that can be used to investigate if charged particles interact with the tenuous atmosphere or possible plumes. 

In this thesis, I investigate how to constrain the properties (density, scale height) of Europa's tenuous atmosphere and plumes, using data from the Galileo in-situ particle detector instruments (PLS, EPD).
First, I analyse and compare the flybys, with a focus on the E12 flyby as it demonstrates the strongest signs of interaction of the atmosphere and the magnetospheric environment. 
Next, I simulate the depletion of energetic protons and oxygen ions 
measured during the E12 flyby. A Monte Carlo particle tracing method is used to simulate the particle trajectories and determine where they are depleted: at the surface or in the atmosphere. 
The simulations show that the depletions of energetic ions observed in the range 80 to 540 keV can be explained by charge exchange with neutral atmospheric particles, but not by impact on the surface alone. This suggests that an atmosphere must have been present during the E12 flyby. The simulations best represent the sublimated component of the atmosphere and favour higher surface densities than the literature, but are not sensitive to the scale height. 
Furthermore, an additional depletion is visible very briefly in the energetic protons with energies from 540 to 1040 keV. This feature could be consistent with localized losses from a Europa plume, although such a solution is not unique. 
Finally, I simulate the detection of plume and atmosphere originating H$_2$O and H$_2$O$^+$ by the neutral particles sensor and the ion sensor part of the upcoming JUpiter ICy moon Explorer. By comparing the signature of atmospheric H$_2$O and H$_2$O$^+$  and plume originating H$_2$O and H$_2$O$^+$ particles, I show that the signature of the plume can, in principle, be recognized in the combined atmosphere-plume signal.

Though the most direct measurements of Europa's tenuous atmosphere and plumes will come from optical imaging or the in-situ detection of neutral particles, this work shows that the  atmosphere and the plumes can influence energetic particle and plasma properties near the moon. The synergy between the optical measurement and the in-situ measurements of neutral and charged particles will allow a detailed characterization of the atmosphere-plume system, a capability that is available both in the upcoming JUICE and Europa Clipper missions, and can be extended to other moons of Jupiter.

\clearpage\null\newpage

\section*{Zusammenfassung\markboth{Zusammenfassung}{Zusammenfassung}}
\addcontentsline{toc}{chapter}{Zusammenfassung}
Europa ist der viertgr{\"o}{\ss}te Mond des Jupiter und es wird angenommen, dass dieser einen eventuell bewohnbaren Ozean aus Wasser unter seiner eisigen Oberfl{\"a}che beherbergt. Beobachtungen mittels Satelliten-Fernerkundung ('Remote sensing')
deuten darauf hin, dass Europa von einer d{\"u}nnen Atmosph{\"a}re umgeben ist. Dar{\"u}ber hinaus zeigen j{\"u}ngste Beobachtungen sowie historische Daten von der Galileo Satellitenmission, dass Wasserdampferuptionen auftreten, die im Inneren von Europa entstehen und 200 km hohe Dampffahnen ('Plumes') erzeugen k{\"o}nnen. Die Entnahme von Proben aus der Atmosph{\"a}re, insbesondere aus den auftretenden Dampffahnen, k{\"o}nnen daher einen Einblick in Europas Inneres bieten, ohne dass es einer Landung bedarf. Aufgrund unzureichender Messungen gibt es allerdings nicht nur gro{\ss}e Unsicherheiten in den Eigenschaften der Atmosph{\"a}re Europas (z.B. Verlustrate und Struktur), sondern auch der Dampffahnen (z.B. Struktur, Herkunft, Lage, Dauer), deren Existenz nicht einmal best{\"a}tigt ist. In den von der Galileo Mission gesammelten 'in-situ' Daten befinden sich ungenutzte M{\"o}glichkeiten um die Eigenschaften der Atmosph{\"a}re oder der Dampffahnen zu erkunden. Dies war die einzigen Mission, die Europa aus n{\"a}chster N{\"a}he erkundet hat. Da Galileo nicht mit einem neutralen Teilchendetektor ausgestattet war, konnten die Teilchen der Atmosph{\"a}re oder der Dampffahnen nicht direkt nachgewiesen werden. An Bord waren jedoch zwei geladene Teilchendetektoren, der energetische Teilchendetektor (EPD) und das Plasmainstrument (PLS), und ein Magnetometer (MAG), mit denen untersucht werden konnte ob die geladenen Teilchen mit der d{\"u}nnen Atmosph{\"a}re oder den m{\"o}glichen Dampffahnen interagieren. 

In dieser Arbeit untersuche ich, wie man die Eigenschaften (Dichte, Skalenh{\"o}he) der d{\"u}nnen Atmosph{\"a}re und der Dampffahnen von Europa mithilfe von Daten aus dem Galileo 'in-situ' Teilchendetektorinstrumenten (PLS, EPD) erforschen kann. Zuerst analysiere und vergleiche ich die Vorbeifl{\"u}ge, wobei ich mich auf den E12-Vorbeiflug konzentriere, da dieser die st{\"a}rkste Wechselwirkung zwischen Atmosph{\"a}re und Magnetosph{\"a}re aufzeigte. Als N{\"a}chstes simuliere ich den Abbau von energetischen Protonen und Sauerstoffionen, die w{\"a}hrend des E12-Vorbeifluges gemessen wurden. Ein Monte-Carlo Verfahren zur Teilchenverfolgung wird angewendet, um die Flugbahnen der Teilchen zu simulieren und um festzustellen, wo diese abgebaut werden - entweder an der Oberfl{\"a}che oder in der Atmosph{\"a}re. Die Simulationen zeigen, dass der Abbau von energetischen Teilchen, die im Bereich von 80 bis 540 keV beobachtete werden, durch Ladungsaustausch mit neutralen Teilchen in der Atmosph{\"a}re erkl{\"a}rt werden k{\"o}nnen, und nicht durch den Aufprall auf die Oberfl{\"a}che allein. Dies deutet darauf hin, dass w{\"a}hrend des E12-Vorbeifluges eine Atmosph{\"a}re vorhanden sein musste. Die Simulationen stellen vor allem die sublimierte Komponente der Atmosph{\"a}re dar und beg{\"u}nstigen h{\"o}here Oberfl{\"a}chendichten als die in der Literatur vorkommenden, und sind zudem unempfindlich gegen{\"u}ber der Skalenh{\"o}he. Dar{\"u}ber hinaus ist ein sehr kurz beobachtbarer zus{\"a}tzlicher Abbau von energetischen Protonen mit Energien von 540 bis 1040 keV sichtbar. Diese Eigenschaft k{\"o}nnte mit {\"o}rtlich begrenzten Verlusten aus einer Dampffahne {\"u}bereinstimmen, obwohl diese L{\"o}sung nicht eindeutig ist. Schlussendlich simuliere ich die Detektion von der aus der Atmosph{\"a}re und Dampffahne stammenden H$_2$O und H$_2$O$^+$ Signatur mit dem neutralen Teilchen- sowie Ionensensor, die beide Bestandteile der kommenden 'JUpiter ICy moon Explorer' Mission sind. Durch den Vergleich zwischen der atmosph{\"a}rischen und der von Dampffahnen stammenden H$_2$O und H$_2$O$^+$ Teilchen-Signatur, zeige ich, dass die Signatur der Dampffahne im gemeinsamen Signal der Atmosph{\"a}re und Dampffahnen erkannt werden kann.

Obwohl die direktesten Messungen von Europas d{\"u}nner Atmosph{\"a}re und Dampffahnen mithilfe von optischer Bildgebung ('Optical imaging') oder 'in-situ' Detektion von neutralen Teilchen gewonnen werden, zeigt meine Arbeit, dass die Atmosph{\"a}re und Dampffahnen die energetischen Teilchen- und Plasmaeigenschaften in der N{\"a}he des Mondes beeinflussen k{\"o}nnen. Die Synergie zwischen optischer Messung und 'in-situ' Messungen von neutralen und geladenen Teilchen erm{\"o}glicht eine sehr genaue Bestimmung des Atmosph{\"a}ren-Dampffahnen Systems. Beide Methoden werden in den kommenden JUICE und Europa Clipper Missionen verf{\"u}gbar sein, und k{\"o}nnen auch auf andere Monde des Jupiter angewendet werden.

\clearpage\null\newpage

\chapter{Introduction and motivation}\label{ch_introduction}

The goal of this thesis is to constrain the properties of Europa's tenuous atmosphere or any water vapour plumes, using data from the Galileo mission's in-situ particle detector instruments. Galileo studied the Jupiter system between 1995 and 2003 and made several flybys of Europa. In this chapter the context of the thesis and the motivation are provided, and, based on this, the main research question will be formulated. 

First, Europa, its tenuous atmosphere and its plumes, will be introduced (Section~\ref{s_europa_object}). Next, since plasma and energetic charged particles are key to this thesis, Europa's magnetosopheric environment and the interaction will be discussed (Section \ref{s_mag_environment}). Special emphasis is put on the interaction of plasma and energetic charged particles with Europa (Section \ref{s_mag_interaction_epd_pls}). Finally the main research question of this thesis will be formulated (Section \ref{s_goal}).

\begin{figure}[h]
  \centering
  \includegraphics[width=0.73\textwidth]{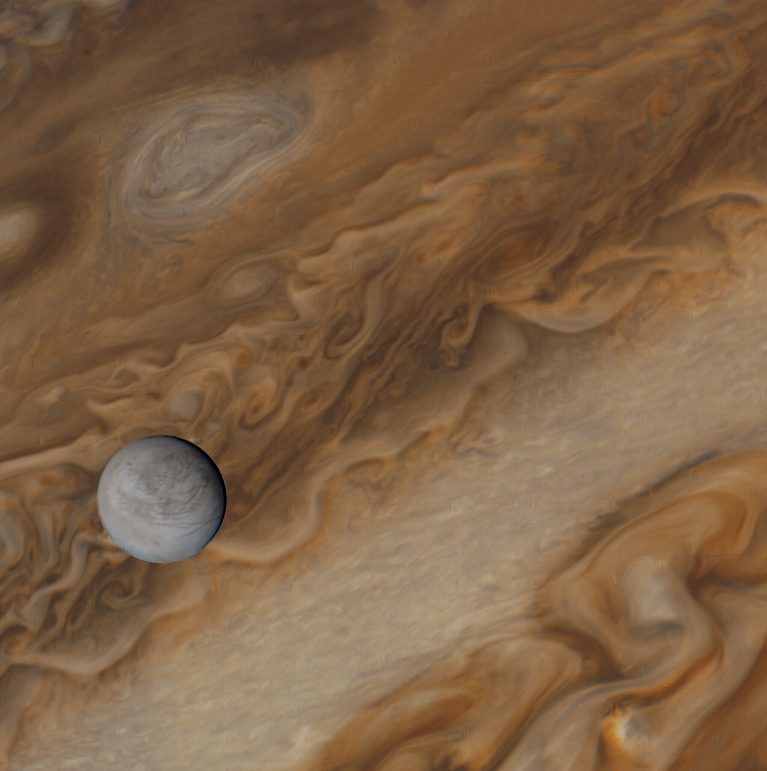}
  \caption{Europa in front of Jupiter, photographed by Voyager 1 on March 3rd 1979. [NASA/JPL-Caltech/Kevin M. Gill, CC BY 2.0].}
  \label{img_europa_jupiter}
\end{figure}

\section{Europa as a planetary object}\label{s_europa_object}

Europa is Jupiter's fourth largest moon. It was discovered in 1610 by Galileo Galilei, together with the other Galilean moons Ganymede, Callisto and Io. Its orbit is located between those of Io and Ganymede.  Europa in particular has captured the attention of the scientific community because it is thought to harbour a subsurface ocean of liquid water, which could be habitable. Europa possesses a tenuous atmosphere and a torus of neutral particles collocated with its orbit. Furthermore plumes of water vapour are thought to erupt on the surface.

This section provides a description of Europa as a planetary object. First Europa's interior, surface and habitability are discussed before moving outwards to Europa's atmosphere and the plumes therein.  Table \ref{tab_properties} provides an overview of general properties of Europa that are relevant to this work.
\begin{table}[h]
\centering
\begin{tabular}{|r|l|}
  \hline
  \textbf{Property} & \textbf{Value and units} \\
  \hline
  \hline
  Mean radius & 1562.09 km \\
  \hline 
  Mass & $4.79982 \times 10^{22}$ kg \\ 
  \hline 
  Escape velocity & 2.025 km/s. From \cite{Singer2013}. \\
  \hline
  Orbital eccentricity & 0.0101 \\
  \hline
  Semi-major axis & 670900 km or $~9.38$ Jupiter radii ($R_j$)\\ 
  & 1 $R_j$ = Jupiter's equatorial radius = 71492 km \\
  \hline
  Orbital period & 3.551810 days \\
  \hline
  Rotation period & Approximately synchronous, \\
  & so the same as the orbital period. \\
  \hline
\end{tabular}
\caption{General properties of Europa relevant to this chapter. From \cite{Pappalardo2009_Schubert}, unless specified otherwise.}
\label{tab_properties}
\end{table}
\subsection{Subsurface ocean and interior}
\label{ss_interior}
\begin{figure}[h]
  \centering
  \includegraphics[width=0.5\textwidth]{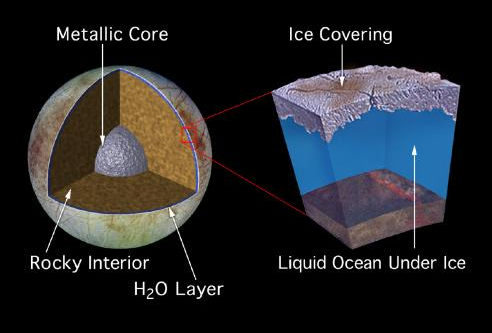}
  \caption{Overview of Europa's interior [https://photojournal.jpl.nasa.gov/catalog/PIA01669].}
  \label{img_interior}
\end{figure}

\begin{figure}[h]
  \centering
  \includegraphics[width=1.0\textwidth]{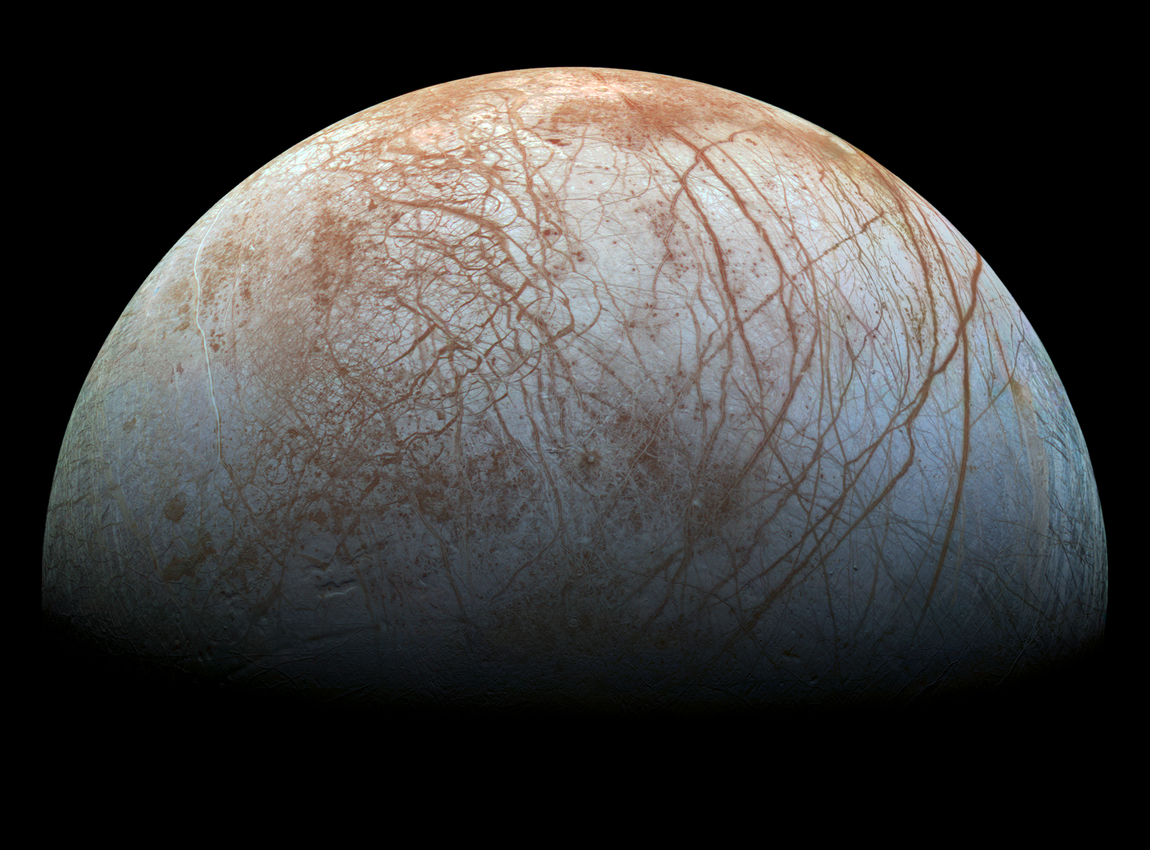}
  \caption{Europa's surface is relatively crater-free. Image obtained by the Galileo mission. [https://www.nasa.gov/jpl/europas-stunning-surface].}
  \label{img_surface}
\end{figure}

Europa is thought to have a differentiated interior, made out of a water ice-liquid outer layer, covering a rocky interior and a metal core. An overview of the interior is shown in Figure \ref{img_interior}. Evidence for a differentiated interior has been obtained from gravity field measurements from Doppler shifts in Galileo's radio signals \citep{Anderson1997,Anderson1998}.
The ice-liquid layer is thought to exist out of an ice shell covering a liquid water ocean. There are multiple signs that indicate that there is a liquid (or at least a warm, ductile, partially melted "slushy" layer) beneath the icy shell, some of these are:
\begin{itemize}
\item Young surface age: Europa's surface is devoid of impact features, compared to objects of similar age, such as the other Galilean moon Callisto (see Figure \ref{img_surface}). Crater densities on the surface of Europa suggest a young surface of age of 40 to 90 million years \citep{Pappalardo2009_Bierhaus}. Such a young age suggest that resurfacing processes are taking place that remove craters over time.
\item Chaos terrain: a terrain type characterized by broken apart pieces of surface ice that have moved with respect to each other, formation of such structures might require ductile layers underneath \citep{Bagenal2004_Greeley}.
\item The occurrence of hydrated materials close to surface features thought to be signs of disruption in the surface: such as ridges or chaos terrain \citep{Bagenal2004_Greeley}.
\item Unusual crater morphology: various unusual properties of Europa's craters such as a relative shallowness and concentric rings around the craters have been interpreted as signatures of a ductile or liquid layer underlying the ice layer \citep{Bagenal2004_Greeley}.
\end{itemize}

Strong evidence for a liquid layer underneath the outer ice shell has been obtained from Galileo magnetometer data. These data suggest that Jupiter's magnetic field induces a magnetic field in Europa. Jupiter's magnetic moment is tilted by about 10 degrees with respect to the orbital plane of Europa. Because of this, Europa experiences a magnetic field that varies in time. The varying component is responsible for the induction of a magnetic field in a conductive layer in Europa's interior. This conductive layer is thought to be a salty liquid water ocean \citep{Khurana1998,Kivelson1999,Zimmer2000,Pappalardo2009_Schubert}.

Gravity field measurements suggest that the total thickness of the combined ice-ocean layer ranges from 80 to 170 km \citep{Anderson1998}. Galileo magnetometer data constrain the thickness of the ice-only layer to less than 7-15 km, depending on the assumed composition of the ice layer \citep{Pappalardo2009_Schubert}. The total volume of water in Europa's ocean ($3 \times 10^9$km$^3$) is thought to exceed the total volume of water on Earth's surface \citep{Bagenal2004_Greeley}.

Europa's ocean is likely kept in a liquid state by tidal heating \citep{Bagenal2004_Schubert}. Europa's orbit is eccentric (see Table \ref{tab_properties}), which causes the tidal forces it experiences during one orbit around Jupiter to vary. This results in friction that is released in the form of heat and also deforms the surface. The eccentricity of Europa's orbit is maintained by the gravitational interaction between Io, Europa and Ganymede.

Europa's subsurface ocean is considered as a potential habitat. The JUICE Yellow Book \citep{JUICE}, which defines the science case for the future JUpiter ICy moons Explorer (JUICE) mission, defines four criteria required for (earth-like) life and states why these are met at Europa: the presence of liquid water for an extended period time, essential elements (S, P, O, N, C, H) and a source of (chemical) energy. The presence of liquid water was discussed previously. It is thought the ocean has been liquid continuously for at least a large part of Europa's history. The required elements have either been observed or their presence is inferred based on detections on other moons or on Jupiter \citep{Pappalardo2009_Carlson,Pappalardo2009_Hand}. Hydrothermal vents on the bottom of Europa's ocean are thought to be a possible source of chemical energy \citep{Kargel2000,Pappalardo2009_Vance}.

The potential habitability of Europa justifies the special interest in this moon. Assessing the habitability of Europa's ocean is complicated by the ice-layer which seals it off completely and prevents any direct study. One possibility to study the contents of the ocean is by sampling Europa's atmosphere. Organic molecules could be transported to the surface and sputtered away from it by impacting heavy energetic ions \citep{Johnson2018}. Europa's plumes could provide a more direct way of accessing the ocean, assuming the water of the plumes originates from it. This emphasizes the importance of studying Europa's atmosphere and plumes in the context of assessing Europa's habitability. 

\subsection{Tenuous atmosphere}
\label{s_atmosphere}
Europa possesses a tenuous atmosphere, which is the product of interaction between its surface and charged particles from Jupiter's magnetosphere . It is thought that sputtering by energetic ions (O$^{n+}$, S$^{n+}$ and H$^+$) is the main process responsible for the formation of the atmosphere \citep{Plainaki2018}. Sputtering mostly releases water molecules. The water subsequently disassociates in OH, H, H$_2$ and O, through processes such as electron impact dissociation. These products react chemically to produce mostly O$_2$ and H$_2$. Molecular oxygen is thought to be the most abundant species in the atmosphere, because it doesn't stick to the surface as much as other species (such as H$_2$O) or escapes Europa's gravity easily (such as H$_2$) \citep{Plainaki2018}. While sputtering is responsible for the extended, large scale height atmosphere, sublimation could also cause an additional low altitude confined atmospheric component with a much smaller scale height \citep{Smyth2006}. 

Published results about the atmosphere are restricted to remote sensing observations with the Hubble telescope, Earth-based telescopes and results obtained during the Jupiter flyby of the Cassini mission \citep{Plainaki2018}. No direct in-situ detections of the atmosphere have ever been published. Electrons from Jupiter's magnetosphere impact neutral atmospheric particles and dissociate and excite them, which leads to Auroral glows that are observable in UV. This allowed for the first detection of the atmosphere \citep{Hall1995}. Hubble UV observations support the hypothesis that the extended atmosphere is mostly composed out of O$_2$ \citep{Hall1995,Roth2016}. The presence of H$_2$ , H$_2$O, SO$_2$ , and Cl has been inferred indirectly \citep{Plainaki2018}.

\begin{figure}[h]
  \centering
  \includegraphics[width=1.0\textwidth]{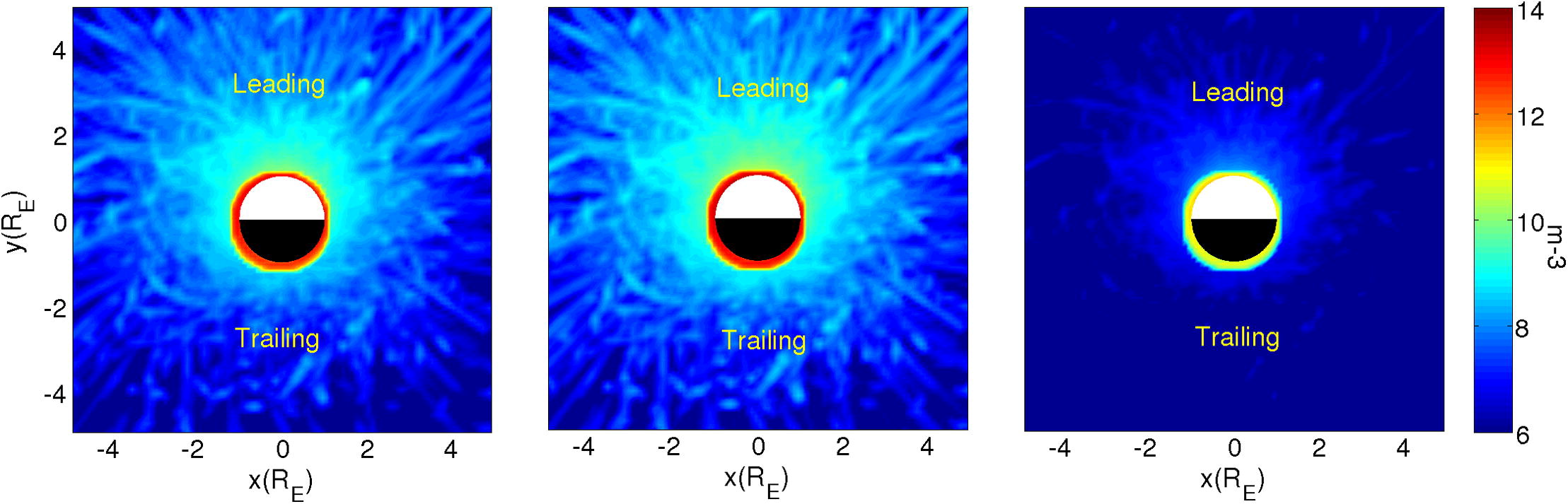}
  \caption{Spatial distribution of O$_2$ released by S$^+$, O$^+$ and H$^+$, including the sublimation component. 
 [Reprinted from \cite{Plainaki2012} with permission from Elsevier].}
  \label{img_atmosphere_plainaki}
\end{figure}

\begin{figure}[h]
  \centering
  \includegraphics[width=0.5\textwidth]{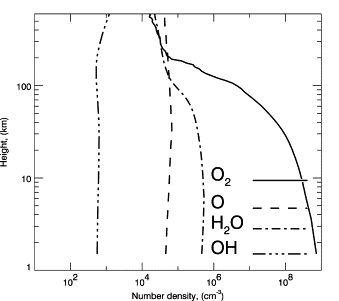}
  \caption{Altitude profile of Europa's atmosphere. This model includes sublimation and sputtering. 
 [Reprinted from \cite{Shematovich2005} with permission from Elsevier].}
  \label{img_shematovich2005}
\end{figure}

The spatial distribution of O$_2$ as modelled by \cite{Plainaki2013} is shown in Figure \ref{img_atmosphere_plainaki}. This model takes into account the release of O$_2$ by sputtering and sublimation. Various models have provided density profiles of the atmospheric constituents, in Figure \ref{img_shematovich2005} the profile of O$_2$, O, H$_2$O and OH provided by \cite{Shematovich2005} is shown.

Large uncertainties exist in properties of the atmosphere, such as the loss rate and structure, due to a lack of adequate measurements \citep{Plainaki2018}.  Hubble observations reported by \cite{Hall1995} suggest an average scale height of 145 km.
Various analytical models approximate the density of the O$_2$ atmosphere as a single component atmosphere, exponentially decaying as a function of altitude. In these models the profile is entirely determined by a scale height and a surface density \citep{Saur1998,Liu2000b,Schilling2007,Schilling2008,Dols2016}. The O$_2$ scale height assumptions in these models range from 145 km \citep{Saur1998,Schilling2007,Schilling2008}, 150 km \citep{Dols2016}, to 175 km \citep{Liu2000b}. Surface density estimates in these models are around $\sim10^7$ particles per cm$^3$ \citep{Saur1998,Liu2000b,Schilling2007,Schilling2008}.
Other analytical models approximate the O$_2$ atmosphere as a two component exponentially decaying atmosphere, the two components being a sputtered and sublimated component, each entirely determined by a scale height and surface density \citep{Rubin2015,Jia2018}. 
The sputtered O$_2$ scale height assumptions range from 145 km \citep{Rubin2015} to 500 km \citep{Jia2018}. Surface densities of the sputtered component vary from $5\times10^4$ particles per cm$^3$ \citep{Rubin2015} to 10$^6$ particles per cm$^3$ \citep{Jia2018}. Scale heights for the sublimated components vary from 20 \citep{Rubin2015} to 100 km \citep{Jia2018}. Densities for the sublimated component vary from $5\times \sim 10^7$ \citep{Jia2018} to $5\times10^8$ \citep{Rubin2015}.
Other models employ a Monte Carlo method to determine the distribution of the atmospheric components \citep{Shematovich2005,Smyth2006,Plainaki2012,Plainaki2013,Vorburger2018}. These models consider both sublimation and sputtered components, \cite{Vorburger2018} shows separate results for sputtering and sublimation. In this model, the atmospheric density for altitudes > $300$ km is dominated by the sublimated component. The sublimated O$_2$ surface density near the surface is $5\times10^7$ particles per cm$^3$ while the sputtered component is below $10^4$  particles per cm$^3$. The scale height of the sublimated component is around 20 km, while the scale height for the sputtered component is around 600 km. \cite{Shematovich2005,Smyth2006} report surface densities of $10^8$ 
particles per cm$^3$, while \cite{Plainaki2013} predicts densities of up to $10^8$ per cm$^3$ under sunlit conditions. The scale heights near the surface in these models are near 20 km. The range of atmospheric parameters is summarized in Table \ref{tab_properties_atmosphere}. By comparing the different previous studies it is clear that the surface density of the sublimated component is expected to exceed that of the sputtered component, while the scale height of the sputtered component is larger than the sublimated component.

\begin{table}[h]
\centering
\begin{tabular}{|r|l|l|}
  \hline
  \textbf{Property} & \textbf{Value} & \textbf{Sources} \\
  \hline
  \hline
  Surface density (single component) [cm$^{-3}$] & $\sim10^7$ & [1-5]  \\
  \hline
   Scale height density (single component) [km] &  145-175 & [1-5] \\
  \hline
  Scale height (sputtered) [km]& 145-600 & [6-8] \\
  \hline 
  Surface density (sputtered) [cm$^{-3}$] & $\sim10^4$ to $\sim10^6$ & [6-8] \\
  \hline
  Scale height (sublimated) [km] & 20-100 & [6-10] \\ 
  \hline
  Surface density (sublimated) [cm$^{-3}$] & $\sim10^7$ to $\sim10^8$ & [6-8,10-12] \\ 
  \hline 
\end{tabular}
\caption{Overview of the range of surface densities and scale heights of Europa's O$_2$ atmosphere reported in previous studies. Sources: [1] \cite{Saur1998}, [2] \cite{Liu2000b}, [3] \cite{Schilling2007}, [4] \cite{Schilling2008}, [5] Dols2016, [6] \cite{Rubin2015}, [7] \cite{Jia2018}, [8] \cite{Vorburger2018}, [9] \cite{Plainaki2012}, [10] \cite{Plainaki2013}, [11] \cite{Shematovich2005} and [12] \cite{Smyth2006}.}
\label{tab_properties_atmosphere}
\end{table}

As the model by \cite{Plainaki2012} shows (Figure \ref{img_atmosphere_plainaki}), the atmosphere of Europa is not spherically symmetric. \cite{Plainaki2013} states that the peak density occurs when the sub-solar point and the leading edge towards the flow of plasma at Europa's orbit fall together. This is because the flux of energetic ions, which is responsible for the extended atmosphere, is largest on the leading edge and because the incidence of solar radiation is what determines sublimation.
The structure of the atmosphere can be expected to vary in time, since the incidence of solar 

The upper part of Europa's atmosphere is referred to as the exosphere. The lower boundary of the exosphere, the exobase, is defined as the altitude at which the average distance travelled by a particle between collisions (mean free path) becomes larger than the scale height. The exobase is species dependant, for the O$_2$ profile in \cite{Shematovich2005} the exobase is estimated to be $\sim 80$ km \citep{Plainaki2018}. In the rest of the thesis I will refer to the entirety of atmosphere and exosphere as Europa's tenuous atmosphere.

It has been argued that eruptions of water vapour originating from Europa's interior ('plumes') contribute to the formation of its atmosphere (see also  Section \ref{s_europa_plumes}). Another topic related to Europa's atmosphere is the Europa torus, a torus of neutral particles located along the moon's orbit, see Section \ref{ss_interaction_EPD}. 
Interaction between Europa, its atmosphere and its magnetospheric environment are described in more detail in Section \ref{s_mag_environment} and \ref{s_mag_interaction}.
\subsection{Water vapour plumes}\label{s_europa_plumes}

Recent Hubble observations indicate that recurring plumes of water vapour occur on Europa. In the absence of published reviews of the plume related papers, this section provides an overview of the existing literature on this topic. First, early arguments for the existence of the plumes and previous observation attempts are discussed. Next, the main findings of the observations by Hubble and Galileo are summarized. 

\subsection*{Early arguments for activity and remote sensing observation attempts}
\label{early}
The occurrence of multiple processes on the surface of Europa that result in the release of vapour and the formation of plumes have been hypothesized by several authors, such as the boiling of water exposed to vacuum when a surface crack is opened by tidal forces \citep{Hoppa1999}, the production of vapour by shear heating when two sides of a crack move with respect to each other \citep{Nimmo2007} or the release of water by exploding bubbles formed when volatiles exsolve under vacuum conditions \citep{Fagents2000,Quick2013}.

No plumes or changes in surface features were observed with optical instruments during the flybys of Galileo, the Voyagers or New Horizons \citep{Hoppa1999,Pappalardo1999,Phillips2000,Fagents2003,Hurford2007,Roth2014a}.
\cite{DeLaFuenteMarcos2000} suggested that a transient bright feature on Europa, observed in 1999 with an Earth-based telescope, could be the result of explosive venting. \cite{Saur2011} discusses observations of UV auroral emissions ($O_I$ 130.4 nm and 135.6 nm emissions) of Europa with the Hubble Space Telescope in 1999. In these observation a surplus of emission near 90$^{\circ}$ west longitude is noticed. The authors argue that this could point to a radially asymmetrical atmosphere and state that active plumes could be one mechanism to explain this feature.

\subsection*{Remote sensing observations by \cite{Roth2014a}}
\label{roth}

\begin{figure}[h]
  \centering
  \includegraphics[width=0.5\textwidth]{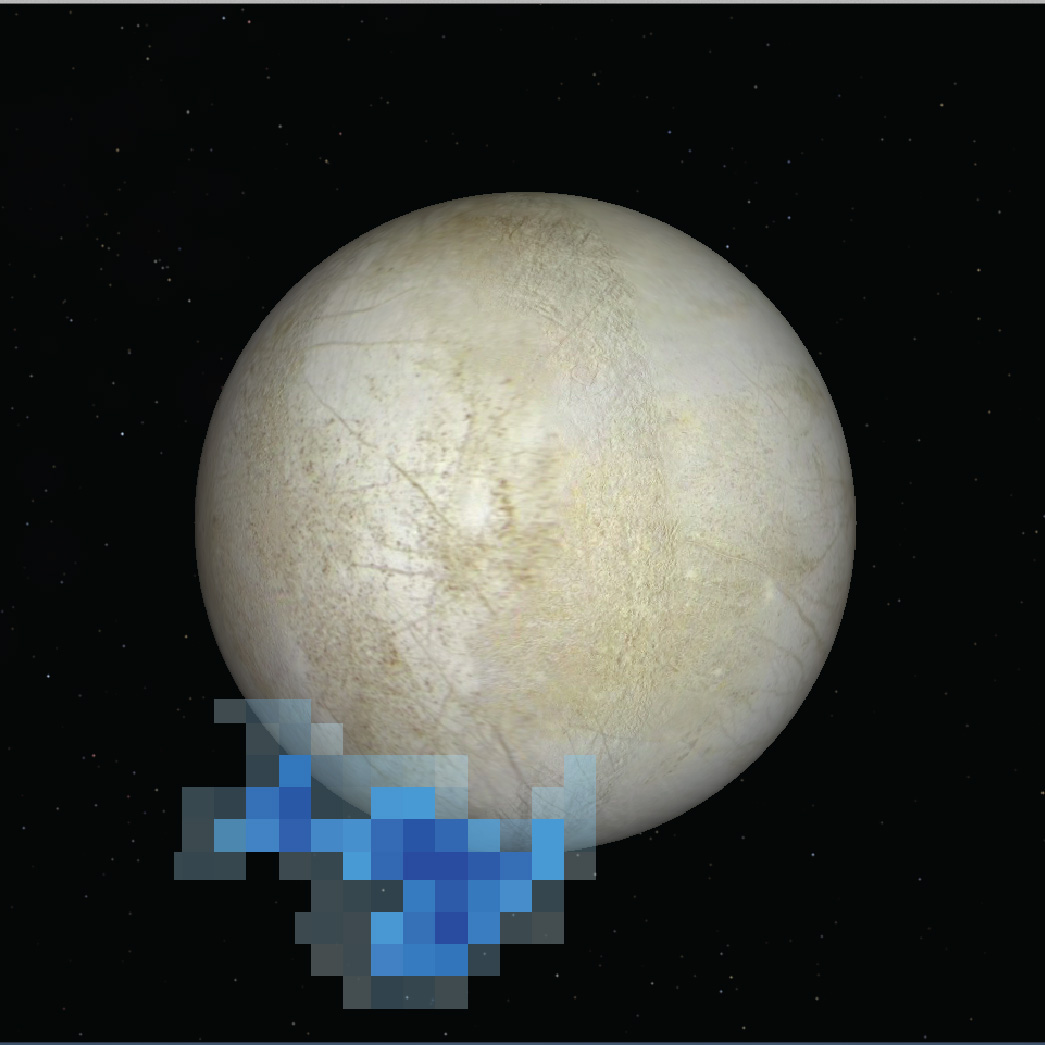}
  \caption{Graphic interpretation showing the position and extent of the emission surplus reported in \cite{Roth2014a} [https://www.nasa.gov/content/goddard/hubble-europa-water-vapor]}
  \label{img_roth}
\end{figure}

The first clear measurement of active plumes on Europa came from \cite{Roth2014a}. They discussed UV auroral emissions observed with the Hubble Space Telescope at Europa, a similar method to \cite{Saur2011}. The authors report the observation of an emission surplus observed near Europa's south pole. The surplus is observed in the oxygen O$_I$ line (130.4 nm). What is different from \cite{Saur2011} is that a surplus is also visible in the Lyman-$\alpha$ line, while no enhancement is noticed in the oxygen O$_I$ line (135.6 nm). The authors argue that this is indicative of electron impact ionization of H$_2$O. Furthermore the surplus is observed off the limb of Europa, so without Europa in the background (see Figure \ref{img_roth}). The surplus was observed for $\sim$7 hours, the duration of the observation window. The observed plume reached altitudes of 200 km, with an uncertainty of 100 km due to the low resolution of the observation. Supersonic velocities of 700 m/s are needed for plume particles to attain this height. The mass flux is estimated to be 7000 kg/s. The observations of \cite{Roth2014a} and \cite{Roth2014b} showed no correlation with Europa's orbital location. Further repeat observations using the same method did not result in additional plume detections \citep{Roth2016,Roth2017}. In total only one out of twenty measurements have resulted in a plume detection. Several reasons for this low detection rate can be thought of. It could be that the event reported in \cite{Roth2014a} was an exceptionally strong event, meaning that the typical plume eruption doesn't generate enough vapour to make detection possible. Another possibility is that the magnetospheric conditions during the successful detection were very different, resulting in a much higher electron flux than usual. 

\subsection*{Remote sensing observations by \cite{Sparks2016} and \cite{Sparks2017}}

\begin{figure}[h]
  \centering
  \includegraphics[width=0.5\textwidth]{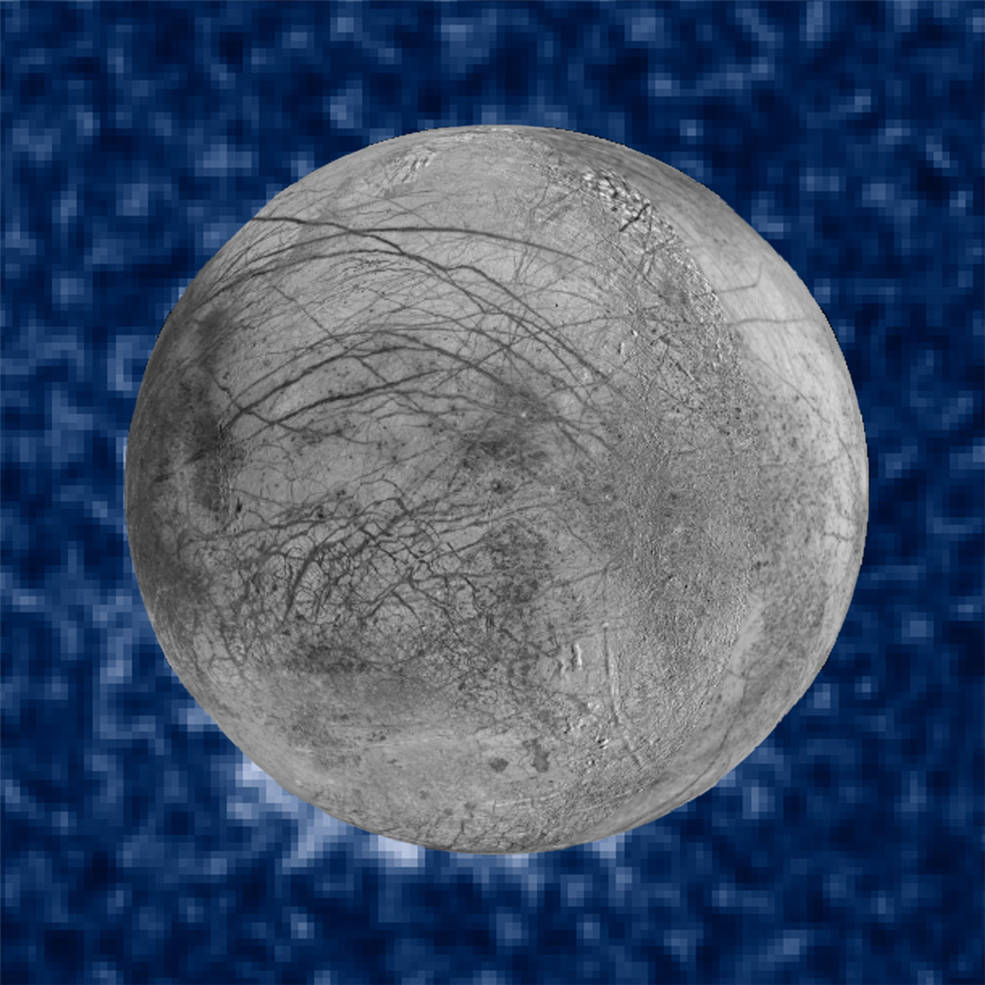}
  \caption{Limb darkening feature (bottom left) observed on January 26, 2014. [NASA/ESA/W. Sparks (STScI)/USGS Astrogeology Science Center]}
  \label{img_sparks}
\end{figure}

\cite{Sparks2016} and \cite{Sparks2017} observed transits of Europa in front of Jupiter. They identify darkening features at the limb of Europa, and argue that these could be caused by plumes (Figure \ref{img_sparks}). The detection rate is higher than the work by \cite{Roth2014a}, the presence of the limb darkening is reported in four out of twelve observations (each last about one hour). No information on the composition of the darkening features can be obtained from these observations. The authors estimate the column density from the observation and combine this with an estimate of the dimensions of the plume to obtain a total mass estimate. The total mass is of the order of $~10^6$kg, which is the same as the mass estimated by \cite{Roth2014a}. The reported plume heights are up to 220 km. The limb darkening featured discussed by \cite{Sparks2017} occurs at the same location on Europa's surface as one of the observations from \cite{Sparks2016}, a location in the vicinity of the crater Pwyll. \cite{Sparks2017} points out that this observation coincides with a thermal anomaly measured by the Galileo Photopolarimeter-Radiometer.
\cite{Trumbo2017} reports ALMA observations of Europa's thermal emissions obtained in 2015, prior to the observation reported in \cite{Sparks2017}. The observation suggests that the thermal anomaly might have cooled down since the Galileo observations.  \cite{McGrath2017} state that the strongest ionospheric signature found in Galileo radio occultation data nearly coincides with a location 50 km above the Pwyll crater. They report a peak electron density of $\sim1.3 \times 10^4$ particles per cm$^3$ at an altitude of $\sim50$ km. The repeated detection of the plume near Pwyll, with Hubble and the coinciding detection of the strongest ionosphere profile and the thermal anomaly lead the authors to suggest that this plume source has been active since the Galileo period.

\subsection*{Arguments from in-situ data by \cite{Jia2018}}

From the point of in-situ data the argument for plume activity has been made previously, in particular for the E12 flyby of Galileo.
\begin{itemize}
\item A significantly raised electron density \citep{Kurth2001} and anomalously strong magnetic fields \citep{Pappalardo2009_Kivelson} were measured during the E12 Galileo flyby. \cite{Pappalardo2009_Kivelson} states that the observed increase in magnetic field could be related to a local source of pickup ions.
\item \cite{Bagenal2015} states that atypical conditions during E12 could be related to active processes on Europa's surface, but also considers the passing of a small, cold, dense blob of plasma originating from Io as another hypothesis.
\end{itemize}
\begin{figure}[h]
  \centering
  \includegraphics[width=0.75\textwidth]{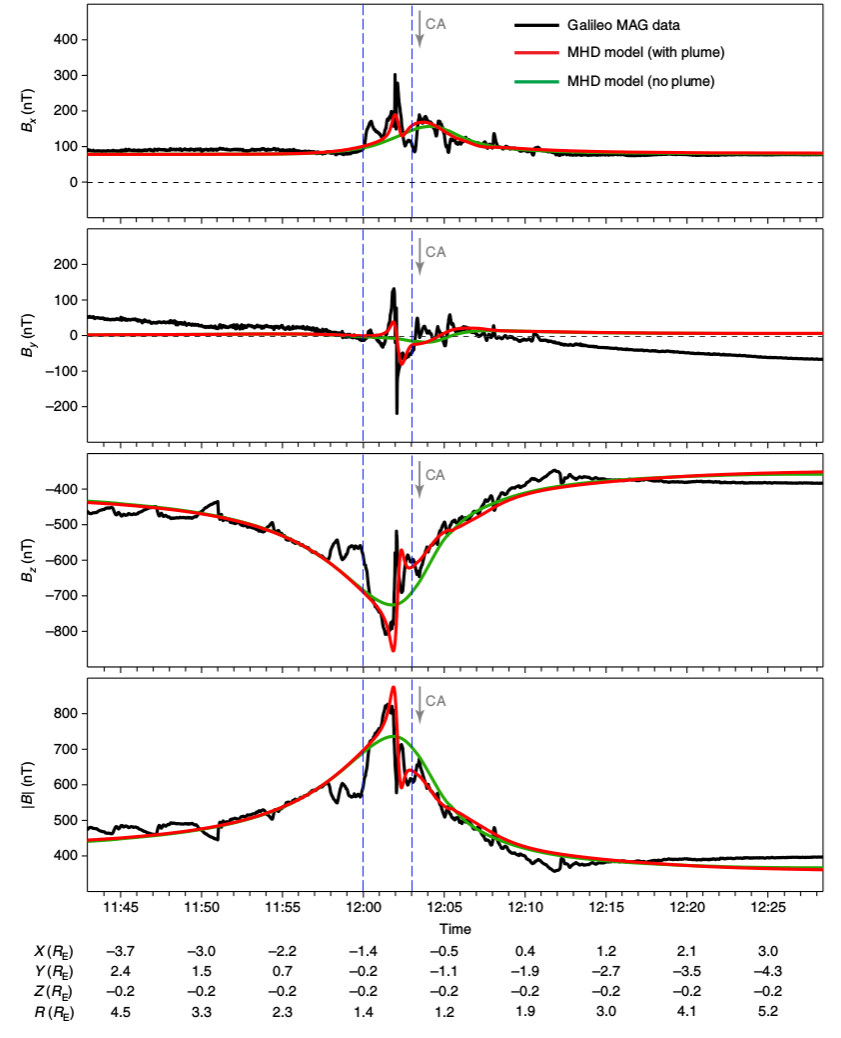}
  \caption{Magnetic field components during the E12 flyby, indicated are the Galileo MAG data and the model by \cite{Jia2018} with and without a plume. [Reprinted by permission from Springer: Nature Astronomy, \cite{Jia2018}, $\copyright$ (2018)]}
  \label{img_jia}
\end{figure}
\cite{Jia2018} implemented a MHD model of Europa's interaction with the corotating plasma to investigate the hypothesis of \cite{Pappalardo2009_Kivelson}. Additionally \cite{Jia2018} presents an electron density enhancement 
not previously reported in \cite{Kurth2001}. \cite{Jia2018} simulates the changes in plasma density and magnetic field components observed during the E12 flyby from 12:00 to 12:03 for two cases: one case where Europa has just a tenuous atmosphere and a second case where a plume is superposed on the atmosphere. The results of the magnetic field simulations are shown in Figure \ref{img_jia}. The authors conclude that the features observed in the data cannot be explained by a globally distributed atmosphere, but that the effect of a plume corresponds to what is seen in the data.
The structure causing the disturbance is about 1000 km in size along Galileo's trajectory. The plume in their model is located at 245 West longitude and 5 South latitude. The plume is modelled analytically, as an exponentially decaying atmosphere with a surface density of $2\cdot 10^9 $cm$^{-3}$ a scale height of 150 km and an angular width of 15$^\circ$. Furthermore the plume is tilted 15$^\circ$ from the radial direction.

Additionally the authors investigate a magnetic perturbation seen during the E26 flyby. Without showing the data or the simulation results they state that this feature cannot be explained by the same plume model as used for the E12 flyby. \cite{Bloecker2016} studies the interaction of the plumes with Europa's magnetospheric environment with an MHD model. 
They state that an exospheric inhomogeneity could have been present during Galileo's E26 flyby.

\subsection*{Overview of plume observations}

\begin{figure}[h]
  \centering
  \includegraphics[width=1.0\textwidth]{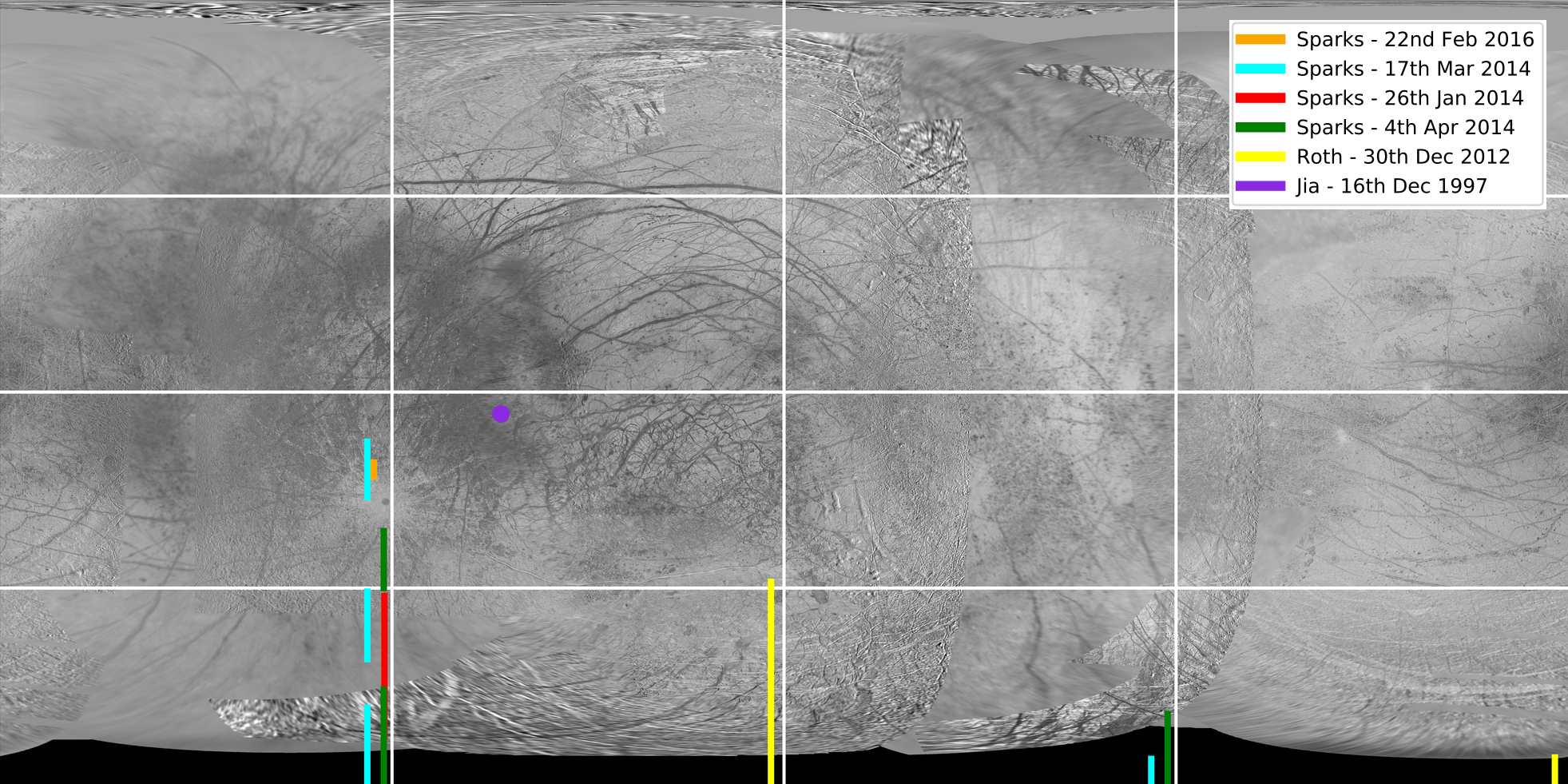}
  \caption{Locations of the plume observations on the surface. The vertical lines represent the latitudinal extent of the observed plumes as reported in \cite{Roth2014a,Sparks2016,Sparks2017}. The location of the plume reported in \cite{Jia2018} is represented by a purple dot.
  [Map of Europa obtained from: USGS]}
  \label{img_europa_map_plumes}
\end{figure}

The recent observations by \cite{Roth2014a,Sparks2016,Sparks2017,Jia2018} suggest that plumes are erupting from Europa's surface, at least in once case from the same location. Current observations can only give approximate constraints on properties such as the column density, height of the plumes and the minimal eruption time.  In Figure \ref{img_europa_map_plumes} the locations of the observations of \cite{Roth2014a,Sparks2016,Sparks2017,Jia2018} are indicated on a map of Europa. The observations are represented by vertical lines of which the size indicates the latitudinal extent spanned by each observation along the limb of Europa at the time of the observation. Table \ref{tab_properties_plumes} presents an overview of the plume properties derived from the observations by \cite{Roth2014a,Sparks2016,Sparks2017,Jia2018}.

\begin{table}[h]
\centering
\begin{tabular}{|r|l|l|l|}
  \hline
  \textbf{Property} & \textbf{\textit{Roth et al}} & \textbf{\textit{Sparks et al}} & \textbf{\textit{Jia et al}} \\
  \hline
  \hline
  Height & $\sim200$ km & $\sim50$ to $\sim220$ km & \\
  \hline 
  Column density & $1.5 \times 10^{20} m^{-2}$  & $0.7 \times 10^{21} m^{-2}$  &\\
  & & to $3.3 \times 10^{21} m^{-2}$ & \\
  \hline
  Surface density & $1.3\times 10^9$ cm$^{-3}$ cm$^{-3}$ & & $2\times 10^9 $cm$^{-3}$ \\ 
  &to $2.2\times 10^9$ cm$^{-3}$ & &  \\ 
  \hline
  Scale height & 200 km & & 150 km\\ 
  \hline 
  Total H$_2$O molecules & $1.3 \times 10^{32}$  & $0.5 \times 10^{32}$ to $1.8 \times 10^{32}$ &\\
  \hline
  Persistence &  > 7 hours   &  > 1 hour & > 3 mins \\
  \hline
  Mass flux & $\sim1000$ kg/s & & \\ 
  \hline
\end{tabular}
\caption{Overview of Europa plume properties, based on \cite{Roth2014a,Sparks2016,Sparks2017,Jia2018}.}
\label{tab_properties_plumes}
\end{table}
\section{Magnetospheric interaction}
\label{s_mag_environment}
Key to this thesis are charged particles, both plasma and energetic particles, that interact with Europa. To understand the populations of charged particles near Europa, it is necessary to introduce Europa's magnetospheric environment, which is done in this section. First, the basic types of magnetospheric interaction with planetary bodies are discussed. Next, Jupiter's magnetosphere as a whole is presented, before focusing on the environment Europa is exposed to directly.

\subsection{Basic types of magnetosphere interaction with planetary bodies}
An overview of the different types of magnetospheric interaction can be made based on properties of the body and the magnetospheric environment. A first way to categorize the types of interaction is according to the relative velocity of the plasma flowing past the object \citep{Kivelson2007,DePater2010}:
\begin{itemize}
\item \textbf{Supersonic:} An example of this interaction is the Earth and the solar wind. Analogous to the formation of a bow wave in front of a boat in water, a bow shock is formed in the solar wind upstream of Earth's magnetosphere. In the case of the bow wave the spreading of the information of the presence of the boat upstream is limited, because the flow moves faster than the waves that can spread the information. This is analogous to the formation of the bow shock, where the relevant waves are waves in a magnetized plasma. 
However, a super-magnetosonic environment does not necessarily imply the formation of a bow shock, as that also depends on the electromagnetic properties of the object which interacts with the magnetized flow, this will be discussed in the following paragraphs.
\item \textbf{Subsonic:} In this case the waves propagating the disturbance caused by the object in the plasma flow are not limited in their propagation, no bow shock is formed upstream of the object. Examples of this kind of interaction are moons embedded in planetary magnetospheres, like Titan, Io or Europa. The relevant plasma population is not the solar wind plasma for these objects, but the plasma corotating with the magnetic field of the central planets. 
\end{itemize}
The interaction can also be categorized according to the electromagnetic properties of the object  \citep{Kivelson2007,DePater2010}:
\begin{itemize}
\item \textbf{Unmagnetized body:}
\begin{itemize}
\item \textbf{Insulating:} An example of this case is Earth's moon, which does not possess an intrinsic magnetic field. Without conductivity of the body and the absence of a conducting ionosphere the Moon cannot influence the flow of the solar wind, the solar wind particles hit the surface of the object and are absorbed by it. A wake is formed behind the moon that lacks particles.
\item \textbf{Conducting:} in the case of a conducting body the plasma will move around the conducting layer, rather than impacting it. The magnetic field lines will drape around the body as a consequence of the diversion of the plasma. What drives this interaction are electric currents imposed by the upstream plasma flow on a conducting layer of the planetary body. In turn, these currents will drive magnetic fields that will react to the upstream disturbance. Examples of this type of interaction are Mars, Venus or Europa, neither of these have a (significant) intrinsic magnetic field, but they do possess a conducting ionosphere. In case of Mars and Venus the relevant plasma population is the solar wind, but in Europa's case it is the plasma from Jupiter's magnetosphere that is corotating with its magnetic field.
The conducting layer can also react to the changes in the magnetospheric environment, driving the formation of an induced magnetic field. Europa's subsurface ocean and Callisto's ionosphere are two cases where reaction to periodic changes in the environmental magnetic field induce magnetic fields.  
\end{itemize}
\item \textbf{Magnetized body:} Bodies with a sufficiently strong intrinsic magnetic field can form a magnetic cavity in which the magnetic field of the body dominates the structure and dynamics of the charged particle populations, referred to as a magnetosphere. Examples of this type are Mercury, Earth, Jupiter, Saturn, Uranus and Neptune but also Jupiter's moon Ganymede.
\end{itemize}
Jupiter is characterized as a magnetized body who's magnetosphere interacts with the supersonic solar wind. Jupiter's magnetosphere is discussed in Section \ref{s_mag_jup} and \ref{ss_magnetosphere_europa_orbit}. Europa is embedded in this magnetosphere and its interaction is characterized as that of an unmagnetized conducting body that is embedded in a sub-sonic flow. The interaction of Europa and its magnetospheric environment will be discussed in more detail in Section \ref{s_mag_interaction}.

\subsection{Jupiter's magnetosphere}
\label{s_mag_jup}

\begin{figure}[h]
  \centering
  \includegraphics[width=1.0\textwidth]{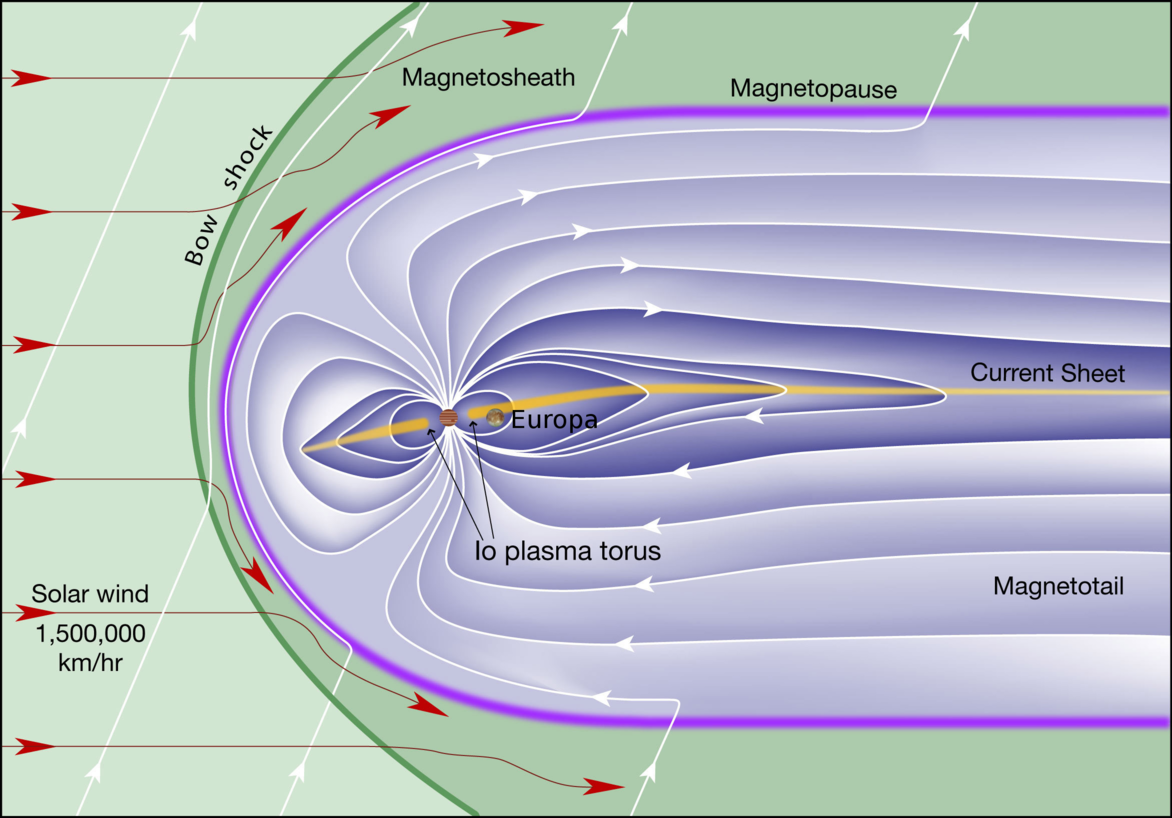}
  \caption{A schematic overview of Jupiter's magnetosphere. Also indicated is Europa's approximate location in the inner magnetosphere. [Schematic: Fran Bagenal and Steve Bartlett http://lasp.colorado.edu/home/mop/files/2012/04/JupMag-8W.jpg Europa: NASA]}
  \label{img_jup_mag_bag}
\end{figure}

Jupiter is a magnetized planet that interacts with the super-magnetosonic solar wind. The region where Jupiter's magnetic field dominates the dynamics and structure of the space environment defines its magnetosphere. A sketch of Jupiter's magnetosphere is shown in Figure \ref{img_jup_mag_bag}. Jupiter's magnetosphere is the largest in the solar system, besides the Sun's. Its magnetopause, the boundary between the interplanetary magnetic field and Jupiter's magnetic field, extends from up to 45 $R_J$ to 100 $R_J$ (1 $R_J$ = 71492 km) in the direction of the Sun \citep{Bagenal2004_Khurana}. Jupiter possesses a strong magnetic field, with equatorial strengths up to $400 000$ nT at the surface \citep{Bagenal2004_Khurana}. It is generated by an internal dynamo. The polarity of the field is opposite to that of Earth's magnetic field, meaning that the field lines flow from the north pole to the south pole. Jupiter's rotation axis is not perfectly aligned with the axis of the magnetic field, the angle between them is approximately 10$^\circ$, meaning that the magnetosphere wobbles as Jupiter rotates with respect to an observer or a moon in a Keplerian orbit within the magnetosphere.

\begin{figure}[h]
  \centering
  \includegraphics[width=0.70\textwidth]{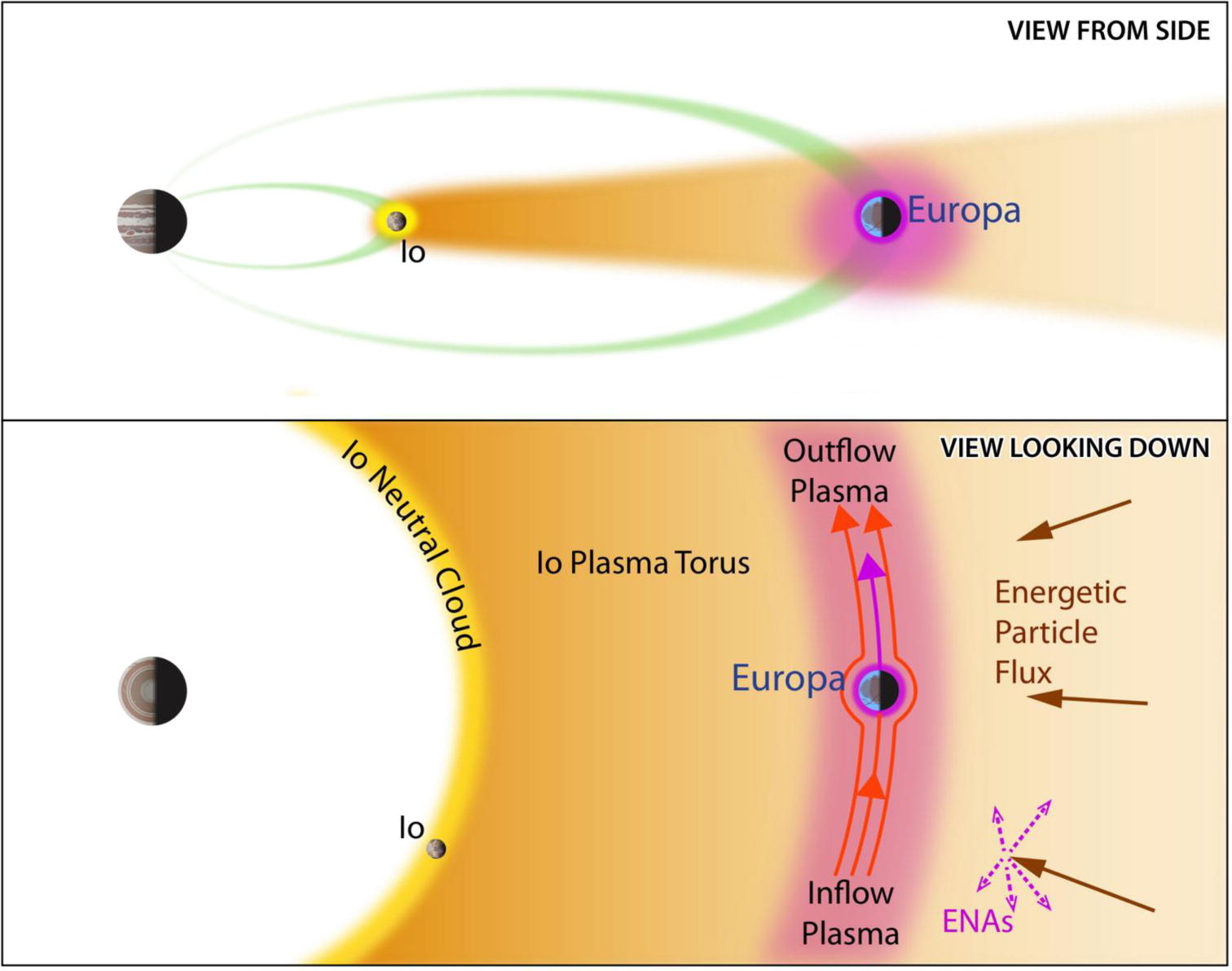}
  \caption{Schematic of the particle environment between Jupiter and Europa. Indicated are Io's neutral torus and the plasma resulting from the ionization of these neutrals. The arrows pointing towards Jupiter indicate that energetic particles are diffusing towards Jupiter. At Europa the flow direction of the plasma is indicated as well as the torus of neutrals associated with Europa's orbit (see Section \ref{s_atmosphere}).
   [Reproduced from \cite{Bagenal2015} with permission from Elsevier.]}
  \label{bag_fig1}
\end{figure}

Most of the plasma in Jupiter's magnetosphere is derived from Io and not from the solar wind \citep{Bagenal2004_Khurana}. Io is located in the inner magnetosphere, at $\sim5.9 R_J$. Neutral particles released from Io become ionized and this results in the formation of Io's plasma torus which is corotating with Jupiter's magnetic field (see also Figure \ref{bag_fig1}). Collisions between the plasma and Io's atmosphere cause the release of 0.6-2.6 ton/s of neutral, atmospheric particles, which form an extended cloud of neutral particles in Io's orbit \citep{Bagenal2015}. This neutral cloud exists mostly out of S and O, with an O/S ratio of 2-4. The neutrals in this cloud are ionized, resulting in the production of heavy ions (various charge states of S and O). The newly formed ions are accelerated to the speed of the bulk plasma flow, which is corotating with Jupiter's magnetic field \citep{Bagenal2004_Kivelson}. The production rate of plasma is about 0.3-1.4 ton/s \citep{Bagenal2015}.

\label{s_mag_jup_eu}

Initially the plasma forms a torus around Io's orbit, rotating together with Jupiter's magnetic field. After which the plasma will flow radially outward because of centrifugal forces, confined to Jupiter's magnetic equator. This plasma is referred to as the plasma sheet. Within the inner magnetosphere the plasma will keep corotating with Jupiter's magnetic field \citep{Bagenal2004_Khurana, Bagenal2014}. This heavy plasma inflates the magnetosphere through the centrifugal force and thermal pressure. The outward motion is a major driver of Jupiter's global dynamics, as it provides the means to transport large quantities of plasma towards the outer magnetosphere, where it can be shed towards the magnetotail.

\begin{figure}[h]
  \centering
  \includegraphics[width=0.60\textwidth]{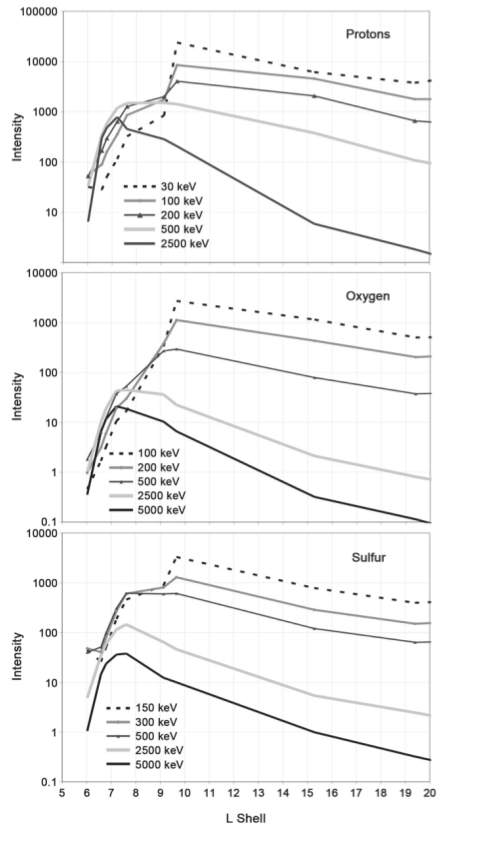}
  \caption{Ion intensity in ions per cm$^2$s sr keV for several energies for protons, sulphur and oxygen, from Galileo Energetic Particle Detector (EPD) data. The x-axis is in units of "L". This is the radial distance of the magnetic field line at the magnetic equator, divided by Jupiter's radius (see Section \ref{ss_trapped}), Europa is located at 9.38 L. [Adapted by \cite{Pappalardo2009_Paranicas} from \cite{Mauk2004}]}
  \label{img_energy_spectra_jup}
\end{figure}

Charged particles are not limited in the thermal, corotating energy range, but extend to much higher, suprathermal energies \citep{Paranicas2000}. After having been transported radially outwards from Io's orbit, charged particles from the magnetosphere are accelerated to these high energies while they diffuse radially inward \citep{Bagenal2004_Khurana}. 
Energetic charged particles have been detected all the way to the magnetopause, and high fluxes of these particles interact with the moons.
Detected species include protons, sulphur and oxygen, with energies > 10 MeV. The intensity of energetic ions in Jupiter's magnetosphere is shown in Figure \ref{img_energy_spectra_jup}.
The number density of energetic particles (energy > 10 keV) is significantly lower than the plasma sheet density \citep{Bagenal2004_Kivelson}. For example, the number of protons between 50-200 keV is approximately a thousand times lower than the total plasma density \citep{Paterson1999,Paranicas2002}. Though the density of energetic particles is significantly lower than that of the plasma, they still play a significant role in the magnetospheric interaction because of their high energy which translates also into high partial particle pressure. Pressure gradients can be especially important for driving current and modifying the local magnetospheric environment, e.g. \cite{Sergis2007}.

\subsection{Jupiter's magnetosphere at Europa's orbit}
\label{ss_magnetosphere_europa_orbit}
Europa is continuously embedded inside Jupiter's magnetosphere, since it orbits Jupiter at 9.38 $R_J$. This region of the magnetosphere is referred to as the inner magnetosphere, a region that extends up to 10 $R_J$ in which the plasma is corotating rigidly with Jupiter \citep{Bagenal2004_Khurana}. At distances >10 $R_J$ corotation breaks down gradually.

The dominant population of plasma at Europa's orbit is the plasma from the Io torus that has moved radially outward, as illustrated in Figure \ref{bag_fig1}. It takes the plasma 10-50 days to move radially outward from Io's plasma torus to Europa \citep{Bagenal2015}. The plasma is corotating with the magnetic field, so the rotation period is about 10 hours. This is much faster than Europa's orbital period (3.55 Earth days). This means that the corotating plasma is overtaking Europa, moving from Europa's leading edge to its trailing edge, with an average velocity of 76 km/s \citep{Pappalardo2009_Kivelson}. 

\begin{figure}[h]
  \centering
  \includegraphics[width=0.75\textwidth]{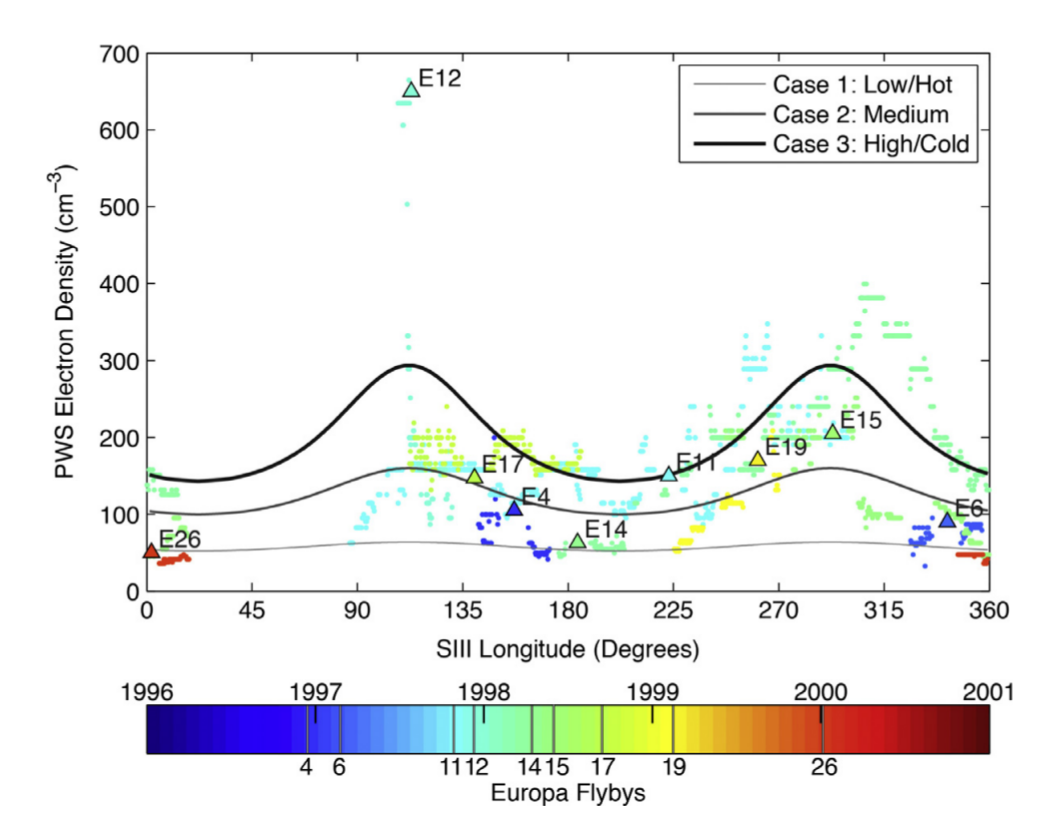}
  \caption{Electron density based on Galileo's plasma wave instrument data (PWS) as a function of longitude. The values were obtained between 8.9 and 9.9 $R_J$ but farther than 2.5 $R_E$ from Europa. E4 to E26 indicate the Galileo flybys of Europa. Longitude is expressed here in the SIII frame, this is a Jupiter centred reference frame that has its z-axis aligned with Jupiter's rotation axis and rotates with the planet. [Reproduced from \cite{Bagenal2015} with permission from Elsevier.]}
  \label{img_bagenal2015}
\end{figure}

\begin{figure}[h]
  \centering
  \includegraphics[width=0.80\textwidth]{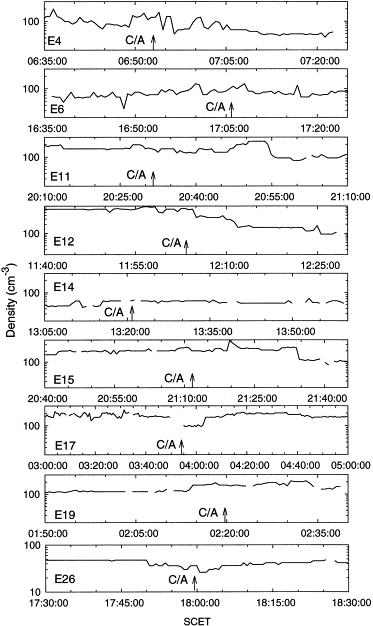}
  \caption{Electron density based on Galileo's plasma wave instrument data (PWS) during the close Europa flybys. [Reproduced from \cite{Kurth2001} with permission from Elsevier.]}
  \label{img_kurth2001}
\end{figure}

In the time period it takes the plasma to move from Io to Europa, collisions occur, which can alter the composition of the plasma; by for example increasing its charge state \citep{Bagenal2015}.
Near Europa, the dominant ions are oxygen and sulphur. Oxygen ions include O$^+$ and O$_2^+$, of which the ratio is not well determined. The dominating sulphur ion is S$^{2+}$ \citep{Bagenal2015}.
Ion densities near Europa vary between 12-170 particles/cm$^3$ \citep{McNutt1981, Bagenal1994, Crary1998, Paterson1999, Frank2001}. Typical electron densities obtained by the plasma wave instrument of the Galileo spacecraft, near Europa's orbit but not closer than 2.5 $R_E$ to Europa, are shown in Figure \ref{img_bagenal2015}. The electron density varies between 50-400 particles/cm$^3$ \citep{Bagenal2015}. It is clear that the E12 flyby is an exceptional case with an extreme density of 600 particles/cm$^3$. Figure \ref{img_kurth2001} shows the electron density during the close flybys of Europa, as measured by the plasma wave instrument. The typical densities near Europa vary 35-300 particles/cm$^3$. Similarly as for the data collected further away, shown in Figure \ref{img_bagenal2015}, E12 represents an extreme case with densities up to 600 particles/cm$^3$. No ion densities have been derived for the E12 flyby. 

Europa orbits around Jupiter in Jupiter's rotational equator. Because of the tilt of the magnetic field, Europa experiences a varying magnetospheric environment. The position of the center of the plasma sheet will move above and below Europa. When Europa is located in the center of the plasma sheet the plasma density will be higher than when it is located above or below it \citep{Bagenal2004_Khurana}. Additionally, the components of the magnetic field will change from Europa's perspective. In a reference frame with the z-axis aligned to Europa's rotation axis the z-component of the field remains mostly constant, while the x-component (along the corotation) and y-component (towards Jupiter) will vary. It is these varying components that are responsible for the creation of Europa's induced magnetosphere, mentioned previously in Section \ref{ss_interior} \citep{Khurana1998,Zimmer2000}. Europa's induced magnetic field will influence the trajectories of charged particles in its vicinity.

\begin{figure}[h]
  \centering
  \includegraphics[width=1.0\textwidth]{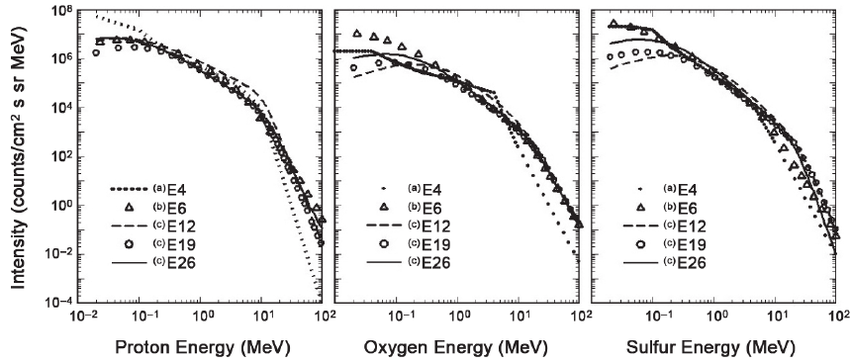}
  \caption{Energy spectra for protons, oxygen ions and sulphur ions at Europa's orbit, from Galileo Energetic Particle Detector (EPD) data. [\cite{Pappalardo2009_Paranicas}]}
  \label{img_energy_spectra}
\end{figure}

At Europa's orbit in Jupiter's magnetosphere energetic charged particles are present. 
Detected species include protons, sulphur and oxygen, with energies > 10 MeV \citep{Pappalardo2009_Paranicas}. \cite{Clark2016} shows that energetic sulphur ions are predominantly multiply charged and energetic oxygen is a mix between singly and doubly charged. In Figure \ref{img_energy_spectra} the energy spectra of energetic ions at Europa's orbit are shown. Below $\sim $0.1 MeV the conditions can differ more than one order of magnitude between flybys, however, above $\sim $0.1 MeV the conditions are stable. Even though the density of the energetic charged particles is low, compared to the plasma, they dominate the energy density \citep{Pappalardo2009_Johnson}.

Finally, in Table \ref{tab_properties_magnetosphere} properties of Europa's magnetospheric environment relevant to this work are summarized.
\begin{table}[h]
\centering
\begin{tabular}{|l|l|l|}
  \hline
  \textbf{Property} & \textbf{Value and units} & \textbf{Source} \\
  \hline
  \hline
  Magnetic field magnitude & $\sim 400$ nT & [1] \\
  \hline 
  Electron density range & 50-400 [$\#$/cm$^-3$] & [2] \\
  \hline 
  Electron temperature & 10-30 [eV] & [2] \\
  \hline 
  Ion density range & 12-170 [$\#$/cm$^-3$] & [3] \\
  \hline 
  Ion charge &  single and multiple & [2] \\
  \hline
  Ion mass & 18.5 amu & [3] \\
  \hline
  Ion temperature & 30-500 [eV] & [2] \\
  \hline 
  Plasmasheet rel. vel. to Europa (range) & 76 (56-86) km/s & [3] \\
  \hline
\end{tabular}
\caption{General properties of Europa's magnetospheric environment relevant to this work. Sources: [1]: \cite{Khurana1997}, [2]:  \cite{Bagenal2015}, [3]: \cite{McNutt1981, Bagenal1994, Crary1998, Paterson1999, Frank2001}}
\label{tab_properties_magnetosphere}
\end{table}

\subsection{Interaction of Europa with its magnetospheric environment}
\label{s_mag_interaction}
This section discusses how Europa interacts with its magnetospheric environment. 
Europa is characterized as a unmagnetized conducting body that is embedded in a sub-magnetosonic flow. This implies that no bow shock is present in front of Europa, but that Europa's conducting layers divert the corotating plasma flowing past it. There are two conducting layers, as explained below: its inferred subsurface, salty water ocean and its ionosphere.

The ionosphere is derived from Europa's tenuous atmosphere that is partly ionized by charge exchange with magnetospheric plasma, electron impact ionization and photo-ionization \citep{Lucchetti2016}. The ionosphere has been detected with radio occultation measurements by Galileo \citep{Kliore1997}. Other indications that Europa is a source of plasma have been obtained with the plasma particle detector PLS on the Galileo mission \citep{Paterson1999}. This dataset will be discussed in more detail in Section \ref{s_mag_interaction_epd_pls}. Additionally, waves in the magnetic field detected downstream of Europa during the E11 and E15 flybys have been identified as signatures of positively and negatively charged pickup ions \citep{Volwerk2001}. Pickup ions are
atmospheric neutral particles that become ionized and are transported in the direction of the corotational flow. 
More pick-up ions seem to be present during the E15 flyby than during the E11 flyby \citep{Volwerk2001}. Furthermore, \cite{Intriligator1982}, \cite{Russell1998} and \cite{Eviatar2005} suggest that Europa is trailed in the direction of the corotational flow by a plume of plasma, based on Pioneer 10 and Voyager 2 plasma analyser data and on Galileo magnetometer and plasma wave data.

\begin{figure}[h]
  \centering
  \includegraphics[width=0.75\textwidth]{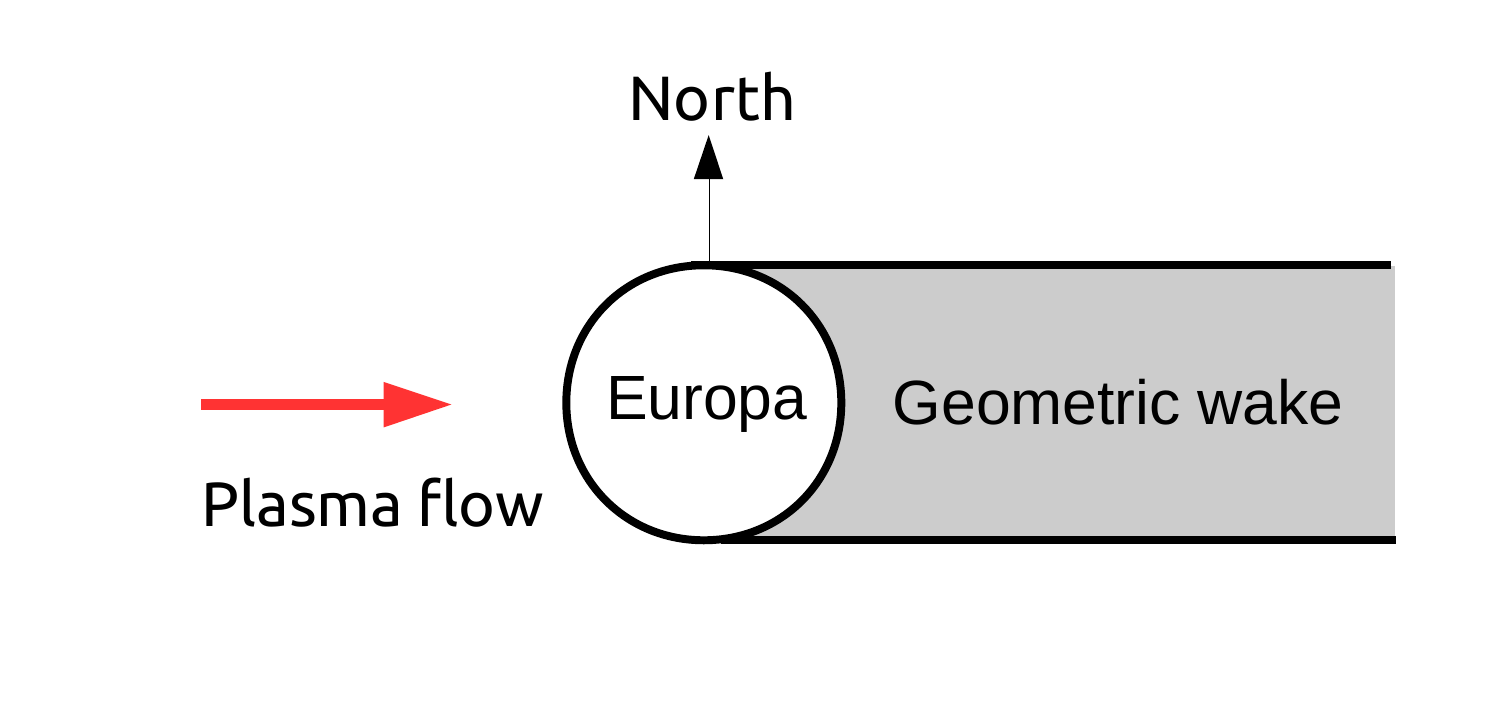}
  \caption{Schematic of Europa's geometric wake with respect to the corotational plasma flow.}
  \label{img_wake}
\end{figure}

The deflection of the corotating plasma by the conducting ionosphere prevents most of the cold plasma flow from hitting the surface \citep{Saur1998}. This deflected flow has been modelled as a fluid passing by a cylindrical object \citep{Ip1996,Dols2016}. 
The diversion of the flow causes the velocity magnitude of the corotational plasma to vary. Just in front of and behind Europa (from the point of view of the plasma flow) this results in a slow down of the corotating plasma \citep{Pappalardo2009_Kivelson}. 
The diversion of the corotational plasma by Europa's ionsosphere has been observed directly by the plasma particle detector PLS \citep{Paterson1999}, more about this in Section \ref{s_mag_interaction_epd_pls}. The region directly downstream of Europa (from the point of view of the corotational plasma) which is within the cross section of Europa, is referred to as the geometric wake (Figure \ref{img_wake}).

When the flow is slowed down the magnetic field is affected too, because the field is frozen-in to the corotating plasma. So when the plasma slows down in front of Europa, magnetic field lines start piling up. This results in a local increase of the magnetic field strength, a phenomenon referred to as magnetic pile-up. Increases in the magnitude of the magnetic field in the vicinity of Europa have been reported for the E12, E14 and E19 flybys that all passed upstream of Europa. The conditions during the E12 flyby are exceptional: one minute before the closest approach 
the field strength reaches $\sim800$~nT, almost double the upstream value \citep{Pappalardo2009_Kivelson}. 
The occurrence of both this extreme magnetic field increase and the extreme plasma density mentioned earlier, suggest an increased production of ions. Which might indicate a high density of neutral particles \citep{Pappalardo2009_Kivelson}.

\begin{figure}[h]
  \centering
  \includegraphics[width=0.5\textwidth]{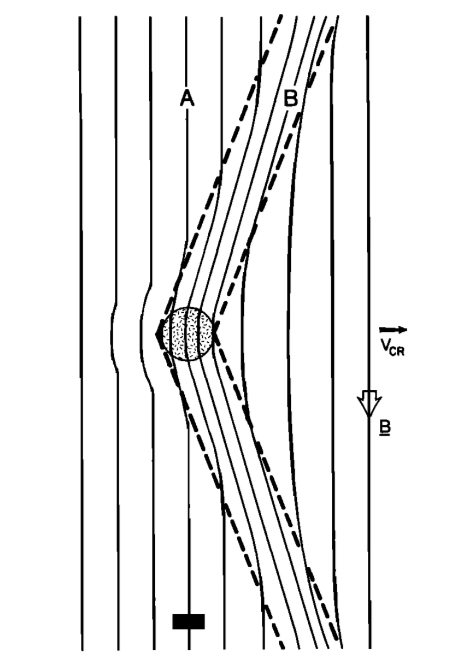}
  \caption{Schematic of Alfv\'{e}n wings. Corotation velocity is to the right ($V_{cr}$). [Reproduced from \cite{Southwood1980}]}
  \label{img_alfven}
\end{figure}

An additional aspect of the slow down is the formation of so called Alfv\'{e}n wings. As already mentioned the magnetic field lines are frozen in the plasma, hence they move together \citep{Pappalardo2009_Kivelson}. The plasma closer to Europa will be slowed down more than the plasma further away, this results in a bending of the magnetic fields lines. This is shown in Figure \ref{img_alfven}. In three dimensional space these Alfv\'{e}n wings can be thought of as cylinders extending along the magnetic field lines. These are called Alfv\'{e}n wings because the perturbation induced near Europa propagates along the magnetic field with the Alfv\'{e}n speed, which is the fundamental speed at which magnetic perturbations are transported.


Europa has a wake with respect to the corotating plasma flow (Figure \ref{img_wake}). Galileo flybys E4, E11, E15 and E17 passed through the geometric wake of Europa. Perturbations in the magnetic field related to these wake passes have been identified for all these flybys \citep{Pappalardo2009_Kivelson}. 
Another important aspect of Europa's interaction with its magnetospheric environment is the induction of a magnetic field in Europa. This phenomenon was studied with Galileo's magnetometer and was previously discussed in Section \ref{ss_interior}.

The interaction of Europa with its magnetospheric environment influences not only the motion of plasma particles around it, as has just been shown, but also the motion of energetic charged particles trapped in Jupiter's magnetosphere. The diversion of the corotational flow affects the electric fields near Europa, which affects the motion of the energetic charged particles. The magnetic field pile-up will on its turn affect the trajectories of charged particles. Furthermore, energetic charged particles are depleted in the vicinity of Europa \citep{Paranicas2000}. Energetic particles do not carry enough energy density to cause any detectable response in the magnetic or electric field, when their distribution is perturbed. As a result of this, energetic particles in particle simulations can be treated as test particles of which the trajectories are sensitive to the perturbations introduced by cold plasma processes and the magnetic field induced by the conductive ocean layer. 

\section{Plasma \& energetic charged particle measurements at Europa by the Galileo mission}
\label{s_mag_interaction_epd_pls}

In this section measurements of lower energy plasma and energetic charged particles that show the interaction with Europa and its particle environment are discussed in more detail.
Focus is in particular on the data gathered by two Galileo instruments, the plasma particle detector PLS and the energetic particle detector EPD. Technical details of these instruments will be discussed in more detail in Section \ref{ss_pls} and \ref{ss_epd}. Overview plots of the data collected by these instruments are included in Appendix \ref{app_overview}.

\subsection{Low energy plasma particles}
Plasma particles at Europa have been studied by the PLS instrument. \cite{Paterson1999} presents PLS data collected during the E4 and E6 flybys.
In preparation of the PLS analysis presented in this thesis, the main findings of \cite{Paterson1999} relevant for this thesis are discussed. Data collected during these flybys are shown in Figure \ref{img_E4_all} and Figure \ref{img_E6_all} in Appendix \ref{app_overview}. The E4 and E6 flybys were the first two Europa flybys. Studies of the PLS data of none of the subsequent Europa flybys has been published.

Fits to the spectra measured two radii upstream, during the E4 flyby, indicate that multiple heavy ion species are present in the corotating plasma. The species that are fitted are: H$^+$, O$_2^+$, O$^+$, S$_2^+$ and S$^+$. The authors argue that pick-up ions originating from Europa are also present, in the downstream region during both E4 and E6, and upstream during E6. The assumed pick-up ion species are H$^+$, H$_2^+$ and H$_2$O$^+$. Europa's wake was shown to contain enhanced ion densities, possibly caused by pick-up ions from Europa. The presence of pick-up ions provide an indirect indication that Europa possesses an atmosphere.

\begin{figure}[h]
  \centering
  \includegraphics[width=0.75\textwidth]{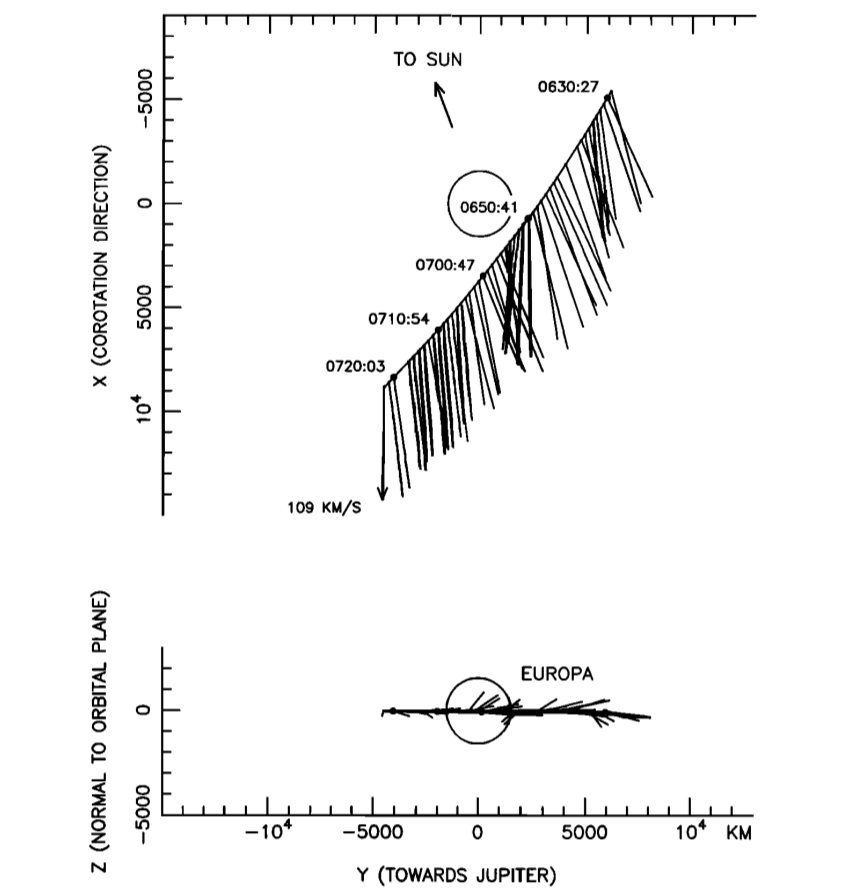}
  \caption{Bulk flow-vectors during Galileo's E4 flyby on December 19, 1996. [Reproduced from \cite{Paterson1999} with permission from John Wiley and Sons. Copyright 1999 by the American Geophysical Union.]}
  \label{img_paterson1999_fig2}
\end{figure}

The bulk velocity vectors of the plasma were determined from the PLS data. A  flow deflection (see Section \ref{s_mag_interaction}) was measured during flybys E4 and E6. This verifies that Europa is a weakly mass loading object, hosting a tenuous exosphere. The deflection during the E4 flyby is shown in Figure \ref{img_paterson1999_fig2}. The temporal extent of the flow deflection is also indicated in the overview plots, Figures \ref{img_E4_all} and \ref{img_E6_all}.
 
\subsection{Energetic charged particles}
\label{ss_interaction_EPD}

Energetic charged particles at Europa have been studied by the EPD instrument. \cite{Paranicas2000} presents energetic ion and electron data collected by the EPD instrument during the E4, E12, E14 and E15 flybys of Europa and physically interprets the data. A variety of features is identified that are linked to interaction of energetic particles with Europa. Here an overview is presented of the most relevant aspects, in preparation for the EPD analysis later in this thesis. 

\begin{figure}[h]
  \centering
  \includegraphics[width=1.0\textwidth]{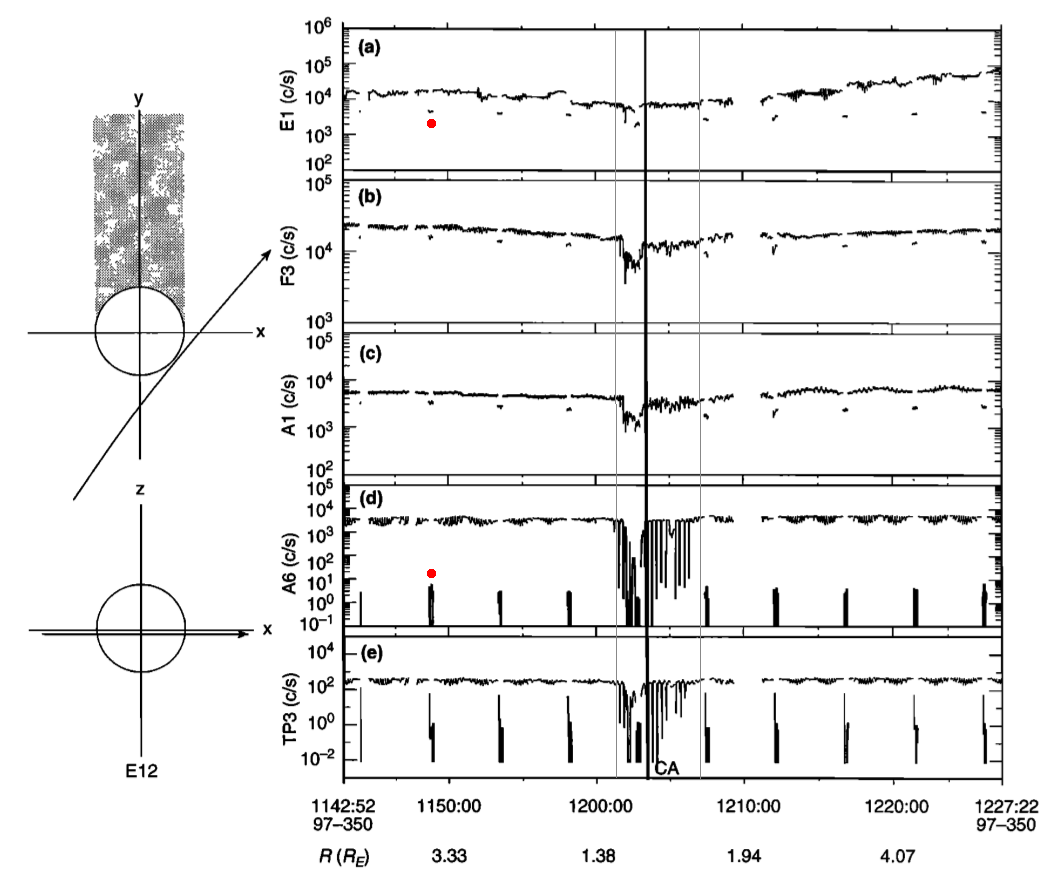}
  \caption{Energetic particle data collected during the E12 flyby. E1 (29-42 keV) and F3 (527-884 keV) are electron channels, A1 (42-180 keV) and A6 (825-2050 keV) are ion channels that do not discriminate between species and TP3 (540-1040 keV) is a proton channel. The two red dots indicate two examples of depletions that occur because the detector has moved behind the radiation shield. The grey vertical lines indicate the extent of the depletion features discussed in the text. [Reproduced from \cite{Paranicas2000} with permission from John Wiley and Sons. Copyright 2000 by the American Geophysical Union.]}
  \label{img_paranicas2000_fig3}
\end{figure}

During the E12 upstream flyby, deep decreases in the energetic ion rate are reported, from 12:01 to 12:07, this period is marked by vertical grey lines in Figure \ref{img_paranicas2000_fig3}. Note that there are two types of depletion visible in this figure: depletions when the detector moves behind the radiation shield (see Section \ref{ss_epd}) and depletions that occur without the shield. The two red dots indicate examples of radiation shield depletions. These measurements allow to determine the noise level in the specific channel of the instrument. Of importance here are the measurements without the shield. An overview of all the E12 particle data can be seen in Figure \ref{img_E12_all} in Appendix \ref{app_overview}. It is argued by \cite{Paranicas2000} that the depletions  of the more energetic ions during this flyby is a consequence of the size of these particles' gyroradius. This is the radius of the circular orbit that these energetic particles perform about the magnetic field lines (the motion of charged particles will be discussed in more detail in Section \ref{s_charged_particle}). Their gyroradius is larger than the altitude of Galileo, hence these particles are lost by collisions with the surface. As an example \cite{Paranicas2000} states that the gyroradius of protons with energies of 750 keV is around 200 km at the closest approach, which is the same as the altitude of Galileo. These particles are measured in the TP3 channel shown in Figure \ref{img_paranicas2000_fig3}. The "bite-out" feature in lower energy ions, visible in the A1 channel in Figure \ref{img_paranicas2000_fig3} just before the closest approach at 12:02, cannot be explained by this effect. Instead it is proposed that this is a consequence of a drift motion that lets electrons drift towards Jupiter and ions away from Jupiter. This drift motion is referred to as the grad-B drift (see Section \ref{sss_gradB}). Due to magnetic field pile-up upstream of Europa charged particles approaching Europa experience a gradient in the magnetic field perpendicular to the field lines, which causes the grad-B drift motion. The drift makes particles drift perpendicular to the gradient in the magnetic field and the field lines. 
The strongest gradient coincides in time with the "bite-out", the particles would thus be drifting away from this region. Furthermore the bite-out in higher electrons is larger than in the lower energy ones, which is to be expected for the grad-B drift, since its effects scales linearly with the particle energy. The hypotheses of \cite{Paranicas2000} have not been verified by modelling.

During the E4 flyby, prior to entering the wake, a depletion of ions is observed of which the gyroradius is not large enough to reach the surface. The E14 flyby is another upstream flyby. During this flyby no signs of interaction with Europa are reported. Finally it should be noted that the other upstream flybys E19 and E26 are not discussed in \cite{Paranicas2000}.

Depletions of energetic electrons and ions are observed in the wake of Europa, during flybys E4, E11 and E15. 
\cite{Paranicas2000} propose that the main mechanism responsible for the depletions observed  is the depletion of trapped energetic particles bouncing up and down along magnetic field lines that come in contact with Europa and impact the trailing side. The authors note that the center of the wake is shifted towards Jupiter (during the E4, E11 and E15 flybys).

\begin{figure}[h]
  \centering
  \includegraphics[width=0.5\textwidth]{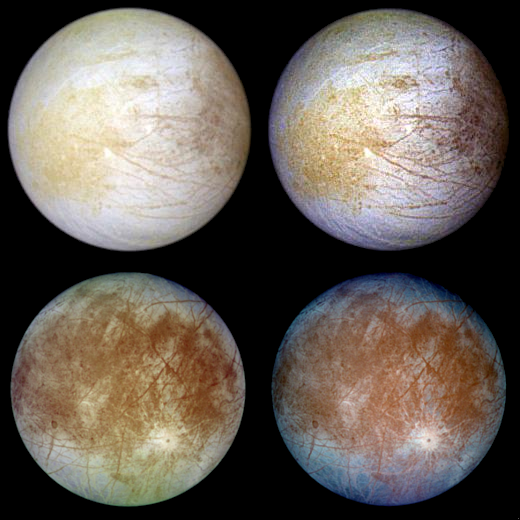}
  \caption{Albedo dichotomy between Europa's leading (top) and trailing (bottom) hemispheres. Left: natural color, right: enhanced color.  [NASA https://photojournal.jpl.nasa.gov/target/Europa].}
  \label{img_dichotomy}
\end{figure}

\cite{Paranicas2001} discusses EPD electron data from E12, E14 and E26. It is argued that energetic electrons (10 keV-10 MeV) primarily impact Europa's trailing hemisphere, explaining the observed difference in albedo between Europa's trailing and leading hemispheres (Figure \ref{img_dichotomy}). 
Electrons primarily impact the trailing side because their bounce period is small compared to the contact time of the magnetic field line their are travelling along with Europa (the ratio of the half bounce time and the contact time is less than one). This is unlike ions whose bounce period is larger than the contact time (ratio exceeds one). Bombardment of electrons is highest at low latitudes.
E26 data for the ions and electrons are discussed in \cite{Paranicas2007}. A reduction of particle flux in both channels is seen corresponding to the direction of Europa, suggesting that these particles have been depleted after direct interaction with the surface. E26 is a high latitude pass at low altitudes. \cite{Paranicas2007} argues that what is seen in the EPD data is the ongoing process of Europa moving with respect to the magnetic field and the particles travelling along the field being depleted.

\begin{figure}[h]
  \centering
  \includegraphics[width=0.75\textwidth]{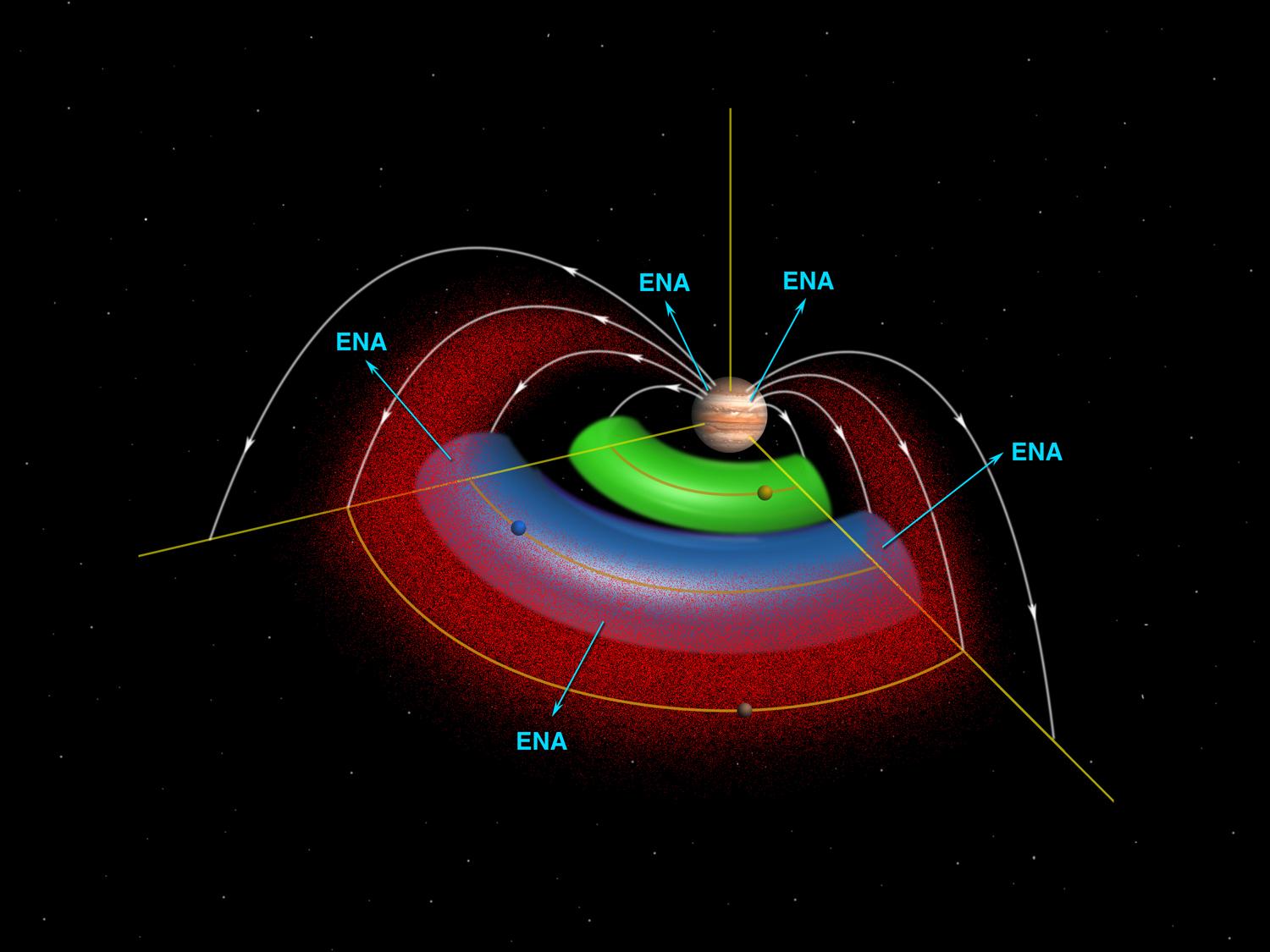}
  \caption{Jupiter and the Europa torus are identified as major sources of Energetic Neutral Atom (ENA) emission by \cite{Mauk2003} using data from Cassini's ENA detector. The red hue indicates the trapped energetic protons thought to be responsible for the ENA generation. The trapped particles responsible for the ENA's are depleted to such an extent that Io's neutral torus does not appear as a signficant source of ENA's.  [NASA https://photojournal.jpl.nasa.gov/catalog/PIA04433].}
  \label{img_torus}
\end{figure}

Part of the neutral particles of Europa's tenuous atmosphere escape Europa's gravity and form a torus of (neutral) particles located along Europa's orbit around Jupiter \citep{Pappalardo2009_Johnson}. The torus provides evidence of the release of material from Europa's surface from the atmosphere, the plumes or a combination of both. Evidence for this torus has been provided by the Ion and Neutral Camera (INCA) instrument on the Cassini mission, that observed the Jovian system when it flew by Jupiter on its way to Saturn \citep{Mauk2003}. INCA detected energetic neutral atoms (ENAs) originating from Europa's orbit around Jupiter (see Figure \ref{img_torus}).
These ENAs are formed when neutral atoms charge exchange with energetic ions trapped in Jupiter's magnetosphere, this is similar to the ENAs predicted at the Io torus \citep{Futaana2015}. Without being influenced by the Lorentz force, the neutralized energetic ions continue as ENA's along a path determined by gravity and their velocity vector at the point of the charge exchange.
Independent evidence for the existence of the neutral torus has been obtained from Galileo EPD data \citep{Lagg2003,Kollmann2016}. A depletion of protons at energies between 80 and 220 keV was observed by \cite{Lagg2003} and \cite{Kollmann2016}, they argue that this depletion is caused by the charge exchange of the energetic protons with neutral particles in the torus. 
\cite{Mauk2004} constrains the total number of neutral atoms and molecules in the torus to $0.6 (\pm 0.25) \times 10^{34}$, while the vertical thickness of the torus is constrained to 1 $R_J$ and its density to 1.6-410 particles/cm$^3$ \citep{Kollmann2016}. The large density range is indicative of systematic uncertainties in the method to invert EPD data rather than a large scale variability. Modelling of atmospheric losses of O and H$_2$ by \cite{Smyth2006} predict densities for Europa's torus that are within the range determined by these observations. In their model the loss is dominated by H$_2$. Recently \cite{Dols2016} has suggested that the atmospheric loss is dominated by O$_2$ losses instead. 
These measurements discussed by  \cite{Lagg2003} and \cite{Kollmann2016} reveal the depletions of energetic ions detected in-situ due to charge exchange with neutrals of the torus. This method could in principle be used to detect Europa's atmosphere, but has never been applied to the close vicinity of Europa.

\section{Main research question of the thesis}
\label{s_goal}

Remote sensing observations indicate that Europa is surrounded by a tenuous atmosphere. Large uncertainties exist in properties of this atmosphere, such as the loss rate and structure, due to the lack of adequate measurements \citep{Plainaki2018}. 
Recent observations by the Hubble telescope and past data from Galileo in-situ instruments hint at the existence of eruptions of water vapour originating from the interior of Europa at six different occasions \citep{Roth2014a,Sparks2016,Sparks2017,Jia2018}. The cloud of material created by these eruptions is referred to as "plumes". Large uncertainties remain in the properties these plumes, as the observations can only approximately constrain properties such as their gas column density, scale height, source location and the minimal eruption time.  Even their existence has not been confirmed by direct measurements. The plumes could offer an opportunity to directly study Europa's interior, by flying a spacecraft through the plumes and taking samples. Therefore, the plumes could offer a "window" to Europa's interior without even having to land on the surface. As the atmosphere contains material originating from the surface, it also offers an opportunity to assess properties of Europa's interior, though the information derived from the atmosphere is less direct than from the plumes. 

The only missions that has studied Europa from close by is the Galileo mission, which was active in the Jupiter system from 1995 to 2003. In this time period it made 12 flybys of Europa \citep{Pappalardo2009_Alexander}. As it was not equipped with a neutral particle detector, no direct detections of the particles of the atmosphere or the plumes could have been made. Amongst its in-situ instruments Galileo carried two charged particle detectors, the energetic particle detector (EPD) and the plasma instrumentation (PLS) \citep{Williams1992,Frank1992}. Data from in-situ instruments on Galileo have been used to show that:
\begin{itemize}
\item the plasma corotating together with Jupiter is diverted near Europa, based on PLS data from the first two flybys \citep{Paterson1999}. This confirms that an ionosphere is present, which is the result of the ionization of atmospheric particles. Furthermore direct signatures of newly formed pick-up ions originating from Europa's atmosphere have been seen in PLS data during the same flybys. PLS data collected during the other flybys have not been published in literature.
\item an extreme plasma density derived from the plasma wave data and changes in the magnetic field measured during the E12 flyby could be caused by a local enhancement of the neutral particle density \citep{Jia2018}, such as a plume.
\item energetic charged particles are depleted in the vicinity of Europa \citep{Paranicas2000}. A torus of neutral gas associated with Europa's orbit depletes energetic ions by charge exchange. This is based on data from the energetic particle detector (EPD) collected inside Europa's orbit, but not in the vicinity of the moon \citep{Lagg2003,Kollmann2016}. The depletions detected close to the moon have not been analysed previously in the context of an atmosphere or plume.
\end{itemize}
This implies that even though the atmosphere and the plumes could not have been detected directly by Galileo instruments, the in-situ data can contain signatures of the interaction of the charged particle environment with the atmosphere, plumes and torus. Furthermore, possible unexploited opportunities might remain in the Galileo in-situ data to detect signatures of the atmosphere or plumes and constrain their physical properties. Thus, 
I formulate the following research question: 
\begin{center}
\textit{"Can the properties (density, scale height) of Europa's tenuous atmosphere or any water vapour plumes be constrained using data from the Galileo in-situ particle detector instruments (PLS, EPD)?" 
}
\end{center}
In addition to this main topic, I will exploit the opportunity to make predictions for future observations of Europa's atmosphere and plumes with in-situ particle detector instruments. The upcoming Europa missions are NASA's Europa Clipper mission and ESA's Jupiter ICy moons Explorer (JUICE) \citep{Grasset2013}, which are scheduled to make 45 and 2 flybys of Europa in the early 2030's, respectively. In this thesis I will focus on JUICE in particular.

The structure of this thesis is as follows. 
First the methodology chosen to address the research question is discussed in Chapter \ref{ch_method}. The approach chosen is to combine the interpretation of spacecraft observations and simulations. Then, in Chapter \ref{ch_comparison}, the Galileo charged particle dataset is analysed. Next, in Chapter \ref{ch_atmosphere}, simulations are used to reproduce the data to characterize the atmosphere and possible plumes. Subsequently, Chapter \ref{ch_prospects}, is dedicated to predictions of future measurements. Final conclusions and recommendations for future work are discussed in Chapter \ref{ch_conclusion}.


\chapter{Research methodology}
\label{ch_method}

A combination of the interpretation of spacecraft observations and simulations is chosen as the approach to identify signatures of Europa's tenuous atmosphere and plumes in the Galileo charged particle data. The first step is to analyze the data to identify signatures of the interaction between Europa's atmosphere or plumes, and the magnetosphere in plasma and energetic particles ({Goal 1}). 
Then, simulations will be used to reproduce the identified signatures and characterize the atmosphere and possible plumes ({Goal 2}).

Goal 1 will be addressed by characterizing and comparing the charged particle data collected by Galileo during the Europa flybys to select the best time period(s) to search for signatures of the atmosphere or any plumes. Goal 2 involves simulating the data, specifically the depletion of energetic charged ions near Europa. These particles are depleted near Europa when they hit the surface or charge exchange with the tenuous atmopshere. 
Therefore, to determine the cause of the depletion it is necessary to simulate the trajectories of the particles. Different scenarios that involve an atmosphere and no atmosphere will be simulated, to characterize the contribution of the atmosphere to the depletions. Then, the simulation results will be compared to the data to determine the contributions of surface impacts or any atmospheric losses, in the measured depletion of the energetic particles. The comparison will allow to constrain atmospheric properties such as density and scale height.

In this Chapter the methodology is explained in more detail. First, in Section \ref{ss_galileo}, the Galileo mission and the charged particle detector instruments onboard this mission (EPD and PLS) will be introduced. 
Then, in Section \ref{s_charged_particle} the motion of the charged particles that can be detected by the instruments will be described and the appropriate method to simulate them. Finally, in Section \ref{s_particle_simulation}, the computer code is discussed that will be used to simulate the trajectories of neutral and charged particles.

\section{Galileo and instrumentation}
\label{ss_galileo}

In this section NASA's past Galileo mission, a mission of particular interest to this thesis, is discussed in more detail, including its most relevant instrumentation.

\begin{figure}[h]
  \centering
  \includegraphics[width=0.5\textwidth]{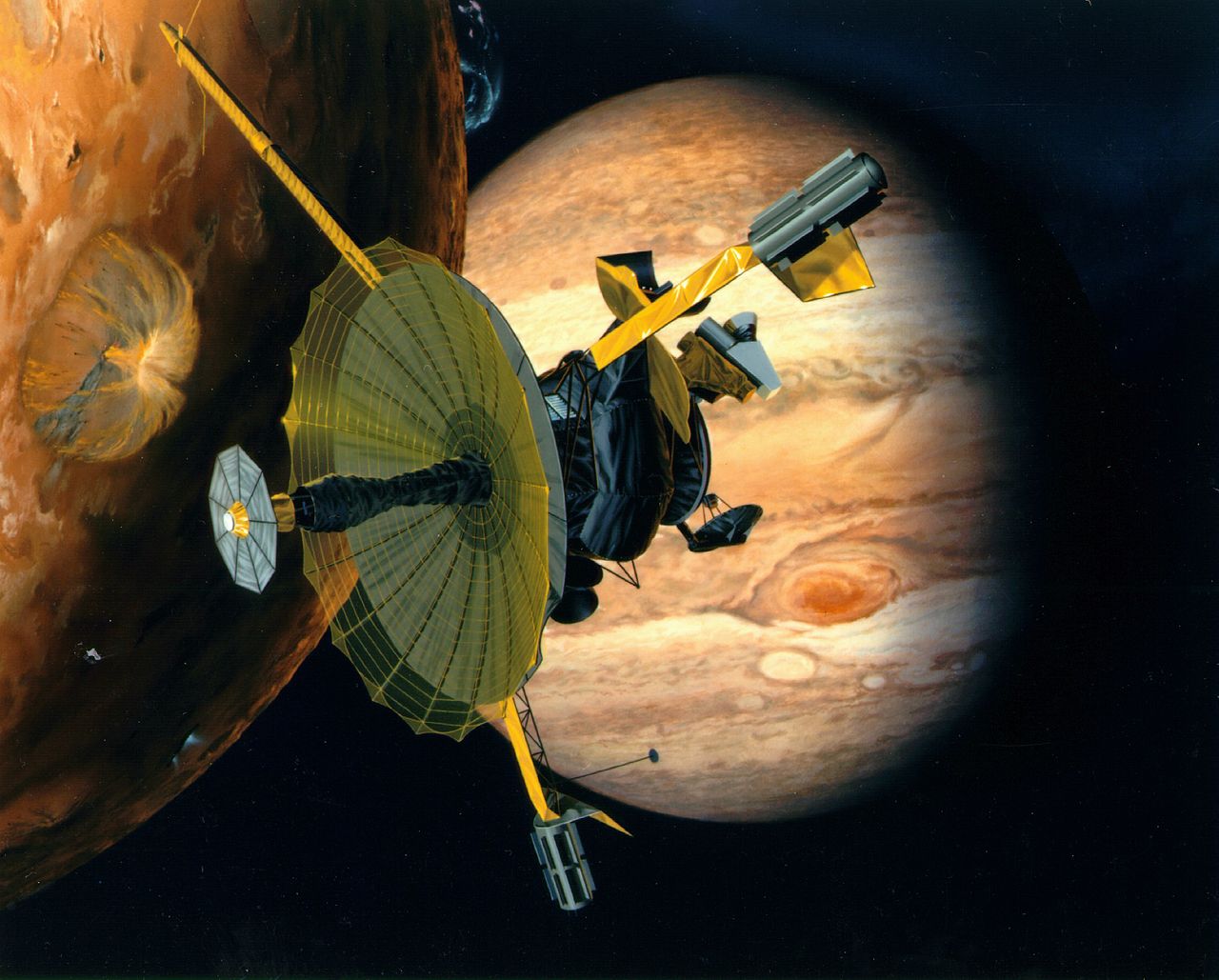}
  \caption{Artist impression of the Galileo mission [NASA https://photojournal.jpl.nasa.gov/catalog/PIA18176]}
  \label{img_galileo}
\end{figure}

The Galileo mission (Figure \ref{img_galileo}) was launched in 1989 with Space Shuttle Atlantis that was active in the Jupiter system between 1995 and 2003. During this period the Galileo mission studied Jupiter and several of its moons. The primary part of the mission ran from December 1995 to December 1997, after which it was extended two times. An object of special interest for the Galileo mission was Europa, to which the first extension of the mission was dedicated: it was named the 'Galileo Europa Mission". A detailed overview of all the Galileo Europa flybys will be presented in Section \ref{s_flybys_overview}. Galileo carried several in-situ and remote sensing instruments. Of particular interest for this work are the following in-situ instruments: the plasma instrumentation (PLS), the Energetic Particle Detector (EPD) and the magnetometer (MAG) (\cite{Frank1992,Williams1992,Kivelson1992}). PLS and EPD are discussed in more detail in the next section. A particular challenge facing the scientists working on the Galileo mission is the failure of the main antenna to unfold. This severally limited the data rate during the whole Galileo mission. After the final mission extension, the spacecraft descended into Jupiter's atmosphere in 2003 (\cite{nasa_galileo}). 

\subsection{Plasma instrumentation (PLS)}
\label{ss_pls}
\begin{figure}[h]
  \centering
  \includegraphics[width=0.5\textwidth]{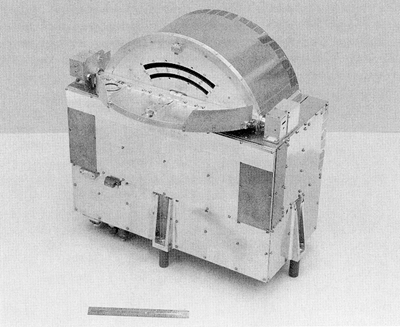}
  \caption{Photo of PLS. [NASA PDS]}
  \label{img_pls}
\end{figure}

\begin{figure}[h]
  \centering
  \includegraphics[width=0.75\textwidth]{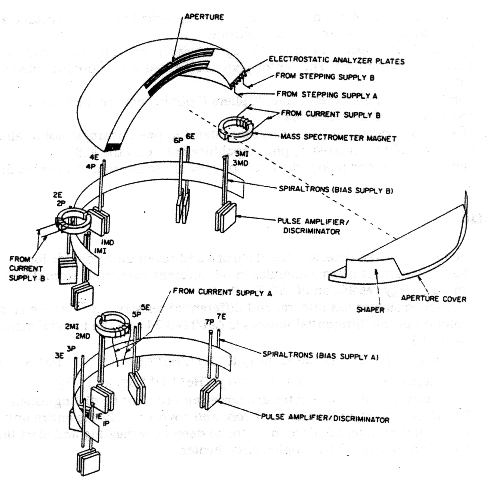}
  \caption{Diagram of PLS [NASA]}
  \label{img_pls_diagram}
\end{figure}

\begin{figure}[h]
  \centering
  \includegraphics[width=0.75\textwidth]{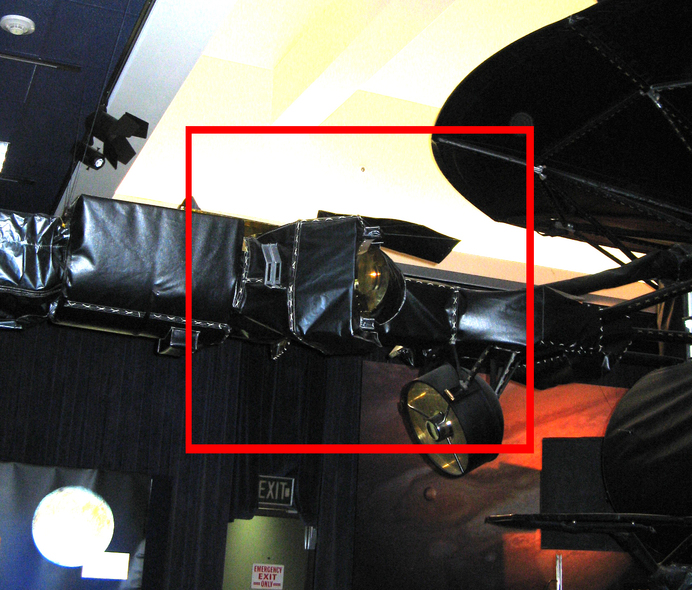}
  \caption{Location of the PLS instrument on the magnetometer boom. Shown is the Galileo model at the Jet Propulsion Laboratory, Pasadena. The main body of the spacecraft is on the right side, the main antenna can be recognized in the top right corner. [N. Krupp]}
  \label{img_pls_galileo}
\end{figure}

For the purpose of studying the plasma environment in-situ, Galileo was equipped with the plasma instrumentation PLS \citep{Frank1992}. PLS was a charged particle detector designed to detect both (positive) ions and electrons. A photo of the instrument is shown in Figure \ref{img_pls}. Figure \ref{img_pls_diagram} shows a diagram of the main components of PLS.
Since this thesis only considers the PLS ion data, only the ion detection aspect of PLS is explained in more detail. Charged particles entered PLS through either of the two apertures that are indicated in the diagram in Figure \ref{img_pls_diagram} and are also clearly visible in the photo shown in Figure \ref{img_pls}. After entering the detector through the aperture the particles passed through the electrostatic analyser (ESA). The electrostatic analyser was made out of two curved plates with a spacing in between. Over the spacing a voltage was applied. Only those particles with the desired energy per charge (E/q) could pass through the spacing between the plates and reach the detectors, without hitting the plates. By varying the field over the space between the plates, PLS could scan over energy per charge. The energy-per-unit charge (E/q) ranges from from 0.9eV to 52 keV.
PLS had 7 ion detectors (1P to 7P in Figure \ref{img_pls_diagram}) that each detected ions coming from a different direction of the sky. PLS was located on the magnetometer boom of Galileo, as is shown in Figure \ref{img_pls_galileo}. Therefore the PLS instrument was rotating about Galileo's rotation axis.
The detectors were oriented in such a way that during one spin (around 20 seconds) of the Galileo spacecraft (approximately) the whole sky was scanned. The detectors were channel electron multipliers (CEM), also referred to as Spiraltrons. CEMs consist out of a tube with a coating that has a high probability of giving secondary electron emissions upon impact by a particle \citep{Hoymork2002_Eklund}. When a particle entered the tube and hits the wall of the tube, secondary electron emission would occur. These secondary electrons were accelerated by the electric field inside the tube, so that they would produce even more secondary electrons when they hit the wall, this is referred to as an 'avalanche' effect. It is the current generated by these secondary electrons that was measured and not the original ion.
PLS data was transmitted in the units of counts per duty time, per energy bin and per direction bin (referred to as sector). Using the calibration data ('chi-factor'), obtained from laboratory experiments, the recorded counts can be converted to differential flux in units of cm$^{-2}$s$^{-1}$sr$^{-1}$eV$^{-1}$.
Due to the malfunction of the deployment of Galileo's main antenna, mentioned before, the PLS data rate was very limited during the entire mission. Hence, for each spin only a limited part of the collected data was transmitted to Earth, resulting in an incomplete coverage of the sky in both energy and direction.

PLS was also equipped with three miniature mass spectrometers, each of these mass spectrometers was equipped with one "Integral" sensor (1MI to 3MI) and one "Differential" sensor (1MD to 3MD), these are indicated in Figure \ref{img_pls_diagram} \citep{Frank1992}. Each mass spectrometer was equipped with an electromagnet that deflects incoming ions. Only particles with the desired mass per charge ratio (M/Q) could enter the MD detectors, allowing to determine the flux of particles per M/Q. The MI detectors were positioned such that they accepted ions that are not deflected by the electromagnet. The mass spectrometer data is not considered further in this thesis due to various technical difficulties associated with this data. 
Finally, in Table \ref{tab_properties_pls} the main properties of PLS relevant to this work are summarized.
\begin{table}[h]
\centering
\begin{tabular}{|r|l|}
  \hline
  \textbf{Property} & \textbf{Value} \\
  \hline
  \hline
  Detectors & 7 \\
  \hline 
  Coverage & $\sim 4 \pi$ \\
  \hline
  Full scan time &  $\sim 20$ s \\
  \hline
  Energy range & 0.9eV to 52 keV \\ 
  \hline
\end{tabular}
\caption{Overview of PLS properties.}
\label{tab_properties_pls}
\end{table}

\subsection{Energetic Particle Detector (EPD)}
\label{ss_epd}
\begin{figure}[h]
  \centering
  \includegraphics[width=0.5\textwidth]{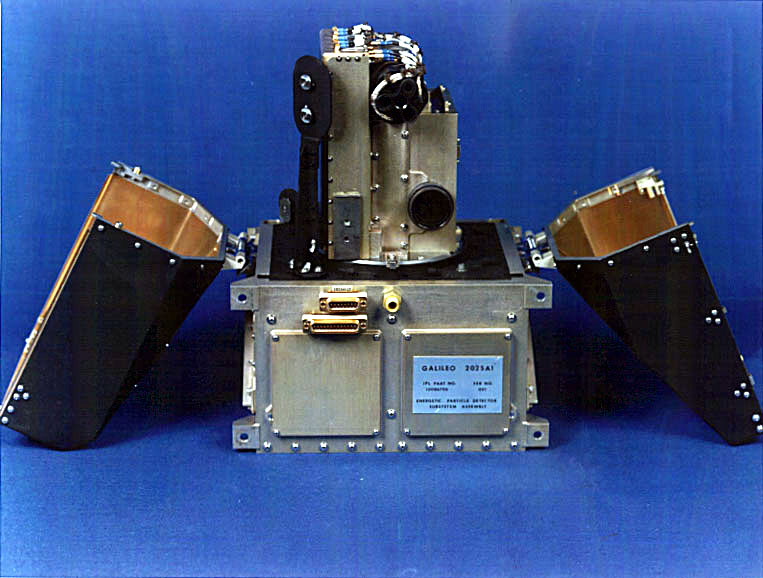}
  \caption{Photo of EPD. [NASA]}
  \label{img_epd}
\end{figure}

The Energetic Particle Detector (EPD), was the Galileo instrumentation tasked to characterize the energetic charged particle environment in Jupiter's magnetosphere \citep{Williams1992}. A photo of EPD is shown in Figure \ref{img_epd}.
EPD could detect charged particles (ions and electrons) with energies much higher than PLS. For ions with atomic number Z > 1 the energy is range is from 20 keV to 55 MeV, while for electrons the energy range is from 15keV to > 11 MeV. For elemental species Helium to Iron the energy range is from 10 keV per nucleus to 15 MeV per nucleus.

\begin{figure}[h]
  \centering
  \includegraphics[width=0.75\textwidth]{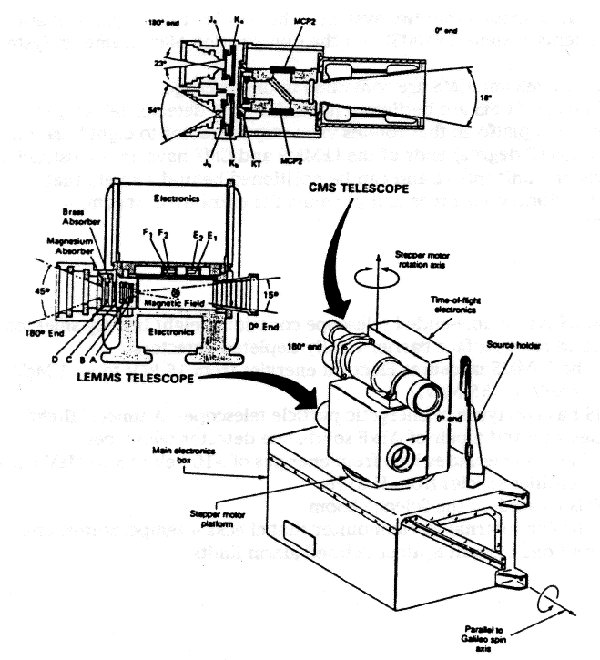}
  \caption{Diagram of EPD. [NASA]}
  \label{img_epd_diagram}
\end{figure}

The particle detectors of EPD were mounted on a rotating platform. This is shown in Figure \ref{img_epd_diagram}.
The axis of the rotating platform was perpendicular to the spin axis of the Galileo spacecraft. During one spin (20 seconds) only one segment of the sky (perpendicular to the spin axis) was scanned. After every spin of Galileo, the motor position of the rotating platform was changed. There were seven motor steps, so in around 140 seconds a spatial coverage of $4\pi$ could be achieved. 
A consequence of this canning procedure is that one particular direction of the sky was only scanned every 140 seconds. In this sense EPD differs from PLS, as PLS had seven detectors that operate simultaneously and that could therefore scan the entire sky in one spin.

Also shown in Figure \ref{img_epd_diagram} is the source holder or foreground shield, behind which the 0 degree ends of the detector could be positioned to estimate the noise level. When the detector passed behind this shield, the particle species of interest could not enter the detector any more. Only the noise signal would remain. The source of noise were particles with extremely high energies that penetrated the instrument shielding and the foreground shield and reached the detectors without passing through the filtering systems that allowed only the particles of interest to pass.
EPD was composed out of two particle detecting telescopes that were mounted on the rotating platform (Figure \ref{img_epd_diagram}).
The two telescopes are the Low Energy Magnetospheric Measurement System (LEMMS) and the Composition Measurement System (CMS). 
Solid State Detectors (SSD) were used to detect the charged particles in both telescopes. A solid state detector is made of a  semi-conducting material. When a charged particles passed through the detector, electric charge was released that was used to detect the particle. The amount of charge depends on the particle species and energy. Both telescopes were bi-directional telescopes, particles could enter the detectors from both sides. The properties of the two entrances of LEMMS are:
\begin{itemize}
\item 0 degree end: Used magnetic deflection to separate ions and electrons. Ion detectors A and B (E > $\sim 20$ keV), and electron detectors E1, E2, F1 and F2 ($\sim 0.52$ MeV > E > $\sim 15$ keV) were on this side.
\item 180 degree end: this end detected the higher energy electrons and ions.
Absorbers were located on this side to filter out particles with a lower energy than the ones of interest. Detector C and D were on this side. Both ions (E >  $\sim 16$ MeV) and electrons (E > $\sim 11$ MeV) were detected.
\end{itemize}
The two sides of CMS were configured as follows: 
\begin{itemize}
\item 0 degree end: a time of flight mass spectrometer was located on this side. The mass spectrometer allowed for the separation of ions in different species. CMS provided energy spectra above around 10 keV per nucleus for elements from helium to iron.
Stop and start signals in the TOF system were generated by, respectively, a foil and a detector (named KT). When charged particles passed through this foil and hit the detector electrons were generated that were detected by Multi Channel Plates (MCP). These are plates of electron
multiplier tubes stacked together.
Detector KT also gave the total energy of the particle, combining this information with the time of flight (thus the velocity of the particle) allows for the determination of the mass per charge (m/q).
A valid detection of a particle in the CMS TOF required a triple coincidence: a start signal, a stop signal and a KT energy pulse. This made CMS TOF less sensitive to penetrating background radiation since penetrating particles would not generate these three coincidences.
The channels of CMS that detected protons (TP) and oxygen (TO) are the most important ones for this work. In particular the data from the TP1, TP2, TP3, TO2 and TO3 channels, which differ in energy (Table \ref{tab_properties_epd}), will be used in this work. The definition of these channels is explained in more detail in Appendix \ref{a_epd_tof}.
\item 180 degree end: this end was not used due to data transmission issues, therefore it will not be discussed here further.
\end{itemize}
Unlike PLS, EPD operated at full resolution, but only during the close moon flybys. The data of this type is referred to as "record mode", which is the only data product that  will be considered in this thesis. Finally, in Table \ref{tab_properties_epd} the main properties of EPD relevant to this work are summarized.
\begin{table}[h]
\centering
\begin{tabular}{|r|l|}
  \hline
  \textbf{Property} & \textbf{Value} \\
  \hline
  \hline
  Coverage & $\sim 4 \pi$ \\
  \hline
  Motor positions & 7 \\
  \hline
  Full scan time &  $\sim 20$s $\times7$ = $\sim 140$ s \\
  \hline
  Energy range ions & 20 keV to 55 MeV \\ 
  \hline
  Energy range electrons & 15keV to > 11 MeV \\
  \hline
  Mass separation & Helium to Iron \\
  \hline
  \hline
  Proton channels & TP1 (80-220 keV), TP2 (220-540 keV) \\
   & TP3 (540-1040.0 keV) \\
  \hline
  Oxygen channels & TO2 (416-816 keV), TO3 (816-1792 keV)\\
  \hline
\end{tabular}
\caption{Overview of EPD properties (in the high resolution mode used during close moon flybys). Also indicated are the TP and TO channels, which are most relevant to this work.}
\label{tab_properties_epd}
\end{table}

\subsection{Magnetometer (MAG)}
\begin{figure}[h]
  \centering
  \includegraphics[width=0.75\textwidth]{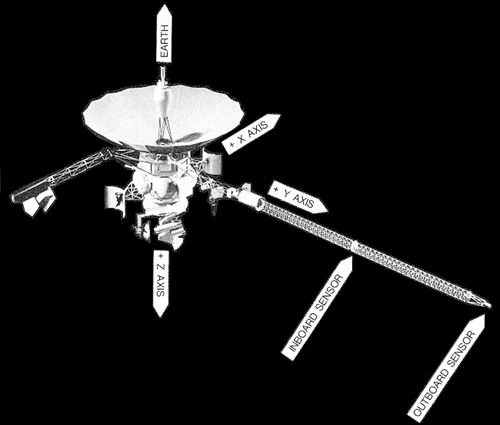}
  \caption{Configuration of the two magnetometers. [NASA]}
  \label{img_mag_diagram}
\end{figure}

Though not the main focus of this project, the magnetometer data is referred to frequently in this work, therefore Galileo's magnetometer instrument is introduced here briefly. Galileo's magnetometer (MAG) consisted of two triaxaial fluxgate magnetometers \citep{Kivelson1992}, that could take up to 30 samples per second. The two magnetometers were located on the boom of the Galileo spacecraft, one at 11.03 meter and the other at 6.87 meter from the spin axis, this is shown in Figure \ref{img_mag_diagram}. The reason for having two magnetometers was to acquire information about the spacecraft generated magnetic field and also to provide redundancy.
Magnetic fields generated by the spacecraft decrease in magnitude with increasing distance from the spacecraft. These fields were characterized by comparing the fields measured by the first and second detector, which were located at different distances from the spacecraft. By assuming the sources are dipoles, their contribution could be removed from the measured magnetic field.
Calibration of the magnetometer was done by measuring the field generated by a known source at the end of the boom.
The magnetometer data was processed onboard. The data was calibrated and converted into a reference frame that is not spinning with the spacecraft, it can therefore be interpreted directly. The final data product are the three components of the magnetic field in the non-spinning fame. The details of the processing and further technical aspects of MAG are described in \cite{Kivelson1992}. In this thesis only the processed data product from PDS is used.

\section{Charged particle motion}
Understanding the motion of charged particles (ions, electrons) under the influence of electric and magnetic fields is important for this thesis. Here the main aspects of this motion are described. The equations provided express particle motion in the non-relativistic limit, which is appropriate for this thesis as only non-relativistic ions are studied.

\label{s_charged_particle}

\subsection{The Lorentz force and gyration}
\label{ss_lorentz}
\begin{figure}[h]
  \centering
  \includegraphics[width=0.85\textwidth]{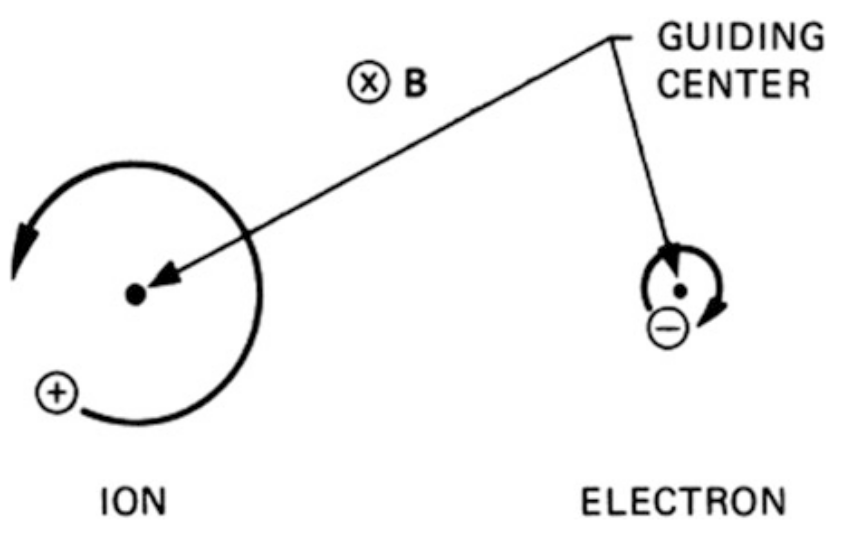}
  \caption{Gyration of a positively charged ion and a negatively charged electron. The magnetic field is into the plane of the figure. Also indicated is the center of the gyration, the guiding center. [Reprinted/adapted by permission from Springer International Publishing: Introduction to Plasma Physics and Controlled Fusion by Francis. F. Chen $\copyright$ (2016)]}
  \label{img_gyration}
\end{figure}

The basic equation expressing the force experienced by a charged particle under the influence of electric and magnetic fields is that of the Lorentz force (Equation \ref{eq_lortentz}). In this equation $m$ is the mass of the particle, $\bm{v}$ is its velocity, $q$ is its charge, $\bm{E}$ is the electric field, $\bm{B}$ is the magnetic field and $t$ is the time. 
\begin{equation}
m\frac{d\bm{v}}{dt} = q(\bm{E}+\bm{v}\times\bm{B})
\label{eq_lortentz}
\end{equation}
If only an electric field is present, the charged particle will be accelerated along the direction of the field. The sign of the acceleration will be determined by the charge, for positive particles in the direction of the electric field and for negative particles in the opposite direction.
The effect of the magnetic field is more complicated. A particle at rest with respect to the magnetic field, will not experience a magnetic force. Also, a particle with its velocity perfectly aligned with the magnetic field will experience no Lorentz force (since the cross product will be zero). Any charged particle with a velocity component perpendicular to the magnetic field, will experience a Lorentz force. Because of the cross product between velocity and magnetic field, the Lorentz force will always be perpendicular to the particle velocity and the magnetic field. This force will cause the charged particle to start gyrating: it will perform a circular orbit around the magnetic field line, but its velocity magnitude (and energy) won't change. The center of the gyration is referred to as the 'gyrocenter' or 'guiding center'. The angle between the magnetic field and the velocity vector of a charged particle is referred to as 'pitch angle'. The gyration for a positively charged ion and a negatively charged electron are shown in Figure \ref{img_gyration}.
This gyrating motion is characterized by two factors: the gyroradius and the gyrofrequency, these are expressed by respectively Equation \ref{eq_gyroradius} and Equation \ref{eq_gyrofrequency}.
\begin{equation}
r_g = \frac{v_{\perp}}{|\omega_g|}  = \frac{m v_\perp}{|q|B}
\label{eq_gyroradius}
\end{equation}
\begin{equation}
\omega_g = \frac{qB}{m}
\label{eq_gyrofrequency}
\end{equation}
Heavier particles of the same velocity have a larger gyroradius but a smaller gyrofrequency. The gyroradius linearly scales with the perpendicular component of the velocity. Larger field strengths lead to higher frequencies but smaller gyroradii. The direction of the gyration is determined by the charge of the particle: it is opposite for ions and electrons, this is shown in Figure \ref{img_gyration}.

The gyration of charged particles is associated with an invariant quantity, referred to as the "first adiabatic invariant". It states that the magnetic moment of a charged particle can be considered a constant if the frequency of the field variations is lower than the gyrofrequency. The magnetic moment $\mu$ is expressed by Equation \ref{eq_first_invariant}. 
\begin{equation}
\mu = \frac{1}{2}\frac{mv_{\perp}^2}{B}
\label{eq_first_invariant}
\end{equation}

\subsection{ExB drift}
\label{sss_ExB}
\begin{figure}[h]
  \centering
  \includegraphics[width=0.85\textwidth]{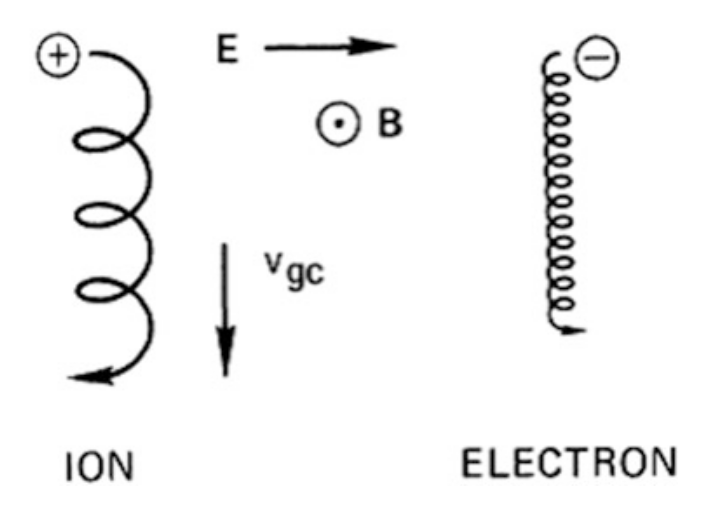}
  \caption{ExB drift for an ion and a electron. $V_{gc}$ is the same as $V_E$ in Equation \ref{eq_ExB}. [Reprinted/adapted by permission from Springer International Publishing: Introduction to Plasma Physics and Controlled Fusion by Francis. F. Chen $\copyright$ (2016)]}
  \label{img_ExB}
\end{figure}

In reality, magnetic fields in solar system or astrophysical environments are not uniform in space and exist in combination with electric fields. In the next subsections the motions resulting from these complicating factors is described. First the case is considered where a magnetic and electric field occur together in a perpendicular configuration. Such a configuration results in the motion of the guiding center of the particle, referred to as the ExB drift. The particle's guiding center will have a net motion in the direction perpendicular to the electric and magnetic field. The drift velocity of the guiding center is expressed in Equation \ref{eq_ExB}.
\begin{equation}
\bm{v}_E = \frac{\bm{E}\times\bm{B}}{B^2}
\label{eq_ExB}
\end{equation}
The ExB drift does not depend on the charge or mass of the particle and is thus the same for ions and electrons. The ExB motion is shown in Figure \ref{img_ExB}.
This drift is the consequence of a varying gyroradius, which is caused by the introduction of the perpendicular electric field. During one half of the gyration the electric field accelerates the particle and thereby increases the gyroradius. During the other half the electric field has a decelerating effect, thereby decreasing the size of the gyroradius.
Thus, during one part of the gyromotion the gyroradius is larger than during the other, the consequence of this is a net motion in the direction perpendicular to the electric and magnetic field.
Newly formed ions in Europa's ionosphere perform an ExB drift motion.

\subsection{Magnetic gradient drift}
\label{sss_gradB}
\begin{figure}[h]
  \centering
  \includegraphics[width=0.75\textwidth]{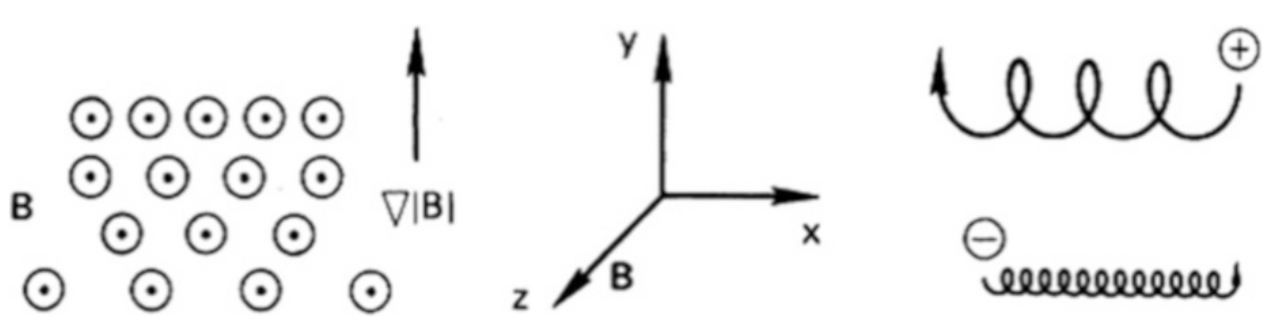}
  \caption{Gradient B drift. [Reprinted/adapted by permission from Springer International Publishing: Introduction to Plasma Physics and Controlled Fusion by Francis. F. Chen $\copyright$ (2016)]}
  \label{img_gradB}
\end{figure}

A drift motion of the guiding center occurs when there is a gradient in the magnetic field strength in the direction perpendicular to the field lines. The drift velocity of the guiding center is expressed by Equation \ref{eq_gradB}. 
\begin{equation}
\bm{v_{\nabla B}} = \frac{m v_{\perp}^2}{2qB^3}(\bm{B}\times\nabla{\bm{B}})
\label{eq_gradB}
\end{equation}
This drift motion is perpendicular to the field lines and the gradient in the field strength. The dependence on charge $q$ indicates that the direction of this drift is different for positively and negatively charged particles. The drift arises due to a variation of the gyroradius, this is illustrated in Figure \ref{img_gradB}. During its gyromotion the charged particle will travel through a region with a higher and a lower field strength. The gyroradius is larger in the region with a lower field strength, but smaller in the region with higher field strength. The effect of this is that the gyro-orbit is not a closed circle any more, which results in a net motion of its guiding center.

\subsection{Magnetic curvature drift}
\label{sss_curvB}
The curvature B drift of the guiding center occurs when there is a curvature in the magnetic field. The drift velocity of the guiding center is expressed by Equation \ref{eq_curvB}. 
\begin{equation}
\bm{v_{R}} = \frac{1}{q}\frac{\bm{F}_{cf}\times{\bm{B}}}{\bm{B^2}}=\frac{mv^2_{||}}{q\bm{B^2}}\frac{\bm{R}_{c}\times{\bm{B}}}{\bm{R_c^2}}
\label{eq_curvB}
\end{equation}
It is a consequence of the centrifugal force $\bm{F}_{cf}=mv^2_{||}\frac{\bm{R}_c}{{R}^2_c}$ experienced by a charged particle moving along a curved magnetic field. Where $v_{||}$ is the velocity along the magnetic field and ${R}_c$ the radius of curvature.

This drift occurs in the direction perpendicular to the radius of curvature ($R_c$) and the magnetic field. The direction of the drift is opposite for particles with positive and negative charge, because of the dependence on charge.

Note that the curvature drift always occurs together with the gradient B drift described previously. This is because a curved magnetic field requires a change in field strength, a field without this gradient in field strength would violate $\nabla \cdot \bm{B} = 0$. Hence a gradient drift perpendicular to the field and the gradient occurs. In planetary magnetospheres the combined curvature drift and gradient B drift causes an azimuthal drift of the gyrocenter, from the perspective of the planet.

\subsection{Magnetic mirror}
\label{sss_mag_mirror}
\begin{figure}[h]
  \centering
  \includegraphics[width=0.75\textwidth]{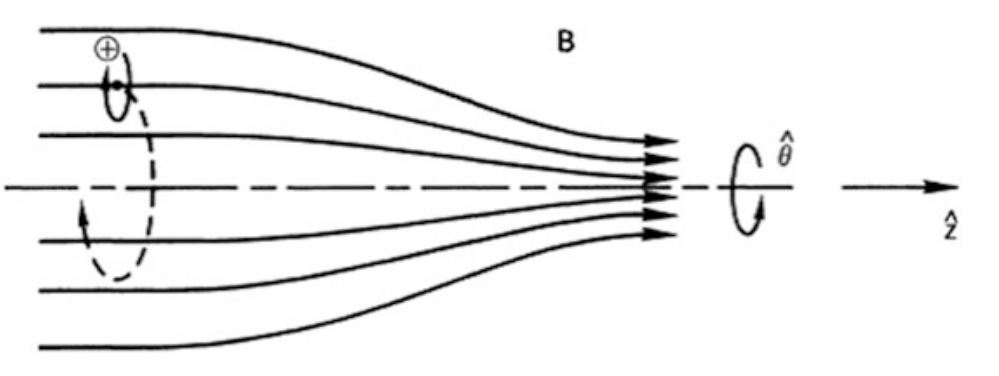}
  \caption{Schematic of a magnetic mirror. In the figure converging magnetic field lines are illustrated by the solid black lines. The horizontal dashed line indicates the axis of symmetry of the field. On the left a positive ion is indicated. A gyro orbit about a field line removed from the axis of symmetry is indicated by the solid circle. The dashed trajectory indicates a gyro orbit about the axis of symmetry of the field.
   [Reprinted/adapted by permission from Springer International Publishing: Introduction to Plasma Physics and Controlled Fusion by Francis. F. Chen $\copyright$ (2016)]}
  \label{img_mirror}
\end{figure}

The magnetic mirror motion of the guiding center occurs when a charged particle enters a region of converging magnetic field lines, which is illustrated in Figure \ref{img_mirror}.

A gradient in the magnetic field occurs along the field lines shown in Figure \ref{img_mirror}. A particle travelling from left to right in this figure will experience an increase in magnetic field strength. The particle moving towards the stronger field experiences a Lorentz force, which is by definition perpendicular to the field. On the left side of the figure, where the magnetic field is homogeneous, the effect of the Lortenz force results in a gyration about the magnetic field. However, when the particle enters the region where the field lines start converging, the Lorentz force will have a component that decreases the velocity component of the particle parallel to the field. Assuming that the first adiabatic invariant is conserved, meaning that the magnetic moment expressed by Equation \ref{eq_first_invariant} remains constant, the perpendicular component of the particle velocity should increase to keep the magnetic moment constant. However, since the Lorentz force cannot increase the total velocity of the charged particle, the parallel velocity of the particle will decrease. Parallel velocity is converted into perpendicular velocity, slowing down the particle. Essentially the pitch angle of the charged particle is increasing. When all parallel velocity has been converted into perpendicular velocity the particle will be reflected away from the converging zone, rather than continue moving towards it.
Particles between two converging field regions will oscillate (bounce) provided that the condition in Equation \ref{eq_mirror_cond} is met.
\begin{equation}
B_m > \frac{B_0}{\sin^2 \alpha_{eq}}
\label{eq_mirror_cond}
\end{equation}
In this Equation $B_m$ is the field at the mirror point, $B_0$ the equatorial field and $\alpha_{eq}$ the equatorial pitch angle. This type of motion is referred to as mirror motion. An example of this is a planetary magnetosphere, regions of converging field lines occur near the magnetic poles of the object and charged particles can mirror between these poles, this is addressed in more detail in the next Section. 

\subsection{Trapped particle motion}
\label{ss_trapped}
Planetary magnetic fields combine all aspects discussed above. Field gradients, curvature and electric fields all lead to the trapping of charged particles due to a motion that can be decomposed in three elements: gyration, mirror motion and azimuthal drift. The energetic charged particles that will be simulated in this thesis are an example of these trapped energetic particles.

Planetary magnetic fields are often described as a dipole magnetic field. In that case the magnetic field strength, at a specific radial location $r$ and latitude $\lambda$ with respect to the center and magnetic axis of the object, is given by Equation \ref{eq_dipole}. In this equation $M_E$ is the magnitude of the magnetic moment. 
\begin{equation}
B = \frac{\mu_0}{4\pi}\frac{M_E}{r^3}(1+3\sin^2 \lambda)^{(1/2)}
\label{eq_dipole}
\end{equation}
In the context of planetary magnetic fields it is often convenient to express the radial position with respect to the object using the L-shell parameter. This parameter is the radial distance of a field line at the magnetic equator $r_{eq}$ divided by the radius of the object $R_{object}$, as shown in Equation \ref{eq_Lshell}. 
\begin{equation}
L = \frac{r_{eq}}{R_{object}} = \frac{r}{R_{object} \cos^2 \lambda}
\label{eq_Lshell}
\end{equation}
On the right hand side of the equation, $r_{eq}$ is expressed as a function of any position (expressed in radial distance $r$ and latitude $\lambda$) along the field line.
Planetary magnetic fields are not perfectly dipolar. For example in the case of Jupiter, the magnetic field is stretched outwards in the magnetic equator by the coroting plasma, this is depicted in Figure \ref{img_jup_mag_bag}. However in such magnetospheres the mirror motion still occurs and does not change qualitatively.

\begin{figure}[h]
  \centering
  \includegraphics[width=1.0\textwidth]{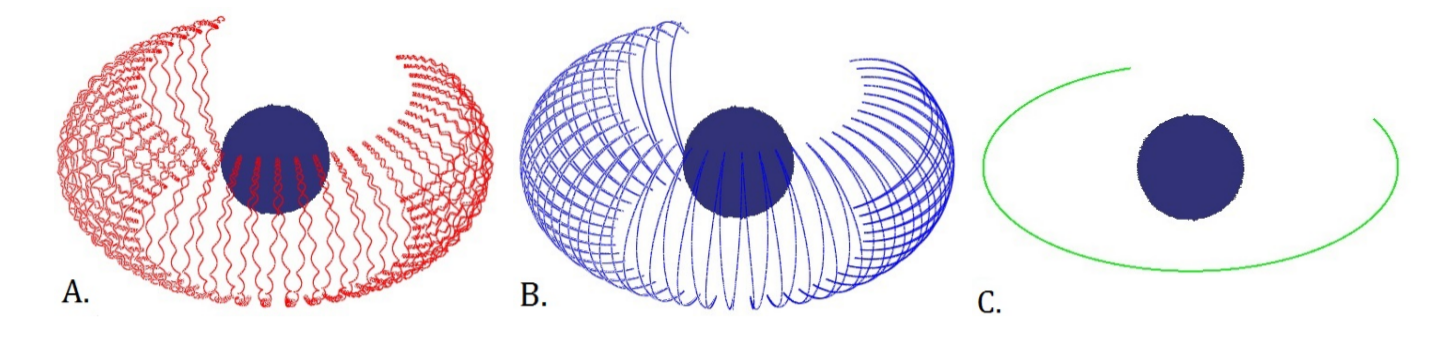}
  \caption{Trapped particle motion, A: gyration, mirror motion and azimuthal drift, B: motion of the gyrocenter, C: azimuthal drift. [From \cite{Kotova2016}]}
  \label{img_trajectories_ak}
\end{figure}

Figure \ref{img_trajectories_ak} illustrates the different aspects of the trapped particle motion. \ref{img_trajectories_ak}A shows the combined effect of the gyration, the mirror motion and the azimuthal drift. \ref{img_trajectories_ak}B shows the motion of the guiding center, that is the result of the mirror motion and the azimuthal drift. \ref{img_trajectories_ak}C shows only the azimuthal drift.

\begin{figure}[h]
  \centering
  \includegraphics[width=0.75\textwidth]{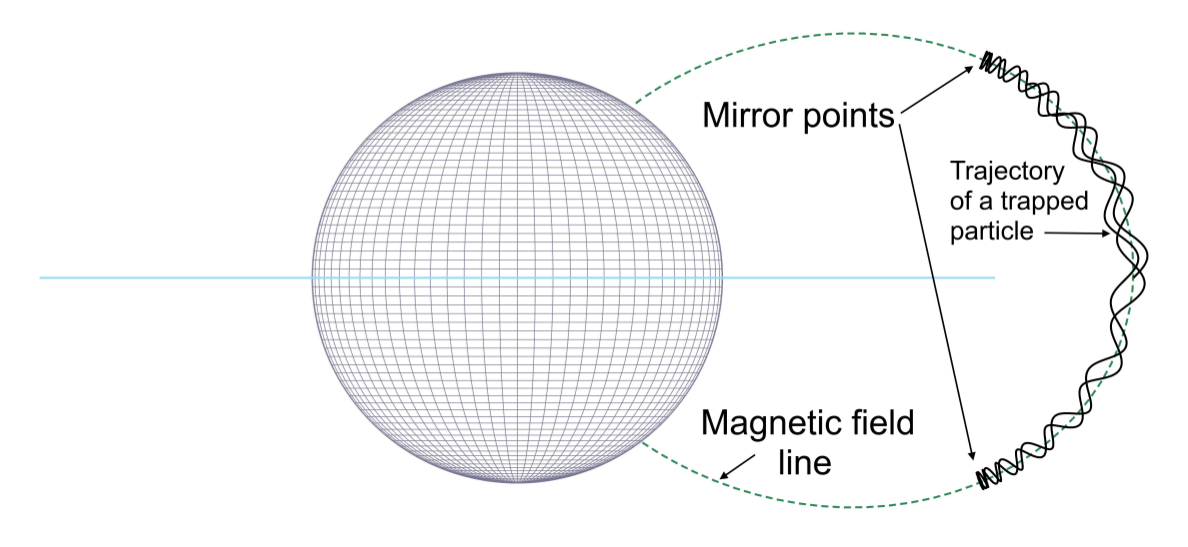}
  \caption{Trapped particle motion. [From \cite{Kotova2016}]}
  \label{img_trapped}
\end{figure}

The mirror motion is illustrated in Figure \ref{img_trapped} for an arbitrary planet. This motion occurs is Jupiter's magnetosphere, trapped energetic charged particles are performing mirror motions between the magnetic poles. 
The latitude of the mirror point is determined by the equatorial pitch angle of the energetic charged particle, which is defined as the angle between the velocity vector of the particle and the magnetic field at the magnetic equator. Equation \ref{eq_pa} expresses the equatorial pitch angle in a dipolar magnetic field.
\begin{equation}
\sin^2 \alpha_{eq} = \frac{B_{eq}}{B_m} = \frac{\cos^6 \lambda_m}{(1+3\sin^2 \lambda_m)^{1/2}}
\label{eq_pa}
\end{equation}
In this equation $B_m$ is the magnetic field at the mirror point and $\lambda_m$ the magnetic latitude of the mirror point. Particles with an equatorial pitch angle of zero and 180$^\circ$ will travel all the way to the magnetic poles. A particle with a 90 degree equatorial pitch angle will stay in the equatorial plane. In Figure \ref{img_trapped} an intermediate case is shown.

The period of the mirror motion or bounce period is expressed by Equation \ref{eq_bounce_period} for a trapped particle in a dipole magnetic field, from e.g. \cite{Baumjohann1997}.
\begin{equation}
\tau_b \approx \frac{L R_E}{(W/m)^{1/2}}(3.7-1.6\sin\alpha_{eq})
\label{eq_bounce_period}
\end{equation}
In this equation L is the L-shell the particle is mirroring in, W the energy of the particle and $R_E$ the radius of the planetary object. The bounce period is larger for more massive particles and for longer field lines (larger L). Particles with a higher energy have a shorter bounce period, since the velocity is higher for these particles. The dependence on equatorial pitch angle is not very strong. This can be explained by the fact that particles with a smaller pitch angle have a higher velocity along and the field, but also have a larger travel distance, and vice versa for particles with a large pitch angle.

Besides the drift motion between the magnetic poles, the guiding centres of the trapped particles are drifting azimuthally because of curvature and gradient B drift. The resulting drift velocity for a trapped particle in a dipole magnetic field is expressed by Equation \ref{eq_drift_az} (see \cite{Baumjohann1997} for a derivation), in which $B_E$ is the equatorial magnetic field on the surface of the planetary object.
\begin{equation}
\langle v_d \rangle \approx \frac{6 L^2 W}{q B_E R_E}(0.35+0.15\sin\alpha_{eq})
\label{eq_drift_az}
\end{equation}
The drift speed depends on the energy per charge of particles. The drift velocity is therefore the same for ions and electrons of the same energy, but in opposite direction. The azimuthal drift increases with energy of the particle, but also with the L-shell. Like the bounce period, the drift period only depends weakly on the equatorial pitch angle.

\section{Charged particle tracing simulation}
\label{s_particle_simulation}
Depletions of energetic ions have been observed in the Galileo Energetic Particle Detector (EPD) data. To determine the origin of these features, either by impact on the surface or charge exchange with the atmosphere, it is necessary to simulate the trajectories of the energetic ions. In this section the Monte Carlo test particle approach that will be used to simulate the particles is discussed. 
This code will also be used to simulate trajectories of neutral and charged particles to make predictions for future measurements.

\subsection{Simulation approach}
\label{s_model}
\begin{figure}[h]
  \centering
  \includegraphics[width=0.5\textwidth]{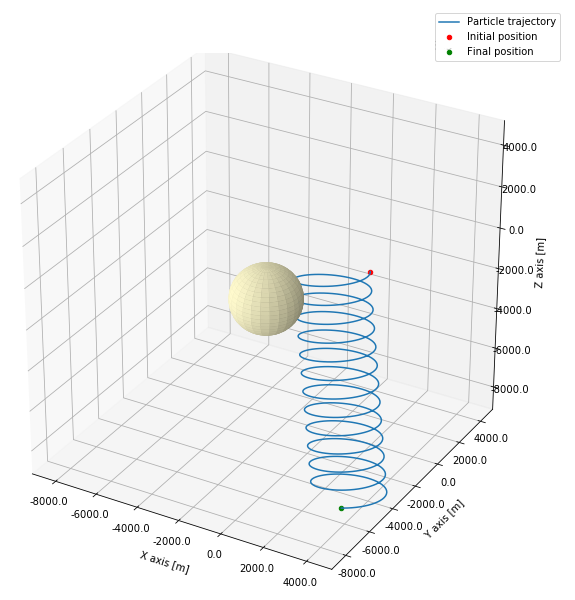}
  \caption{An example of the results of the particle tracing code: a singly charged oxygen trajectory near Europa, the energy of the particle is 1.7 MeV.}
  \label{img_trajectory}
\end{figure}
Depending on the application, the motion of charged particles can be simulated in different ways. Describing the gyration or the mirror motion is not always needed. Therefore, the periodic aspect of the particle motion can sometimes be averaged so that only the motion of its guiding center (Figure \ref{img_trajectories_ak}B and C) needs to be followed. For this thesis a full solution of the Lorentz force is selected as the most relevant approach. 
The goal is to simulate losses of the energetic particles that interact with the surface or atmosphere and for this application the gyration cannot be neglected. The gyroradii of the particles that will be simulated are of the same size or larger than relevant geometric scales, such as the altitude of the spacecraft flyby. 
Thus from an orbital altitude that is smaller than the gyroradius of certain particles, depletions of the these particles should be observed. This effect is indeed observed in the Galileo energetic particle detector (EPD) data, as was discussed in Section \ref{ss_interaction_EPD}. If the gyroradius were to be neglected in the simulation, these losses would not occur in the simulation. Therefore the full motion needs to be described.

The charged particle tracing code used in this thesis is a continuation of the code originally developed for the Master's thesis project described in \cite{Huybrighs2015}. It is entirely developed in the programming language Python and makes use of the NumPy library and the irfpy.util library developed by Dr. Yoshifumi Futaana at the Swedish Institute of Space Physics \footnote{https://irfpy.irf.se/projects/util}. First a summary is provided of the main aspects of the code as described in \cite{Huybrighs2015}. For each aspect it is indicated by 'ions' and or 'neutrals' to which type of simulation it refers to specifically.
\begin{itemize}
\item \textbf{Test Particle (ions \& neutrals):} a test particle type of simulation is employed, this means that individual particles are traced as they propagate under the influence of the relevant forces (gravity or the Lorentz force). The effect that the particles have on their environment is not taken into account. This could be important in particular when dealing with charged particles, their motion can affect the electric and magnetic fields which affect the Lorentz force. However, for the simulation of energetic charged particles it can be assumed that they do not influence the fields, as was discussed in Section \ref{s_mag_interaction}.
Ionospheric plasma can also influence the fields (in Section \ref{s_mag_interaction} it was shown how Europa's ionospheric plasma influences the flow of the corotational plasma). MHD or hybrid simulation could in principle be employed to provide self-consistent fields from the plasma
dynamics. The test particle method has a key advantage over these methods. The results of the simulation depend on multiple parameters. With the test particle method each parameter's contribution can be determined by running simulations with increasing complexity. Furthermore it is less computationally intensive.
The simplest simulation that will be considered is the case with homogeneous fields and no atmosphere. This case will be compared with simulations taking into account, a flow deflection, Europa's dipole and an atmosphere. This way the contribution of each physical effect on the losses of energetic charged particles can be characterized. In a self-consistent simulation it would be more difficult to disentangle the contribution of the different effects to the losses. The models used to simulate the atmosphere, the flow deflection and the induced dipole are discussed later in this section.
\item \textbf{Monte Carlo (ions \& neutrals):} the code generates particles with certain initial conditions according to the Monte Carlo method. This means that code generates a large number of particles of which their initial velocity vector is determined by a certain distribution, which determines the probability that certain initial conditions occur. The choice of the distribution depends on the context. 
\item \textbf{Particle tracing (ions \& neutrals):} the code determines the three-dimensional trajectories of charged particles by numerically integrating Newton's law of
gravitation (first term of the right hand side of Equation \ref{eq_newton_grav}) for neutral particles or the Lorentz force (second term on the right hand side of Equation \ref{eq_newton_grav}) for charged particles.
\begin{equation}
m\frac{d\bm{v}}{dt} = -\frac{GM}{\bm{r}^2}\hat{\bm{r}}+q(\bm{E}+\bm{v}\times\bm{B})
\label{eq_newton_grav}
\end{equation}
As was discussed previously, a full solution of the Lorentz force is the most appropriate for this work. It is assumed that the particles do not exert any forces on each other.
\item \textbf{Leap frog integration (ions \& neutrals):} for the numerical integration of Newton's law of gravitation and the Lorentz force the method of leap frog integration is used. Equations \ref{eq_leap_frog_x} and \ref{eq_leap_frog_v} express how, respectively, the position and velocity are progressed from time step i to time step i+1 in this method.
\begin{equation}
\bm{x}_{i+1} = \bm{x}_i + \bm{v}_{i+1/2}\Delta t
\label{eq_leap_frog_x}
\end{equation}
\begin{equation}
\bm{v}_{i+1} = \bm{v}_i + \bm{f}_{i+1/2}\Delta t
\label{eq_leap_frog_v}
\end{equation}
The equations are implemented as follows. First, the particle position is progressed with Equation \ref{eq_leap_frog_x}. The progression of position depends on the velocity at step i+1/2. For the first time step
(namely, the progression from i = 0 to i =1/2) the velocity at i+1/2 is calculated with Equation \ref{eq_leap_frog_v}, by
calculating the force $f_{i+1/2}$ at i+1/2 using the position at step i and using half a time step $\Delta t$. The force can be either the Lorentz force or gravity. For the subsequent time step velocity is progressed from v at i+1/2 to v at i+3/2 with Equation \ref{eq_leap_frog_v} and using the time step size $\Delta t$.
\item \textbf{Super particles (ions \& neutrals):} the code has the functionality to treat particles as super particles. This means that each particle in the simulation represents a number of particles with the same initial conditions that are travelling together (see for example \cite{Holmstrom2007}). This aspect of the code is only used in the simulations for predictions of future measurements (Chapter \ref{ch_prospects}).
\end{itemize}
For further technical details on the code the reader is referred to the work in \cite{Huybrighs2015}. Various changes have been made to the code since the publication of \cite{Huybrighs2015}, of particular importance to this thesis are the following aspects:
\begin{itemize}
\item \textbf{Backtracing (ions): } a crucial new component added to the code for this thesis is the backtracing functionality. This means that rather than calculating the trajectories of particles forward in time, they are calculated backwards in time (see for example \cite{Futaana2003,Futaana2010}). This is achieved essentially by incorporating a negative timestep in the existing method. This functionality is very useful when determining the origin of particles that have reached a spacecraft at a certain position. In the case of forward calculations one would have to trace many particles from a wide range of positions to determine which ones reach the spacecraft. In the case of backtracing calculations one would only have do such a computation for one position: that of the spacecraft. Thus for this type of simulation, the backtracing method requires less computational resources, which justifies its use.
\item \textbf{Particle losses (ions):} energetic charged particles in the simulation can be lost in two ways: by impacting on the surface or by charge-exchanging with atmospheric neutral particles. The loss due to impact is implemented by removing particles from the simulation as soon as they obtain a radial position smaller than Europa's radius.
Losses due to charge exchange are implemented following the method described on \cite{particleincell2011}. The method is based on Equation \ref{eq_charge_exchange_p}, which expresses the probability $P$ that charge exchange occurs for a particle located at a certain location where the number density is given by $n$.
\begin{equation}
P = 1 - \exp(-n\sigma g dt)
\label{eq_charge_exchange_p}
\end{equation}
In this equation $\sigma$ is the collision cross section, which is a measure for the probability of charge exchange between the particle of interest and an atmospheric particle, $g$ expresses the relative velocity of the particle, and $dt$ expresses the time step used to integrate the particle trajectory. Charge exchange is then said to occur if a random generator, that generates numbers between 0 and 1, generates a number larger than the charge exchange probability $P$. In this work this is implemented by calculating $P$ for every integration step. For the calculation of the relative velocity it is assumed that the atmospheric particles are at rest. Then for each of these steps a random number is generated. The first integration step for which the random number is larger than $P$, is then considered as the point where charge exchange occurs. The particle that charge exchanges is then discarded in the simulation. Since the simulation is a Monte Carlo approach, a large number of particles is used and therefore a statistical analysis of the particle losses by charge exchange is possible.
The atmospheric models used to determine the atmospheric density at every position are discussed in the next section.
\item \textbf{Electric and magnetic fields (ions):} assumptions for the electric and magnetic fields are required to calculate the Lorentz force. These fields can be set as homogeneous fields, but it is also possible to use analytical equations that express the fields at every position. An example is the analytical model in \cite{Ip1996} to express the electrical field near Europa affected by the diversion of the corotating plasma. The models used in this thesis are discussed in the next section. 
\end{itemize}

\subsection{Model inputs}
In this section an overview is provided of analytical models that are used as inputs for the simulations of the (energetic) charged particle trajectories.

\subsection*{Analytical expressions for the atmosphere and plume}

\begin{figure}[h]
  \centering
  \includegraphics[width=0.75\textwidth]{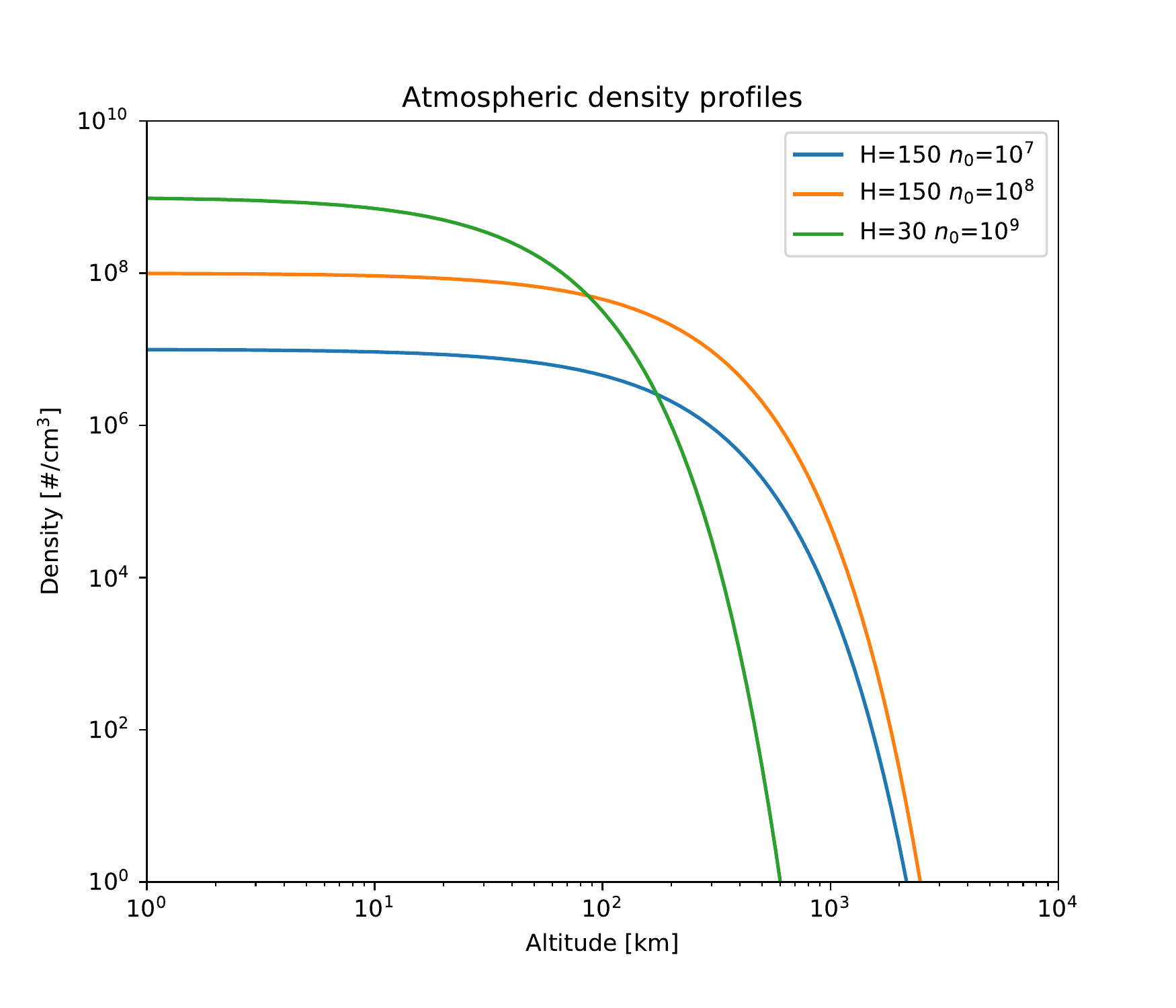}
  \caption{Several atmospheric profiles, using the model from \cite{Saur1998}}
  \label{img_saur_atm}
\end{figure}

The analytical expression from \cite{Saur1998} is used to determine the density of Europa's atmosphere as a function of radial position $r$. The model is expressed by Equation \ref{eq_atm_saur} which depends on a surface density $n_0$ and scale height $H$. $r_{sat}$ expresses the radial dimension of the planetary object.
\begin{equation}
n = n_0(r_{sat}/r)^2 \exp{[(r_{sat}-r)/H]}
\label{eq_atm_saur}
\end{equation}
This equation is derived in \cite{Saur1998} as a solution of the spherical continuity equation for radial outflow. Some examples of atmospheric profiles calculated with this model are shown in Figure \ref{img_saur_atm}. 

\cite{Jia2018} uses Equation \ref{eq_plume_jia} to express the density of a plume as a function of radial position $r$ and angular distance $\theta$.
\begin{equation}
n_p = n_{p0} \exp{[(r_{sat}-r)/H_p]} \exp{(-\theta/\theta_p)^2}
\label{eq_plume_jia}
\end{equation}
It is similar to Equation \ref{eq_atm_saur} in the sense that it expresses an exponentially decaying density, determined by a surface density $n_{p0}$ and scale height $H_p$. The difference is a factor that gives the plume an angular width $\theta_p$. We use this model in the simulation to determine the density of a plume.

\subsection*{Induced magnetic field}
\label{ss_induced_dip}

\begin{figure}[h]
  \centering
  \includegraphics[width=0.75\textwidth]{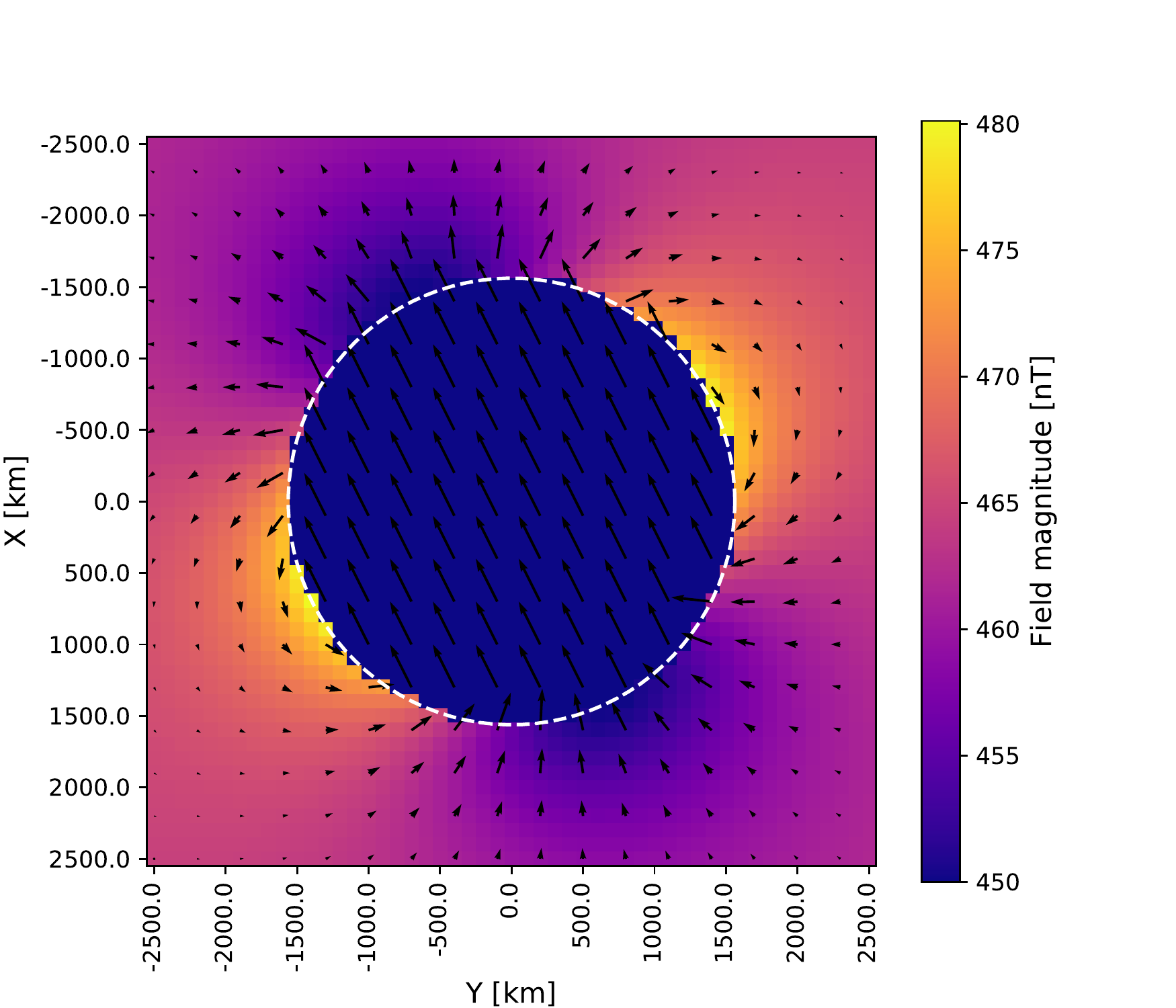}
  \caption{Europa's induced magnetic field during the E12 flyby according to the analytical model from \cite{Zimmer2000}. The arrows indicate the directions of the magnetic field and the colour the magnitude of the field, in nT. The figure represents Europa's equatorial plane. The positive y-axis is towards Jupiter and the positive x-direction is in the direction of the corotational plasma. For this example it is assumed that the varying magnetic field components are: 50 nT in direction of Jupiter, 100 nT in the direction of the corotational plasma and 450 nT in the direction in the plane. The magnitude of the field is 463 nT.}
  \label{img_induced_dip}
\end{figure}

The induced dipole of Europa can be expressed as a dipole magnetic field, as shown by \cite{Zimmer2000}. An example output of the model from \cite{Zimmer2000} is shown in Figure \ref{img_induced_dip}.

Equation \ref{eq_B_sec} expresses the magnetic field components at a position $\bm{r}$. Here $\bm{B}_{sec}$ refers to the induced magnetic field.
\begin{equation}
\bm{B}_{sec} = \frac{\mu_0}{4 \pi}[3(\bm{r}\cdot\bm{M})\bm{r}-r^2 \bm{M}]/r^5 
\label{eq_B_sec}
\end{equation}
This equation depends on the magnetic moment $\bm{M}$, which is expressed by Equation \ref{eq_mag_moment}.
\begin{equation}
\bm{M} = -\frac{4 \pi}{\mu_0}Ae^{i \phi} \bm{B}_{prim}\bm{r}_m^3/2 
\label{eq_mag_moment}
\end{equation}
$\bm{M}$ on its turn depends on $\bm{B}_{prim}$, which is the varying component of Jupiter's magnetic field, as experienced by Europa (see Section \ref{ss_interior}). The radius of the moon is expressed by $r_m$. The parameters A and $\phi$ are, respectively, the amplitude and the phase lag, of the induced dipole moment relative to $\bm{B}_{prim}$. The value of A is between 0.7 and 1 and the value of $\phi$ is zero, according to \cite{Zimmer2000}. It will be assumed that A is 1 for the rest of this work. Inside the radius of the conductor ($r$ < $r_{Europa}$A$^{-1/3}$) the magnetic field is given by Equation \ref{eq_B_sec_in}.
\begin{equation}
\bm{B}_{sec} = [-B_{prim_x}, -B_{prim_y}, 0]
\label{eq_B_sec_in}
\end{equation}

\subsection*{Flow deflection}
\label{s_model_flow}
\begin{figure}[h]
  \centering
  \includegraphics[width=0.72\textwidth]{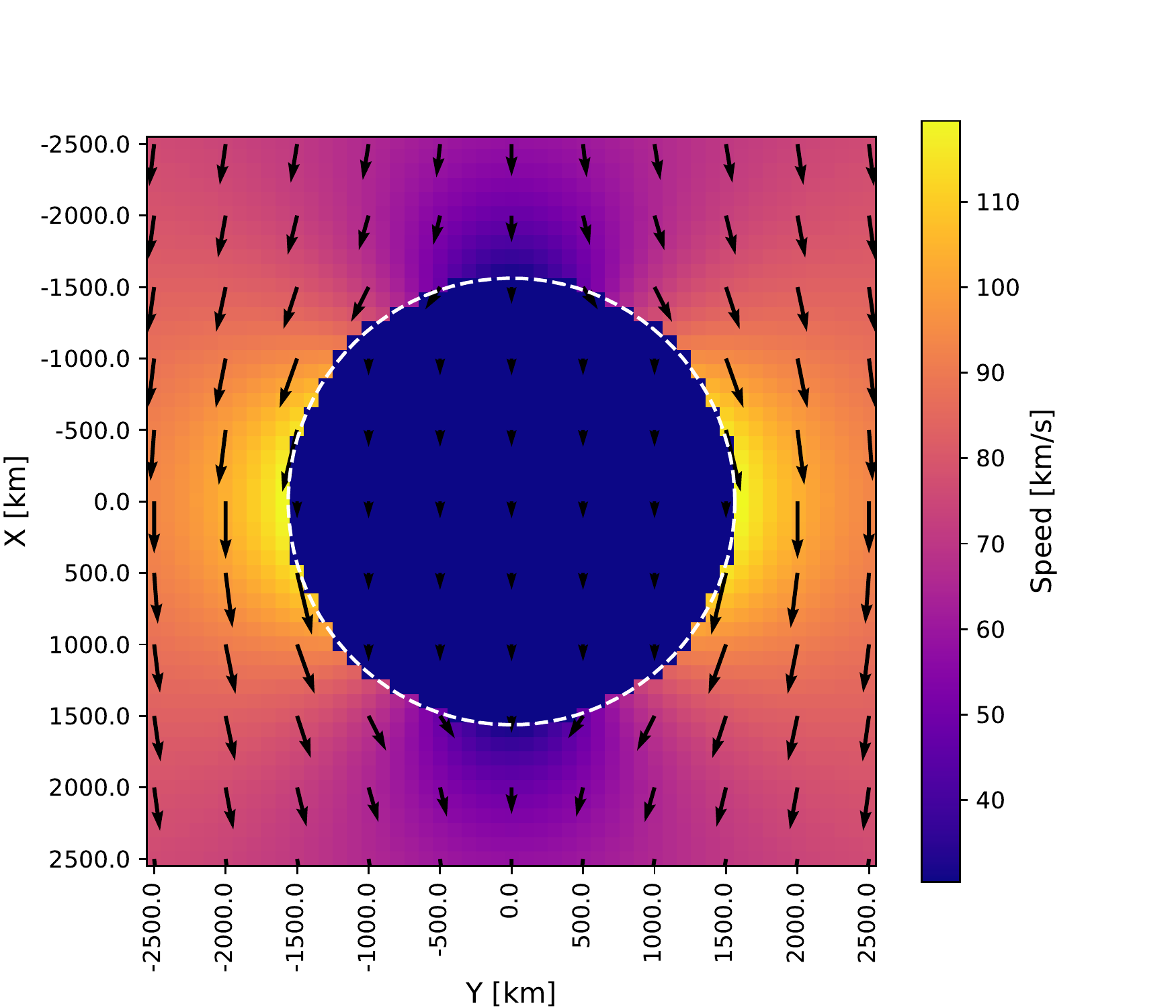}
  \caption{The analytical model of the deflection of the corotating plasma by Europa's ionosphere, from \cite{Ip1996}. The arrows represent the direction of the flow, where the arrow length scales with the speed. The colors represent the speed. The figure represents Europa's equatorial plane. The positive y-axis is towards Jupiter and the positive x-direction is in the direction of the corotational plasma. For this example it is assumed that the upstream flow velocity is 76 km/s and that it slows down to 40\% of the upstream velocity.}
  \label{img_ip_flow}
\end{figure}

To express the deflection of the corotational plasma around Europa the analytical model from \cite{Ip1996} is used, which was mentioned previously in Section \ref{s_mag_interaction}. The resulting flow field of the model is shown in Figure \ref{img_ip_flow}.

The model by \cite{Ip1996} is expressed by Equations \ref{eq_ip_vx} to \ref{eq_ip_vy_iono}. The model expresses the velocity vectors, $V_x$ and $V_y$, with Equations \ref{eq_ip_vx} and \ref{eq_ip_vy}, outside of the Alfv\'{e}n wing disturbance ($r$>$R_c$). Inside of the disturbance ($r$<$R_c$) $V_x$ and $V_y$ are expressed by Equations \ref{eq_ip_vx_iono} and \ref{eq_ip_vy_iono}.
\begin{equation}
V_x = -2(1-\alpha)V_0(R_c/r)^2xy/r^2
\label{eq_ip_vx}
\end{equation}
\begin{equation}
V_y = V_0 + (1-\alpha)V_0(R_c/r)^2(1-2y^2/r^2)
\label{eq_ip_vy}
\end{equation}
\begin{equation}
V_x = 0 
\label{eq_ip_vx_iono}
\end{equation}
\begin{equation}
V_y = \alpha V_0
\label{eq_ip_vy_iono}
\end{equation}

The diversion of the flow (Figure \ref{img_ip_flow}) causes the velocity magnitude of the corotational plasma to vary. Just in front of and behind Europa (from the point of view of the plasma flow) this results in a slow down of the corotating plasma. 
On the flanks, with respect to the plasma flow direction, the flow is accelerated. Upstream the flow direction is diverted away from Europa, while downstream the flow returns to its original direction.

The model treats the disturbance of the flow by Europa as a cylinder, not as a sphere, thereby taking into account the Alfv\'{e}n wings. The model assumes that the corotational flow upstream of Europa, $V_0$, and the disturbed components $V_x$ and $V_y$ are in the same plane. The deflection of the flow is thus two dimensional. The equations in this model do not vary along the height of the cylindrical disturbance.
Apart from  $V_0$ the model is determined by the parameters $R_c$ and $\alpha$. $R_c$ is the radius of the disturbance, which is the radius of Europa plus the height of the ionosphere. The other important parameter is $\alpha$, the interaction parameter, which has a value between 0 and 1. It expresses the percentage the upstream velocity $V_0$ slows down within the disturbance cylinder. 

\begin{figure}[h]
  \centering
  \includegraphics[width=0.72\textwidth]{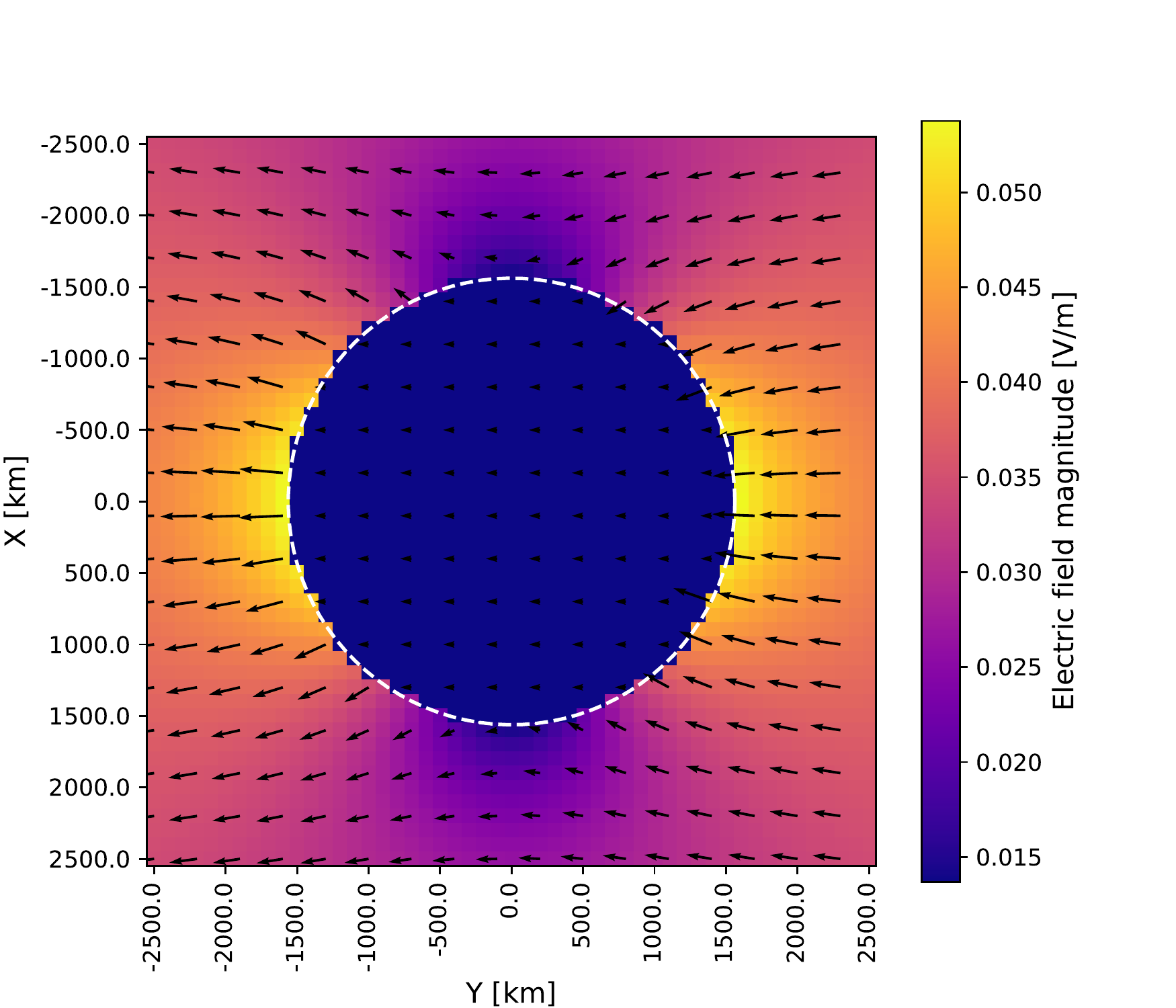}
  \caption{Electric field during the E12 flyby from Equation \ref{eq_E} using the velocity field from Figure \ref{img_ip_flow}. The arrows indicate the directions of the field and the colour the magnitude. The magnetic field components are: 50 nT in direction of Jupiter, 100 nT in the direction of the corotational plasma and 450 nT in the direction in the plane.} 
  \label{img_e_flow}
\end{figure}

The electric field resulting from the flow velocity and the magnetic field in the frame of Europa is expressed by Equation \ref{eq_E}, the result is shown in Figure \ref{img_e_flow}.
\begin{equation}
\bm{E} = -\bm{V} \times \bm{B}
\label{eq_E}
\end{equation}

\clearpage


\section{Selected instrumentation and simulation method}
\label{s_method_concl}

In this chapter the methodology that I will use to address the main research question was discussed. Data analysis and simulations will be combined to find signatures of Europa's atmosphere and plumes in the Galileo charged particle data.

In Chapter \ref{ch_comparison} the first goal will be addressed: characterize and compare the charged particle data collected during the Europa flybys to select the best time period(s) for the search for signatures of the atmosphere or any plumes. Galileo charged particle data is available for two instruments: the plasma particle detector PLS and the energetic charged particle detector EPD.

The second goal, the simulations of the data, will be discussed in Chapter \ref{ch_atmosphere}.
The depletions of energetic ions will be simulated and compared to the data to determine if any atmospheric contribution to the measured losses occurred. I select a test particle tracing method in which the trajectories of individual particles are modelled to determine where they are lost. Even though this method doesn't determine the effect of the charged particles on the fields self-consistently, it has a key advantage over self-consistent methods, it allows to study the individual effect of different physical effects. In this thesis simulations with increasing physical complexity will be discussed. This way the contribution of each physical effect on the losses of energetic charged particles can be characterized. In a self-consistent simulation it would be more difficult to disentangle the contribution of the different effects to the losses.  Because the gyroradius of the energetic particles is comparable to or larger than Europa it has to be taken into account when studying particle losses. Therefore, a full solution of the gyromotion is opted for, rather than a guiding center approximation. 
The charged particle tracing code will also be used in Chapter \ref{ch_prospects} to make predictions for future measurements.

\chapter{Comparison of the Galileo Europa flybys using charged particle data}\label{ch_comparison}
Previous research shows that the charged particle data obtained by the Galileo plasma detector (PLS) and energetic particle detector (EPD) contain signatures of the interaction between Europa and the charged particle environment. However, no comprehensive overview of the data form a limited number of flybys is available. Therefore, in this chapter the charged particle data collected during the Europa flybys will be characterized and compared to identify candidate signatures of the atmosphere or any plumes. These candidate signatures will be simulated with the particle tracing code in the next chapter.
First, in Section \ref{s_flybys_overview} a comparison will be made of the geometries of the Europa flybys. Then, in Section \ref{s_data_analysis}, the data analysis is discussed. Next an overview is provided of the flybys and a comparison is made, in Section \ref{s_comparison}. Subsequently, in Section \ref{s_candidate}, several criteria to select the most interesting flyby(s) are formulated. Finally, the selected candidate(s) will be discussed in more detail in Section \ref{s_candidate_e12}.

\section{Geometric overview of the Galileo flybys of Europa}
\label{s_flybys_overview}

\begin{figure}[h]
  \centering
  \includegraphics[width=1.0\textwidth]{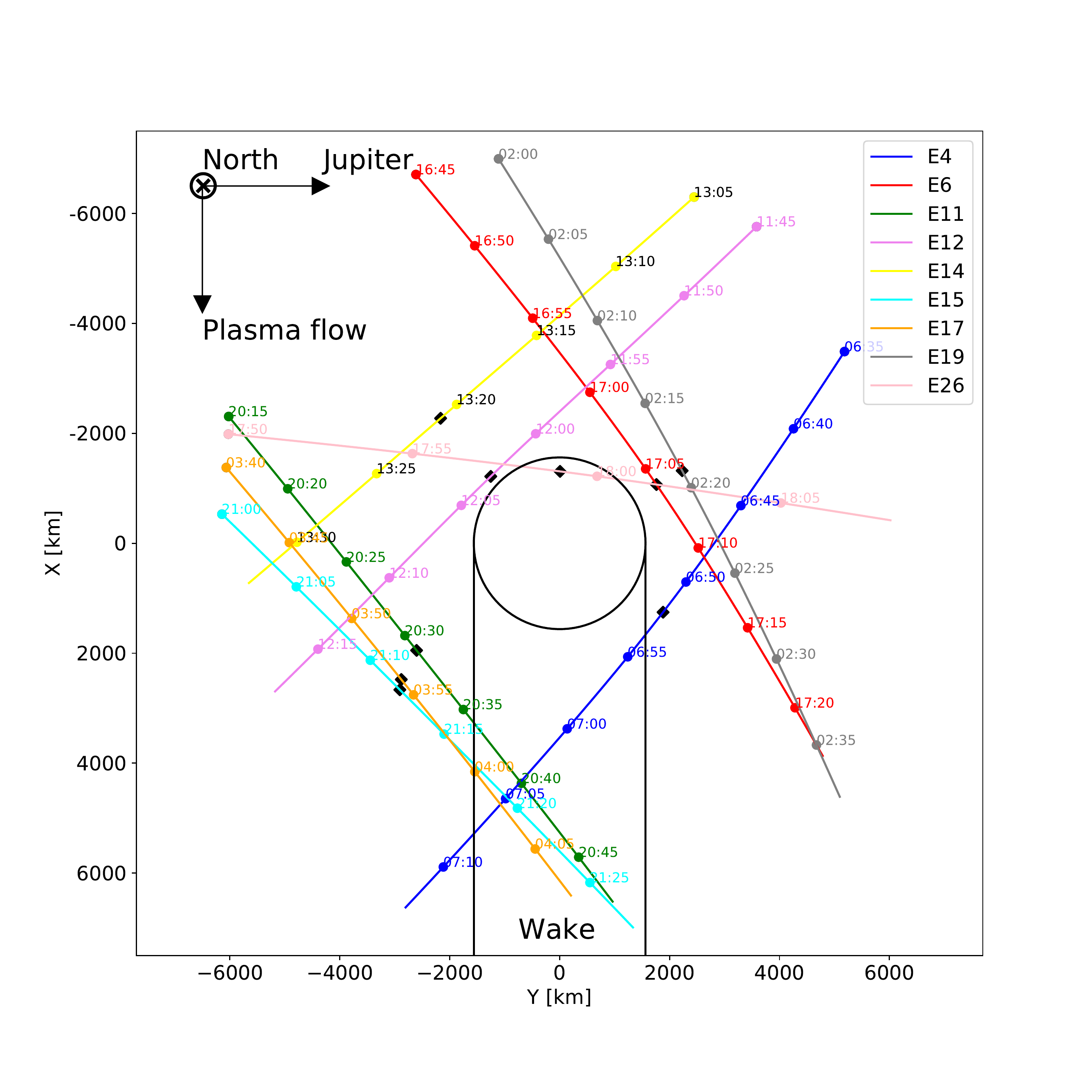}
  \caption{Europa flybys of Galileo. Positive y-axis is towards Jupiter, positive x-axis is in the direction of the corotational flow. The z-axis (out of the image) is along Europa's rotation axis. Black diamonds indicate the closest approach. Two black lines indicate Europa's geometric wake. Each flyby is represented by the positions during the flyby that were within 5 $R_E$.}
  \label{img_orbit_xy}
\end{figure}

\begin{figure}[h]
\centering
\includegraphics[width=0.70\textwidth]{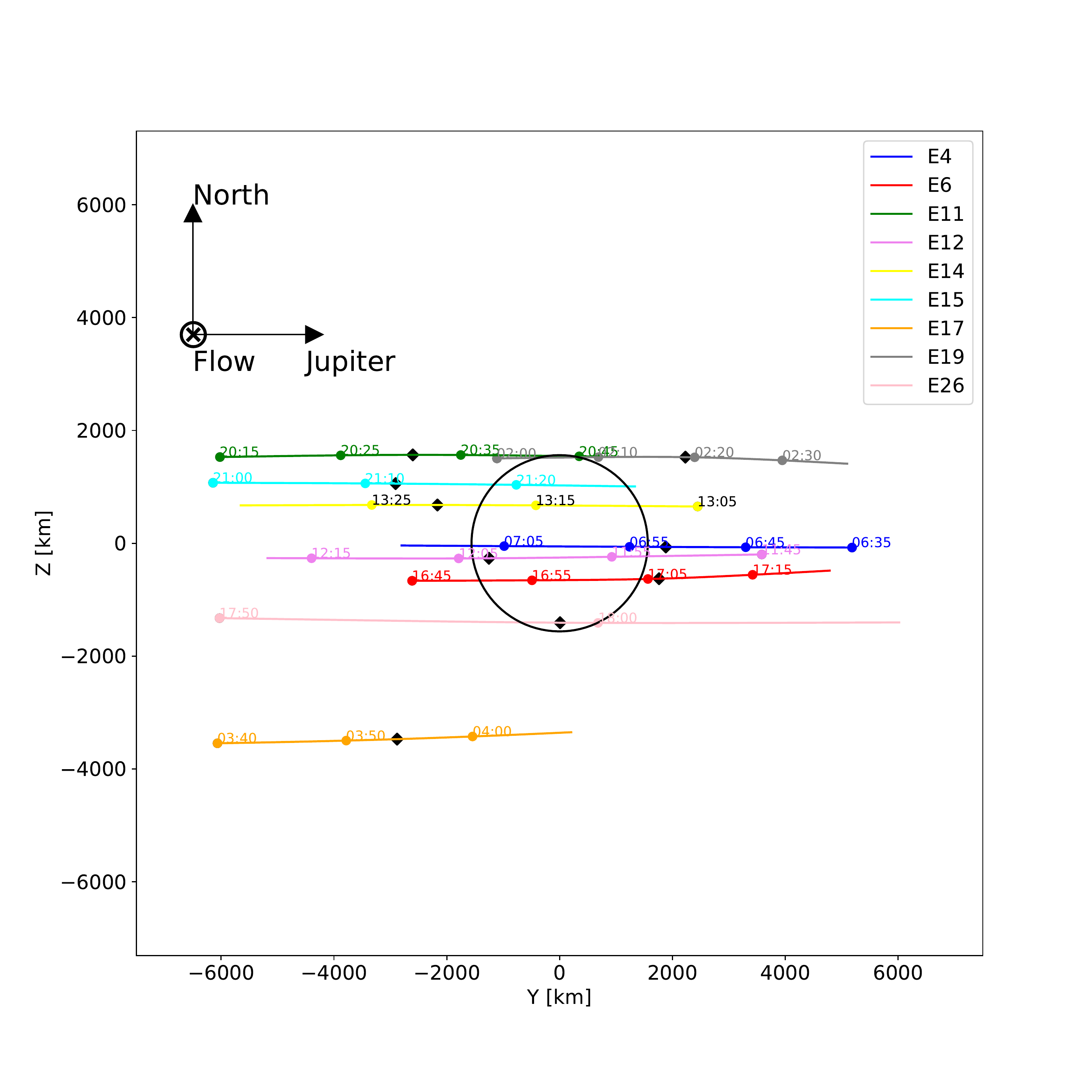}
\caption{Galileo flybys of Europa, XZ cut of Figure \ref{img_orbit_xy}}
\label{img_orbit_xz}%
\end{figure}

\begin{figure}[h]
\centering
\includegraphics[width=0.70\textwidth]{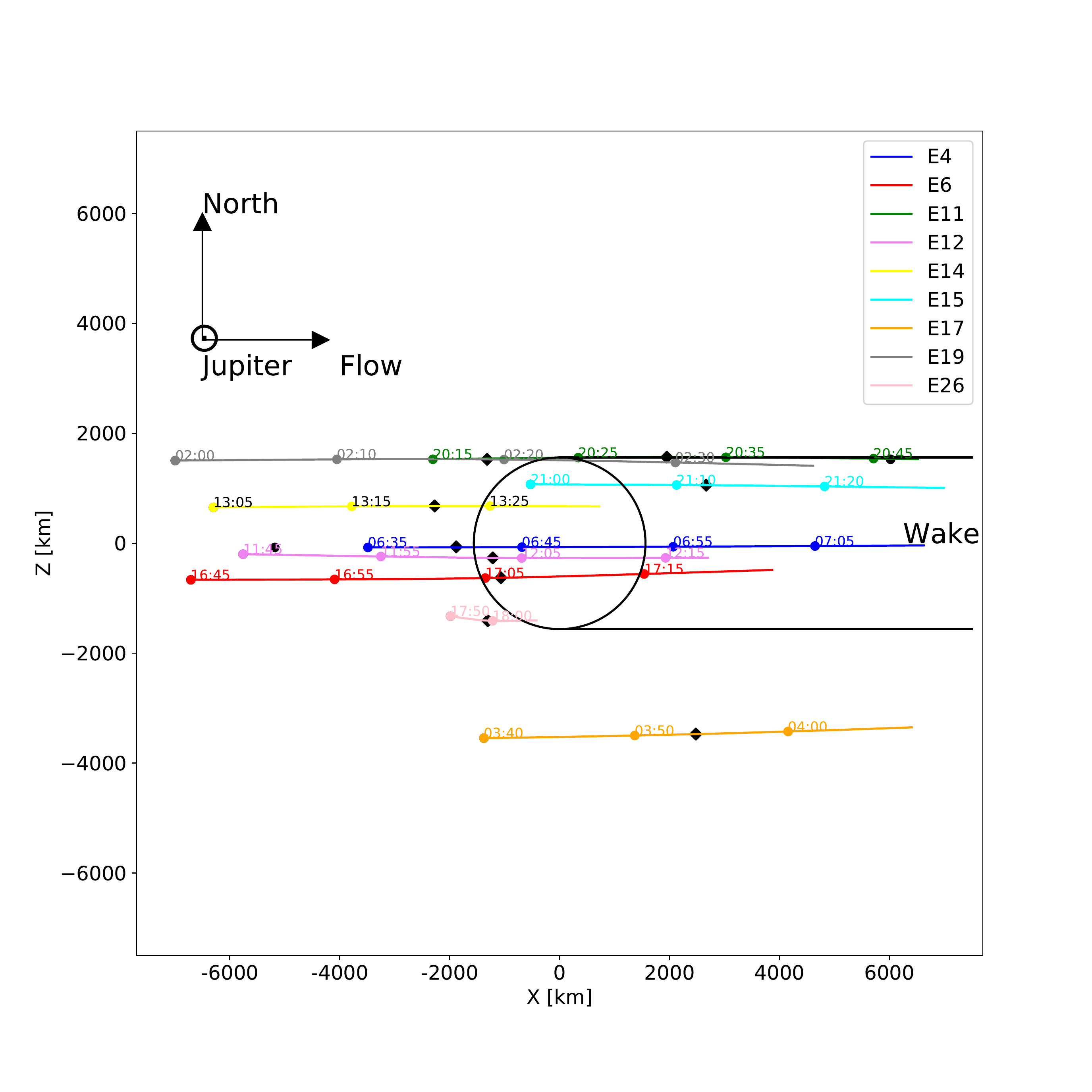}
\caption{Galileo flybys of Europa, YZ cut of Figure \ref{img_orbit_xy}.}
\label{img_orbit_yz}
\end{figure}

For the interpretation of the data it is relevant to know properties of the flybys such as their geometry with respect to Europa and its wake or the position of the flyby in Jupiter's magnetosphere. 
The Galileo flybys that passed by Europa are called E4, E6, E11, E12, E13, E14, E15, E16, E17, E18, E19 and E26. In the names of the Galileo orbits the letter refers to the object that Galileo passed by closest and the number refers to index in the overall number of Galileo orbits (for example G2 means that the second Galileo orbit passed by Ganymede). In the rest of this section E13, E16 and E18 will be excluded since no EPD or PLS data is available for these flybys. Figures \ref{img_orbit_xy} to \ref{img_orbit_yz} provide an overview of the flybys with respect to Europa. The y-axis points towards Jupiter, the positive x-axis points in the direction of the corotational plasma flow and the z-axis is along Europa's rotation axis. 

\begin{figure}[h]
  \centering
  \includegraphics[width=1.0\textwidth]{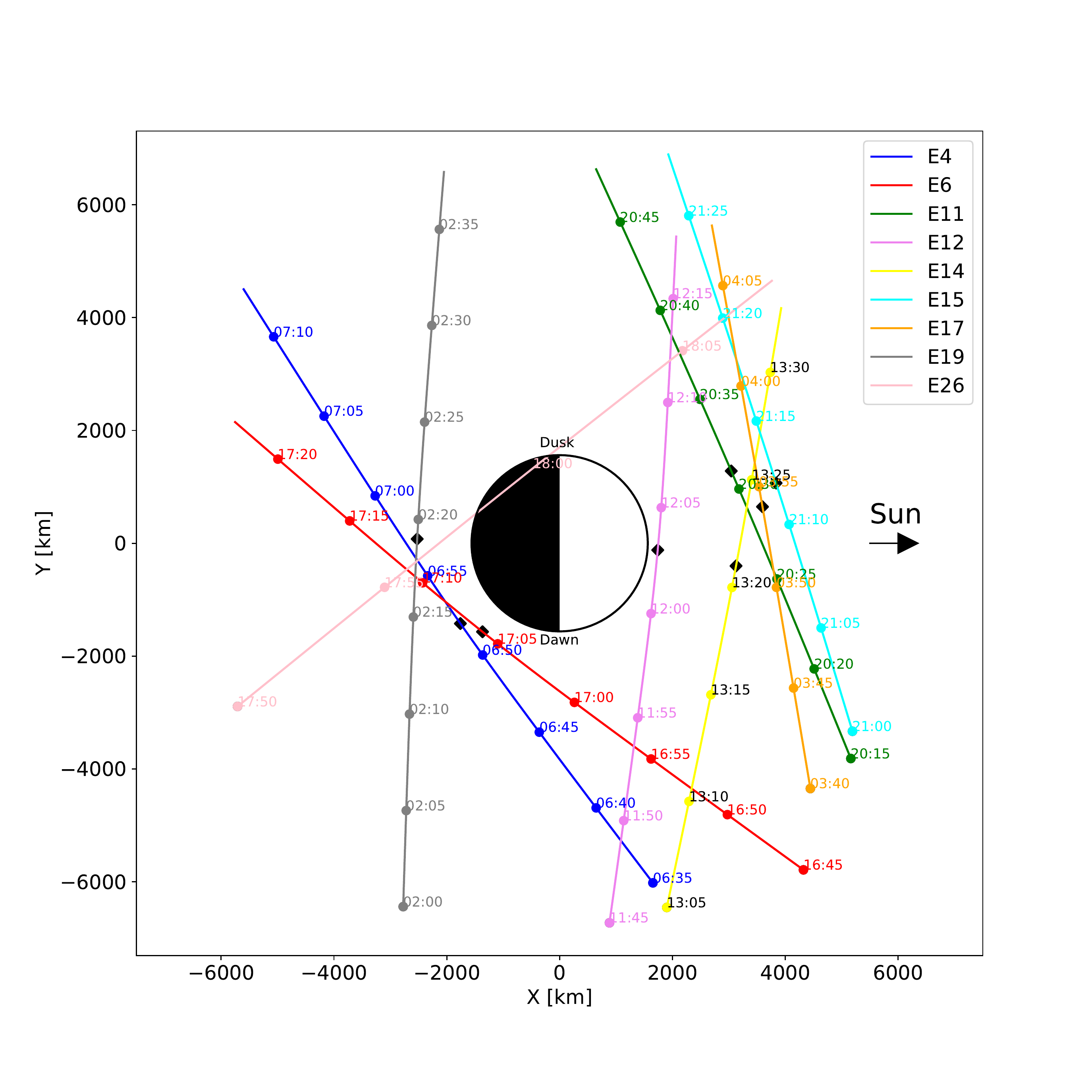}
  \caption{Europa flybys of Galileo. Positive x-axis points towards the sun, y is along Europa's day-night terminator and the z-axis is along Europa's rotation axis. Black diamonds indicate the closest approach to Europa. Each flyby is represented by the positions during the flyby that were within 5 $R_E$.}
  \label{img_orbit_europa_lt}
\end{figure}

\begin{figure}[h]
  \centering
  \includegraphics[width=0.75\textwidth]{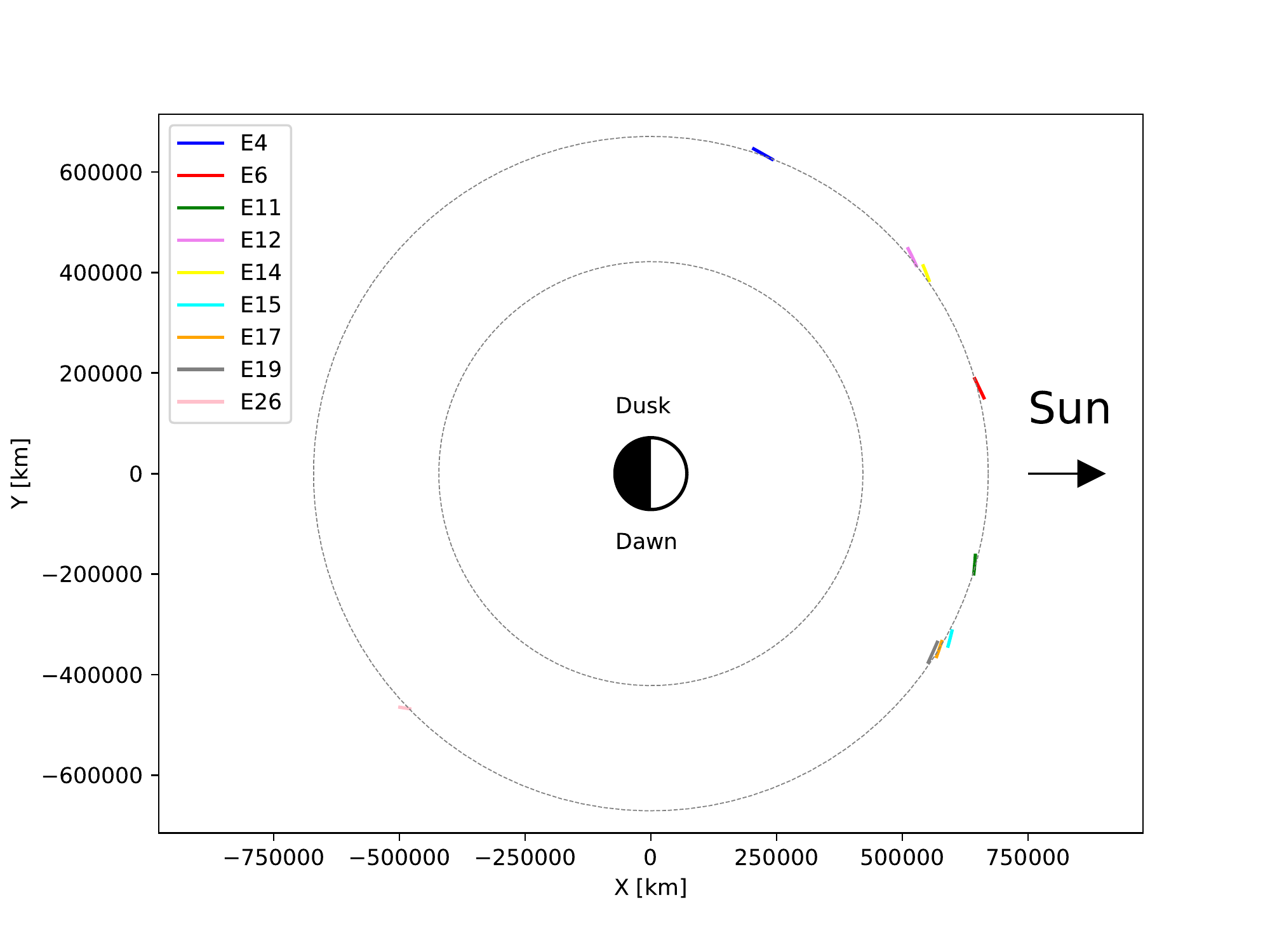}
  \caption{Location of the Europa flybys in a reference frame with respect to Jupiter. The black arrow indicates the direction of the sun. The y-axis is along Jupiter's day-night terminator. The z-axis (out of the image) is along Jupiter's rotation axis. The inner grey circle and the outer grey circle indicate, respectively, the orbits of Io and Europa. Each flyby is represented by the positions during the flyby that were within 5 $R_E$.}
  \label{img_orbit_lt}
\end{figure}

\begin{figure}[h]
  \centering
  \includegraphics[width=0.60\textwidth]{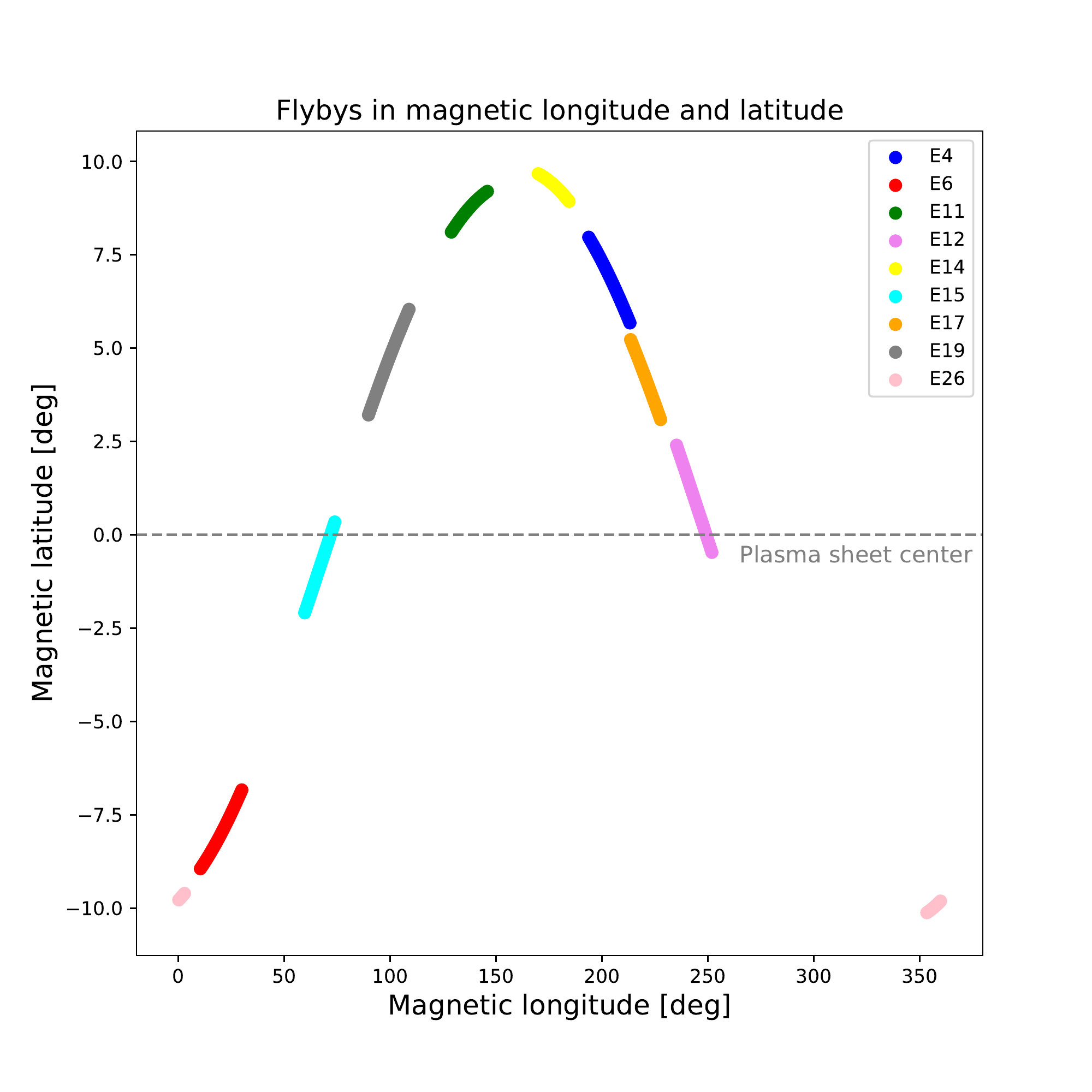}
  \caption{Flyby coordinates in a magnetic frame. The x-axis represents magnetic longitude and the y-axis magnetic latitude. The dashed grey line indicates the magnetic equator, the location where Galileo is in the center of the plasma sheet. Each flyby is represented by the positions during the flyby that were within 5 $R_E$.}
  \label{img_orbit_magnetic}
\end{figure}

\begin{figure}[h]
  \centering
  \includegraphics[width=1.0\textwidth]{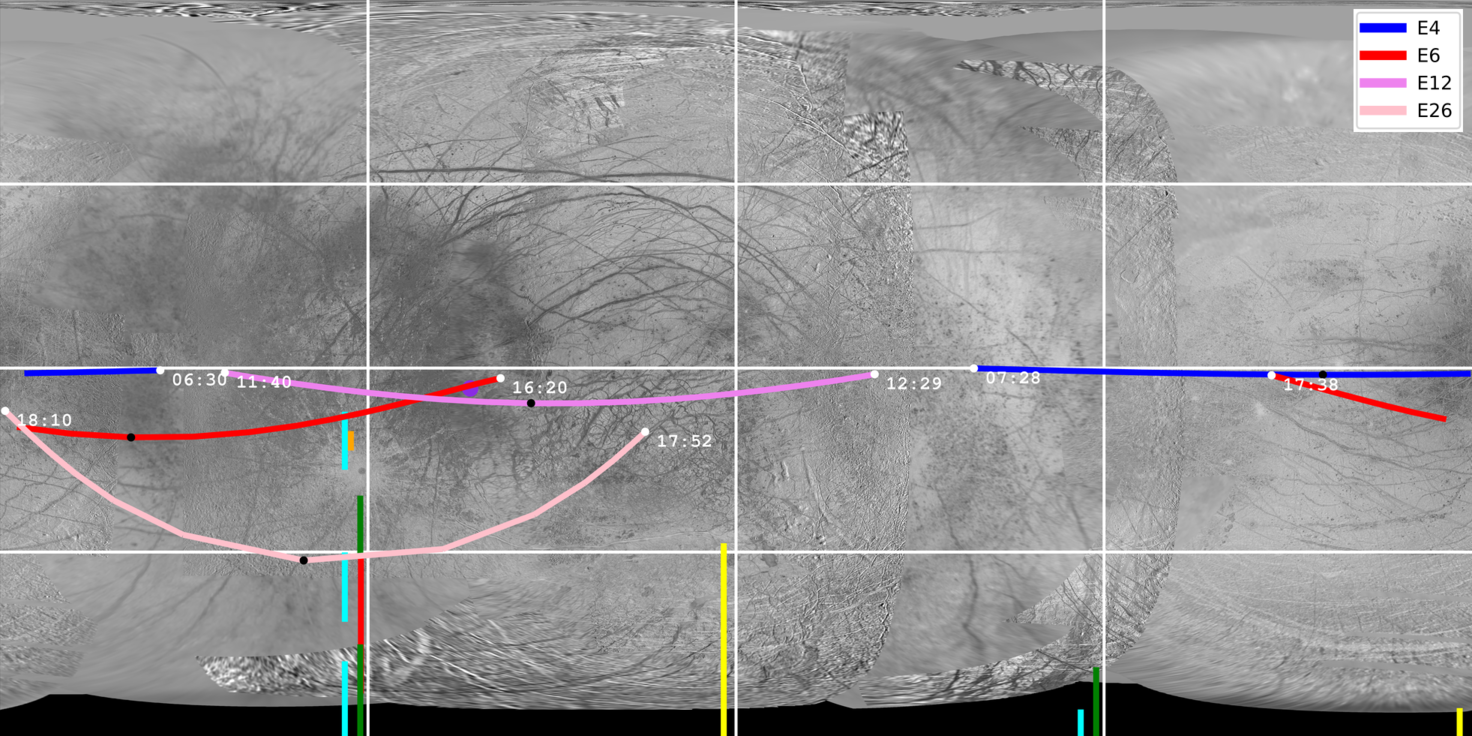}
  \caption{Map of Europa with projection of the four closest Galileo flybys. The beginning and end time of the plotted trajectories are marked with white dots. Black dots mark the closest approach. The locations of the observed plumes from Figure \ref{img_europa_map_plumes} are also shown. Map of Europa obtained from: USGS.}
  \label{img_map_flybys}
\end{figure}

Figure \ref{img_orbit_europa_lt} shows the flybys with respect to Europa in a sun fixed frame. The difference to Figure \ref{img_orbit_xy} is that the x-axis is pointing towards the sun. These figures allows to determine which flybys happened on Europa's dayside or nightside.
Figure \ref{img_orbit_lt} shows the location of the flybys in a reference frame with respect to Jupiter. In this frame the x-axis points towards the sun. The z-axis (out of the image) is along Jupiter's rotation axis. The y-axis is along Jupiter's day-night terminator. 
Figure \ref{img_orbit_magnetic} shows the coordinates of the flybys in a magnetic frame, allowing to separate the flybys in magnetic latitude and longitude.
Figure \ref{img_map_flybys} shows a map of Europa with the projection of the four closest flybys on the surface.
\clearpage
Table \ref{tab_properties_orbits} gives an overview of all the flybys and summarizes their main properties.
\begin{table}[h]
\centering
\begin{tabular}{|r|l|l|l|l|l|l|l|}
  \hline
  & \textbf{CA} & \textbf{CA} & \textbf{CA} & \textbf{Loc.} & \textbf{LT (J)} & \textbf{LT (E)} & \textbf{Mag. lat.} \\
 & \textbf{[date]} & \textbf{[time]} & \textbf{[km]} & & \textbf{[hr]} & \textbf{[hr]} & \textbf{[deg]} \\
  \hline
  \hline
  E4 & 19 Dec. 1996 & 06:52:58 & 688.1 & wake & 16.6-16.8 & 7.0-21.4 &  5.7 to 8.0  \\
  \hline 
  E6 & 20 Feb. 1997  & 17:06:10 & 582.3 &  & 12.8-13.1 & 8.4-22.6  & -8.9 to -6.8  \\
  \hline
  E11 & 6 Nov. 1997 & 20:31:44 & 2039.3 & wake & 10.8-11.1 & 9.6-17.6 & 8.1 to 9.2  \\
  \hline
  E12 & 16 Dec. 1997 & 12:03:20 & 196.0 &  & 14.5-14.7 & 6.5-16.6 & -0.5 to 2.4  \\ 
  \hline
  E14 & 29 Mar. 1998 & 13:21:05 & 1649.1 &  & 14.3-14.5 & 7.1-15.1 & 8.9 to 9.7  \\
  \hline
  E15 & 31 May 1998 & 21:12:56 & 2519.5 & wake & 10.0-10.1 & 9.8-17.0 & -2.1 to 0.3  \\
  \hline
  E17 & 26 Sep. 1998 & 03:54:20 & 3587.4 & wake & 9.8-10.0 & 9.0-16.3 & 3.1 to 5.2  \\
  \hline
  E19 & 1 Feb. 1999 & 02:19:50 & 1444.4 &  & 9.7-10.0 & 4.4-19.2 & 3.2 to 6.0  \\
  \hline
  E26 & 3 Jan. 2000 & 17:59:43 & 348.4 & upstr. & 2.8-3.0 & 1.8-15.4 & -10.1 to  \\
   &  &  &  & polar & & &  -9.6  \\
  \hline
\end{tabular}
\caption{Overview of Galileo flybys of Europa. Loc. refers to the location of the orbit with respect to Europa. LT refers to local time, with J referring to Jupiter and E referring to Europa. The properties in the last three columns correspond to when the spacecraft is within 5 $R_E$ from Europa. Column 2, 3 and 4 are obtained from \cite{Pappalardo2009_Kivelson}.}
\label{tab_properties_orbits}
\end{table}

\clearpage
\section{Data analysis}
\label{s_data_analysis}
Technical details of the data processing of EPD and PLS data are documented in Appendix \ref{a_pls_processing}. Therein it is described how the major challenges related to the analysis of PLS data have been dealt with, these are: the poor spatial and time coverage resulting from the malfunction of the main antenna and the noise introduced in the data by penetrating radiation (see Section \ref{ss_galileo}). 

\begin{figure}[p]
  \centering
  \includegraphics[width=1.0\textwidth]{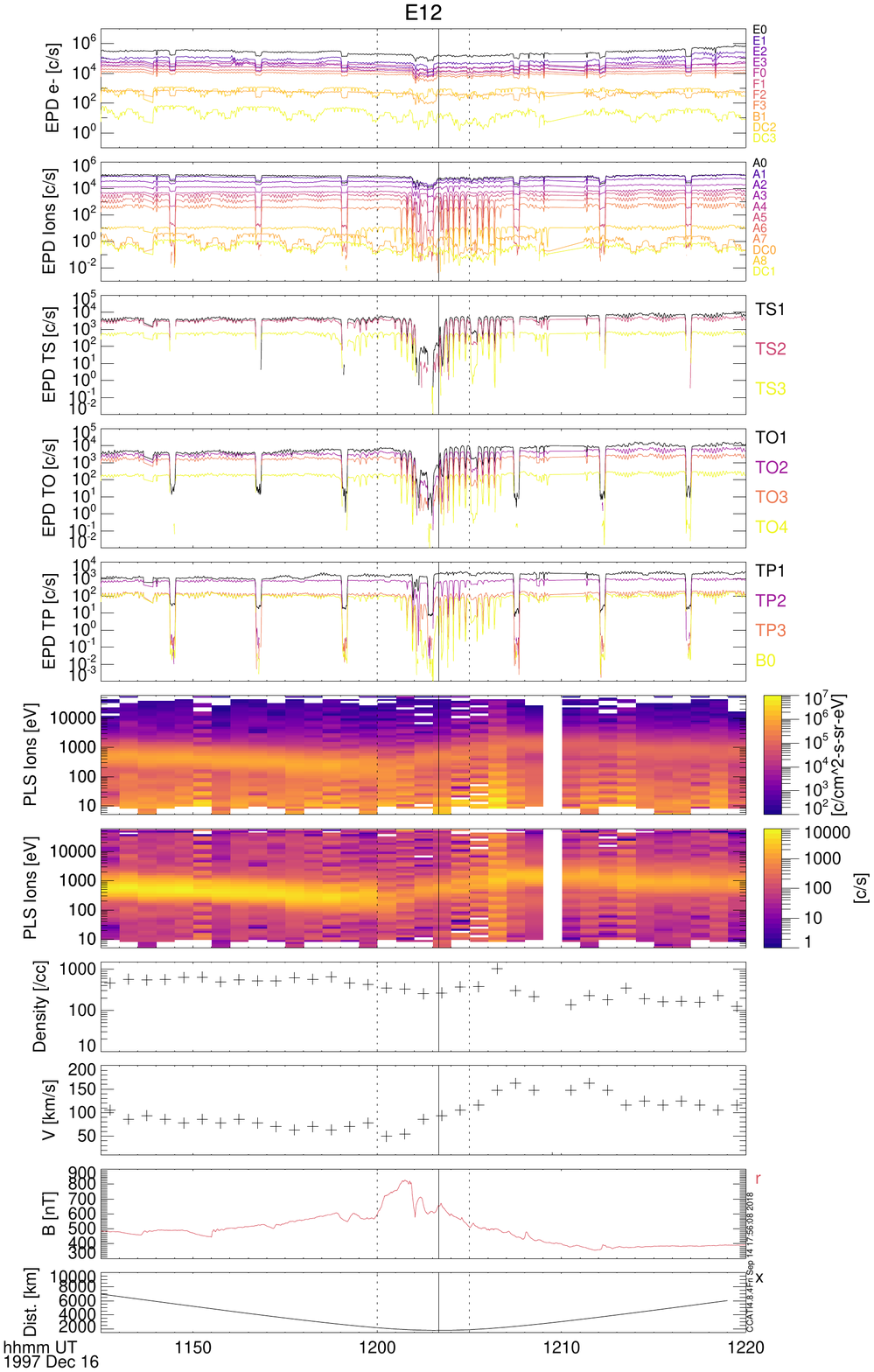}
  \caption{E12 flyby data. From the top: EPD data (panel 1-5), PLS data (panel 6-9), MAG data (panel 10) and altitude (panel 11). The dotted lines indicate the region during which the direction of the corotational plasma changes according to the analysis of the PLS data (this chapter). \cite{Jia2018} reports a change in magnetic field direction between the first dotted line and the closest approach (solid line), possibly related to a plume.}
  \label{img_E12_all_comp}
\end{figure}

Overview plots of the EPD, PLS, and MAG data, collected during the Europa flybys of Galileo, are provided in the figures in Appendix \ref{app_overview}. For convenience of the reader the overview plot for the E12 flyby is also shown in this Chapter in Figure \ref{img_E12_all_comp}. 
Data products that are presented are the count rates in the EPD E, A, TS, TO and TP channels (panel 1-5 from the top). For PLS the omnidirectional energy spectrum in units of counts/s and differential flux, the plasma density and the speed of the corotational plasma are presented (panel 6-9 from the top). For MAG the magnitude of the magnetic field is shown (panel 10). Finally also the distance to Europa is shown (panel 11). 
These plots are provided for all flybys that are listed in Table \ref{tab_properties_orbits}, except the E17 flyby. For this flyby only sporadic PLS measurements are available and no EPD data, therefore it is not considered further.

\section{General comparison of the flyby data}
\label{s_comparison}
In this section a general comparison of the data collected during the flybys is made, the comparison is divided in three parts: magnetic field data, plasma particle data and energetic charged particle data.

\subsection{Magnetic field measurements}
\begin{figure}[p]
  \centering
  \includegraphics[width=0.6\textwidth]{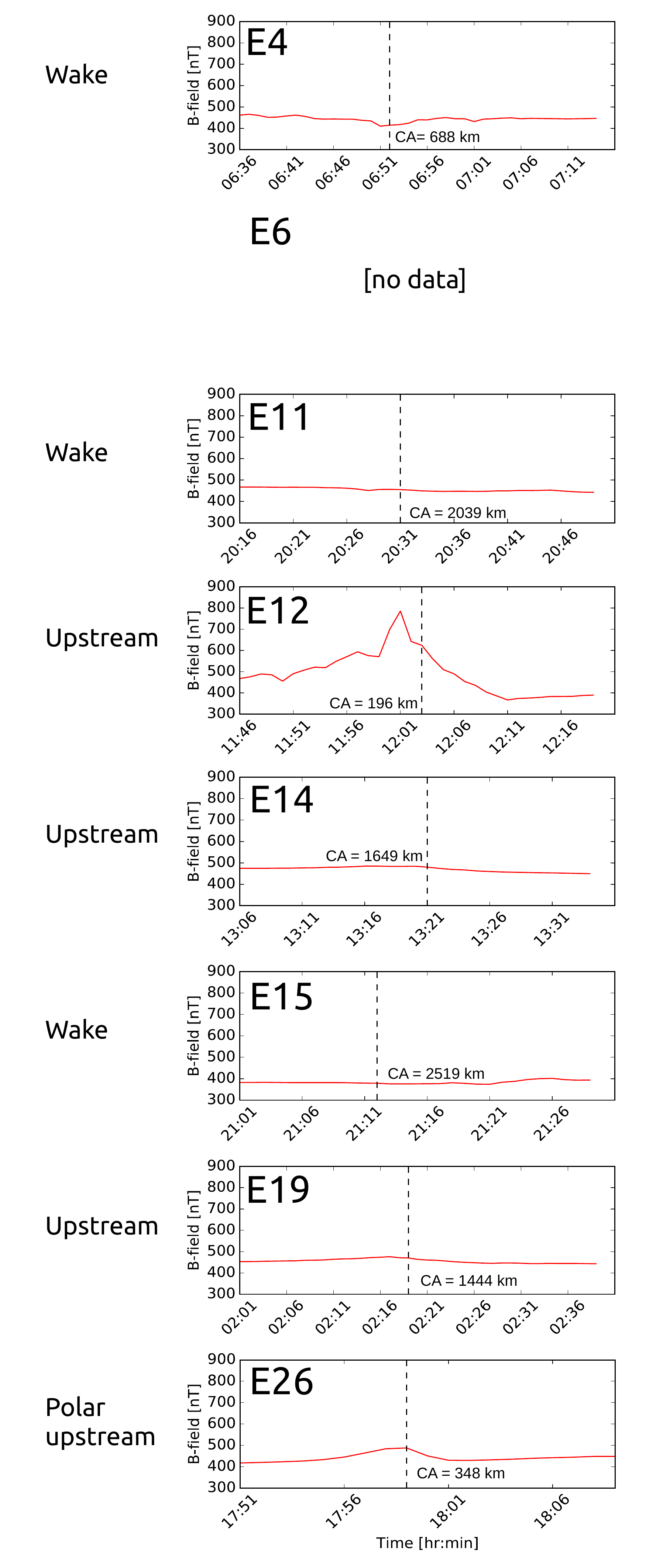}
  \caption{Magnetic field magnitude for different Galileo flybys. CA refers to the closest approach, of which the corresponding time is indicated by the dashed vertical line. Data is obtained from https://pds.nasa.gov/. One minute time averaging is applied. The corresponding geometry of the flybys can be seen in the Figures of Section \ref{s_flybys_overview}.}
  \label{img_mag_overview}
\end{figure}

\begin{figure}[h]
  \centering
  \includegraphics[width=1.0\textwidth]{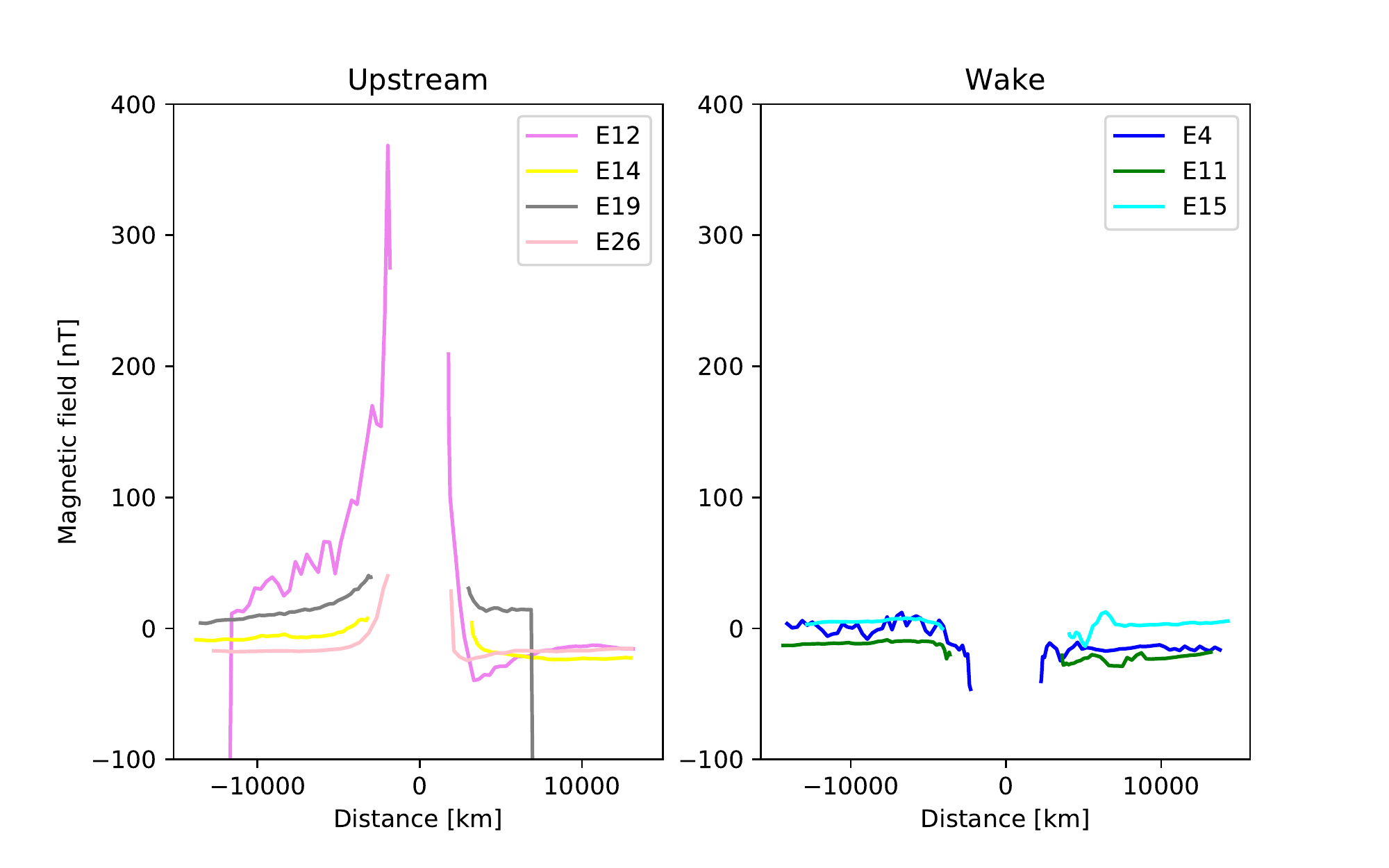}
  \caption{Magnetic field during the Europa flybys, the field as predicted by the VIP4 model of Jupiter's magnetic field has been subtracted, to make the local contribution by Europa visible. The field is represented as a function of distance with respect to Europa, negative distance representing measurements taken prior to the closest approach and positive distance measurements taken after closest approach. The gaps represent radial distances over which no measurements were taken because Galileo did not approach Europa close enough. The flybys on the left passed upstream of Europa, the flybys on the right passed through Europa's wake.}
  \label{img_mag_vip4}
\end{figure}

A first comparison of the flybys can be made based on the magnetic field magnitude, this is shown in Figure \ref{img_mag_overview}. From this Figure it is clear that the E12 flyby stands out.
Figure \ref{img_mag_vip4} shows the difference between the measured field and the VIP4 model of Jupiter's magnetic field. VIP4 is a commonly used empirical model of Jupiter's magnetic field that has been fitted to Voyager and Pioneer magnetic field measurements and observations of Jupiter's aurora \citep{Connerney1998}. From Figure \ref{img_mag_vip4} the local contribution due to Europa can be determined, it is clear that during all the flybys there is some magnetospheric interaction, as the field is not unperturbed. Figure \ref{img_mag_vip4} also shows that the magnetic field is piling up during flybys E12, E14, E19 and E26, providing an indirect indication of the existence of Europa's atmosphere during these flybys. These four flybys pass upstream of Europa, where the corotational flow slows down. Note that the E6 flyby is not included, as no magnetic field data is available for this flyby.

\subsection{Plasma observations}
\begin{figure}[h]
  \centering
  \includegraphics[width=0.75\textwidth]{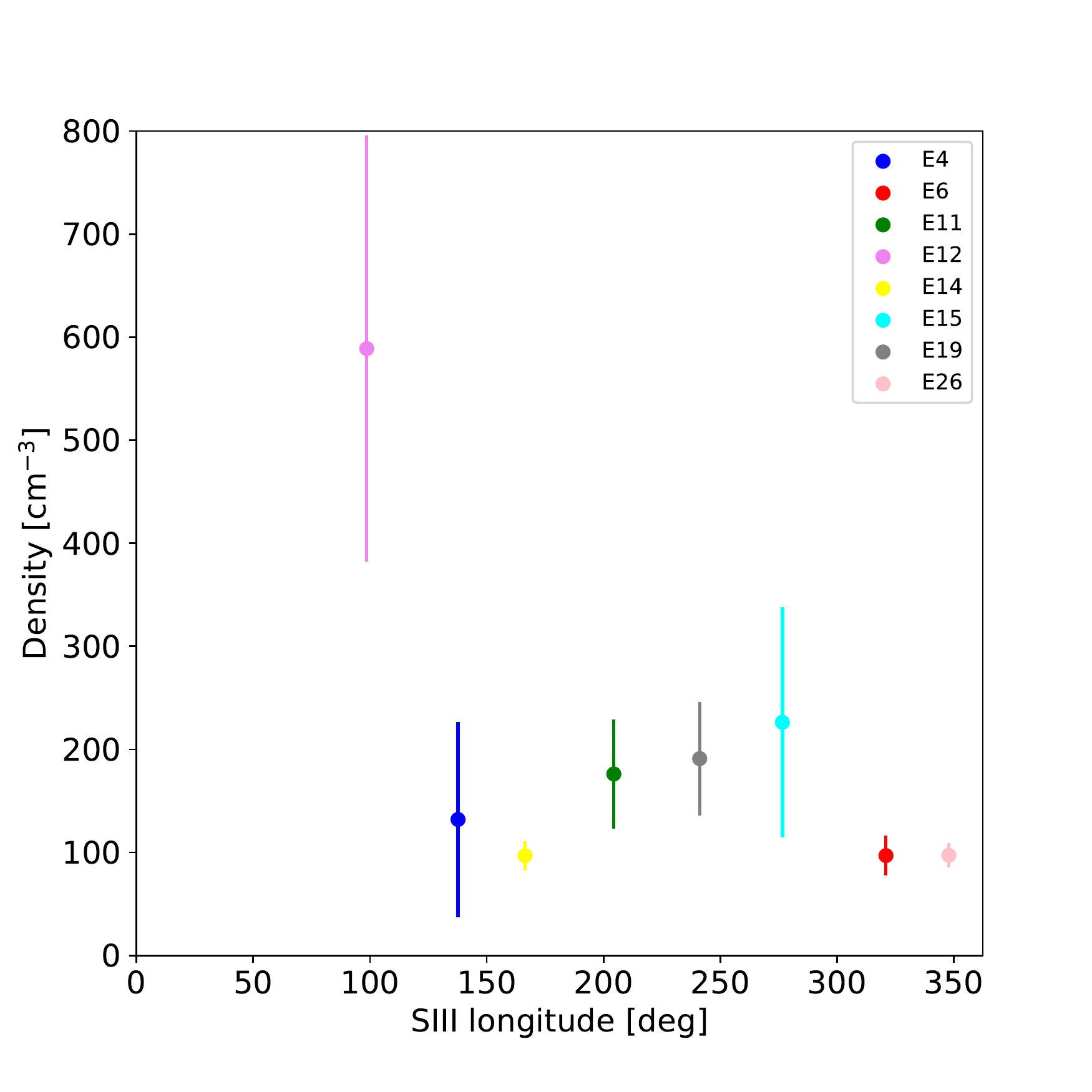}
  \caption{Plasma number density from PLS data in the vicinity of Europa as a function of longitude in the System III reference frame. Only values for positions between 2.5 and 5 $R_E$ are considerd, to enable comparison with Figure \ref{img_bagenal2015}. The error bar represents the standard deviation of density as measured in this interval.}
  \label{img_pls_dens_long}
\end{figure}

A comparison can also be made of the flybys from the point of view of the plasma data. Previously, \cite{Bagenal2015}, has provided an overview of the electron density near Europa's orbit (shown in Figure \ref{img_bagenal2015}). In Figure \ref{img_pls_dens_long} the ion density derived from PLS data as part of this work is shown. The same trend can be seen as in Figure \ref{img_bagenal2015}: E12 is clearly standing out. Data presented in \cite{Kurth2001} from the close flybys has also indicated the electron density is highest during the E12 flyby, see Figure \ref{img_kurth2001}.

The interaction of Europa's ionosphere with the corotational plasma results in changes of the velocity of the corotational plasma flowing past Europa (this was described previously in Section \ref{s_mag_interaction}). In the overview plots in Appendix \ref{app_overview} the PLS energy spectrum is shown in panel 5 and 6 as seen from the top. The corotational plasma can be recognized as the band of increased counts (panel 5 from the top) or differential flux (panel 6 from the top) around 1000 eV, in most flybys. 
The shifts in energy of the corotational plasma correspond to changes in velocity. Changes in the corotational plasma velocity are most clear during the E12 flyby (Figure \ref{img_E12_all_comp}), possible weaker decelerations are visible in the E4 flyby as a deceleration near the closet approach (06:53, Figure \ref{img_E4_all}), in the E6 flyby as a acceleration near closest approach (17:06, Figure \ref{img_E6_all}) and in the E15 flyby as a deceleration at the start of the wake entry (21:17, Figure \ref{img_E15_all}). It is clear however, from visual inspection, that the spatial extent and change in energy during E12 is much larger than what is observed during other flybys. This feature is not simply an effect of the E12 flyby being the closest flyby. Deceleration is already occurring at 11:45, when Galileo is about 4 Europa radii away from Europa, other flybys have passed upstream of Europa at similar distances without observing any deceleration.

E4 and E6 have been identified as flybys during which the flow of the plasma is deflected by Europa according to \cite{Paterson1999}, as was discussed in Section \ref{s_mag_interaction_epd_pls}. From the data shown in Figures \ref{img_E4_all} and  \ref{img_E6_all} in Appendix \ref{app_overview} these changes in direction are not visible. Since the techniques used for \cite{Paterson1999} are not elaborated in that work, I developed a new method to try to identify changes in corotational velocity (the method described here is discussed in more detail in Appendix \ref{a_pls_processing}). I do this by separating the data either in 8 azimuthal bins or 7 polar bins, in a spacecraft reference frame that is not spinning, but constant with respect to the physical background. With this method I have not been able to reproduce the changes in direction reported for E4 and E6 by \cite{Paterson1999}. However, I was able to identify a change in direction of the plasma during the E12 flyby, from 12:00 to 12:05 (these results 
will be discussed in more detail in Section \ref{s_candidate}). My method did not show any deflections during other flybys. I consider that this is due to the strength of the corotational signal. During the E12 flyby the corotational plasma signature is very strong, while it is more difficult to distinguish during E4 or E6 for example. An inherent limitation of my method is that it stacks all azimuth bins together to represent the data in polar angles and vice versa to represent the data in azimuth angles, thereby rendering weaker features of deflection invisible. Furthermore noise was not removed in this analysis method, which could further complicate making weaker features visible.

\subsection{Energetic particle observations}
\begin{figure}[h]
  \centering
  \includegraphics[width=1.0\textwidth]{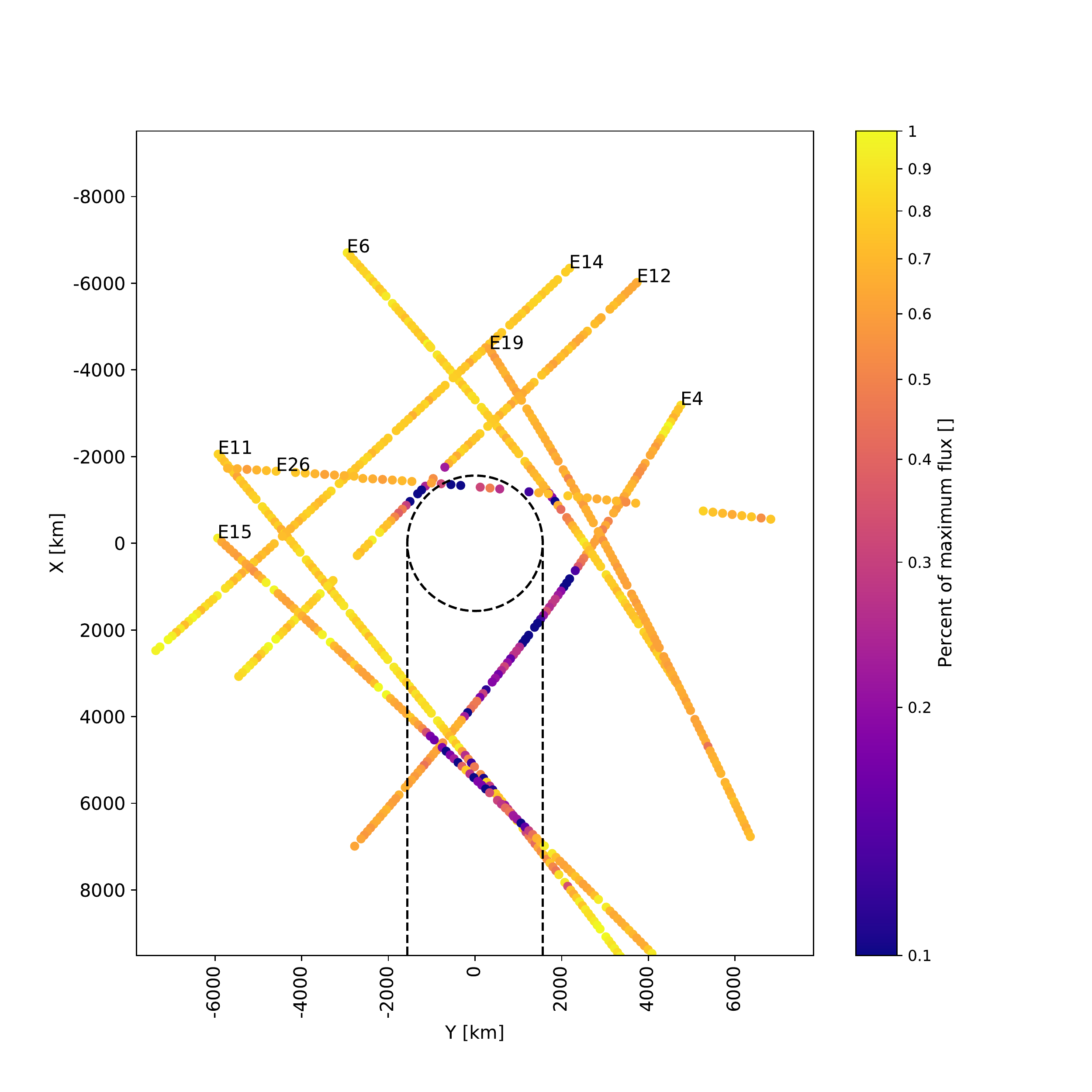}
  \caption{Depletions of energetic protons (220 to 550 keV, EPD TP2 channel). As every 16 data points have the same position tag, the minimum value of the 16 data points available per position tag is shown. The positive x-axis points in the direction of the corotational plasma, the y axis points towards Jupiter.}
  \label{img_depletions_xy}
\end{figure}

Here the flybys are compared from the point of view of the energetic particles. Previously, depletions in the number of energetic ions and electrons have been identified near Europa by \cite{Paranicas2000,Paranicas2007}, both upstream during E12 and E26, and in Europa's wake during E4, E11 and E15 (as was described in Section \ref{s_mag_interaction_epd_pls}). First I will separately discuss the depletions that occur in the wake and outside of it. Then, I will focus on the features that are relevant for the simulation effort in the next chapter. In Figure \ref{img_depletions_xy} an overview of the depletions of energetic protons (220 to 550 keV, TP2 channel) is shown. Depletions that correspond to the wake (E4, E11 and E15) and during the closest three upstream flybys (E6, E12 and E26) can be recognized.

First, I will focus on the depletions that occur in the wake. In the comparison plots provided in Appendix \ref{app_overview} the depletions reported in previous studies of energetic electrons (> 15 keV, E and F channels), energetic ions (>22 keV, A channels) and energetic protons (80 to 1040 keV, TP channels) in Europa's wake during E4 (Figure \ref{img_E4_all}), E11 (Figure \ref{img_E11_all}) and E15 (Figure \ref{img_E15_all}) can be recognized, confirming the findings of \cite{Paranicas2000}. The time when Galileo was in the geometric wake is indicated by dashed vertical lines in the respective figures. 
Depletions of energetic electrons, ions and protons in Europa's wake during E4 flyby have been identified previously by  \cite{Paranicas2000}. Furthermore \cite{Paterson1999} reported an increase in plasma density in the wake. From Figure \ref{img_E15_all} it is clear that there is also an increase in the plasma density during E15's wake (panel 8). However, no such increase in density is visible during the E11 flyby. As discussed in \cite{Paterson1999} the increase in ion density could be related to pick-up ions originating from Europa's atmosphere. As E11 and E15 crossed Europa's wake at similar radial distances, these results could point to a variability of Europa's atmospheric density. However, other parameters are varying too: the effect of electron impact ionization, which generates the pick up ions, is likely higher during E15 than E11 because of the higher electron density during E15 (see Figure \ref{img_bagenal2015}). E15 occurs when Europa is near the magnetic equator (see Figure \ref{img_orbit_magnetic}) and thus when Europa is in the center of the plasma sheet, where the plasma density is expected to be higher.
During the wake passage of the E11 and E15 flyby, depletions in electrons with energies from 15 keV and higher (E and F channels) can be observed. At the same time depletions of protons with energies of 80 to 1040 keV (TP channels) can be distinguished during E11 and E15, while depletion features of sulphur ions of 0.96 to 9.92 MeV (TS2-TS3 channels) and oxygen ions of 416 to 1792 keV (TO2-TO3 channels) are present during E11 only, but less pronounced. This is an effect of the gyroradius and bounce period of the particles. Electrons have small gyroradii compared to the size of Europa (<10 km for electrons in the E-F channels and a magnetic field strength of 400 nT). Furthermore their bounce period (<30 seconds for electrons in the E-F channels and a magnetic field strength of 400 nT) is much shorter than the ion bounce period (> 140 seconds for protons in the TP channels and a magnetic field strength of 400 nT), they bounce multiple times along the field line, before they convect over Europa's surface. Therefore many of them are depleted on Europa's surface before they can reach the wake, leading to the observed depletions of energetic electrons in the wake. Europa itself is not a source of energetic electrons. More massive particles, like sulphur and oxygen have larger gyroradii and can enter the wake, also because of their longer gyroperiod they do not necessarily interact with the surface. E4 differs from E11 and E15 in the sense that there are large bite-outs in the flux of energetic protons (80-1040 keV, TP channels), oxygen (416-1792 keV, TO2-TO3 channels) and sulphur (0.512-9.92 MeV, TS channels), before the closest approach (discussed later) and during the geometric wake passage.

Next, I will discuss depletions that do not occur in Europa's wake. Energetic ion depletions are visible near the closest approach to Europa, not only in energetic ions during E12 (11:58-12:10, Figure \ref{img_E12_all_comp}) and E26 (17:55-18:05, Figure \ref{img_E26_all}) as reported by previous studies, but also during E4 (06:49-06:53, Figure \ref{img_E4_all}), E6 (17:02-17:13, Figure \ref{img_E6_all}) and E19 (02:20, Figure \ref{img_E19_all}). The EPD data collected during the E14 flyby (Figure \ref{img_E14_all}) does not contain any clear signatures of interaction with Europa. Therefore this flyby will not be considered further in this section.
Depletions during E4, E6, E12 and E26 are observed near the closest approach for energetic ions (> 22 keV, A channels), energetic protons (80-1040 keV, TP1 to TP3), energetic oxygen (416 to 1792 keV, TO2-TO3 channels) and energetic sulphur (0.512-9.92 MeV, TS1 to TS3). During E19 only depletions in energetic sulphur ions (0.96-9.92 MeV, TS2-TS3) are observed. These depletions of energetic ions occurring near the closest approach in E4, E6, E12, E19 and E26, are unrelated to Europa's wake.

\begin{figure}[h]
  \centering
  \includegraphics[width=1.0\textwidth]{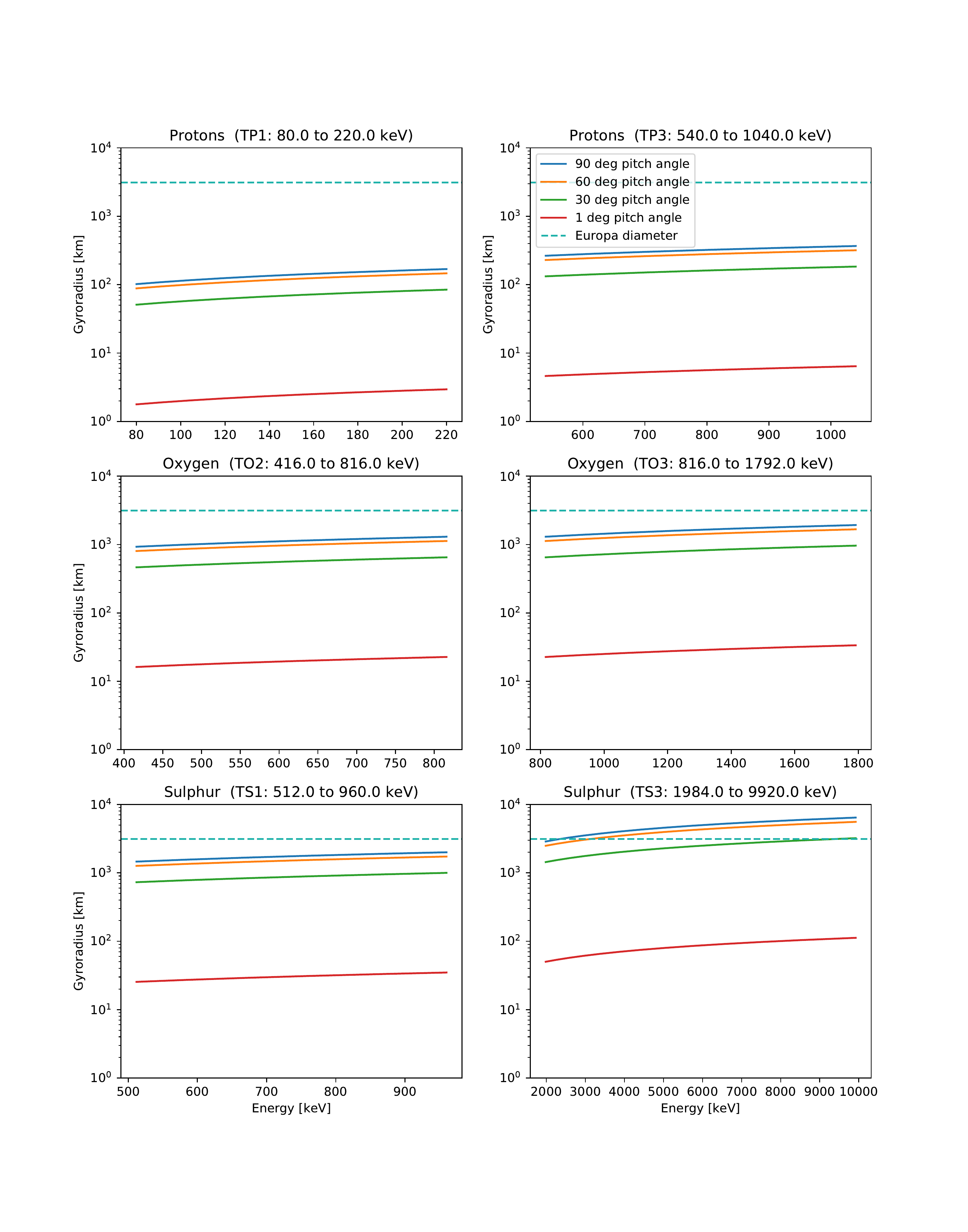}
  \caption{Gyroradius of singly ionized protons, oxygen and sulphur versus energy of the particle, for different pitch angles. Equation \ref{eq_gyroradius} was used to calculate the gyroradius, assuming undisturbed upstream magnetic field conditions that are typical  for the flybys considered in the text (B = 400 nT). The energy range is for two EPD proton channels (TP1 and TP3), two oxygen channels (TO2 and TO3) and two sulphur channels (TS1 and TS3).}
  \label{img_gyro_tp_to_ts}
\end{figure}

As previously stated by \cite{Paranicas2000} during E12 depletions occur at larger distances from Europa than the gyroradius of the particles (see Section \ref{ss_interaction_EPD}). These depletions could be caused by charge exchange with neutral particles from Europa's atmosphere or plumes. I observe such depletions also in the TP1 and TP2 channel, for example at 12:02, which was not reported previously by \cite{Paranicas2000}. The gyroradius in the energy range TP1 spanned by TP2 (50-130 km) is a factor 5.4 and 2.1 larger than the altitude at 12:02 (270 km), for a field magnitude of 800 nT. I also identify depletions, at larger distances from Europa than the gyroradius, during the E4, E6 and E26 flybys, near the closest approach. The E4 flyby closest approach occurs at 688.1 km. This distance is at least a factor two larger than the expected gyroradius of the depleted protons (80-1040 keV, channels TP1 to TP3). This can be see in Figure \ref{img_gyro_tp_to_ts}, in which the gyroradii of several relevant species are shown. The gyroradius of sulphur and oxygen is larger than the altitude at closest approach. During E6 the gyroradius of particles in the proton channels (80-1040 keV, TP1 to TP3) is smaller than the altitude (582.3 km). During E26, which occurs at 348.4 km altitude, depletions are visible in energetic protons (80-540 keV, TP1 and TP2). The gyroradius of both is smaller than the spacecraft altitude (348.4 km). During E19, which occurs at 1444.4 km, a depletion can be seen of energetic sulphur (0.512-9.92 MeV, TS channels). The gyroradius of these particles is comparable to, or larger than the altitude.

\section{Candidate orbit selection}
\label{s_candidate}
In this section I will compare the flybys and determine which are the best candidate(s) for finding signatures of Europa's atmosphere and plumes. Based on Section \ref{s_europa_plumes}, \ref{s_mag_interaction}, \ref{s_mag_interaction_epd_pls} and \ref{s_comparison} I can formulate the following types of features in plasma and energetic particle data, that could be indicative of an atmosphere or plume.
\newline

\noindent
For PLS:
\begin{itemize}
\item \textbf{Criterion 1: }An enhancement of the ion density near Europa, because an increased production of neutral particles will result in an increased ion density.
\item \textbf{Criterion 2: }A deflection of the corotational plasma near Europa, because such a deflection requires a conductive layer. In the case of Europa this is the ionosphere, which is the product of ionized neutral particles.
\item \textbf{Criterion 3: }A change in the plasma velocity magnitude, could be indicative of interaction of the corotational plasma with Europa's ionosphere.
\end{itemize}
For EPD:
\begin{itemize}
\item \textbf{Criterion 4: }Depletions of energetic particles near Europa, occurring at altitudes larger than the gyroradius of the particles involved, because these could be caused by charge exchange with neutral gas.
\end{itemize}
I also take into account the available magnetometer data:
\begin{itemize}
\item \textbf{Criterion 5: }An increase of the magnetic field magnitude upstream Europa, because this indicates the flow of the plasma is decelerated by Europa's ionosphere.
\end{itemize}
The criteria are evaluated in Table \ref{tab_criteria_flybys}. I currently cannot evaluate E26 for criteria 1-3. The software that deals with the missing data points and noise removal of PLS data, produces increases in the count rate over all energies near the closest approach. I consider these features to be unphysical, as they do not occur in the raw data. It is therefore not possible to investigate the density and velocity of the plasma near the closest approach for this flyby.
\begin{table}[h]
\centering
\begin{tabular}{|l|l|l|l|l|l|l|l|l|l|}
  \hline
  \textbf{Criterion} & \textbf{E4} & \textbf{E6} & \textbf{E11} & \textbf{E12} & \textbf{E14} & \textbf{E15} & \textbf{E19} & \textbf{E26}\\
  \hline
  \hline
  Criterion 1 (density) & \textcolor{green}{yes} [1] & \textcolor{green}{yes} [1] & \textcolor{red}{no} & \textcolor{green}{yes} [2] & \textcolor{red}{no} & \textcolor{green}{yes} & \textcolor{red}{no} &    \\
  \hline 
  Criterion 2 (deflection) & \textcolor{green}{yes} [1] & \textcolor{green}{yes} [1] & & \textcolor{green}{yes} & & & &    \\
  \hline
  Criterion 3 (vel. mag.) & \textcolor{green}{yes} & \textcolor{green}{yes} & \textcolor{red}{no} & \textcolor{green}{yes} & \textcolor{red}{no}  &  \textcolor{red}{no} & \textcolor{green}{yes} & \\
  \hline
  Criterion 4 (depletion) & \textcolor{green}{yes} & \textcolor{green}{yes} &  \textcolor{red}{no} & \textcolor{green}{yes} & \textcolor{red}{no} &  \textcolor{red}{no}&  \textcolor{red}{no} & \textcolor{green}{yes}    \\
  \hline 
  Criterion 5 (pile-up) & n.a. & no data & n.a. & \textcolor{green}{yes} & \textcolor{green}{yes} & n.a. & \textcolor{green}{yes} &  \textcolor{green}{yes} \\
  \hline 
\end{tabular}
\caption{Evaluation of the criteria for all the flybys. External sources are: [1] \cite{Paterson1999} and [2] \cite{Jia2018}. N.a. specifies the criterion is not applicable to the flyby. Blank fields indicate that it was not possible to evaluate the criterion for that specific flyby.}
\label{tab_criteria_flybys}
\end{table}

It is clear from Table \ref{tab_criteria_flybys} that all the flybys meet at least one of the criteria. It indicates that that the data of multiple flybys could potentially be used to study Europa's atmosphere. However, the E12 flyby stands out clearly. This is because of the following reasons:
\begin{itemize}
\item It has the highest plasma density, as shown by \cite{Jia2018}
\item It features a unique slow-down region upstream
\item Energetic ions are depleted at altitudes larger than their gyroradius and thus may have been depleted by an atmosphere.
\item The highest magnetic field strength near Europa was observed during this flyby
\item \cite{Jia2018} has argued that a plume and atmosphere could have been present during this flyby.
\end{itemize}
Therefore I consider the E12 flyby the most interesting candidate to find signatures of Europa's atmosphere and plumes in the Galileo charged particle detector data.
 
\section{Selected candidate: E12}
\label{s_candidate_e12}
\begin{figure}[h]
  \centering
  \includegraphics[width=0.62\textwidth]{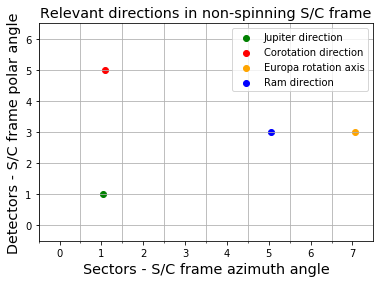}
  \caption{Coordinates, azimuth and polar angle, in the non-spinning spacecraft frame of several physically interesting directions. The determination of these directions is discussed in Appendix \ref{a_pls_processing}}
  \label{img_e12_pls_fov}
\end{figure}

Having selected E12 as the most interesting flyby to contain atmospheric or plume signatures, I will now discuss the data collected during this flyby in more detail.
First I will take a closer look at the deflection of the corotational flow during the E12 flyby. I have developed my own method to identify changes in corotational velocity. I do this by separating the data either in 8 azimuthal bins or 7 polar bins, in a spacecraft reference frame with a fixed inertial pointing. In this frame the expected polar angle and azimuth angle of physically interesting directions, such as that of the corotational plasma can be determined. The results are shown in Figure \ref{img_e12_pls_overview}. The two panels on the top show the omnidirectional data as presented in Appendix \ref{app_overview}. The third panel from the top shows the data separated in 7 polar angles versus time. The fourth panel shows the data separated in 8 azimuthal bins versus time. Figure \ref{img_e12_pls_fov} shows the coordinates of several relevant physical directions in this non-spinning reference frame that apply for the duration of the data. The used method is discussed in more detail in Appendix \ref{a_pls_processing}.

\begin{figure}[p]
  \centering
  \includegraphics[width=1.0\textwidth]{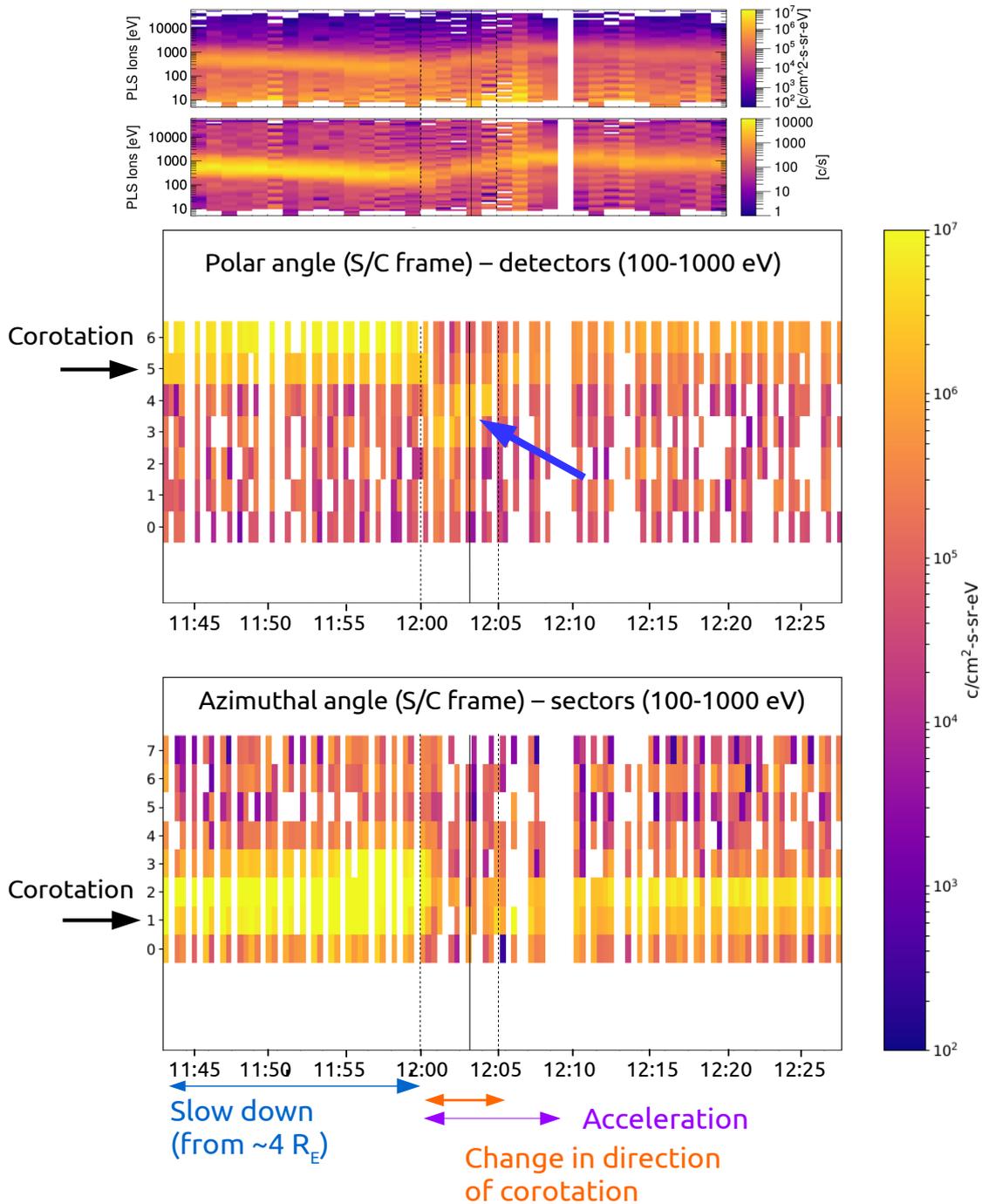}
  \caption{Overview of PLS data collected during the E12 flyby. The top two panels show the omnidirectional energy spectrum (as shown in Figure \ref{img_E14_all}). The next two panels show the data binned in azimuth and polar angles, in a non-spinning spacecraft frame. The black arrows indicates the expected direction of the corotation plasma. Physical interpretation of the directions can be done with Figure \ref{img_e12_pls_fov}. The big blue arrow indicates the change in direction of the corotational plasma.}
  \label{img_e12_pls_overview}
\end{figure}

The expected direction the corotational plasma is indicated by black arrows in \ref{img_e12_pls_overview}, generally the peak of the differential flux corresponds to these directions. However between 12:00 and 12:05 there is a change in direction of the corotational plasma. The polar angle is changing from the normal corotating direction, towards the rotation axis of Europa. According to the model by \cite{Jia2018} of this flyby, the largest changes in magnetic field occur between 12:00 and 12:05 for both the plume and the non-plume scenario they compare (see Figure \ref{img_jia}). This period corresponds to the period where I see changes in the direction of the corotational plasma. In \cite{Jia2018} the change in the field is strongest for the z-component. As \cite{Jia2018} does not show the velocity vectors, further comparison is not possible.

\begin{figure}[h]
  \centering
  \includegraphics[width=1.0\textwidth]{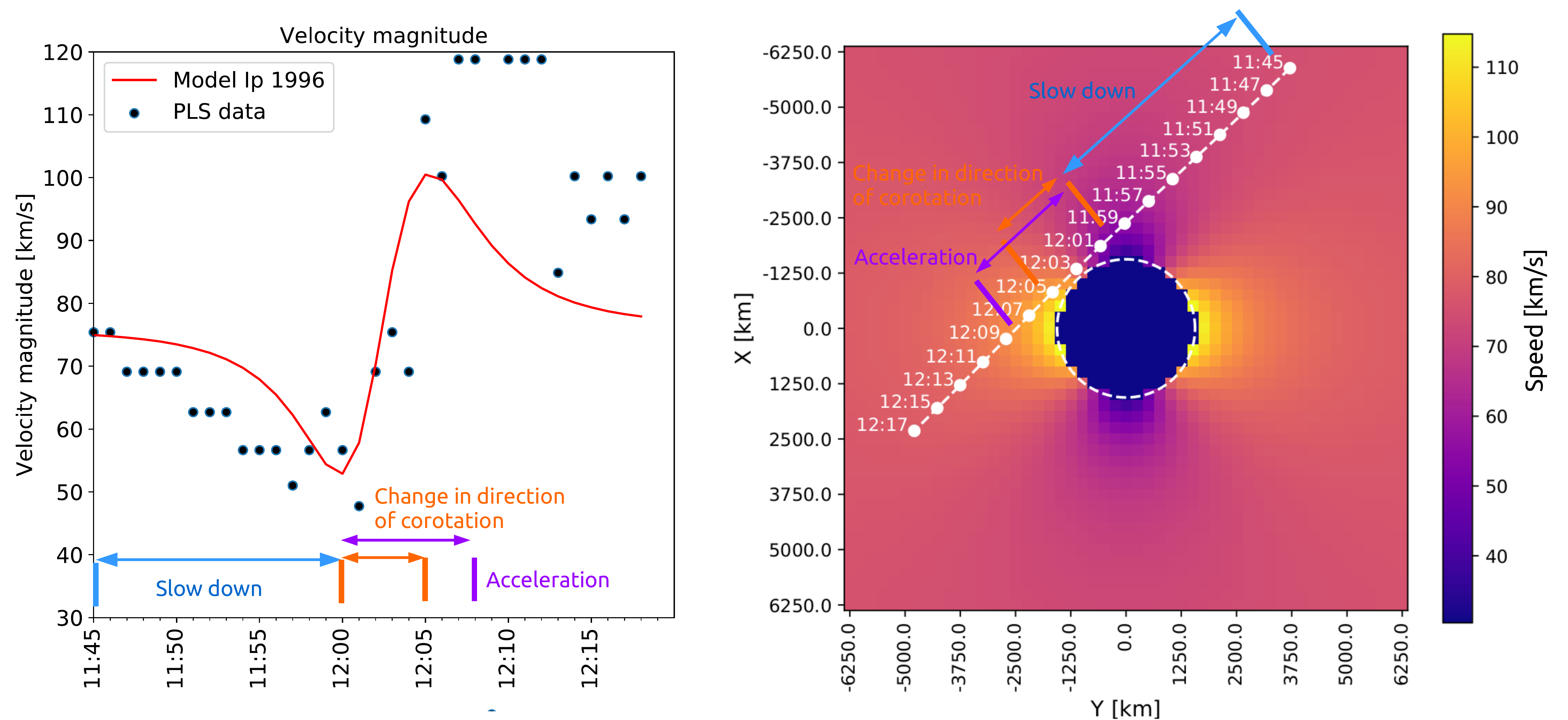}
  \caption{Left: velocity magnitude during the E12 flyby (data) and a fit with the model by \cite{Ip1996} (Section \ref{s_model_flow}). Here I set the $\alpha$ parameter to 0.4 and $R_c$ to the radius of Europa. Right: velocity magnitude in the same model with the spacecraft trajectory.}
  \label{img_e12_overview}
\end{figure}

Next, I discuss the magnitude of the velocity measured during the E12 flyby. I have developed a method to estimate the magnitude of the corotational plasma velocity (Appendix \ref{a_pls_processing}). The resulting velocity magnitude is shown in the left panel of Figure \ref{img_e12_overview}. Also in the left panel in this figure the velocity magnitude is compared with the analytical model by \cite{Ip1996}. This model reproduces the trend observed in the data, up to 12:05. A phase of slow-down and acceleration can be distinguished. After 12:05 the model predicts a stronger deceleration than observed. The right panel of Figure \ref{img_e12_overview} shows the velocity magnitude in the model together with the trajectory. The observed slow down corresponds to the area upstream of Europa where the corotating plasma is expected to be slowed down by the model. The acceleration corresponds to the region on the flanks of Europa where the plasma is expected to accelerated. The region where the plasma changes direction can be found in between them. This flow pattern will also influence energetic particles and will be an input in the simulations discussed in the next chapter.
\begin{figure}[h]
  \centering
  \includegraphics[width=1.0\textwidth]{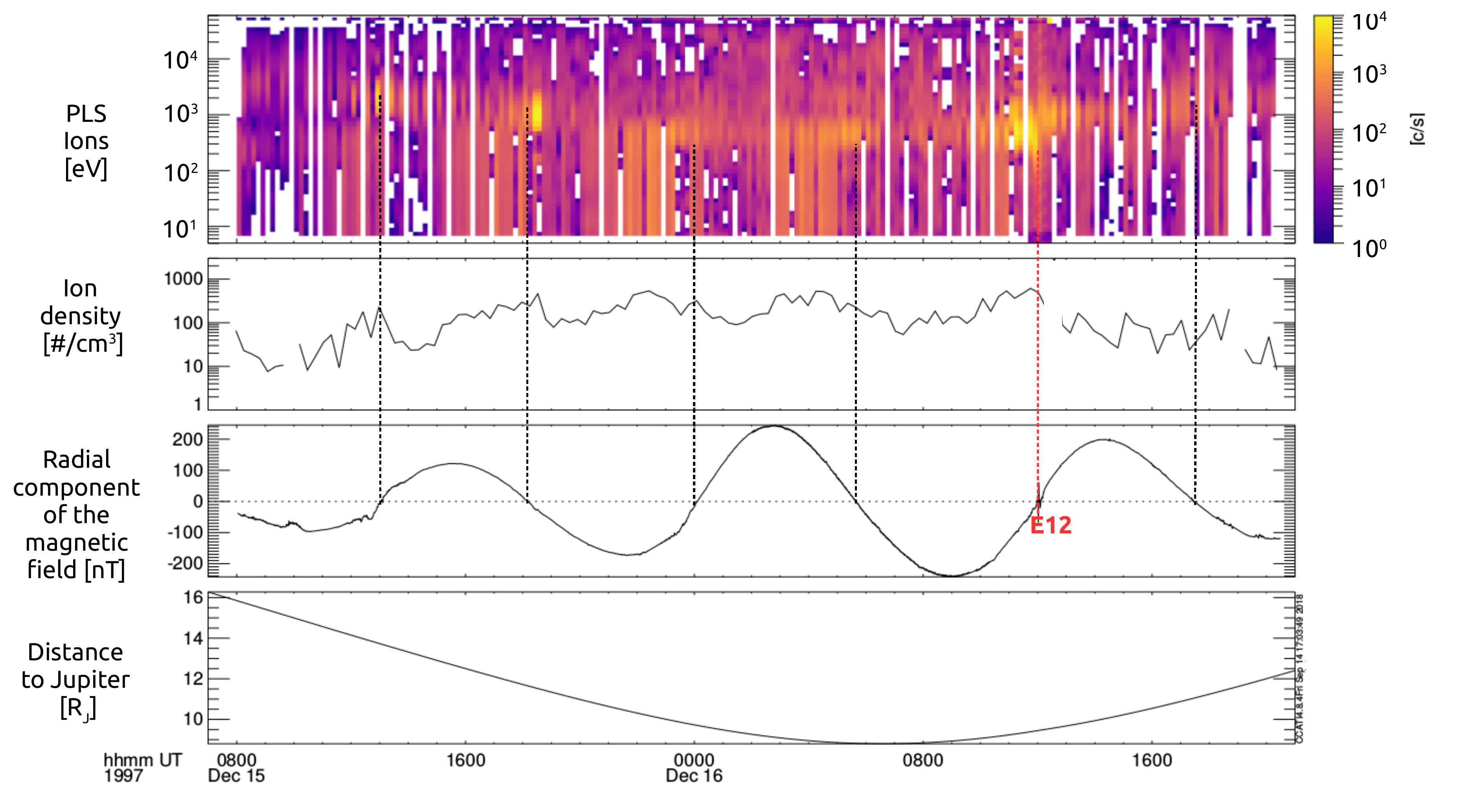}
  \caption{Top panel: all PLS data available directly adjacent to the E12 flyby,
10 min average. Second panel: PLS ion density, 10 min average, energies < 200 eV 
have been ignored for the density calculation.}
  \label{img_e12_crossings}
\end{figure}

It should be noted that during the E12 flyby the density of the plasma is already high at the beginning of flyby and decreases towards the end. This is visible in both PWS data, Figure \ref{img_bagenal2015} and PLS data, Figure \ref{img_E12_all_comp}). This raises the question if the high density of the corotational plasma is a local density enhancement or if a general disturbance of the magnetospheric plasma occurs. Investigating this question is hampered by the unavailability of data at same time resolution as available during 11:45 and 12:20. Some sparse measurements were made outside of this time frame. In Figure \ref{img_e12_crossings} a ten minute average of the available data is shown. As stated in Section \ref{s_flybys_overview} E12 is a flyby that occurs at low magnetic latitude and thus when Europa is in the densest part of the plasma sheet. In Figure \ref{img_e12_crossings} the radial component of the magnetic field is shown to indicate when the previous and next such plasma sheet crossings should occur. At these crossings it is clear that the count rate around the corotational energy (near 1000 eV) increases, resulting in increases of the plasma density, corresponding to what is expected to occur at a plasma sheet crossing. From Figure \ref{img_e12_crossings} it is clear that the plasma density during the E12 flyby is comparable to the previous plasma sheet crossings, suggesting that the high density observed upstream of E12 is related to a change in conditions of the plasma sheet that lasts longer than the E12 flyby. 

\begin{figure}[p]
  \centering
  \includegraphics[width=1.0\textwidth]{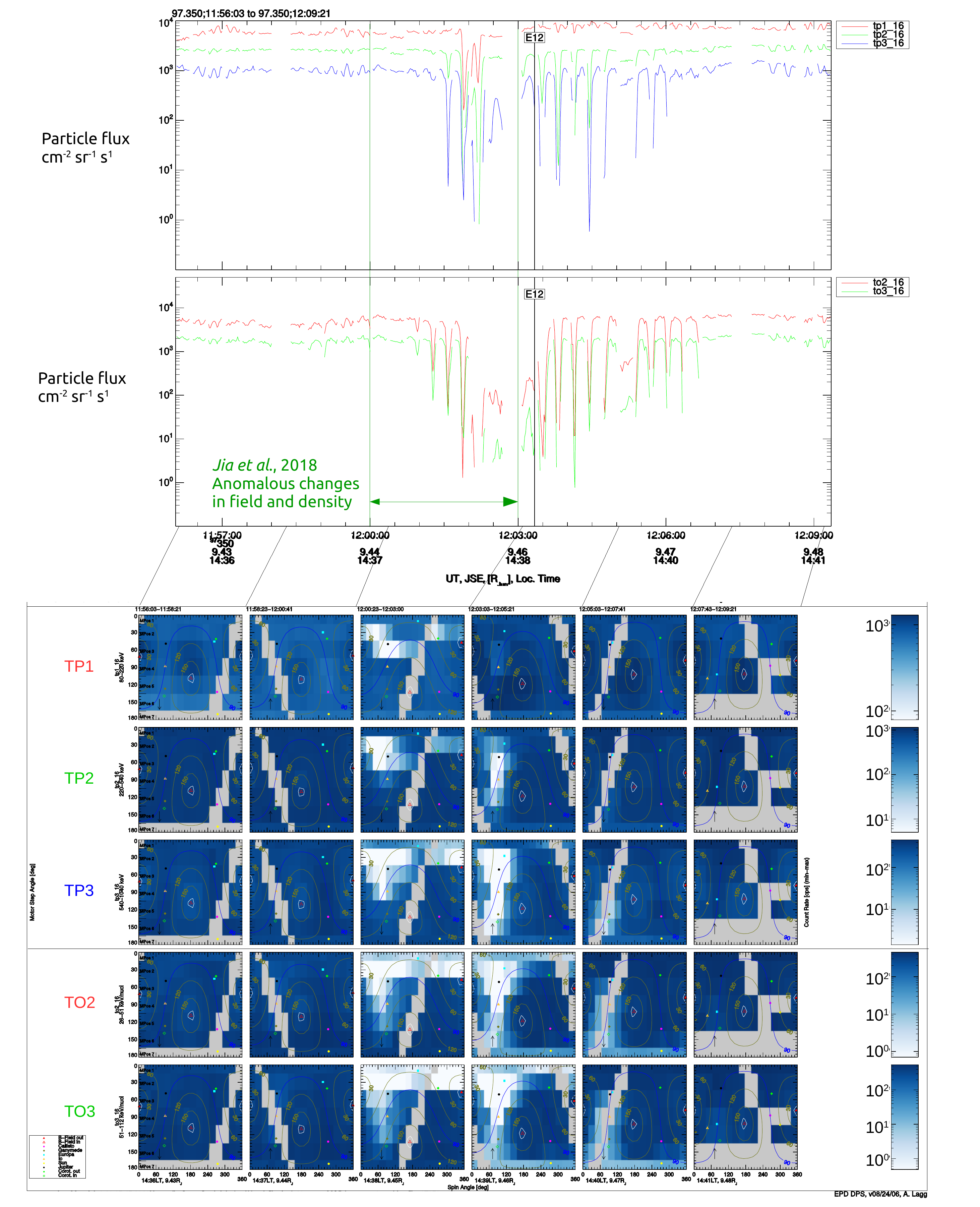}
  \caption{Overview of the EPD TP and TO data during the E12 flyby. The top two panels show the line plots of the particle flux in three proton channels (80 to 1040 keV, TP1 to TP3) and two oxygen channels (416-1792 keV, TO2 and TO3). The bottom panels represent the same data, but separated in six figures that each represent the field of view of the EPD instrument for one full sky scan. Each skymap is obtained during approximately 140 seconds. Indicated by contour lines in the sky maps is the angle with respect to the magnetic field. Additional details about the construction of the skymaps can be found in the text.}
  \label{img_combined_fov_line}
\end{figure}

Finally I discuss the EPD data collected during the E12 flyby. From Figure \ref{img_E12_all_comp} it can be determined that the A channels (energetic ions) are heavily affected by noise. This can be seen by looking at the data when the instrument moves behind the foreground shield (see Section \ref{ss_epd}). For example at 11:49 the residual signal is $\sim 50$\% of the total signal. For the proton (80-1040 keV, TP1-TP3), oxygen (416-1792 keV, TO2-TO3) and sulphur (0.96 to 9.92 MeV, TS1-TS3) channels the signal is reduced by at least two orders of magnitude. This indicates that the effect of noise is not significant. It is to be expected that the T channels are not significantly affected by noise, as a triple coincidence system is used for these channels (see Section \ref{ss_epd}). Therefore I will focus specifically on the energetic proton channels (80 to 1040 keV, TP1 to TP3) and oxygen channels (416-1792 keV, TO2 and TO3) in the simulation effort discussed in the next chapter. 
An overview of the energetic proton (80-1040 keV, TP1-TP3) and energetic oxygen (416-1792 keV, TO2-TO3) data for the time period near the closest approach is shown in Figure \ref{img_combined_fov_line}. The top two panels show the line plots of the flux in the channels. Compared to Figure \ref{img_E12_all_comp} the scans behind the foreground shield are not shown. The bottom panel shows the same data, but separated in six subfigures that show the field of view of the EPD instrument. It is important to explain how the skymaps are made. Each skymap is constructed over a period of $\sim 2$ minutes (Table \ref{tab_properties_epd}). Every data point in each skymap is collected at a different point in time. The horizontal axis of each individual skymap represents the azimuth of the instrument field of view. Each row of data in a skymap represents data collected during a different spin of Galileo. The order in which the data points are collected is from left to right. The vertical axis represent the polar angle in the instrument field of view. After every spin the motor position of the EPD instrument was changed, such that a different polar angle could be scanned (Section \ref{ss_epd}). Thus each row of a skymap also shows the data collected during a different motor position of the instrument. Indicated by contour lines are the angles with respect to the magnetic field (iso pitch angle contours). The depletion features depend on the direction that the instrument is looking at. Not all directions are depleted equally. The depletions extend further in time in channels with higher energy, as is clear from comparing energetic protons of 540-1040 keV (TP3) with energetic protons of 80-220 keV (TP1) or energetic ions of 816-1792 keV (TO3) with energetic ions of 416-816 keV (TO2). This is because the gyroradius of the higher energy particles is larger. A larger gyroradius particle can easier reach the surface (or atmosphere near Europa) and can thus be depleted from a larger distance.

\begin{figure}[h]
  \centering
  \includegraphics[width=0.75\textwidth]{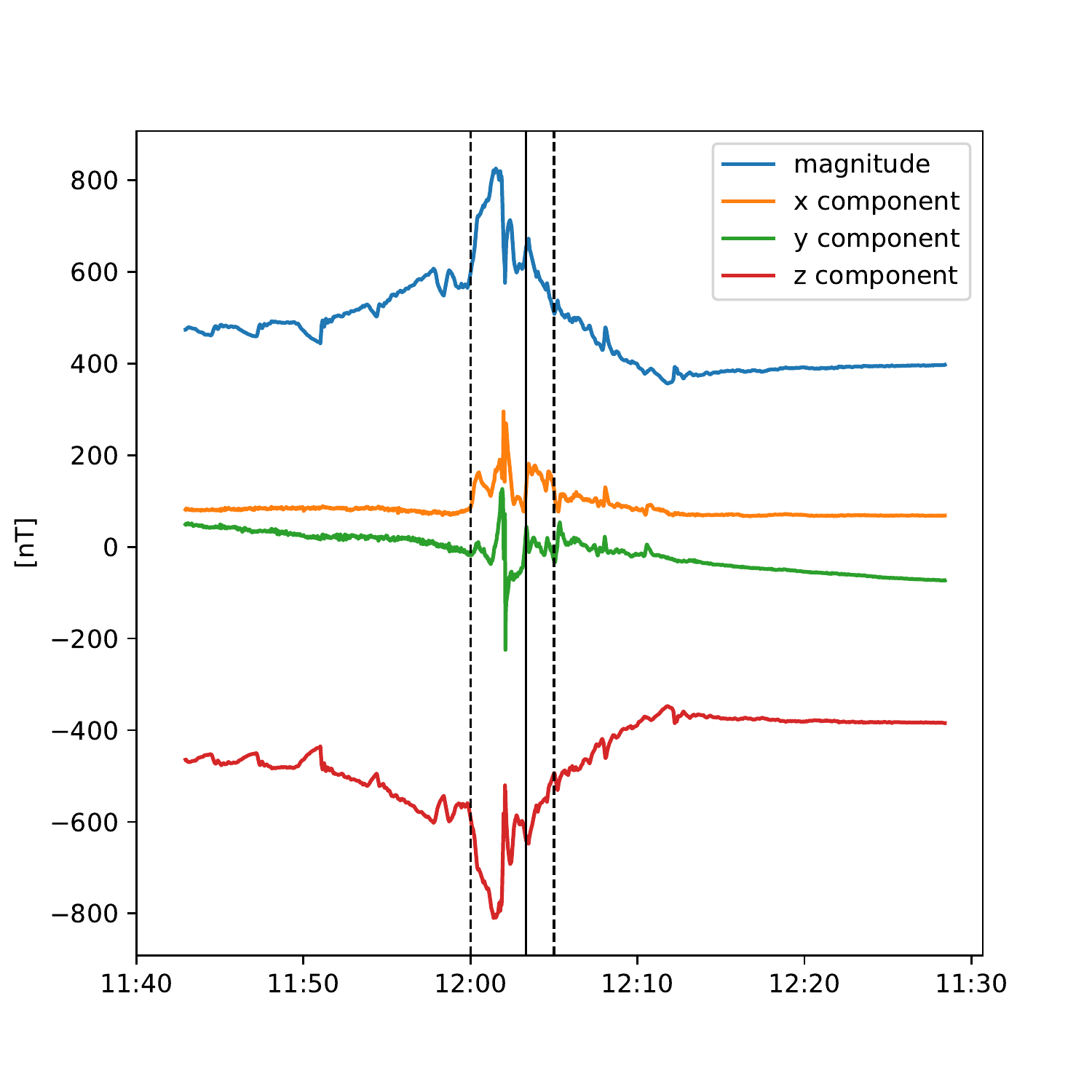}
  \caption{Components of the magnetic field during the E12 flyby in the IAU Europa frame. In this figure the x component is in the direction of the corotational plasma, y is towards Jupiter and z is along Europa's rotation axis. The dotted lines indicate the region during which the direction of the corotational plasma changes according to the analysis of the PLS data. \cite{Jia2018} reports a change in magnetic field direction between the first dotted line and the closest approach (solid line), possibly related to a plume.}
  \label{img_components}
\end{figure}

Finally, in Figure \ref{img_components} the components of the magnetic field during the E12 flyby are shown. This is relevant for the next chapter.

\clearpage

\section{Main findings and implications for the simulation}
In this chapter an overview was made of the charged particle and magnetic field data collected during the Europa flybys. During all flybys signatures of interaction of Europa with its magnetospheric environment are visible. I evaluated all flybys for five criteria to select the candidate(s) that are most relevant to find signatures of Europa's atmosphere and potential plumes in the charged particle data. Based on this comparison I selected the E12 flyby as the best candidate. This flyby stands out because of its high plasma density, magnitude of the local change in the magnetic field, magnitude of the slow down of the plasma and the extent of the region over which the slow-down occurs. Furthermore \cite{Jia2018} has indicated that a plume and atmosphere could be present during this flyby and interact with the corotational plasma. I conclude that signs of interaction of the magnetospheric environment with Europa's atmosphere-derived ionosphere are stronger than during any other flyby. A detailed analysis of the charged particle data collected during this flyby was made. I formulate several conclusions:
\begin{itemize}
\item The number of energetic ions (EPD A channels), protons (EPD TP channels), oxygen ions (EPD TO channels) and sulphur ions (EPD TS channels) is depleted near Europa.
\item The observed depletions depend on particle energy, particle composition and spatial direction.
\item Depletions of energetic protons (80-540 keV, TP channels) occur at altitudes larger than their gyroradius. These depletions could be caused by charge exchange with neutral gas.
\item The A channels (energetic ions) are heavily affected by noise, while the T channels are not. The T channels are thus most suitable for the modelling effort in the next chapter.
\item The velocity and direction of the corotational plasma are changed near Europa. The trend in the velocity magnitude can be reproduced by the model in \cite{Ip1996}.
\item The high plasma densities observed upstream during this flyby could be be related to a change in global magnetospheric conditions.
\end{itemize}
These findings will be used in the next chapter, in which I attempt to model the depletion of energetic ions.

\chapter{Characterisation of Europa's tenuous atmosphere during the E12 flyby}\label{ch_atmosphere}

In the previous chapter depletions in the number of energetic ions near Europa have been identified that occur at altitudes larger than the particles' gyroradius. These depletions could be caused by charge exchange with neutral particles from Europa's tenuous atmosphere. Therefore, in this chapter I present simulations of the trajectories of the energetic particles to reproduce the depletions. The simulations  will be compared to the Galileo Energetic Particle Detector (EPD) data to constrain the surface density and scale height of Europa's tenuous atmosphere. The presence of a potential plume will also be investigated. First, in Section \ref{s_sim_setup} the setup of the simulation is discussed. Then, in Section \ref{s_sim_results} the results are presented. After that, in Section \ref{s_data_compare}, the simulation is compared to the data to constrain the atmospheric properties. In Section \ref{s_plume_e12} it is investigated if a plume signature could be present.

\section{Energetic particle simulation setup}
\label{s_sim_setup}
In this section I discuss the setup of the simulation and describe the different assumptions that have been made. For these simulations the code described in Section \ref{s_particle_simulation} is used.
The simulation results have to meet the following requirements regarding the depletion features (from Chapter \ref{ch_comparison}):
\begin{itemize}
\item \textit{The simulations should be conducted for different ion species to account for the species-dependent depletion.} Here I will simulate protons (80-1040 keV) and oxygen ions (416-1792 keV) , for which there are reliable data available (TP1 to TP3 and TO2-TO3 channels). Data on depletions of energetic sulphur ions (TS channels) are also available, but there are no charge exchange cross sections available for sulphur ions on O or O$_2$ that cover the energy range of interest, which are needed to simulate potential losses of energetic sulphur ions by Europa's atmosphere. I assume that the energetic charged particles are singly charged.
\item \textit{The simulations should be conducted for different particle energies to account for the energy-dependent depletion features.} To meet this requirement I will simulate the depletions for different energy channels (TP1 to TP3 and TO2-TO3). I use the Monte Carlo method and launch particles in the energy range of the EPD channels. Every energy channel is binned in several steps and the contribution of each bin is weighted according to the energy spectrum derived from the data (Figure \ref{img_energy_spectra}).
\item \textit{The simulation should be able to discriminate between depletions in different physical directions, as the data show that the depletions depend on the pointing of the EPD instrument.} To address this I will simulate the depletions for different pitch angles. The pitch angles are shown by the contour lines in Figure \ref{img_combined_fov_line}. In addition to the pitch angles I will also simulate the depletions for every azimuth angle for selected pitch angle ranges.
\end{itemize}
Based on Section \ref{s_mag_interaction} and the analysis of the E12 flyby in Chapter \ref{ch_comparison} the impact of the following aspects of the magnetospheric environment need to be investigated.
\begin{itemize}
\item \textit{The flow of the corotational plasma is deflected around Europa.} To simulate the flow deflection I will use the analytical model by \cite{Ip1996} as discussed in Section \ref{s_particle_simulation}. I will use the fitted model shown in Figure \ref{img_e12_overview}. The magnitude of the upstream corotational flow velocity is assumed to be 76 km/s, based on the speed determined from the PLS data (Figure \ref{img_E12_all}). The resulting flow field is shown in Figure \ref{img_ip_flow} and the electric field, calculated using Equation \ref{eq_E}, in Figure \ref{img_e_flow}.
\item \textit{Europa possesses an induced magnetic dipole.} I will simulate the dipole using the analytical equations from \cite{Zimmer2000}, as discussed in Section \ref{s_particle_simulation}. The resulting dipole field is shown in Figure \ref{img_induced_dip}.
\item \textit{A plume could be present during the E12 flyby}. The density contribution of the plume to the neutral particle environment is simulated using the analytical expression discussed in Section \ref{s_particle_simulation}.
\end{itemize}
To test the contribution of these effects I will consider four simulation cases:
\begin{itemize}
\item \textbf{Case 1: homogeneous fields}. I assume a homogeneous magnetic field. The components of the field are taken from the MAG data at 11:40 upstream of Europa: [50, -10, -450] in nT. When I simulate the period from 12:00 to 12:06 I use the field at 12:00: [100, 0, -600] in nT. The homogeneous electric field is calculated using Equation \ref{eq_E}. There is no disturbance on the fields by Europa. Furthermore, no atmosphere is present, thus the only loss process of energetic particles in this case are impacts on the surface.
\item \textbf{Case 2: Case 1 + atmosphere}. The difference with Case 1 is that an atmosphere is included in this case. The atmosphere is described by an analytical model that depends on scale height and surface density (see Equation \ref{eq_atm_saur}). Here I investigate an O$_2$ atmosphere that consists out of one atmospheric component that is described by surface density and scale height. In this case losses can occur both due to surface impacts and charge exchange with the atmosphere. The selection of charge exchange cross sections is discussed in Section \ref{ss_cross_section}.
\item \textbf{Case 3: Case 2 + dipole + flow deflection}. This case differs from Case 2 by the inclusion of an induced dipole and a flow deflection. The dipole and the flow deflection are described above. In this case, by modifying the trajectories of the energetic ions, flow and magnetic perturbations may form forbidden access regions near Europa which "mimic" regions of energetic ion depletions, without however actual losses taking place.
\item \textbf{Case4: Case 1 + plume}. This case differs from Case 1 in the sense that a plume has been included. It is simulated using Equation \ref{eq_plume_jia}.
\end{itemize}
By comparing Case 1 and 2, it can be determined if (part of) the depletion observed in the data can be attributed to atmospheric losses. The addition of Case 3 will help to determine the importance of the induced dipole and the flow deflection on the depletion features. Case 1-3 will be discussed in Section \ref{s_data_compare}). Additionally the effect of the inclusion of a plume (Case 4) is discussed in the dedicated Section \ref{s_plume_e12}.

\cite{Paranicas2000} suggested that magnetic drifts could contribute to the formation of depletions observed during the E12 flyby (as was described in Section \ref{s_mag_interaction_epd_pls}). My trajectory simulation method involves a numerical integration of the Lorentz force and can thus in principle determine the effect of gradients on the trajectories. However, a description of the magnetic field that includes the gradients observed in the data is required, which are: the magnetic field pile up, the peak in field strength and the sudden change of direction. Because the magnetic field data is only available along the trajectory, the cannot be used for this, as the data does not tell us what the field looks like away from the spacecraft trajectory. Thus a MHD or hybrid model of the interaction of the plasma with the atmosphere and possible plumes is required. In the absence of such a model my simulations are not able to determine the contribution of the magnetic drifts to the formation of the depletion features.

\subsection{Charge exchange cross sections}
\label{ss_cross_section}
The method through which I simulate charge exchange is discussed in Section \ref{s_particle_simulation}. For this method charge exchange cross sections of the energetic particle on the atmospheric particles are required. These cross sections define the probability that charge exchange occurs between two particles.

\begin{figure}[h]
  \centering
  \includegraphics[width=1.0\textwidth]{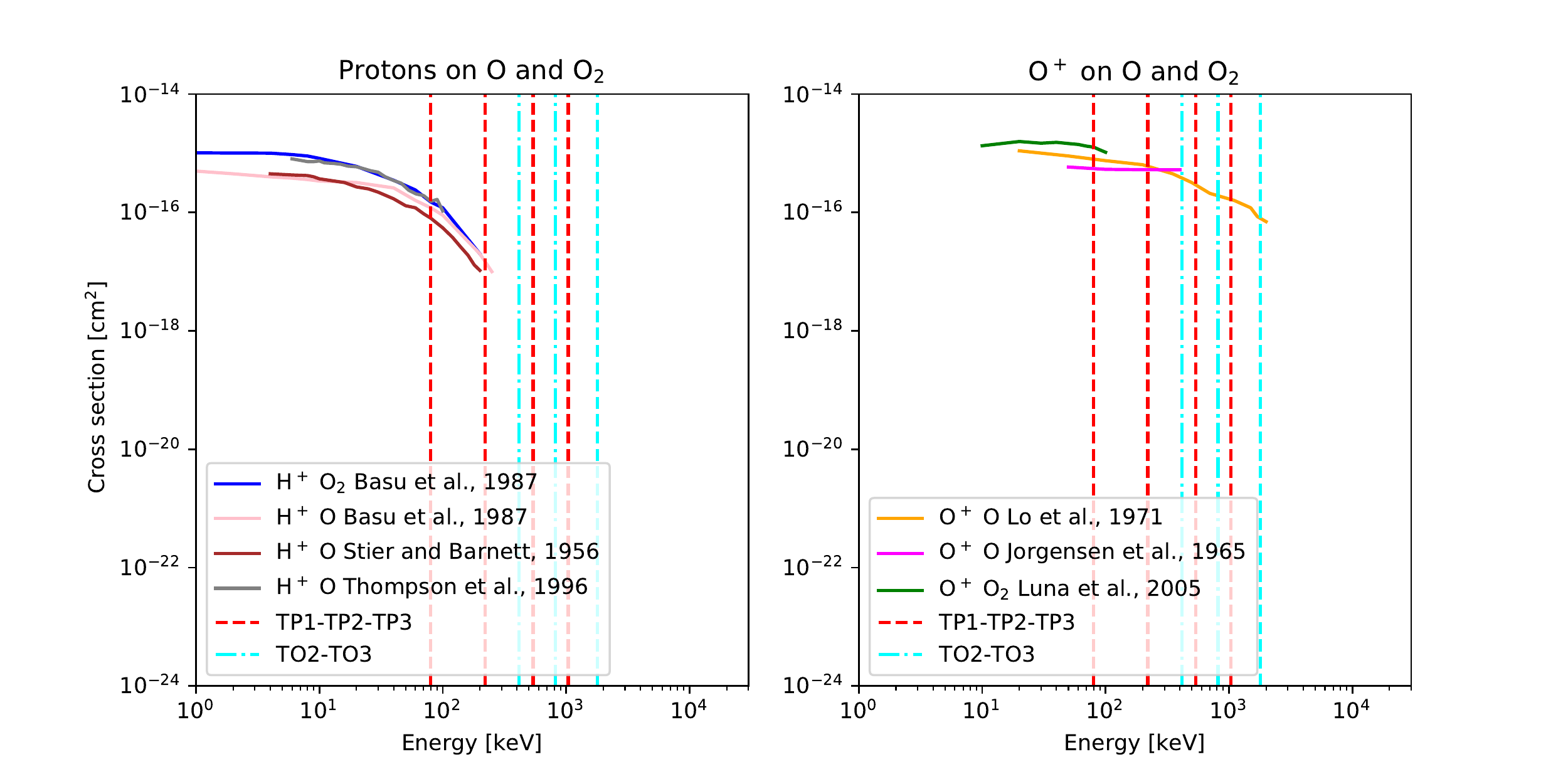}
  \caption{Charge exchange cross sections versus energy, for (left) H$^+$ on O and (right) O$_2$, and O$^+$ on O and O$_2$. The dashed red vertical lines indicate the energy range of the three adjacent EPD proton channels: TP1 (80-220 keV), TP2 (220-540 keV) and TP3 (540-1040.0 keV). The dashed cyan lines indicate the two adjacent EPD oxygen channels: TO2 (416-816 keV) and TO3 (816-1792 keV). Original sources are \cite{Stier1956,Basu1987,Thompson1996} for protons and \cite{Jorgensen1965,Lo1971,Luna2005} for oxygen.}
  \label{img_cross_sections_all}
\end{figure}

In Figure \ref{img_cross_sections_all} charge exchange cross sections from laboratory experiments, are shown for H$^+$ on O and O$_2$, and O$^+$ on O and O$_2$. Specifically shown are those cross section profiles that cover at least the lower ends of the relevant EPD channels. What is immediately clear from these figures is that the cross sections are not constant over the energy range covered by the different EPD channels. The cross sections decrease with increasing energy, thus particles of a higher energy are less likely to charge exchange with atmospheric particles. The decrease is particularly strong for protons. In the TP1 channel the cross section decreases with approximately one order of magnitude.

\begin{figure}[h]
  \centering
  \includegraphics[width=0.65\textwidth]{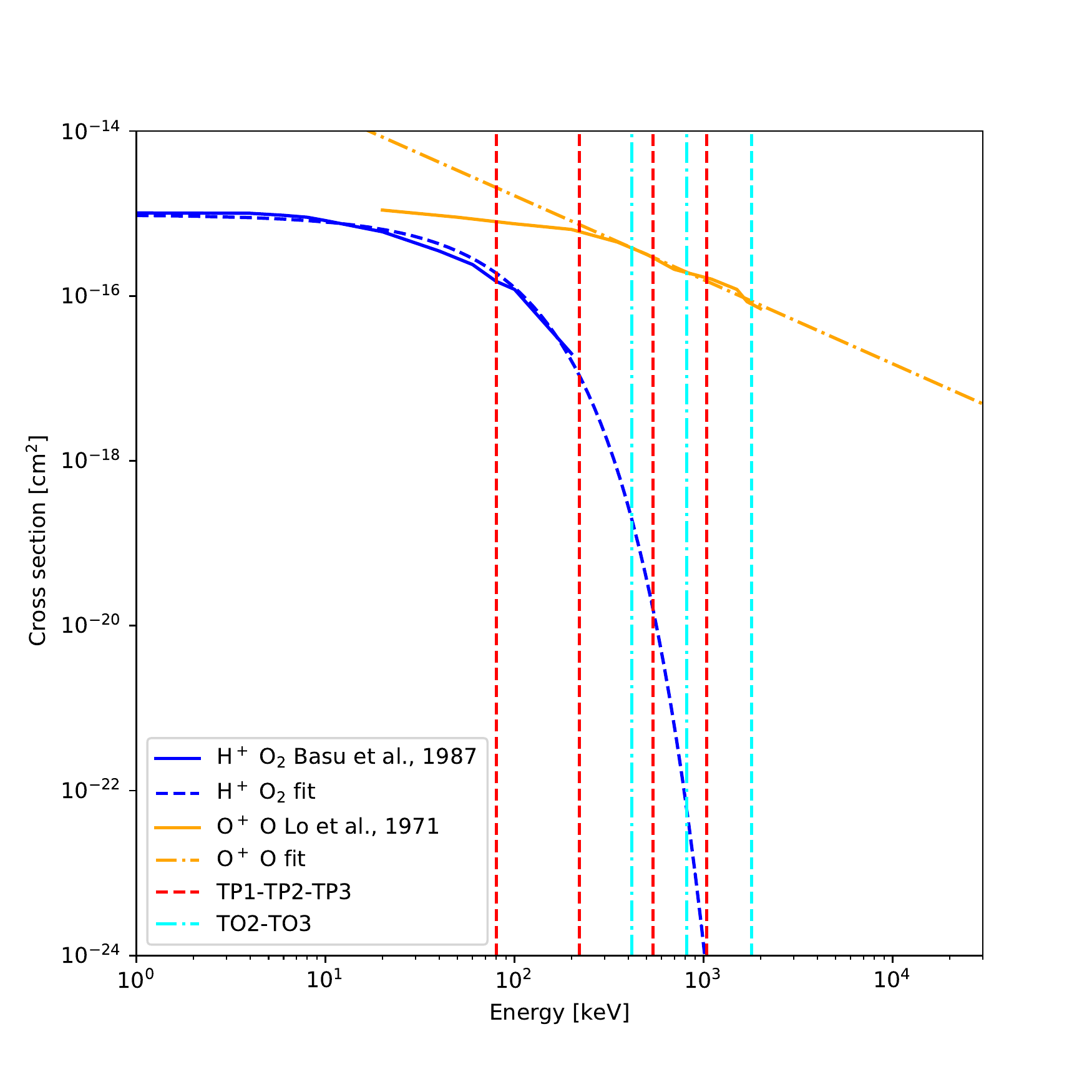}
  \caption{Selected charge exchange cross sections versus energy, for H$^+$ on O$_2$ and O$^+$ on O. The solid lines show the cross sections as determined by \cite{Basu1987} and \cite{Lo1971}. The fits used for the simulations in this work are also shown. The dashed red vertical lines indicate the energy range of the three adjacent EPD proton channels: TP1 (80-220 keV), TP2 (220-540 keV) and TP3 (540-1040.0 keV). The dashed cyan lines indicate the two adjacent EPD oxygen channels: TO2 (416-816 keV) and TO3 (816-1792 keV).}
  \label{img_cross_sections}
\end{figure}

In the left panel of Figure \ref{img_cross_sections_all} both the cross sections for H$^+$ on O and O$_2$ are shown. As the most abundant component of Europa's extended atmosphere is thought to be O$_2$ (see Section \ref{s_atmosphere}), H$^+$ on O$_2$ is more important here. It is clear that the four profiles show the same trend and differ less than one order of magnitude. The profile for protons on H$^+$ on O$_2$ is comparable to, or higher than the profiles for H$^+$ on O.
As no cross sections are available for energies above the upper limit of the TP1 channel, I have fitted the H$^+$ on O$_2$ profile presented in \cite{Basu1987}. 
The resulting fit is shown in Figure \ref{img_cross_sections}. Similar trends for the charge exchange cross section have been predicted for H$^+$ on H and H$^+$ on H$_2$O \citep{Johnson1990}. 

In the right panel of Figure \ref{img_cross_sections_all} both the cross sections for O$^+$ on O and O$_2$ are shown. No profile is available of O$^+$ on O$_2$ that covers the energy range of TO2 or TO3. From the available profiles I determine that they do not differ more than one order of magnitude where they cover the same energies. The profile for O$^+$ on O$_2$ being the largest from the available energies. Here I will use the profile of O$^+$ on O from \cite{Lo1971} as an approximation of the  O$^+$ on O$_2$ cross section, because \cite{Lo1971} covers the energy range of TO2 and TO3. In the energy range between 1 and 100 keV, the O$^+$ on O$_2$ profile by \cite{Luna2005} and the O$^+$ on O from \cite{Lo1971} have the same trend, the difference being that the profile values of the O$^+$ on O are lower for all energies. I therefore assume that the profile by \cite{Lo1971} is a reasonable approximation, though it might underestimate the probability of charge exchange. In the simulation I use a linear fit of this profile, which is shown in Figure \ref{img_cross_sections}.

\subsection{Boundary conditions}
\begin{figure}[h]
  \centering
  \includegraphics[width=1.0\textwidth]{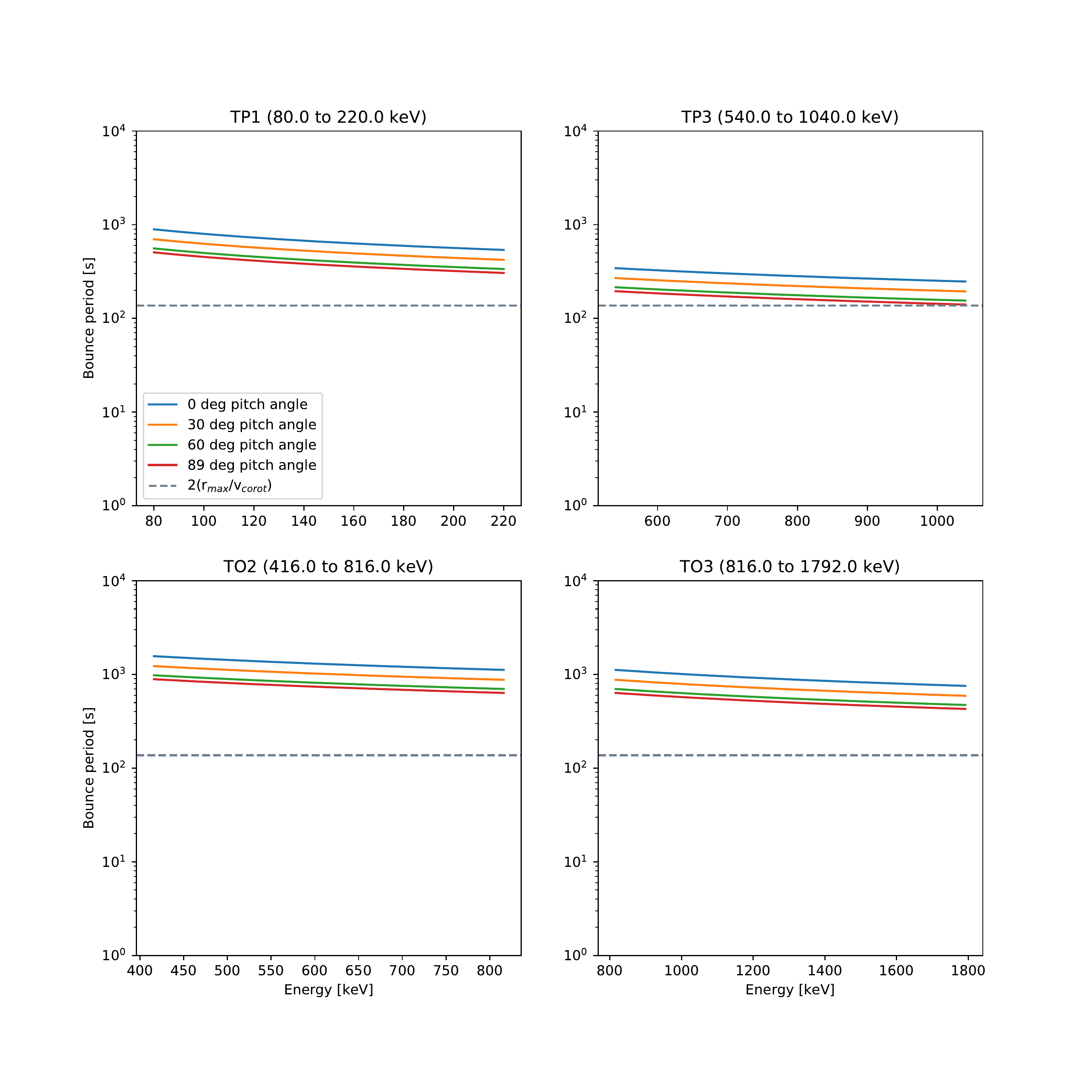}
  \caption{Bounce period of the particles types and energies used in the simulation, calculated using Equation \ref{eq_bounce_period}. The horizontal dashed lines indicate the bounce period of a particle that would just re-enter the simulation box. This distance is calculated by dividing the maximum radial distance of the spacecraft trajectory which is used in the simulation (r$_{max} \approx$ 5000 km), by the corotational velocity (v$_{corot}$) and multiplying by two.}
  \label{img_bounce_period}
\end{figure}

Another important assumption is made regarding the bounce motion of the particles. I assume that energetic particles that enter the simulation box and exit it north or south along their motion along the magnetic field do not re-enter the simulation box after mirroring at high latitudes. 
In Figure \ref{img_bounce_period} the bounce period of the relevant EPD channels is shown for different pitch angles of the particles. The bounce period is calculated using Equation \ref{eq_bounce_period}. Indicated in these figures by the horizontal dashed line is the bounce period of a particle that would just re-enter the simulation box. This period is calculated by dividing the size of the simulation box by the corotational velocity (v$_{corot}$). The size of the simulation box is defined as the double of the maximum radial distance of the spacecraft trajectory which is used in the simulation (r$_{max} \approx$ 5000 km). From Figure \ref{img_bounce_period}  it can be deduced that during the bounce period of the considered particles, the motion of the corotational plasma and thus the motion of Europa with respect to the magnetic field exceeds the size of the simulation box. Thus, the particles have a single chance to interact with Europa and they do not have to be traced outside the simulation box boundaries. This reduces the simulation time significantly. Additionally, because the size of the simulation box is small with respect to the radius of the field lines, I neglect the gradient of the magnetic field that the energetic charged particles experience when they move to lower or higher latitudes.

I neglect the azimuthal drift described in Equation \ref{eq_drift_az}. It depends on energy and not on mass, thus the largest azimuthal drift occurs for the highest energy particles  considered in the simulation, which corresponds to the upper end of the TO3 channel. The slowest particles in the azimuthal sense, in this simulation, are the 90$^\circ$ pitch angle particles, that travel at the corotational speed with respect to Europa. For those particles the drift velocity is around 6 km/s. This is small with respect to the corotational velocity of 76 km/s, thus I neglect it. 

\clearpage

\subsection{Overview of the input models and parameters}
\label{ss_parameters}

Finally, an overview is provided of all the simulation inputs in Table \ref{tab_sim_settings}.

\begin{table}[h]
\centering
\begin{tabular}{|r|l|}
  \hline
  \textbf{Property} & \textbf{Value} \\
  \hline
  \hline
  Reference frame & x: direction of the corotational plasma \\
  & y: direction of Jupiter \\
  & z: along Europa's rotation axis \\
  \hline 
  Magnetic field [nT] & [100, 50, -450]\\
  & when simulating 12:00-12:06 [100, 0, -600]\\
  & from Figure \ref{img_components}  \\
  \hline 
  Induced magnetic field & \cite{Zimmer2000}, Equations \ref{eq_B_sec} to \ref{eq_mag_moment} \\
  A & 1 \\
  $\phi$ & 0 \\
  \hline 
   Electric field & Equation \ref{eq_E} \\
   Undisturbed electric field [V/m] & [0, -0.0342, -0.0038] (using field at 11:40) \\
  \hline
   Flow deflection & \cite{Ip1996}, Equations \ref{eq_ip_vx} to \ref{eq_ip_vy_iono} \\
  $\alpha$ & 0.4 \\
  R$_c$ & 1.0 R$_E$ \\
  V$_0$ [km/s] & 76 (in positive x direction) \\
  \hline
  Atmospheric model & \cite{Saur1998}, Equation \ref{eq_atm_saur} \\
  Atmosphere species & O$_2$ \\
  Surface density [cm$^{-3}$] & $10^9$ \\
  Scale height [km] & 100 \\ 
  \hline
  Energetic ion species & H$^+$ and O$^+$ \\
  \hline  
  Spectrum  of energetic charged particles & \cite{Pappalardo2009_Paranicas}, Figure \ref{img_energy_spectra} \\
  \hline
  Charge exchange cross section & H$^+$ on O$_2$: \cite{Basu1987} and  \\ 
   & O$^+$ on O: \cite{Lo1971}, Figure \ref{img_cross_sections} \\ 
  \hline
\end{tabular}
\caption{Overview of the input settings for the simulation. Justification for the assumed parameters can be found in Section \ref{s_sim_setup}.}
\label{tab_sim_settings}
\end{table}

\section{Simulations of energetic particle depletions}
\label{s_sim_results}
In this Section I will discuss the results of the simulations of the energetic ions during the E12 flyby, for the first three simulation cases discussed in Section \ref{s_sim_setup}. For the atmosphere a surface density of 10$^9$ cm$^{-3}$ and a scale height of 100 km are used as a reference case.

\begin{figure}[h]
  \centering
  \includegraphics[width=1.0\textwidth]{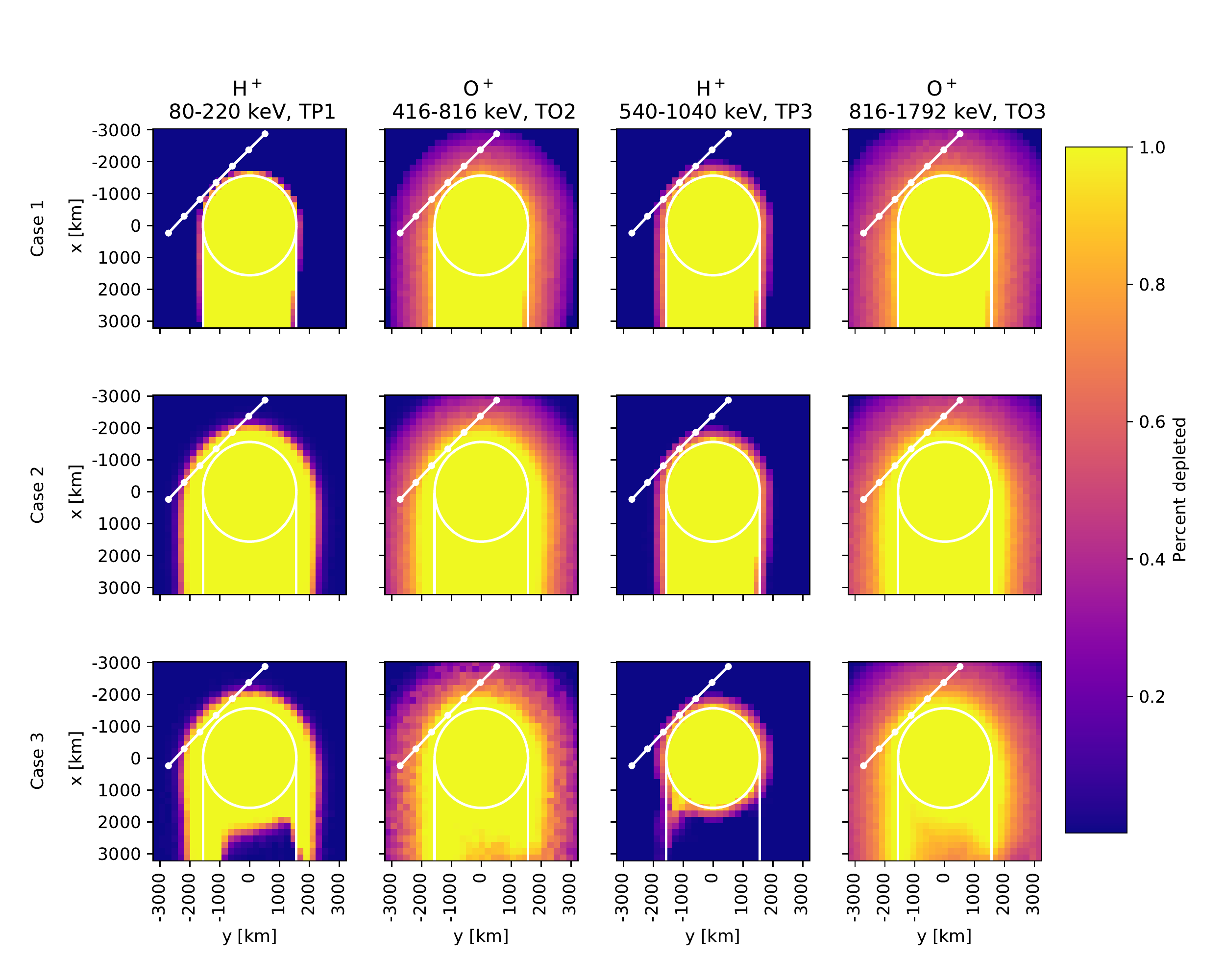}
  \caption{Spatial distribution of the depletion (90$^\circ$ pitch angle) in Europa's equatorial plane. The positive vertical axis points in the direction of the corotational plasma, the positive horizontal axis points towards Jupiter.   The white dotted line indicates the E12 flyby trajectory, white dots are separated by two minutes. The first dot corresponds to 11:57, the last one to 12:09. The white vertical lines indicate the geometric wake of Europa, the white circle indicates Europa's radius.}
  \label{img_depl_map_E12}
\end{figure}

The simulation results are presented in three different types of plot. The first one, Figure \ref{img_depl_map_E12}, shows the spatial distribution of the depletion for the particles with a 90$^\circ$ pitch angle. For this pitch angle the gyroradius of the particles will be the largest and therefore the distance from Europa at which they can be depleted is also the largest. The depletions are shown for energetic protons in EPD proton channels TP1 (80.0 to 220.0 keV) and TP3 (540.0 to 1040.0 keV), and for singly charged oxygen in oxygen channels TO2 (416.0 to 816.0 keV) and TO3 (816.0 to 1792.0 keV), to visualize the effect of the energetic particle energy on the depletion. For reference the E12 trajectory is also indicated.

\begin{figure}[h]
  \centering
  \includegraphics[width=1.0\textwidth]{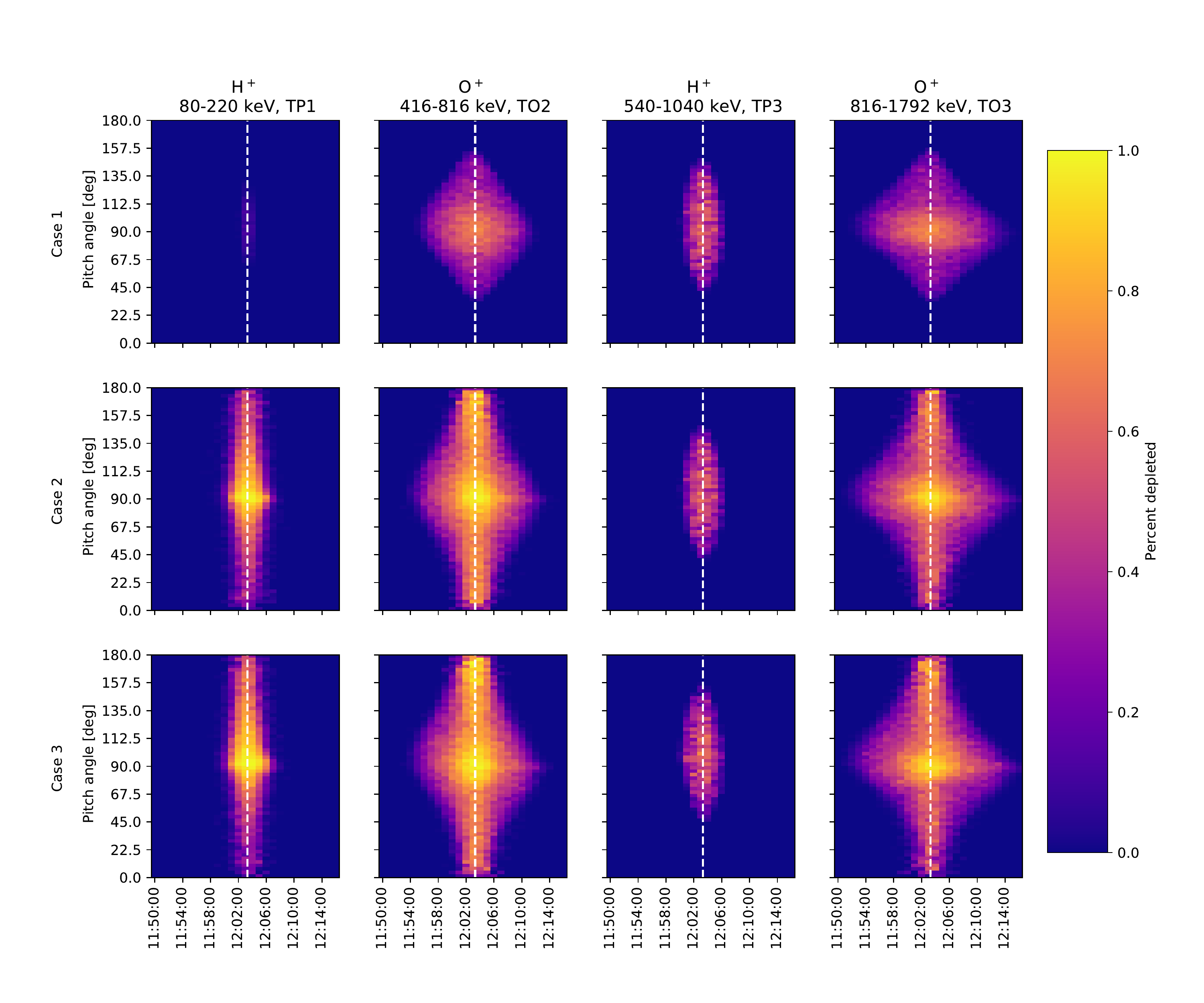}
  \caption{Pitch angle dependence of the depletion along the E12 trajectory. The dashed vertical line indicates the closest approach.}
  \label{img_pitch_angle_E12}
\end{figure}

\begin{figure}[h]
  \centering
  \includegraphics[width=1.0\textwidth]{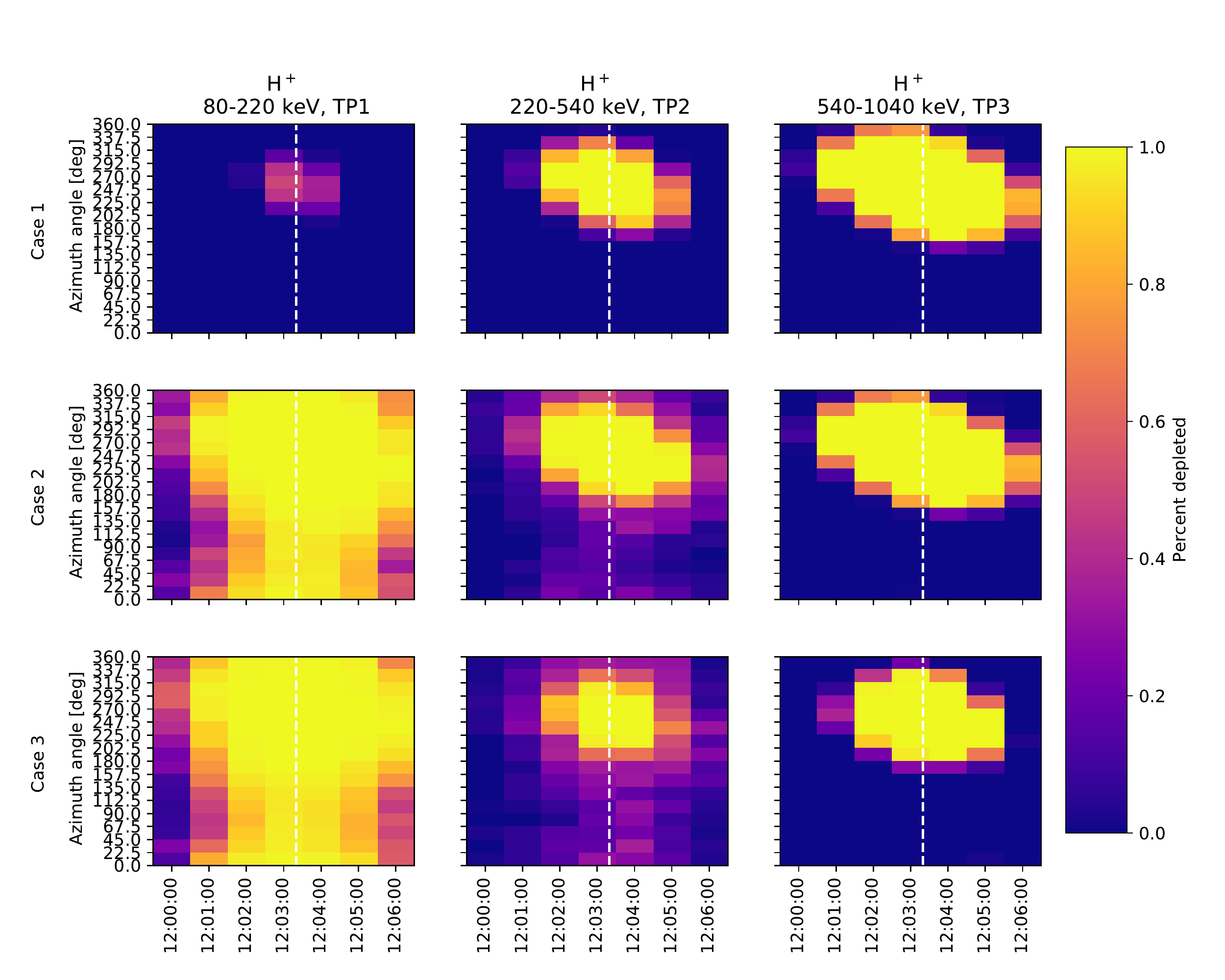}
  \caption{Directional dependence of the depletion for the 90$^\circ$ pitch angle, proton channels TP1, TP2 and TP3. The dashed vertical line indicates the closest approach.}
  \label{img_direction_tp}
\end{figure}

\begin{figure}[h]
  \centering
  \includegraphics[width=1.0\textwidth]{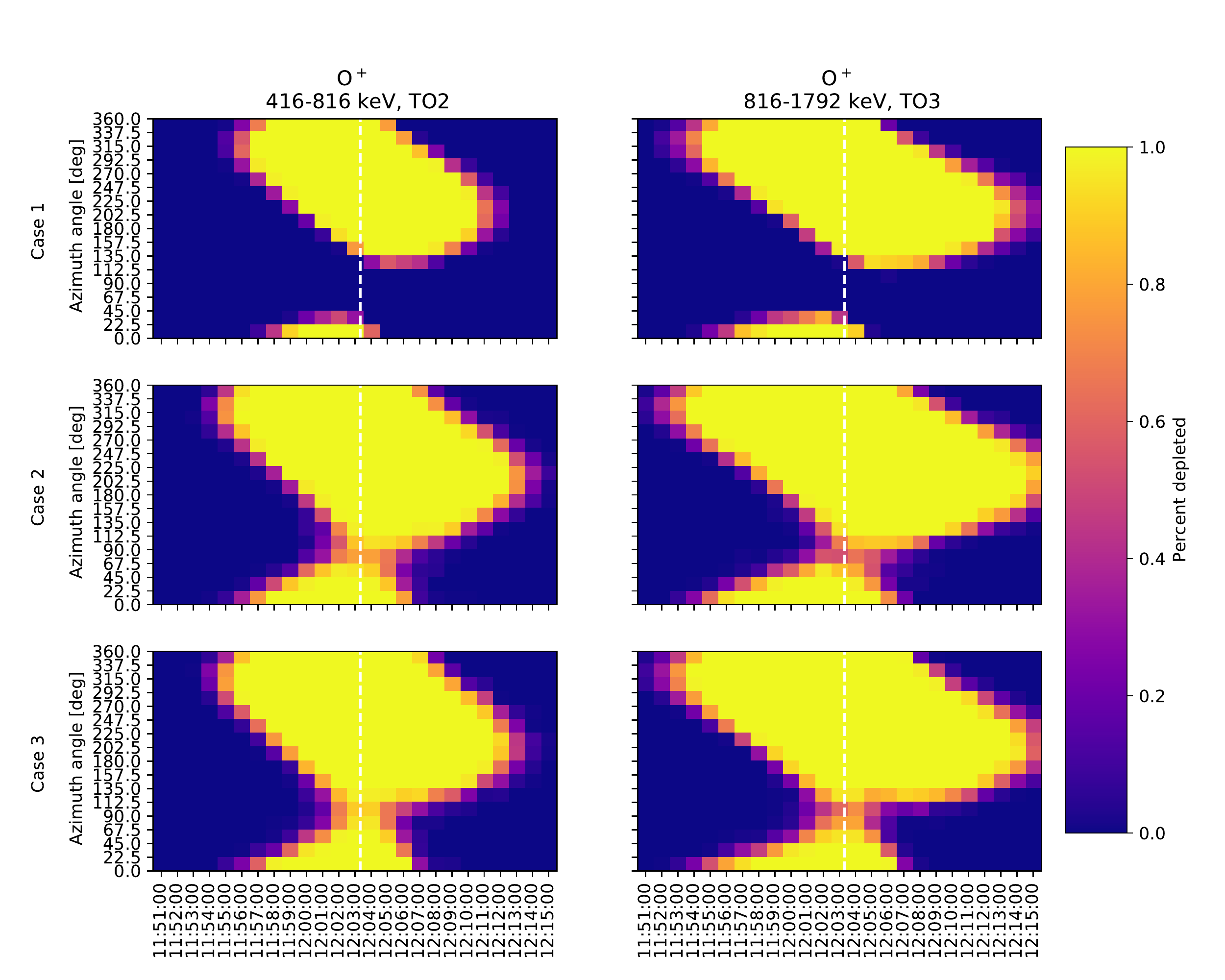}
  \caption{Directional dependence of the depletion for the 90$^\circ$ pitch angle, oxygen channels TO2 and TO3. The dashed vertical line indicates the closest approach.}
  \label{img_direction_to}
\end{figure}

Next I visualize the directional dependence of the energetic particle depletion:
first as a function of pitch angle and then as a function of azimuth with respect to the magnetic field line for a specific pitch angle. In the second type of plot, Figure \ref{img_pitch_angle_E12}, I show the variation as a function of pitch angle. The same EPD channels as Figure \ref{img_depl_map_E12} are shown. In the pitch angle plot it is not possible to differentiate between particles with the same pitch angle, but with a different azimuth angle with respect to the magnetic field. This is addressed in the third type of plot, Figure \ref{img_direction_tp}
 and \ref{img_direction_to}, in which the depletion is shown as a function of azimuth angle for the 90$^\circ$ pitch angle.  This azimuth angle is a measure for the gyrophase of the particles. The 0$^\circ$ azimuth angle corresponds to the positive y-axis in Figure \ref{img_depl_map_E12}. In this case I show the simulation results for the proton energy channels TP1, TP2 and TP3, and the oxygen energy channels TO2 and TO3. The vertical axis (azimuth) in Figure \ref{img_direction_tp}
 and \ref{img_direction_to} reproduces the pitch angle contour lines shown in Figure \ref{img_combined_fov_line}.

Case 1 in Figure \ref{img_depl_map_E12} shows that the number of energetic protons (80 to 1040 keV) and O$^+$ (416 to 1792 keV) can be depleted by impact on Europa's surface.
The depletion region extends beyond the radius of Europa, because of the size of the particles' gyroradii. Even though the guiding center of the particle does not pass through Europa, the particle is still depleted. This is because the distance between the gyrocenter and the surface is smaller than the gyroradius. The size of the gyroradius is determined by the mass and energy of the particles (see Equation \ref{eq_gyroradius}). The dependence on energy can be seen by comparing TP1 with TP3, and TO2 with TO3. The dependence on mass is visible by comparing the proton and the oxygen channels. It should be noted that the extent of the depletion will not keep increasing with increasing gyroradius. For very high energies the gyroradius of the particle can become larger than the size of the object (see TS3 in Figure \ref{img_gyro_tp_to_ts}), in that case the particle can gyrate around the object without impacting its surface. From the trajectory plotted in Figure \ref{img_depl_map_E12} it is also clear that the measured depletion depends strongly on the altitude of the spacecraft flyby.

Case 2 in Figure \ref{img_depl_map_E12} shows that the number of energetic protons (80-220 keV) and oxygen particles (416 to 1792 keV) can be depleted by charge exchanging with neutrals from Europa's atmosphere. The atmosphere increases the effective impact area of Europa, as can be seen in TP1, TO2 and TO3 but not in TP3. In TP3 no noticeable difference occurs between Case 1 and 2. This is because the charge exchange cross section in TP3 is significantly lower than in TP1, resulting in no significant atmospheric depletion.

Case 1 in Figure \ref{img_depl_map_E12} shows that the number of energetic protons (80-220 keV) and oxygen particles (416 to 1792 keV) is depleted in Europa's wake.
The width of the depletion region along the horizontal axis is the same as the width of the depletion along the upstream-downstream terminator. Case 2 in Figure \ref{img_depl_map_E12} shows that in the cases where the effective impact area of particles is increased by the atmosphere, the width of the depletion increases accordingly in the wake. Case 3 in Figure \ref{img_depl_map_E12} shows  that the inclusion of the dipole and the flow field changes the depletion downstream, while the depletion upstream is not affected. The dipole is not the major control factor for the difference downstream, as it affects upstream and downstream approximately equally. This can be seen in Figure \ref{img_induced_dip}. From this figure it is also clear that the maximum change in the field magnitude due to the dipole is 15 nT, which is only around three percent of the undisturbed magnetic field magnitude. This change will not result in a significant change of the trajectories, especially since the travel time of ions from locations along the E12 flyby to Europa is a few seconds, a fraction of the gyroperiod. The ion hits Europa and its tenuous atmosphere within such a short period that deflections by the field perturbations have no effect. As Figure \ref{img_orbit_magnetic} shows, E12 occurred when Europa was at the magnetic equator. At this point the induced dipole is weakest because the magnetic field components perpendicular to the rotation axis of Europa are smallest (see Section \ref{ss_induced_dip}). Thus the effect of the dipole could be more important during other flybys which occur at higher magnetic latitudes.
What causes the upstream-downstream asymmetry is the flow deflection. The flow deflection modifies the electric field (see Figure \ref{img_e_flow}), which modifies ExB drift motion of the energetic charged particles, such that they can enter the wake downstream of Europa. This implies that for modelling the depletion features upstream, the effect of the dipole and the flow deflection is small and can be neglected. Thus, for the E12 flyby the inclusion of the dipole and flow field are not critical. However, for modelling the depletion features in the wake of Europa it is important to include the effect of the flow deflection, otherwise the size of the depletion region would be overestimated.

Case 1 in Figure \ref{img_pitch_angle_E12} shows that the depletion of the number of energetic protons (80-220 keV) and oxygen particles (416 to 1792 keV)  varies as a function of the pitch angle. Pitch angles of 90$^\circ$ are depleted less than the pitch angles at 0$^\circ$ or 180$^\circ$. This is because for particles of the same energy, charge and composition, 90$^\circ$ pitch angle particles have the largest gyroradius (see Figure \ref{img_gyro_tp_to_ts}) and thus the most access to Europa's surface. This applies as long as the gyroradius is smaller than the size of the object. Non-90$^\circ$ pitch angle particles avoid Europa by moving north or south of the moon.

Case 2 in Figure \ref{img_pitch_angle_E12} shows that the number of energetic protons (80-220 keV) and oxygen particles (416 to 1792 keV) is depleted over more pitch angles than in Case 1, but only for those energy ranges for which the charge exchange cross section is non-negligible. TP3 and TO3 are depleted less at higher pitch away from 90$^\circ$ than, respetively TP1 and TO2. I attribute this to the decrease in charge exchange cross section, which is less dramatic for TO3 than for TP3. Case 3 in Figure \ref{img_pitch_angle_E12} shows that the difference in depletion between Case 2 and Case 3 is not significant. 
As previously explained, the ion hits Europa and its tenuous atmosphere within such a short period that deflections by the field perturbations have no effect. Thus the flow field does not affect the depletion strongly. I therefore consider the assumption of setting the ionosphere radius $R_c$ in the flow deflection model to Europa's radius justified.

Case 1 in Figure \ref{img_direction_tp} and \ref{img_direction_to} shows that the depletion is a function of the azimuth angle (gyrophase), it can be seen that the depletion can depend on azimuth.
Case 2 in Figure \ref{img_direction_tp} and \ref{img_direction_to} shows that the inclusion of an atmosphere causes the range of azimuth angles over which depletions occur to increase. Case 3 in Figure \ref{img_direction_tp} and \ref{img_direction_to} shows that the difference in depletion between Case 2 and Case 3 is not significant, this confirms the conclusion formulated for Figure \ref{img_depl_map_E12}: the induced dipole and the flow deflection do not affect the depletion significantly during an upstream flyby like E12.

\section{Comparison with data}
\label{s_data_compare}
In this section I compare the simulations with the data to determine the scale height and surface density of the atmosphere in Case 2 and 3 that are required to explain this observed depletion. 

\subsection{Proton channels (TP1, TP2 and TP3)}
First I compare the proton data to the output of the simulations in which the depletion is determined as a function of azimuth and time (such as those presented in Figure \ref{img_direction_tp}). These simulations show the depletion for the 90$^\circ$ pitch angle as a function of time and gyrophase (azimuth of the particles with respect to the magnetic field). I run these simulations for a range of surface density and scale height values. I consider densities in the range of 10$^6$ cm$^{-3}$ to 10$^{10}$ cm$^{-3}$ and scale heights from 50 to 200 km. For every resulting combination of density and scale height I test the simulation results for the following six criteria:
\begin{itemize}
\item \textbf{Criterion 1:} Start of the depletion in TP1 data and the corresponding percentage 
\item \textbf{Criterion 2:} End of the depletion in TP1 data and the corresponding percentage 
\item \textbf{Criterion 3:} Start of the depletion in TP2 and the corresponding percentage 
\item \textbf{Criterion 4:} End of the depletion in TP2 and the corresponding percentage 
\item \textbf{Criterion 5:} Start of the depletion in TP3 and the corresponding percentage 
\item \textbf{Criterion 6:} End of the depletion in TP3 and the corresponding percentage 
\end{itemize}
Each criterion is assumed to be met if in at least one direction (azimuth) the percentage of depletion is reached at the time set by the data. The numerical values for the criteria are shown in Table \ref{tab_sim_criteria}. These times are determined from the data shown in Figure \ref{img_combined_fov_line}, by visual inspection. Each criterion represents the time steps in the simulation that corresponds best to the start and end of the depletion in the data. 
\begin{table}[h]
\centering
\begin{tabular}{|r|l|l|}
  \hline
  \textbf{Criterion} & \textbf{Time} & \textbf{Percentage} \\
  \hline
  \hline
  1 & 12:01-12:02 & 0.9 \\
  \hline 
  2 & 12:04-12:05 & 0.2 \\
  \hline   
  3 & 12:01-12:02 & 0.6 \\
  \hline 
  4 & 12:06 & 0.5 \\
  \hline 
  5 & 12:01-12:02 & 0.9 \\
  \hline   
  6 & 12:05-12:06 & 0.9 \\
  \hline 
  \end{tabular}
\caption{Definitions of the criteria defined in the text.}
\label{tab_sim_criteria}
\end{table}
The uncertainty of the time determination is approximately one minute. Since one scan of the full sky by EPD takes about 140 seconds, a single direction of the sky is observed once only every 140 seconds. Therefore, if a depletion is observed for a certain direction during one full scan of the sky, but not during the previous full scan of the sky, it cannot be excluded that the depletion occurred in between the scans. Thus an uncertainty of about one minute occurs.

The percentage of depletion at the start of the depletion is obtained by normalizing the flux shown in the top panel in Figure \ref{img_combined_fov_line} by the average of the first minute of data far from Europa. The percentage at the end of the depletion is determined by normalizing with the last minute of data. I assume that the flux in the depleted channel is not influenced by noise, since the TP and TO channels are considered insensitive to background noise, as described in Section \ref{ss_epd}. Therefore, I assume that the observed depletions are entirely attributed to impacts on the surface and charge exchange with atmospheric particles. 

I use the same simulation settings as given in Table \ref{tab_sim_settings}, except for the magnetic field. For these TP channel simulations I use the magnetic field measured at 12:00, rather than at 11:40. I consider these field measurements more appropriate since this simulation covers the time frame 12:00 to 12:06. The field components at 12:00 are: [$100\cdot 10^{-9}$, 0, $-600 \cdot 10^{-9}$] nT in the same frame as specified in Table \ref{tab_sim_settings}, from Figure \ref{img_components}.

\begin{figure}[h]
  \centering
  \includegraphics[width=1.0\textwidth]{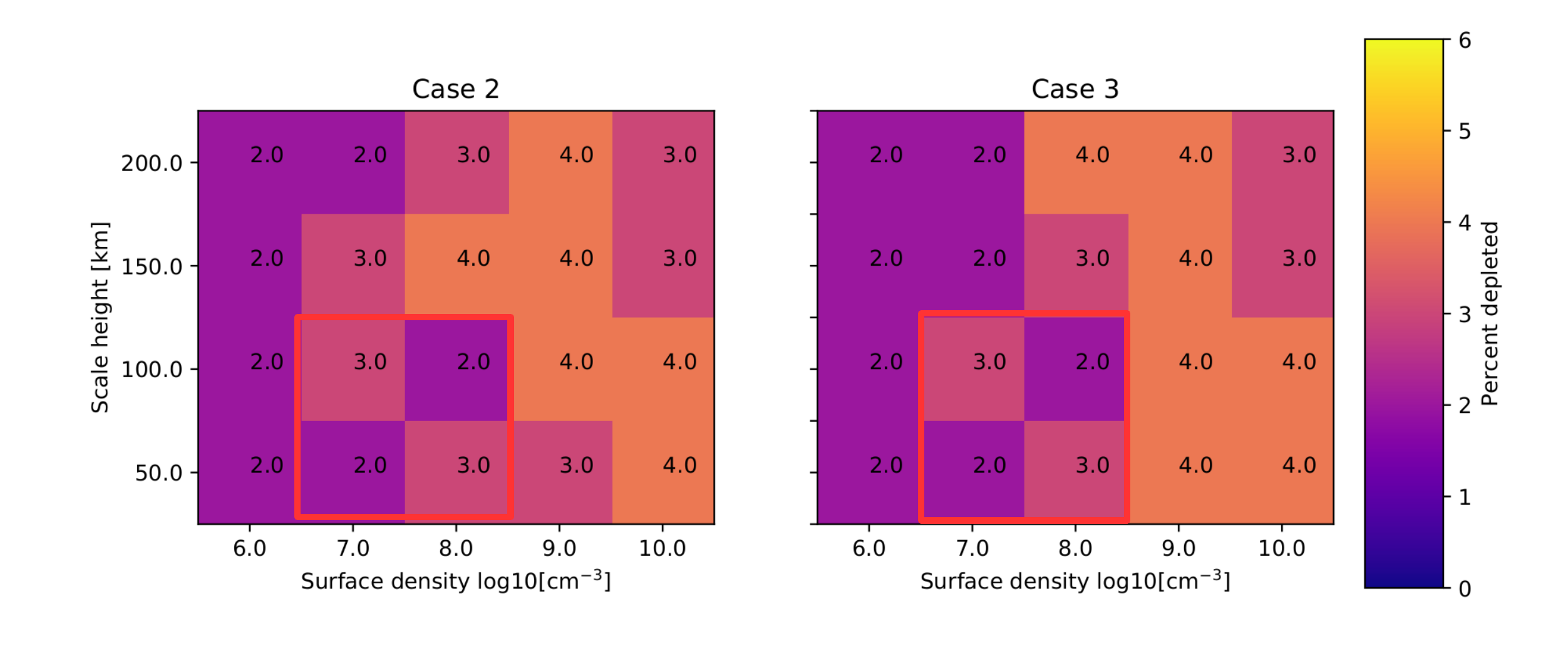}
  \caption{Number of criteria from Table \ref{tab_sim_criteria} that are met for each combination of scale height and surface density. Left: Case 2. Right: Case 3. The red boxes indicate the density and scale height range of the sublimated atmosphere from literature (Section \ref{s_atmosphere}).
  }
  \label{img_solutions_atm}
\end{figure}

The tests are done for Case 1, 2 and 3. In Case 1 the maximum number of solutions is limited to two, only the last two criteria are met. These are the start and end of the depletion in TP3. For none of the combinations of scale height and surface density TP3 is affected, this is because of the low charge exchange ratio in this channel (see Figure \ref{img_cross_sections}). In Case 2 and 3 solutions can be found for all of the first four criteria. Figure \ref{img_solutions_atm} shows the number of criteria that are met for each combination of surface density and scale height. From this figure it is clear that no solution can be found satisfying all the criteria at the same time. Also indicated in Figure \ref{img_solutions_atm}, by a red box, is the scale height and density range of the sublimated component of Europa's atmosphere from literature.
At lower altitudes (<300 km) the density of the sublimated component should exceed that of the sputtered component \citep{Vorburger2018}. Therefore, the sublimated component is possibly the most important source of depletions. The importance of the sublimated component will be discussed in more detail later in this section.
The solutions favour higher densities than the literature, but are not sensitive to the scale height (Figure \ref{img_solutions_atm}). 
Panel two in Figure \ref{img_solutions_atm} shows that the inclusion of the dipole and the flow deflection (Case 3) changes the simulation results slightly, but not significantly. Reasons for this have been discussed in Section \ref{s_sim_results}.

\begin{figure}[h]
  \centering
  \includegraphics[width=1.0\textwidth]{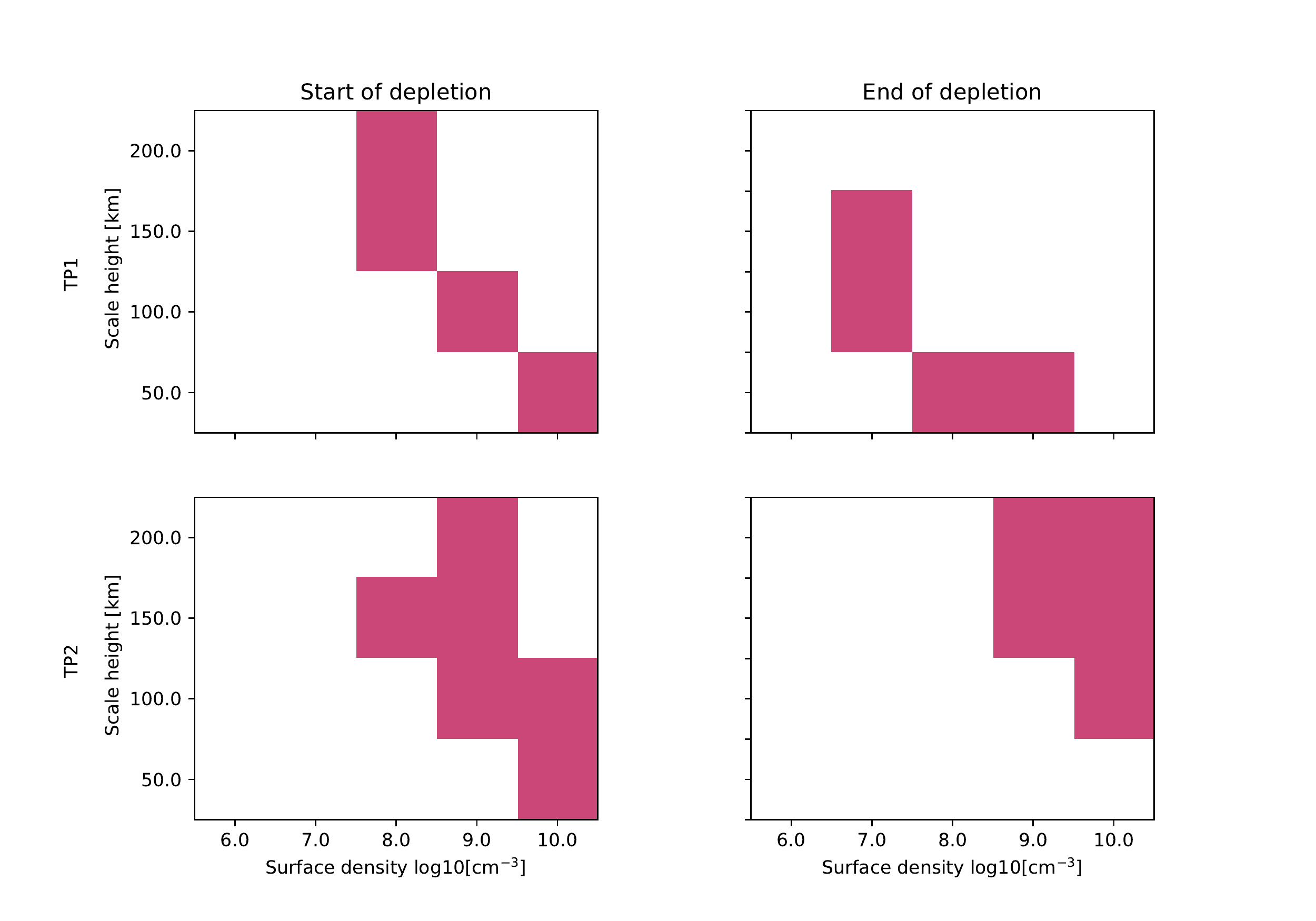}
  \caption{Solutions for criteria 1 to 4, in Case 2. Top row: criterion 1 and 2. Bottom row: criterion 3 and 4.}
  \label{img_solutions_atm_sep}
\end{figure}

Figure \ref{img_solutions_atm_sep} separates the solutions for criteria 1-4 in Case 2. For TP1 there is no solution in common between the start and end of the depletion, though some of the solutions are adjacent. For TP2 there are three common solutions between the start and stop. The start of TP1 and TP2 share three possible solutions, while the end of TP1 and TP2 share none. 
The difference in solutions between the start and stop might indicate that the atmosphere is not homogeneous and that there is an asymmetry between the inbound and outbound part of the flyby. As the model by \cite{Plainaki2012} shows (Figure \ref{img_atmosphere_plainaki}), the atmosphere of Europa is not spherically symmetric. \cite{Plainaki2013} states that the peak density occurs when the sub-solar point and the leading edge (towards the corotational flow) fall together.
During the E12 flyby the sub solar point is near the closest approach, as shown in Figure \ref{img_orbit_europa_lt}. Thus, it can be expected that the highest atmospheric densities occur on the leading side, nearer to the start of the depletion. The solutions for the start of the TP1 dropout indeed prefer higher densities than the stop, the same trend is not visible for TP2.

Since I used the magnetic field conditions at 12:00, which is closest to the start of the depletion feature, I expect that the simulation should reproduce the start conditions better than the end of the depletion. A further implication of using the magnetic field at 12:00 is that the size of the gyroradius is overestimated. In the data it is clear that the magnetic field increases after 12:00 up to 800 nT (Figure \ref{img_components}), which would result in a 30\% decrease of the gyroradius. In particular around 12:02 when the peak of the magnetic field strength occurs, the simulation will differ from the actual depletion. A smaller gyroradius would mean that certain particles will not hit the surface or reach the denser parts of the atmosphere, affecting the depletion feature. Near 12:02 it is thus possible that the contribution of surface impact to the total depletion is overestimated in the current simulation.

\begin{figure}[h]
  \centering
  \includegraphics[width=1.0\textwidth]{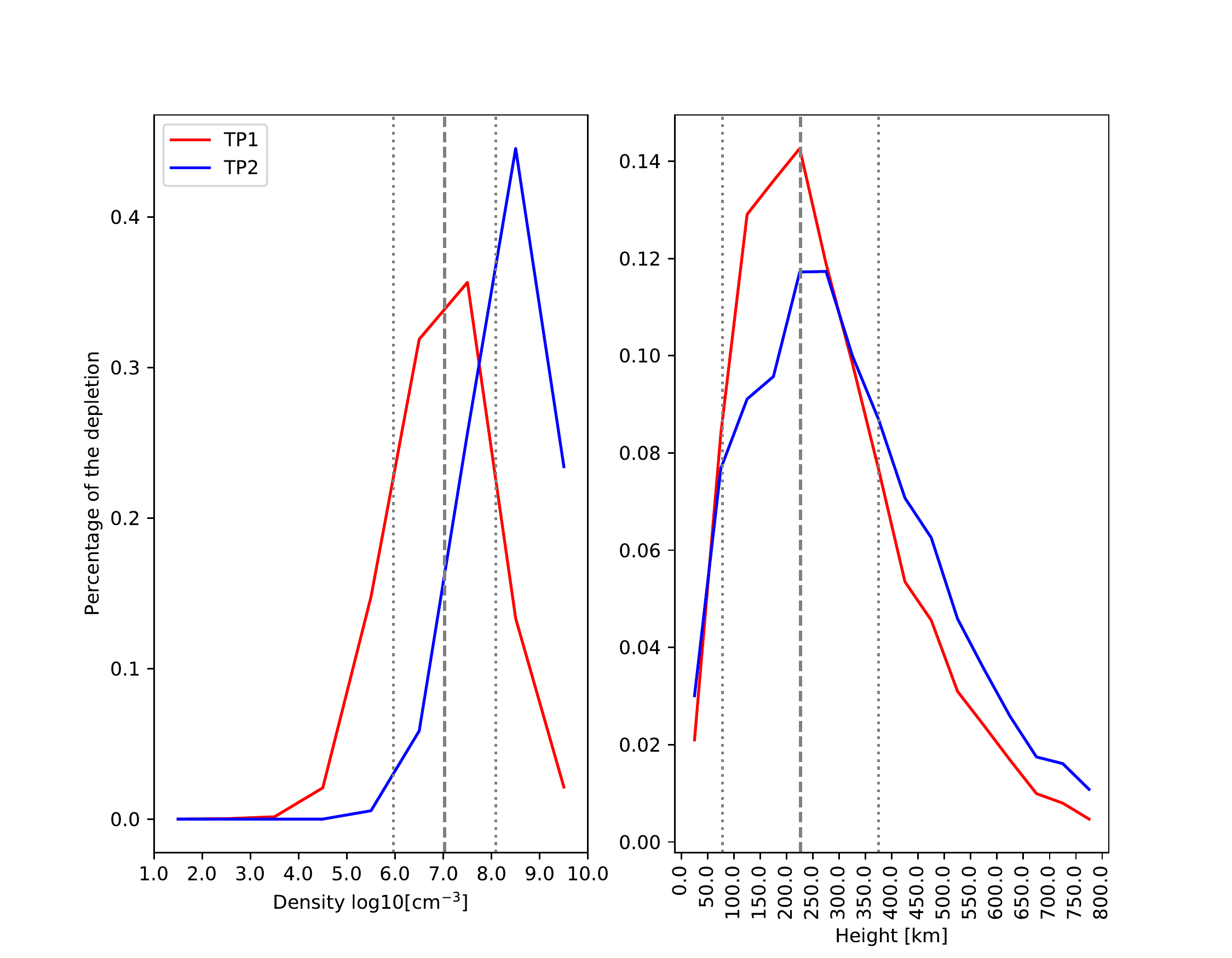}
  \caption{Histogram of density and altitude at which particles are depleted by charge-exchange with the atmosphere. Based on the solutions shown in Figure \ref{img_solutions_atm_sep}. TP3 is not shown as it is not sensitive to the combinations of density and scale height considered here. The mean of TP1 is indicate by the dashed vertical grey line, while the standard deviation for TP1 is indicated by the dotted grey lines.}
  \label{img_solutions_hist}
\end{figure}

Figure \ref{img_solutions_hist} shows a histogram of the density and altitude at which particles are lost due to charge exchange, based on the solutions in Figure \ref{img_solutions_atm} and \ref{img_solutions_atm_sep} for TP1 and TP2. Solutions for TP3 are not included in the histogram, since this channel is not sensitive to charge exchange for the range of densities and scale heights that are considered. To determine the altitude and density range over which most depletions occur in the TP1 simulations, the standard deviation of the density and altitude histograms of TP1 are shown in Figure \ref{img_solutions_hist}.
The standard deviation of the distribution for TP1 shows that in the simulations most particles are lost between altitudes of 50-400 km and between densities of 10$^7$ and $10^9$ cm$^{-3}$. This range of parameters is the range that is best constrained by the TP1 simulation results, and provides constraints for models of the atmosphere.

\begin{figure}[h]
  \centering
  \includegraphics[width=0.75\textwidth]{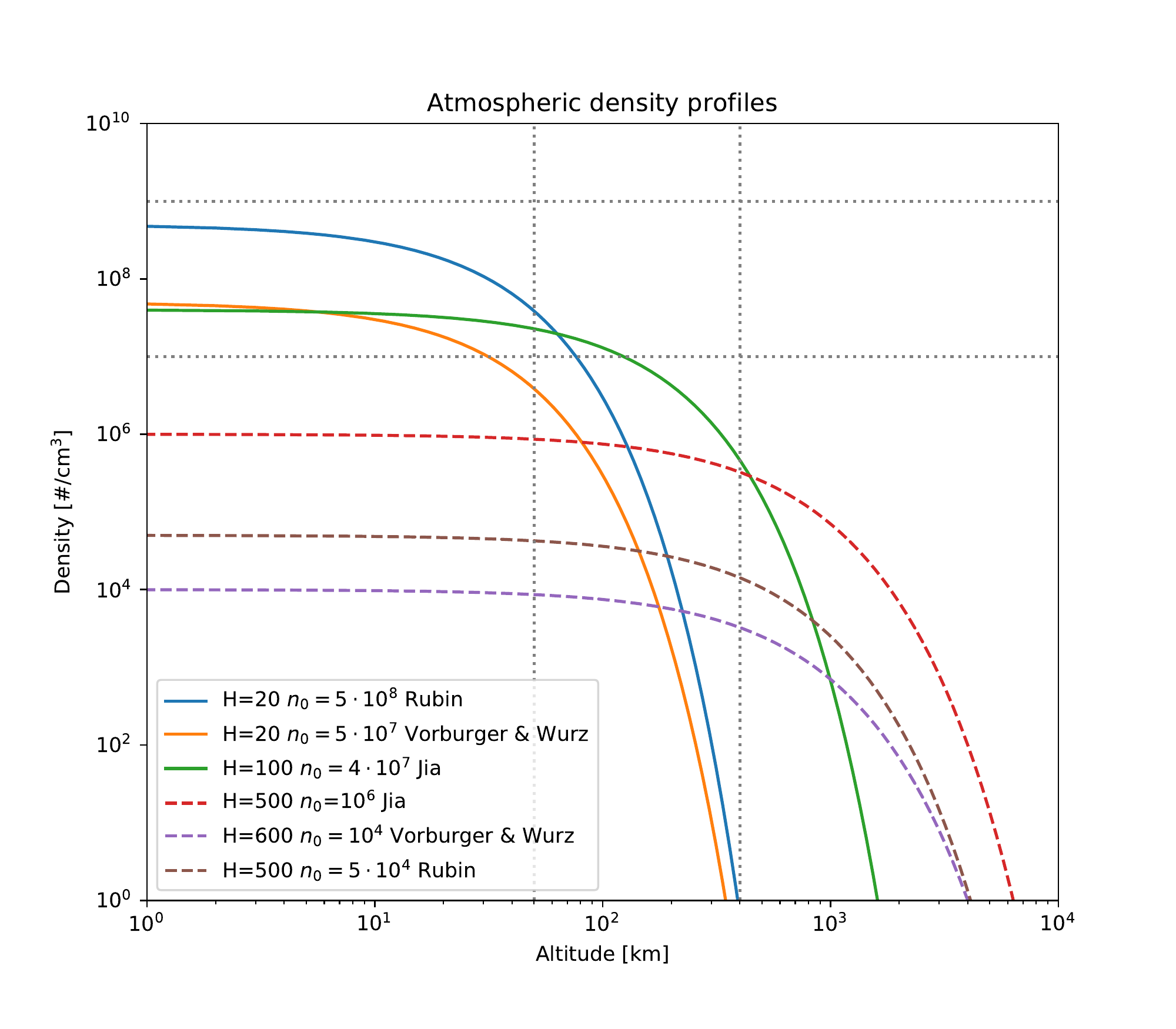}
  \caption{The solid coloured lines indicate several sublimated exospheric profiles, using the scale height and surface density values from \cite{Rubin2015,Jia2018,Vorburger2018} and Equation \ref{eq_atm_saur}. The dashed coloured line indicate the sputtered profiles from \cite{Rubin2015,Jia2018,Vorburger2018} The horizontal dotted grey lines indicate the standard deviation of the histogram in the left panel of Figure \ref{img_solutions_hist}, while the vertical grey lines indicate the standard deviation of right panel in Figure \ref{img_solutions_hist}.}
  \label{img_atm_jia}
\end{figure}

As described in Section \ref{s_atmosphere} the tenuous atmosphere is expected to have a sublimation component that is confined to lower altitudes than the extended sputtered atmosphere. Here I will compare the  the parameter range in which most particles are depleted in the simulations with sputtered and sublimated atmospheric profiles. The profiles are based on scale heights and surface densities from \cite{Rubin2015}, \cite{Jia2018} and \cite{Vorburger2018}. The atmospheric profiles and parameter range for TP1 are shown in Figure \ref{img_atm_jia}.
It is clear that none of the sputtered profiles pass through the parameter space in which most depletions occur in the TP1 simulation. Of these sputtered profiles, the one from \cite{Jia2018} is the sputtered profile with the highest surface density and scale height available in previous studies (see Section \ref{s_atmosphere}). Some of the sublimated components pass through the parameter space of highest surface density, or are close to it. Thus, the TP1 simulation results represent the sublimated component best, rather than the sputtered component. Determined by eye, the distribution of the depletion per density for TP2 in Figure \ref{img_solutions_hist} has a higher mean than the distribution for TP1, but a similar spread. Thus for TP2 depletions are mostly of the sublimated profile as well.

As was shown in Figure \ref{img_solutions_atm} the simulation results favour high surface densities (up to $10^{10}$). This is higher than the range of surface densities for the sublimated component in previous studies ($10^7$-$10^8$ cm$^{-3}$, see Section \ref{s_atmosphere}). Previous studies, e.g. \cite{Vorburger2018}, assume surface temperatures of $\sim130$ K based on the upper end of the observed surface temperature range obtained from Galileo measurements, which ranges from 80 to 130 K \citep{Spencer1999}. Thus, because sublimation is a function of surface temperature, significant changes in the surface density or scale height of sublimated components would require higher surface temperatures, outside of the observed temperature range. However, it is also possible that a systematic error remains in the simulation approach, caused by the choice of single exponential atmospheric profiles. Future simulations should consider two separate atmospheric components: a sputtered and a sublimated component. Also the scale height range of the simulation should be extended to lower scale heights, since previous studies suggest a scale height < 50 km for the sublimated component. 

\subsection{Oxygen channels (TO2 and TO3)}
In the data of the TO2 and TO3 channels depletions are visible between 12:01:00 and 12:06:30. Figure \ref{img_direction_to} shows that in Case 1 depletions of 100\% extend in TO2 from 11:57 to 12:10 and in TO3 from 11:55 to 12:12. The depletions in the Case 1 simulation extend well outside the approximately two minute uncertainty of the data. This suggests that not all relevant physical effects have been taken into account. I consider that this could be attributed to (a) the increase of the magnetic field near Europa and/or (b) multiple charge state of the oxygen particles. Both of these factors decrease the size of the gyroradius and thereby the effective impact area of the particles. However, since comparison of the simulations with proton (TP) measurements show that the field-pile-up effects may be important only around 12:02 I suggest that the larger inconsistencies for the oxygen simulations are indicative of a higher charge state.

The TO2/TO3 depletion of 100\% is visible in the simulation from an altitude of 1367 km altitude. The uncertainty in the data gives an altitude of 590-350 km for the start of the depletion, which is a significant difference.
Assuming that the change in altitude corresponds approximately to the reduction in size of the gyroradius, an estimate can be made of the contribution of the increase in field strength and multiply charged oxygen ions. The magnetic field magnitude in the simulation is approximately 450 nT. In the data of the E12 flyby the field at 1367 km is about 600 nT (Figure \ref{img_components}). This is a reduction of about 25\%. Since the gyroradius depends linearly on the magnetic field strength (Equation \ref{eq_gyroradius}), the change of the magnetic field can at most explain a decrease in size of the gyroradius by a 25\%. This means that to explain the observed reduction in size of the gyroradius requires a change of charge as well, which has to be between 1.74-2.94. This suggests that the energetic ions could have a higher charge state than the single charge I assumed, this is consistent with the results from \cite{Clark2016} which states that the charge of energetic oxygen is a mix between singly and doubly charged.

\section{Possible plume signatures during the E12 flyby}
\label{s_plume_e12}
\cite{Jia2018} has suggested that a plume could have been active during the E12 flyby.  Therefore, in this section I discuss case 4: the effect of a plume on the depletion feature. 

\begin{figure}[p]
  \centering
  \includegraphics[width=1.0\textwidth]{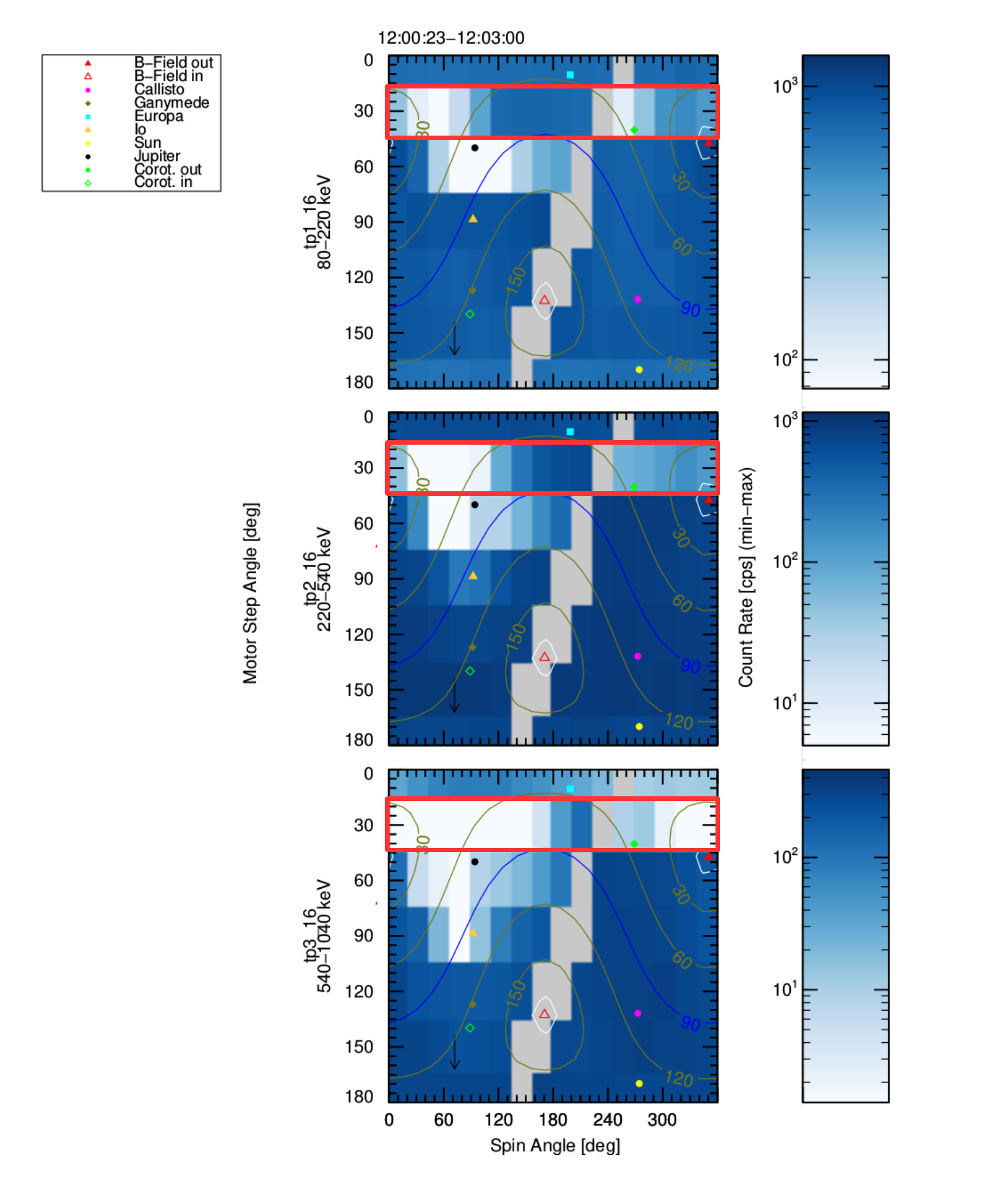}
  \caption{Detail of Figure \ref{img_combined_fov_line}. Count rate sky map for the energetic proton channels TP1, TP2 and TP3 between 12:00:23 and 12:03:00. The red box indicates the time period when the extended depletion in TP3 is visible. Each horizontal row represents one spin of the spacecraft. Each horizontal line starting shows one motor position. The sequence in which the measurements have been obtained is from the bottom row to the top row in each panel. The contour lines represent the pitch angle of the detected particles.}
  \label{img_1203_feature}
\end{figure}

\begin{figure}[h]
  \centering
  \includegraphics[width=1.0\textwidth]{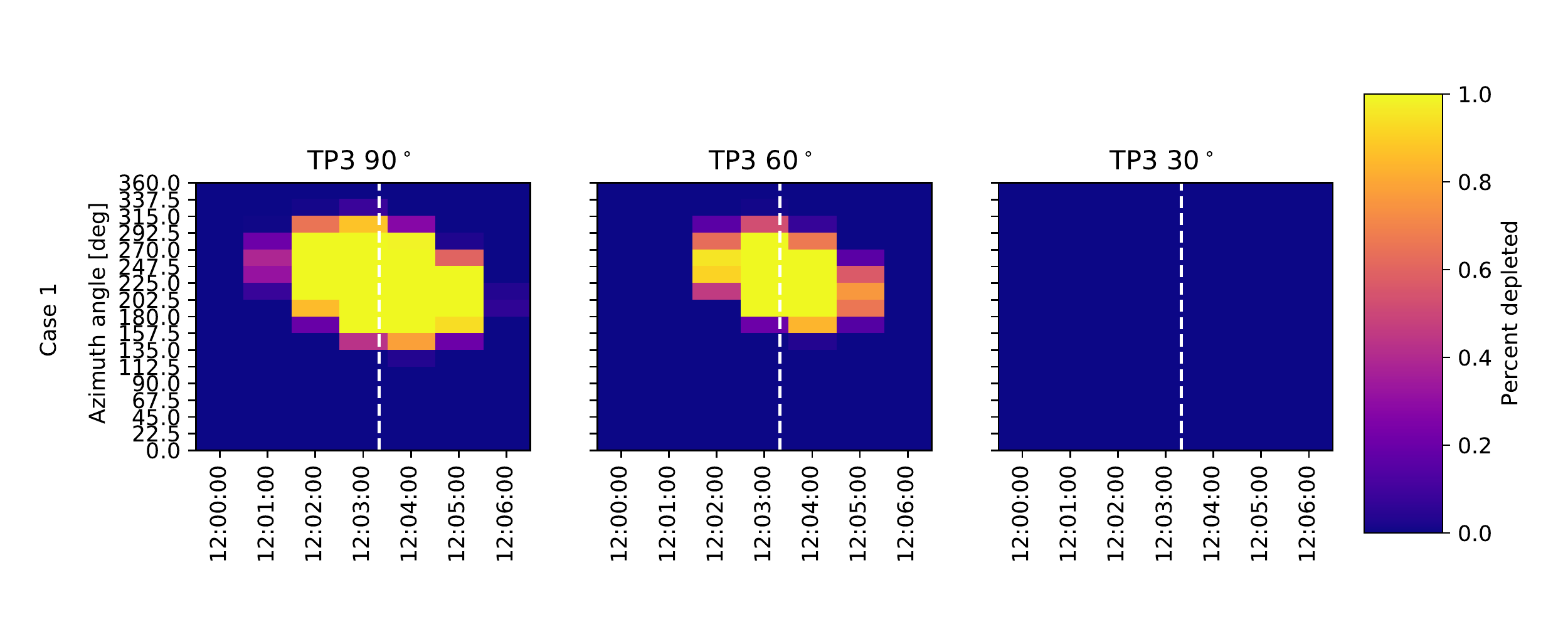}
  \caption{Simulated percentage of depletion in the energetic proton channel TP3 (540-1040 keV) versus time and azimuth with respect to the magnetic field line, for three pitch angles: 90, 60 and 30$^\circ$. The dashed vertical line indicates the closest approach.}
  \label{img_tp3_pitch}
\end{figure}

First I consider the EPD data presented in Figure \ref{img_combined_fov_line} again. \cite{Jia2018} reports a sudden peak in electron density at 12:02 and anomalous changes in the magnetic field over the period from 12:00 to 12:03, both associated with the plume. The EPD data that corresponds best to this period are shown in the third column of the second panel, which covers the time period from 12:00:23 to 12:03:00. A detail of this data is shown in Figure \ref{img_1203_feature}. Depletions can be observed during this period in all channels. A remarkable feature is visible in the TP3 channel. Between 12:02:20 and 12:03:00 the available data suggest that the 30 and 60 degree pitch angle protons are depleted over the entire azimuth range (indicated by the red box in Figure \ref{img_1203_feature}). At the previous time steps (the rows below the red box) the depletions only occur between 0 and 180 degree azimuth. The simulation results shown in Figure \ref{img_tp3_pitch} indicate that depletions over all azimuth directions do not happen in TP3 for Case 1. For the 30 degree pitch angle, no depletion should occur at all, yet a depletion is observed. An atmosphere with surface densities exceeding $10^{14}$ cm$^{-3}$ are required to deplete TP3 beyond the feature in Case 1 for the 30$^\circ$ pitch angle. 
Such a density is several orders of magnitude larger than the typical atmospheric surface density inferred from simulations and remote sensing observations. Such a density would also lead to broader and stronger ion losses that would easily be visible in other EPD channels (TP, TP2, etc.) and is therefore considered unrealistic. Thus, a globally symmetric atmosphere cannot explain the depletion feature visible in TP3 just before 12:03.

\begin{figure}[h]
  \centering
  \includegraphics[width=1.0\textwidth]{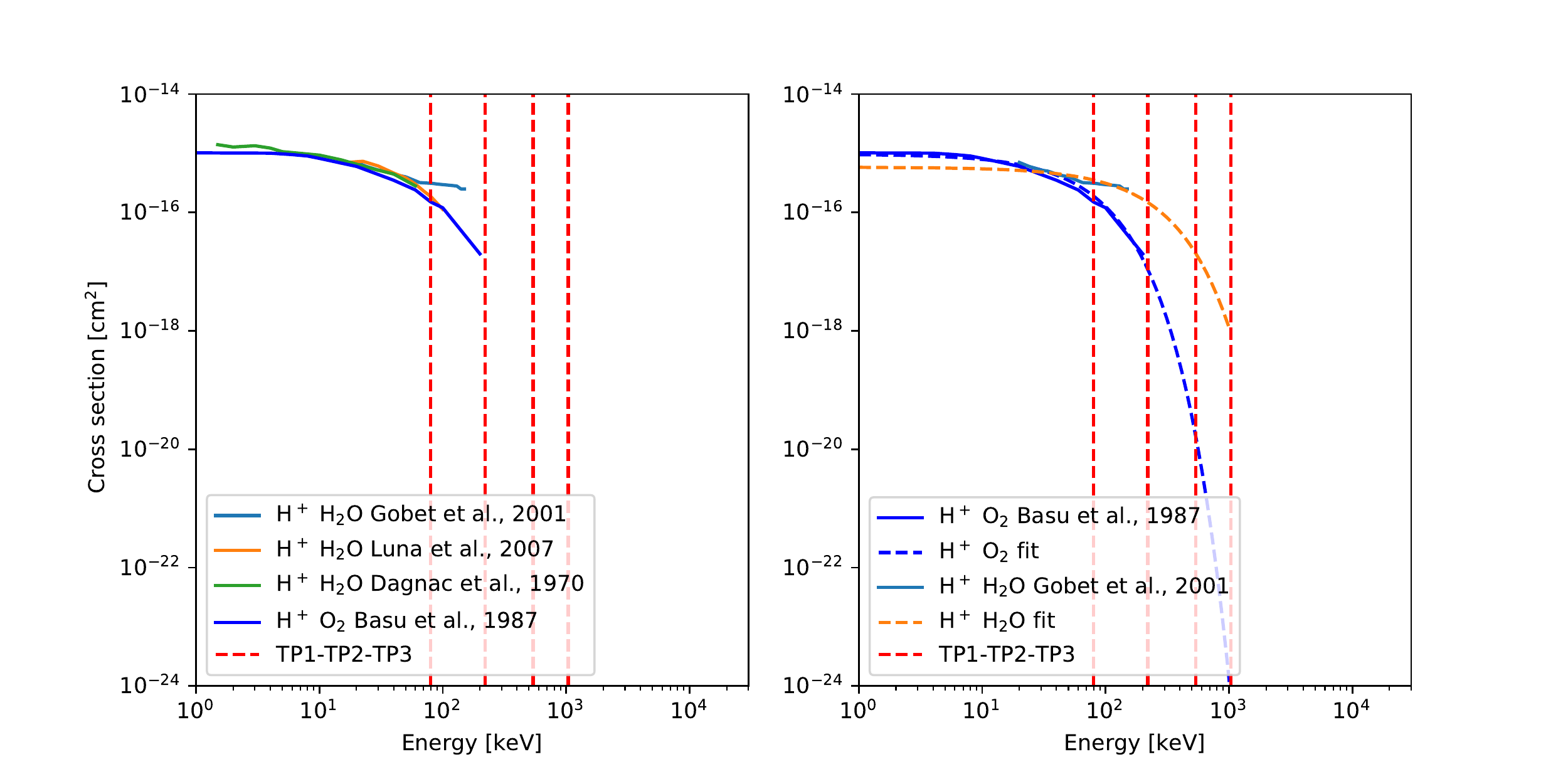}
  \caption{Charge exchange cross sections versus energy of H$^+$ on H$_2$O. For comparison the charge exchange cross section of H$^+$ on O$_2$ from \cite{Basu1987} is also shown. The dashed vertical red lines indicate the energy range of the EPD proton channels: TP1 (80-220 keV), TP2 (220-540 keV) and TP3 (540-1040.0 keV). Original sources for the charge exchange cross sections: \cite{Dagnac1970,Gobet2001,Luna2007}.}
  \label{img_cross_sections_h2o}
\end{figure}

\begin{figure}[h]
  \centering
  \includegraphics[width=1.0\textwidth]{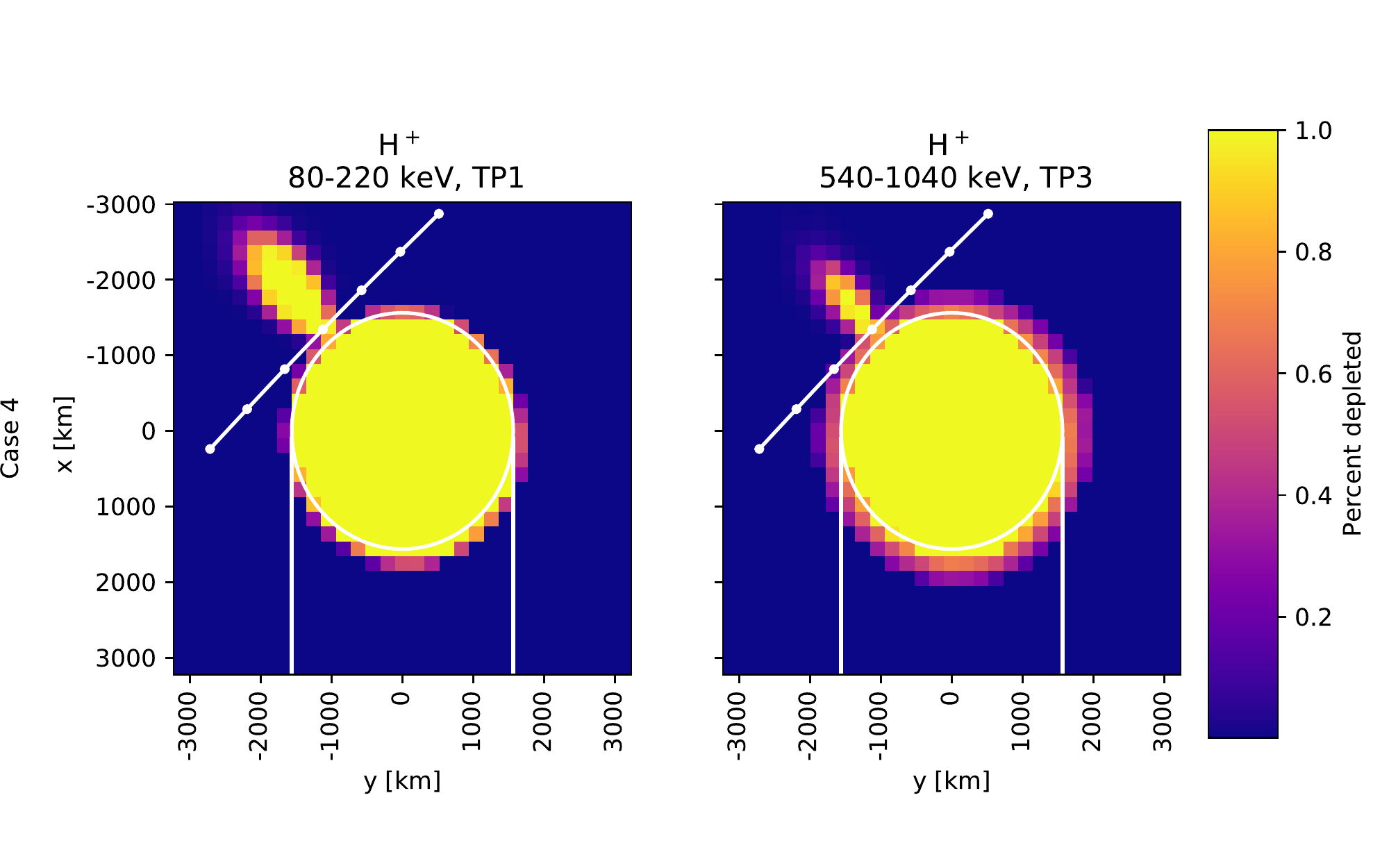}
  \caption{Spatial distribution of the depletion (60$^\circ$ pitch angle) in a plane at -10$^\circ$ latitude. The positive vertical axis points in the direction of the corotational plasma, the positive horizontal axis points towards Jupiter. The depletions are shown for energetic protons in the TP1 (80-220 keV) and TP3 channel (540-1040 keV). The causes of depletion are impact on Europa's surface and charge exchange with a plume (Case 4). The plume is located at 130$^\circ$ longitude and $-10^\circ$ degree latitude in the IAU Europa frame. The plume is modelled using Equation \ref{eq_plume_jia}, using $H_p$ = 150 km, $\theta_p$  = 5$^\circ$ and $n_{p0} = 10^{15}$cm$^{-3}$. The charge exchange cross section by \cite{Gobet2001} is shown in the right panel of Figure \ref{img_cross_sections_h2o}. The white dotted line indicates the E12 flyby trajectory, white dots are separated by two minutes. The first dot corresponds to 11:57, the last one to 12:09. The white vertical lines and the white circle indicate the geometric wake of and Europa's radius in the equatorial plane. }
  \label{img_tp1_tp3_depl_map}
\end{figure}

\begin{figure}[h]
  \centering
  \includegraphics[width=1.0\textwidth]{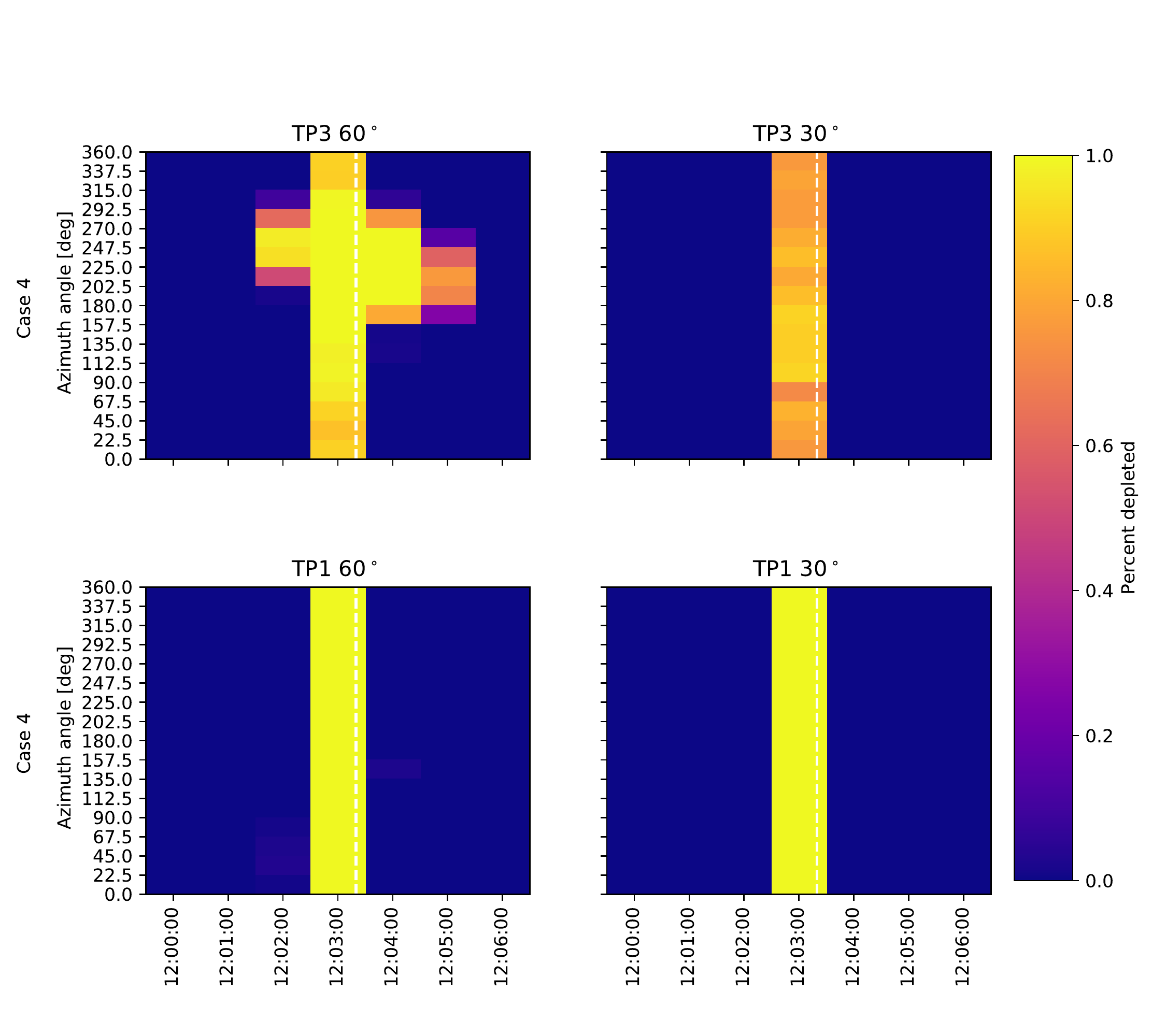}
  \caption{Case 4: depletion of energetic protons in the TP3 channel (540-1040 keV) caused by impact on Europa's surface and charge exchange with a plume located at 130$^\circ$ longitude and $-10^\circ$ degree latitude in the IAU Europa frame. The plume is modelled using Equation \ref{eq_plume_jia}, using $H_p$ = 150 km, $\theta_p$  = 5$^\circ$ and $n_{p0} = 10^{15}$cm$^{-3}$. The charge exchange cross section by \cite{Gobet2001} is shown in the right panel of Figure \ref{img_cross_sections_h2o}. The dashed vertical line indicates the closest approach.}
  \label{img_tp1_tp3}
\end{figure}

Next I consider if the localized plume of \cite{Jia2018} could explain the depletion feature just before 12:03 in the TP3 data. I use the analytical model from Equation \ref{eq_plume_jia} to simulate the density distribution of the plume. I assume that the plume has a scale height of 150 km (as in \cite{Jia2018}) and an angular width $\theta_p$ of 5 degree (narrower than \cite{Jia2018}). I assume a surface density of $10^{15}$cm$^{-3}$, which is significantly larger than the assumption made in \cite{Jia2018}, but on the same order of magnitude as in \cite{Roth2014a} (see Table \ref{tab_properties_plumes}). I neglect the assumed tilt of the plume from \cite{Jia2018} for simplicity. The position of the plume is 130 degree longitude and -10 degree latitude in the IAU Europa frame, while the plume from \cite{Jia2018} is located at 155 degree latitude in the same frame. I assume a $H^+$ on H$_2$O charge cross section. Several of these cross sections are shown in Figure \ref{img_cross_sections_h2o}. As can be seen in this figure in two of three cases the cross section is not significantly different from the $H^+$ on O$_2$ cross section, while the curve derived by \cite{Gobet2001} shows a higher cross section at higher energy. 
For this simulation I will use \cite{Gobet2001}. The results of the simulation are shown in Figure \ref{img_tp1_tp3_depl_map} and \ref{img_tp1_tp3}.

The feature in the data is only visible for the duration of two scans, thus for about 40 seconds. The absence of any depletion between 180 and 360 degree azimuth in the scans immediately before and after suggest that the feature occurs indeed very briefly. In the simulations the time steps are separated by one minute, thus the  12:03 feature should only be visible for one time step. The simulations in Figure \ref{img_tp1_tp3} show that such a narrow feature can indeed be created using the analytical plume model from \cite{Jia2018}, by assuming a narrower angular width.

The plume location I selected makes the peak of the plume related depletion occur at 12:03. Moving the plume away from the assumed position either 5 degree in longitude or latitude, would affect this timing. Using the position from \cite{Jia2018} would result in a peak in depletion not before 12:06. 

Figure \ref{img_tp1_tp3_depl_map} shows that the size of the depletion feature in TP1 is more extended than in TP3, for the $60^\circ$ pitch angle. The difference is explained by the decrease of the charge exchange cross section in TP3. From Figure \ref{img_tp1_tp3} it is clear that for the same plume feature, TP3 is depleted less than TP1 along the E12 trajectory. This in particular clear for the 30$^\circ$ pitch angle plot. However in the data, just before 12:03 the TP3 channel is depleted over more azimuth angles than TP1, as can be seen in Figure \ref{img_1203_feature}. This feature is not reproduced by my simulation. I suggest the absence of a magnetic field gradient drift in the simulation as a possible explanation. The magnetic field strength peaks at 12:02, and immediately afterwards, between 12:02 and 12:03, there is a strong change in magnetic field direction (see Figure \ref{img_components}). Previously \cite{Paranicas2000} suggested that the gradient drift contributes to the strong depletions in EPD energetic ion and electron channels around 12:02 (see Section \ref{ss_interaction_EPD}). The magnetic drift motion is important when there is a gradient in the magnetic field on the scale of the gyroradius, such that the gyroradius can vary over one gyration (see Section \ref{sss_gradB}). 
The scale of the magnetic field peak and change in direction is about 600 km along the Galileo trajectory. This is larger than the gyroradius of the particles in TP1 or TP3 (< 300 km), thus both TP1 and TP3 could be affected by a gradient drift. 
As Equation \ref{eq_gradB} shows, the drift speed increases linearly with energy, the effect of the gradient on TP3 should be larger than TP1. The gradient drift could thus help explain why TP3 is depleted more than TP1.
The gradient drift could have an effect too on the previously discussed solutions for the start and end time of the depletion feature, however the effect will be the largest near 12:02 where the gradient is the most extreme.
To determine the contribution of the gradient drift a simulation using a model of the field that includes the peak of magnetic field strength and the change in direction is required.

\section{Main results and recommendations for future work}
I simulated the depletion of energetic protons and oxygen ions due to surface impact and charge exchange during Galileo's E12 flyby. The results show that these energetic ions are depleted by Europa's surface. The depletion region can extend beyond the radius of Europa, because of the size of the gyroradius of the particles involved. The spatial distribution of the depletion thus depends on the energy, pitch angle, mass of the particles, which each affect the gyroradius. Charge exchange with neutral particles is another possible loss process, charge exchange increases the effective impact area of the moon. The simulation shows that the observed depletion strongly depend on the looking direction of the instrument and the flyby trajectory. The effect of the flow deflection on the depletion of energetic ions for upstream flybys is negligible, at least for E12. The gyroradius of particles is similar to the spacecraft altitude, hence the distance they cover during their gyromotion is much larger than the distance covered because of the flow. However, for wake flybys the effect of the flow deflection is important. The dipole is weak during the E12 flyby, hence its effect is not significant.

The depletion of energetic protons of energies between 80 and 220 keV (TP1) in the data is more than 90\%. This exceeds the calculated depletion in my simulation of the atmosphere-less case, which suggests that additional losses caused by charge exchange with atmospheric particles are be occurring. I have simulated the depletion in all the energetic proton channels for a range of surface densities and scale heights. I have compared the extent in time and percentage of the resulting depletions to determine which combinations of scale height and density fit best to the data. The simulations show that the depletions of energetic ions observed in the range 80 to 540 keV can be explained by charge exchange with neutral atmospheric particles, not by impact on Europa's surface alone. This suggests that an atmosphere must have been present during the E12 flyby. The simulation results best represent the sublimated component of the atmosphere. The solutions favour higher surface densities (10$^8$-10$^{10}$ cm$^{-3}$) than the literature (10$^7$-10$^8$ cm$^{-3}$), but are not sensitive to the scale height.
Differences between the solutions for the inbound and the outbound part of the flyby suggest that an asymmetry of the atmosphere could exist. Including Europa's dipole and a flow deflection in the simulation gives slightly different results, but does not significantly alter the conclusion.

I have also modelled the extent of the depletion in the oxygen channels TO2 and TO3. In the case with no atmosphere the extent of the simulation exceeds that of the depletion in the data of these channels. This suggests that other effects need to be taken account that reduce the gyroradius. As the maximum increase of the magnetic field is known, I propose that an increase of the charge above one is required to explain the reduction of the gyroradius. Charge states exceeding two are possible.

A depletion feature is seen in the energetic protons between 540 and 1040 keV, slightly before 12:03. This feature is not consistent with a global atmosphere. I demonstrate that a narrow depletion feature in this channel can be explained by a dense, narrow plume. However, the simulation efforts cannot reproduce the depletion signature measured at the same time in the lower energy range 80 and 220 keV, which is less depleted in the data. Future simulation efforts taking into account the additional field perturbation effects near Europa should be conducted, to determine to which extent they contribute to the depletion and if they could be responsible for this discrepancy.

For future studies to model the depletions of energetic ions near Europa in the context of determining atmospheric properties, I suggest the following:
\begin{itemize}
\item Simulate the depletions using a model of the magnetic field that reproduces the magnetic field pile up, the peak in magnetic field strength near 12:02 and the change in direction of the field just after that, as observed in the data. All these effects will influence the depletions, the increase in field strength will reduce the gyroradius of particles and the gradient in the field could cause a gradient drift. The field could be simulated using an MHD model of Europa's interaction with the magnetospheric plasma, such as \cite{Jia2018}.
\item Simulate multiply charged energetic oxygen ions to determine if they explain the observed extent of the depletion in the data, which is smaller than in the simulation. Multiple charge reduces the size of the gyroradius and could reduce the area over which energetic oxygen is depleted.
\item Use more realistic atmospheric profiles, which are not spherically symmetrical. They should include multiple components at the same time: a low altitude confined atmospheric component and an extended atmosphere. The analytical approach from \cite{Jia2018} could be used for this.
\item Simulate the case where a plume and one or more atmospheric components
occur at the same time. 
\item Use a more realistic plume model and perform a wider investigation of the plume parameter space (plume location, angular width, density and tilt) in combination with an atmosphere to investigate the contributions of each.
\item Simulate the depletions in the TS channels, when charge exchange cross sections become available.
\end{itemize}

\chapter{Predictions for future missions}
\label{ch_prospects}

In Chapter \ref{ch_atmosphere} it was recommended that future simulations of the losses of energetic ions should use a more realistic plume model. In this chapter\footnote{Partial results of this chapter were published in advance in the following contribution: Huybrighs, H.L.F., Futaana, Y., Barabash, S., Wieser, M., Wurz, P., Krupp, N.,  Glassmeier, K.H., Vermeersen. B., 2017, On the in-situ detectability of Europa's water vapour plumes from a flyby mission. Icarus 289, 270-280.} a start is made for this task by simulating the density distribution of the plume. These simulation results will be used to make predictions for future missions to Europa. In particular I investigate the in-situ detection of particles of Europa's plume with the in-situ instruments onboard of the future JUpiter ICy moon Explorer (JUICE) mission. I consider both H$_2$O and H$_2$O$^+$ particles, that could be detected, respectively, by the Neutral gas and
Ions Mass spectrometer (NIM) and the Jovian plasma
Dynamics and Composition analyser (JDC) instruments, both part of JUICE's Particle Environment Package (PEP). Previously, in \cite{Huybrighs2015} it was demonstrated that the signal of H$_2$O and H$_2$O$^+$ particles from the plume exceeds the noise level of both the  instruments, therefore these particles are detectable. \cite{Huybrighs2015} recommended that the signal of the plume particles should be compared to the signal of atmospheric H$_2$O and H$_2$O$^+$ to determine if the plume and atmospheric particles can be separated. That topic is addressed in this chapter. I use an updated version of the non-collisional Monte Carlo particle tracing software from \cite{Huybrighs2015} to simulate the detection of plume-originating H$_2$O and H$_2$O$^+$ particles in the presence of atmospheric H$_2$O and H$_2$O$^+$.

First, in Section \ref{s_future_missions}, I will briefly introduce two future space missions that are set to visit Europa: JUICE and Europa Clipper. Next, in Section \ref{s_plume_sim}, I will provide a brief overview of existing plume models. After that, in Section \ref{s_plume_sim_setup}, I discuss the set-up of my simulations of future measurements. The results of these simulations will then be discussed in Section \ref{s_plume_sim_results}.

\section{Future missions to Europa}
\label{s_future_missions}
This section introduces two missions that are currently planned to visit Europa in the early 2030's: JUICE and Europa Clipper.

\subsection{JUpiter ICy moons Explorer (JUICE)}
\label{s_juice}
\begin{figure}[h]
  \centering
  \includegraphics[width=0.5\textwidth]{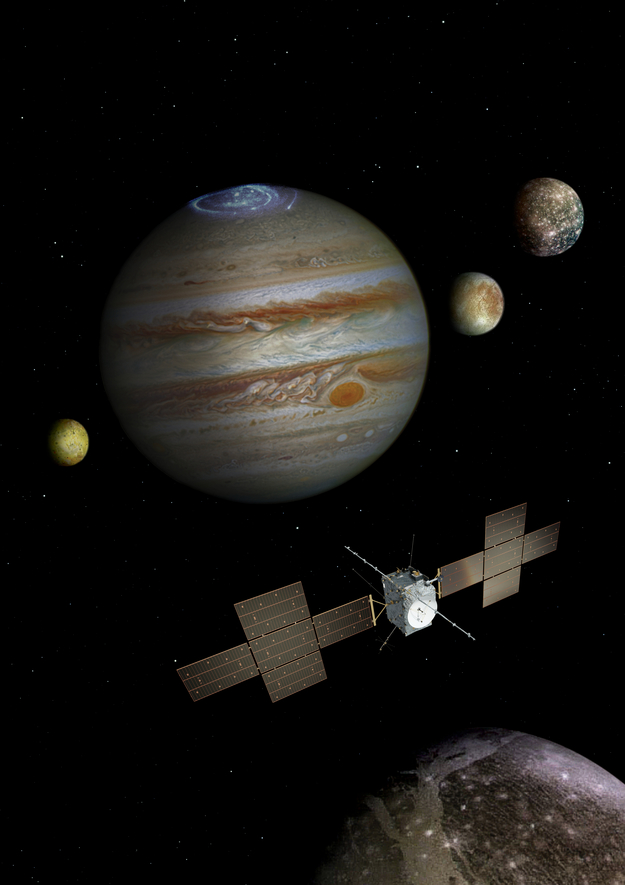}
  \caption{Artist impression of the JUICE mission. [ESA/ATG medialab/NASA/J. Nichols (University of Leicester)/University of Arizona/JPL/DLR http://sci.esa.int/juice/59341-juice-mission/]}
  \label{img_juice}
\end{figure}

The JUpiter ICy moons Explorer (JUICE) is a future ESA mission that is scheduled to launch in 2022 and arrive in the Jupiter system in 2029 (Figure \ref{img_juice}). The mission was selected in 2012 as ESA's first L-class mission. It is tasked to address two themes from ESA's Cosmic Vision 2015-2025:
\begin{itemize}
\item Theme 1: What are the conditions for planet formation and the
emergence of life?
\item Theme 2: How does the Solar System work?
\end{itemize}
The primary focus of JUICE is the moon Ganymede, which the spacecraft will enter orbit around in the later phases of the mission. Before entering Ganymede orbit, two Europa flybys, as well as 12 Gaymede and 13 Callisto flybys, are planned. The Europa flybys are scheduled for 2031 and will approach Europa up to a distance of 400 km. JUICE will carry 11 instruments, in-situ and remote sensing \citep{Grasset2013}. Of special interest for this thesis is the Particle Environment Package (PEP) a package of six particle sensors to measure the neutral and charged particle environment in-situ. Later in this chapter I will simulate the detection of plume particles by two sensors part of PEP. The first one is the Jovian plasma
Dynamics and Composition analyser (JDC), an ion time of flight ion mass
spectrometer. It provides the energy and directional distribution of ions with energies in the range 1 eV - 41 keV and it can achieve a mass resolution (M/$\Delta$M) up to 30. 
The second instrument is the the Neutral gas and Ions Mass spectrometer (NIM). This instrument, based on an earlier version for lunar research \cite{Wurz2012},  provides the density of low energy neutral and ionised particles (energies < 10 eV). Furthermore it has a very high mass resolution (M/$\Delta$M > 1100).

\subsection{Europa Clipper}
\label{s_clipper}

\begin{figure}[h]
  \centering
  \includegraphics[width=0.5\textwidth]{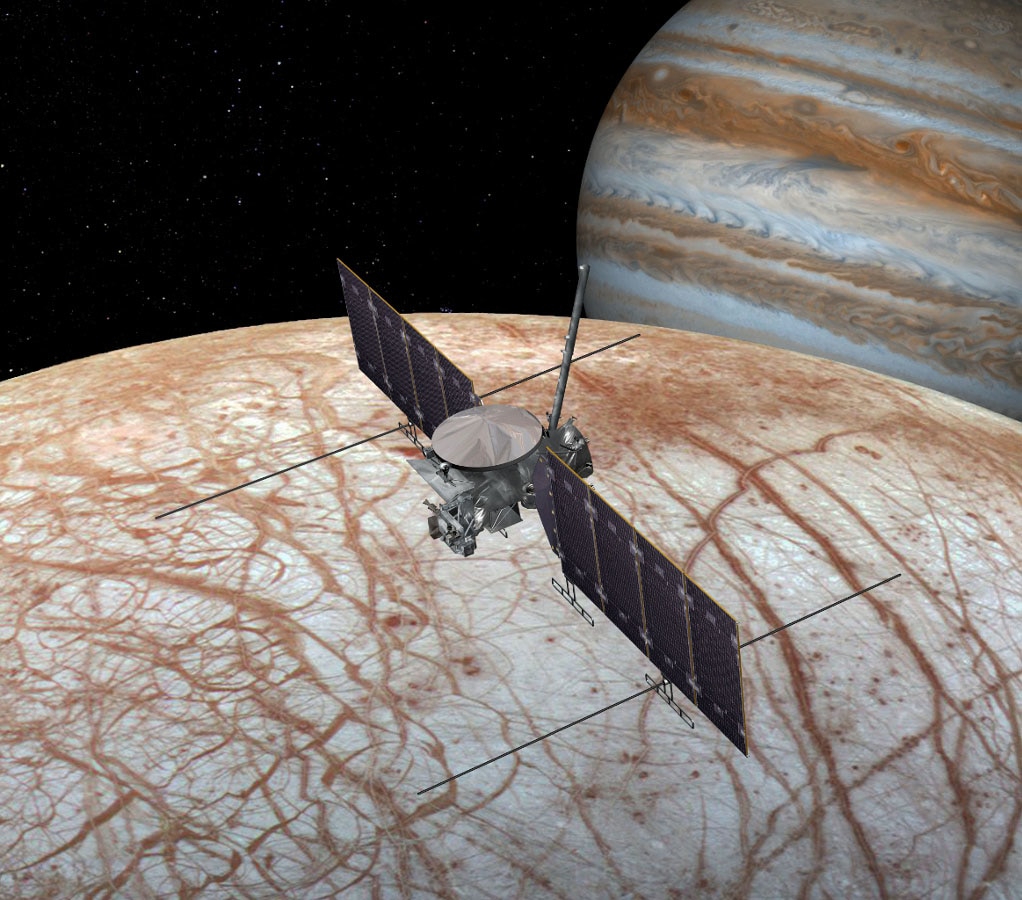}
  \caption{Artist impression of the Europa Clipper mission. [NASA]}
  \label{img_clipper}
\end{figure}

Europa Clipper is a future Europa focused NASA mission (Figure \ref{img_clipper}). It is tasked to investigate Europa to assess whether it could harbour conditions suitable for life. Clipper is scheduled to launch in the 2020's. Depending on the launch date and vehicle that will eventually be chosen, it might be active in the Jupiter system around the same time as JUICE is scheduled to be. Currently 45 flybys of Europa at altitudes from 2700 to down to 25 km and some additional flybys of Ganymede and Callisto are planned.
Clipper will carry nine instruments for in-situ and remote sensing. Instruments that directly characterize Europa's charged and neutral particle environment are the Plasma Instrument for Magnetic Sounding (PIMS) and Mass Spectrometer for Planetary Exploration (MASPEX). PIMS comprises three Faraday cups and measures the plasma mass density and flow velocity. MASPEX is a high-resolution time of flight mass spectrometer for neutral particles.
\section{Overview of Europa plume models}
\label{s_plume_sim}

Several Europa plume models have been published that describe the distribution and propagation of neutral particles of the plumes. Prior to the first observations of Europa's plumes, \cite{Fagents2000} and \cite{Quick2013} used an analytical model to describe the ballistic trajectories of plume particles and investigated if dark regions adjoining several types of surface features are in fact plume deposits. \cite{Lorenz2015} uses an analytical expression that depends on the pressure at the plume vent and its dimensions, to estimate the maximum altitude that can be reached by a particle of a certain size. 
Numerical models are also available. \cite{Southworth2015} models the dust produced by a Europa plume. Trajectories of the plume particles are calculated by integrating the equation of motion. The effect of both Europa's and Jupiter's gravity is taken into account. \cite{Teolis2017} uses a Monte Carlo exospheric model for Europa. The exospheric profile of a wide variety of constituents (for a plume case and case without a plume) is described to prepare for future measurements. \cite{Berg2016} studies in detail the propagation of dust and gas from the source of the plume. The distribution of velocity, density and temperature in the plume are determined using an advanced Direct Simulation Monte Carlo (DSMC) model.

The interaction of the plumes with the magnetospheric environment has also been studied. \cite{Jia2018} uses an MHD model to investigate the interaction of a plume with the corotational plasma. \cite{Bloecker2016} uses an MHD model to investigate the effect of atmospheric inhomogeneities on the formation of the Alfv\'{e}n wings. Both of these models use a analytical equation to describe the plume (see Equation \ref{eq_plume_jia}).

In this work I use a non-collisional Monte Carlo particle tracing model to determine the distribution of plume originating neutrals (H$_2$O) and ions (H$_2$O$^+$). Such an approach has been us previously for plumes on Enceladus (\cite{Burger2007,Smith2010,Tenishev2010,Dong2011,Hurley2015}), Io (\cite{Glaze2000,Doute2002}) and Europa (\cite{Teolis2017}). A key advantage of using this method over collisional methods, such as the DSMC code used in \cite{Berg2016} is that the non collisional model is significantly faster and requires much less computer resources. 

\section{Europa plume simulation setup and assumptions}
\label{s_plume_sim_setup}
The simulation software I use is a product of the continued development of the code described in \cite{Huybrighs2015}, of which the main aspects were discussed in Section \ref{s_particle_simulation}. The neutral particles in my simulation are launched using initial conditions that are determined by a Maxwellian velocity distribution. The spread of this distribution depends on the temperature. I assume a source temperature of 230 K, based on the estimate provided by the supplementary material in \cite{Roth2014a}. Even though water of this temperature will be frozen, the simulation is not very sensitive to changes in the temperature of at least 100 K, thus I consider it a reasonable assumption to use this value. In addition to the thermal velocity I assume that the particles have an initial bulk velocity of 460 m/s that is perpendicular to the surface at the source. With these initial conditions the plume will attain a height of approximately 100 km, when neglecting the thermal spread of the particles. This is comparable to the value of 200 km obtained by \cite{Roth2014a}.

Each particle in the simulation is a super particle representing multiple particles travelling together (Section \ref{s_particle_simulation}). After launch a particle is traced under influence of Europa's gravity only. When a particle's altitude becomes less than Europa's radius, thus when it effectively impacts, it is removed from the simulation.

I assume a mass flux of 1 kg/s, which is three orders of magnitude less than the 1000 kg/s reported by \cite{Roth2014a}. The single event reported by \cite{Roth2014a} could be an exceptionally dense release, as it was only observed once out of twenty observations. Plume eruptions with a much lower mass flux could be more frequent. Additionally, to demonstrate the feasibility of detecting plume originating particles convincingly, it is better to investigate a case with low mass flux rather than a high mass flux. 

The neutral particle detector NIM measures the density of particles. The product of the measurement is a count rate. Equation \ref{eq_nim_counts} from \cite{Huybrighs2015} expresses how the count rate $c$ for five second integration time is related to the number density $\rho$ in cm$^{-3}$ through the conversion factor X, which is equal to 5. 

\begin{equation}
c = \rho \times X
\label{eq_nim_counts}
\end{equation}

I use the distribution of H$_2$O to estimate the generated flux of H$_2$O$^+$ particles. I generate H$_2$O$^+$ particles by applying an ionization rate to the neutral particle distribution. I assume that the ionization rate is homogeneous over the production region. The major sources of ionization of H$_2$O in H$_2$O$^+$ are electron impact ionization and charge exchange with ionospheric plasma. Photoionization is negligible compared to the previous two sources \citep{Pappalardo2009_Kivelson,Lucchetti2016}. The total rate of ionization for an electron density of 110 cm$^{-3}$ is $3.65\times$10$^{-6}$ s$^{-1}$.
The trajectories of the generated ions are determined by numerically integrating the Lorentz force (Equation \ref{eq_lortentz}). I assume a uniform magnetic field and a constant velocity of the corotational plasma. The electric field is determined from these using Equation \ref{eq_E}. Therefore the electric field will also be uniform and perpendicular to the flow velocity and the magnetic field. The resulting motion of the particles is the ExB drift, as shown in Figure \ref{img_ExB}.

To determine the atmospheric contribution of H$_2$O I use the H$_2$O atmospheric profile from \cite{Shematovich2005}, as shown in Figure \ref{img_shematovich2005}. This profile predicts that the density of H$_2$O between 100 and 600 km is about 10$^4$ cm$^{-3}$. These results are comparable to those in \cite{Smyth2006}, which predicted densities of the same order of magnitude for the same altitude range. To generate atmospheric H$_2$O$^+$ pick-up ions I apply the previously mentioned ionization rate to the H$_2$O distribution resulting from the profile from \cite{Shematovich2005}. From the simulation of the ion trajectories I determine the flux of the particles, which is the quantity measured by the ion sensor JDC. The differential flux of particles and the count rate are related through Equation \ref{eq_count_rate}, from \cite{Huybrighs2015}. In this equation $j$ is the differential flux per energy E and direction $\Omega$, $\Delta t$ is the measurement time and $\epsilon$ accounts for additional efficiencies. More details about the instrument simulation are provided in \cite{Huybrighs2015}.
\begin{equation}
r(E, \Omega) = j(E, \Omega)G E \Delta t \epsilon
\label{eq_count_rate}
\end{equation}
The simulation inputs are summarized in Table \ref{tab_sim_plume_settings}.

\begin{table}[h]
\centering
\begin{tabular}{|r|l|}
  \hline
  \textbf{Property} & \textbf{Value} \\
  \hline
  \hline
  Reference frame & x: direction of the corotational plasma \\
  & y: direction of Jupiter \\
  & z: along Europa's rotation axis \\
  \hline
  Plume & \\
  Source temperature [K] & 230, from \cite{Roth2014a}\\
  Mass flux [kg/s] & 1\\
  Bulk speed at source [m/s] & 460 m/s\\
  \hline
  Magnetic field [nT] & [0, 0, -415], Table \ref{tab_properties_magnetosphere}\\
  \hline 
   Electric field & Equation \ref{eq_E} \\
   in [V/m] & [0, -0.0315, -0.0038] \\
   \hline
  Corotational plasma velocity [km/s] & [76, 0, 0], Table \ref{tab_properties_magnetosphere}\\
  \hline
    Ionization rate [s$^-1$] & $3.65\times$10$^{-6}$ from \cite{Pappalardo2009_Johnson}\\
     & and \cite{Lucchetti2016} \\
  \hline
    Ion density [cm$^{-3}$] & 110, Table \ref{tab_properties_magnetosphere}\\
  \hline
  NIM & \\
   Density to count rate factor & 5 [5s integration time]\\
  \hline
  JDC & \\
   Geometric factor [cm$^2$ sr eV/eV] & $5.58\times10^{-4}$\\
   $\epsilon$ [] & 0.035\\
   \hline
   
\end{tabular}
\caption{Overview of the input settings for the simulation. These parameters are justified in Section \ref{s_plume_sim_setup}.}
\label{tab_sim_plume_settings}
\end{table}

\section{Simulations of detecting Europa plume particles}
\label{s_plume_sim_results}
\begin{figure}[h]
  \centering
  \includegraphics[width=1.0\textwidth]{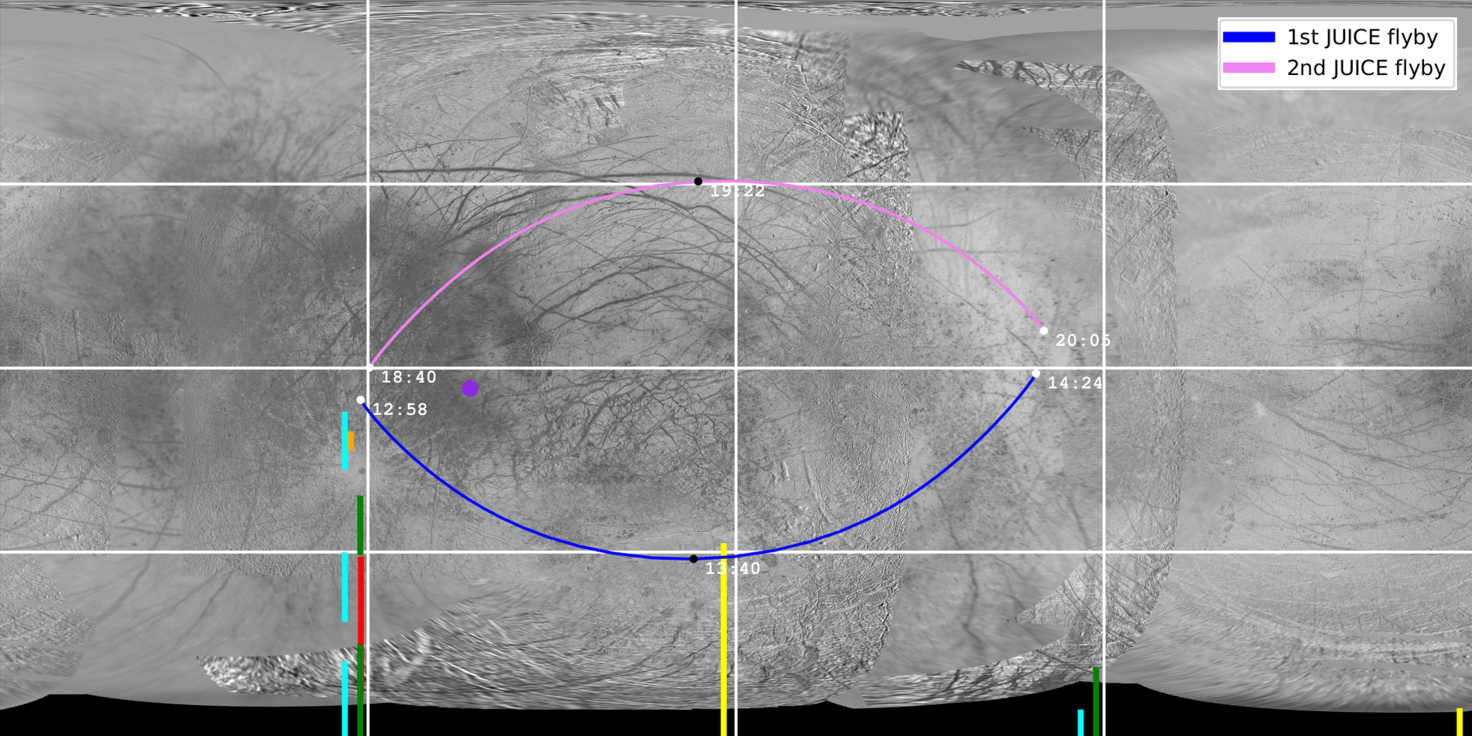}
  \caption{Map of Europa with projection of the two JUICE flybys. The beginning and end time of the plotted trajectories are marked with white dots. Black dots mark the closest approach. The locations of the observed plumes from Figure \ref{img_europa_map_plumes} are also shown. Map of Europa obtained from: USGS.}
  \label{img_map_flybys_juice}
\end{figure}

\begin{figure}[h]
  \centering
  \includegraphics[width=0.50\textwidth]{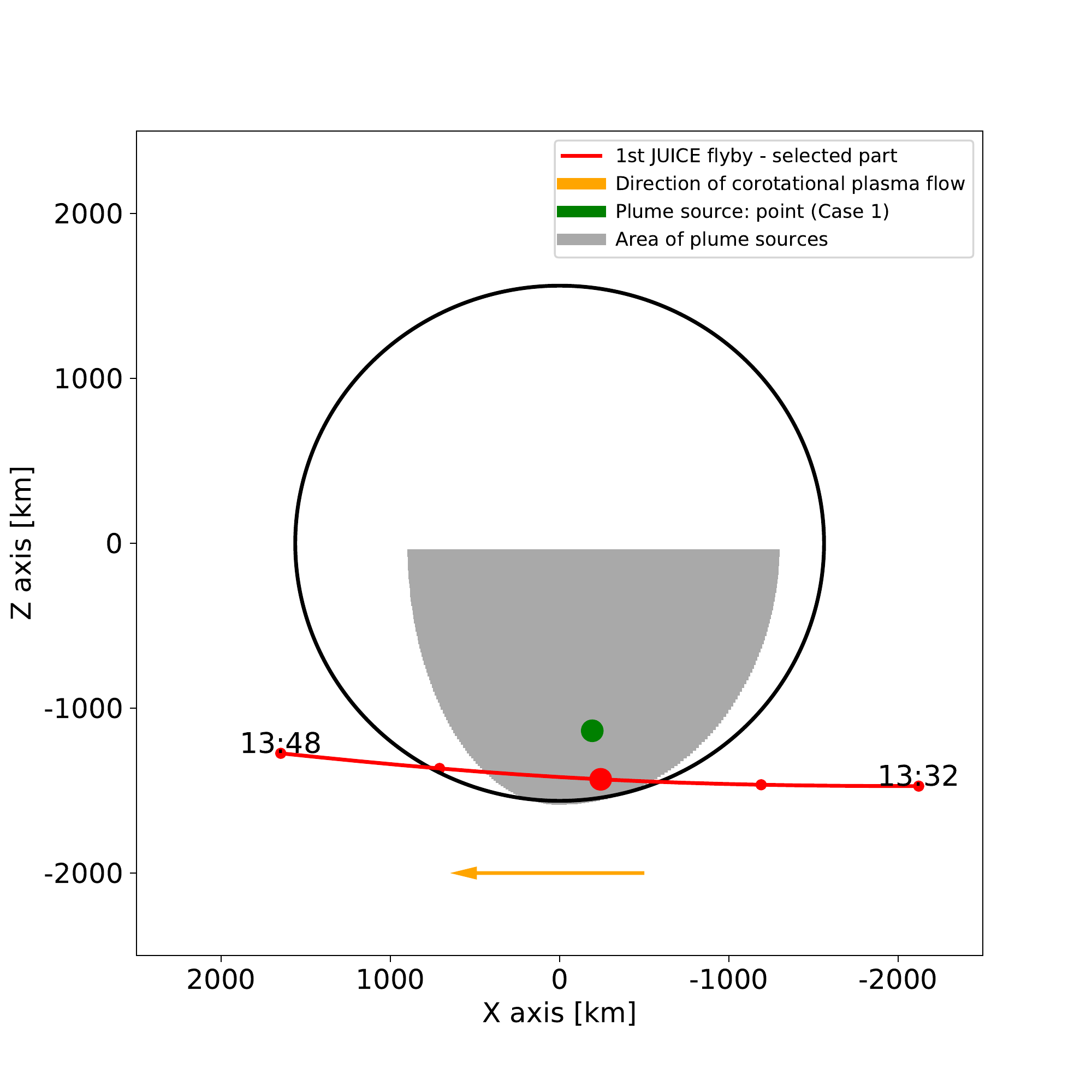}
  \caption{Geometry of the first JUICE flyby. The red line indicates the trajectory of the first JUICE flyby. The big red dot indicates the closest approach. The green dot indicates the point on the surface directly below the closest approach. The grey area indicates an area over which plume sources will be placed. The z-axis is along Europa's rotation axis and the positive x-axis is pointing in the direction of the corotational plasma. An earlier version of this figure was published in \cite{Huybrighs2017} as Figure 1.}
  \label{img_schematic_orbit}
\end{figure}

Previously, \cite{Huybrighs2015} showed that the signal of H$_2$O and H$_2$O$^+$ originating from the plume exceeds the noise level of the respective instruments, during the first JUICE flyby of Europa. As recommended by \cite{Huybrighs2015} I will compare the atmospheric signal of H$_2$O and H$_2$O$^+$ with the plume originating signal. I will not reproduce any of the results of \cite{Huybrighs2015} here. To compare the atmospheric and plume signature I consider the following simulation cases:
\begin{itemize}
\item \textbf{Case 1:} the only source of H$_2$O is Europa's atmosphere, the density is given by the profile from \cite{Shematovich2005}.
\item \textbf{Case 2:} Case 1 + a plume originating from a single point source directly below the closest approach during the first JUICE flyby.
\item \textbf{Case 3:} Case 1 + a plume originating from a 1000 km crack source below the closest approach. The crack is centred on the projection of the closest approach on the surface and aligned along the meridional going through that point. Such a surface crack corresponds to the surface feature referred to as 'Lineae' (\cite{Pappalardo2009_Doggett}). This case is included because the geometry of the plume source cannot be identified in the existing observations due to their low spatial resolution. Furthermore, \cite{Spitale2015} has suggested that the plume sources at Enceladus could be elongated curtains, rather than point sources. The total mass flux of through the crack is the same as through the point source in Case 2.
\item \textbf{Case 4:} Case 2, but the JUICE orbit is lowered by 300 km. In this case the closest approach is located at 100 km altitude instead of 400 km.
\end{itemize}
To investigate the detection of plume particles I select the first JUICE flyby of Europa, since this flyby passes over the regions where plumes have already been observed (Figure \ref{img_map_flybys_juice}), which shows the ground track of the JUICE flybys. Figure \ref{img_schematic_orbit} shows the first flyby with respect to Europa. 

\begin{figure}[h]
  \centering
  \includegraphics[width=1.0\textwidth]{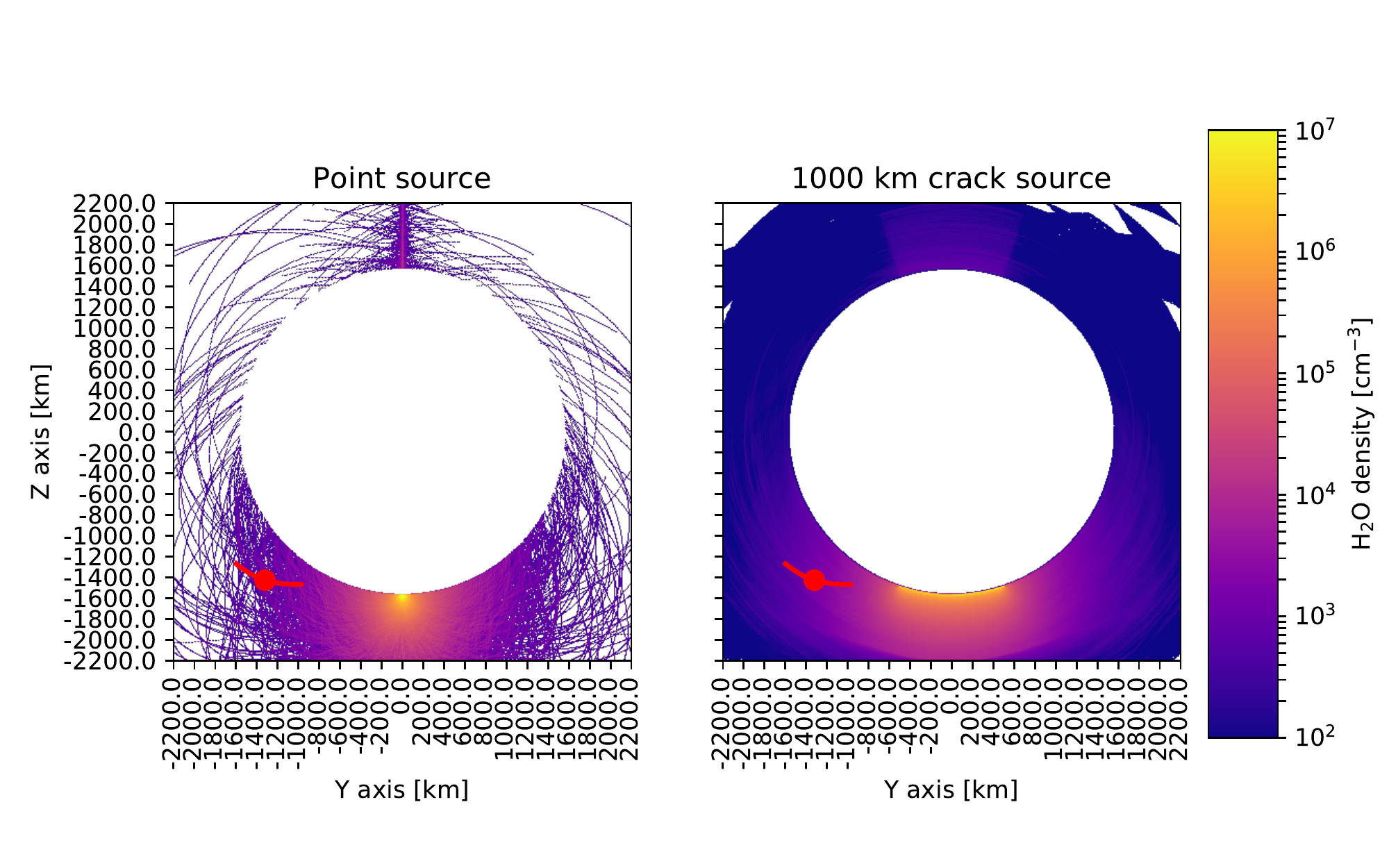}
  \caption{H$_2$O density resulting from a single point source on Europa's south pole (left) and a 1000 km crack centred on the south pole (right). The crack is simulated by placing 200 duplicates of the point source along a 1000 km line along the y-axis, while keeping the total mass flux constant between the two cases. The z-axis is along Europa's rotation axis, the positive y-axis points towards Jupiter. An earlier figure of these simulation results was published in \cite{Huybrighs2017} as Figure 2 and 4.}
  \label{img_density}
\end{figure}

To simulate the Case 3 I place 200 point sources next to each other along a 1000 km line on the surface. To reduce computation time I use the result from Case 2, duplicate it 200 times and place the duplicates along a 1000 km line on the surface. The mass flux is scaled such that the total mass flux of the crack is the same as the point source. The trajectory in Case 4 is obtained by using the same JUICE trajectory as in Case 2, but a constant altitude of 300 km is subtracted from the planned altitude at each point. Figure \ref{img_density} shows the density distribution resulting from a point source and a source crack, in this case located at the south pole of Europa.

\begin{figure}[h]
  \centering
  \includegraphics[width=1.0\textwidth]{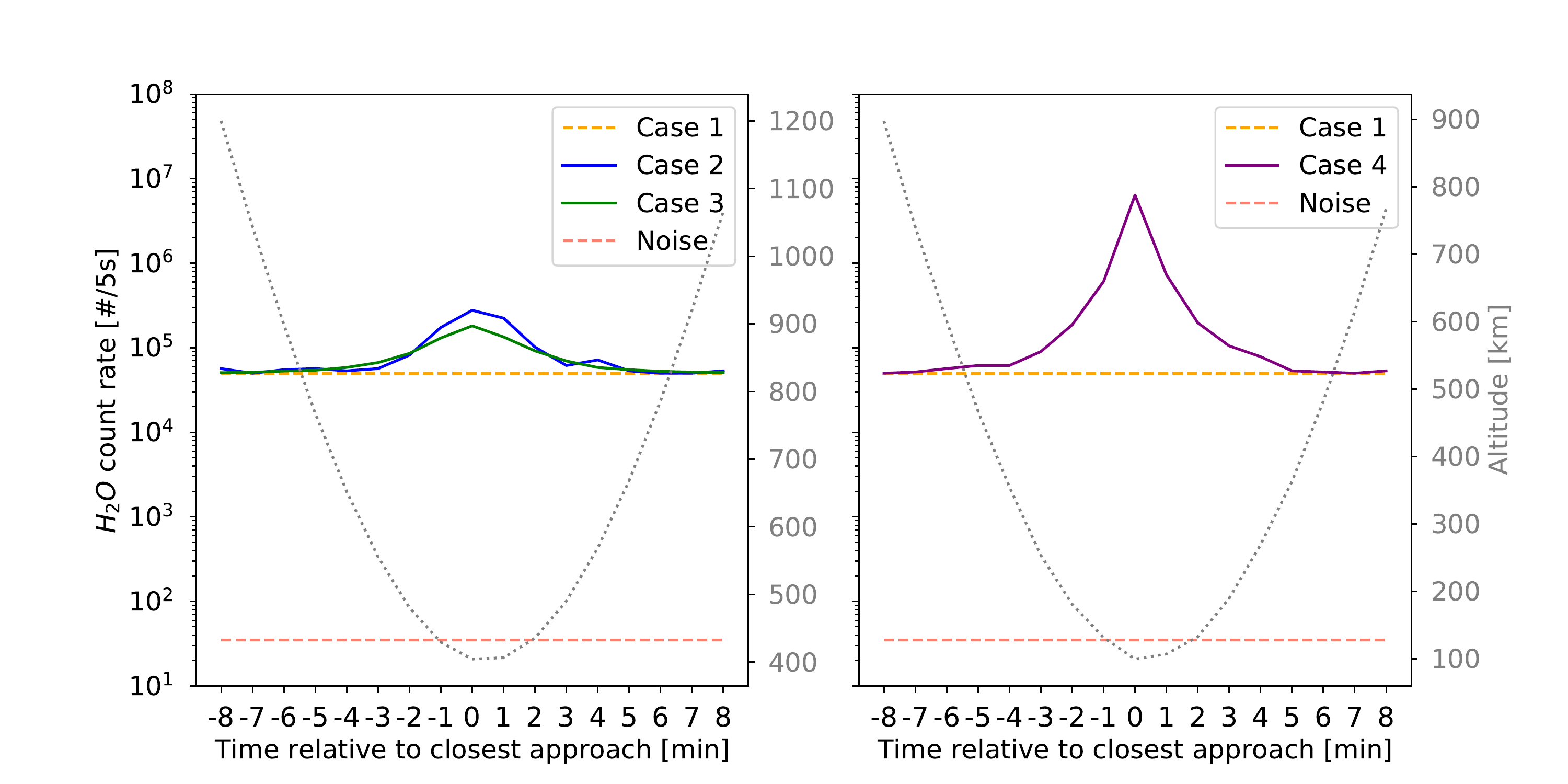}
  \caption{Simulated count rate of H$_2$O for the neutral particle detector NIM during the first JUICE flyby as a function of time with respect to the closest approach. Left: Case 1, 2 and 3. right: Case 4. The dashed orange line indicates the atmospheric profile of \cite{Shematovich2005} which predicts an atmospheric H$_2$O density of around 10$^4$ for altitudes between 100 and 600 km altitude. The dashed grey line shows the altitude of JUICE, which is indicated on the right axis of each panel. The dashed red line indicates the noise level of the instrument. An earlier figure of the simulation results in the left panel was published in \cite{Huybrighs2017} as Figure 3 (right panel).}
  \label{img_nim_solutions}
\end{figure}

In Figure \ref{img_nim_solutions} the simulated count rate of H$_2$O that the neutral particle detector NIM would observe in Case 1-4, is shown. This is the count rate that would be detected if both atmospheric and plume-originating H$_2$O is present. The density contribution of the plume to the particle environment is similar to Figure \ref{img_density}, except that the source of the plume has been moved to directly below the closest approach for the simulations shown in Figure \ref{img_nim_solutions}. Additionally, the atmospheric profile by \cite{Shematovich2005} has been superposed on this. This density profile is almost constant over the range of 100 to 600 km. Therefore, because this is the maximum that density that will be encountered, I represent it as a horizontal line in Figure \ref{img_nim_solutions}.
As previously determined by \cite{Huybrighs2015}, the signal strength exceeds the instrumental background in all cases. Assuming that the integration time of the instrument is about 5s and that penetrating radiation causes 35 counts over the integration time, the signal to noise ratio would reach about 5700 in Case 2. This suggests that the signal to noise ratio of minor species in the plume could also have a signal to noise ratio exceeding one. In Case 4 the signal to noise ratio will increase to around 286000, thus strongly improving the ability to discriminate minor species. These results also show that the introduction of a plume to the environment creates a local increase in the H$_2$O density that can be separated from the atmospheric signal, in Case 2, 3 and 4. The difference in Case 2 and 4 is about one order of magnitude. Lowering the orbit by 300 km dramatically increases the difference between the plume signal and the atmospheric signal by two orders of magnitude.

\begin{figure}[h]
  \centering
  \includegraphics[width=1.0\textwidth]{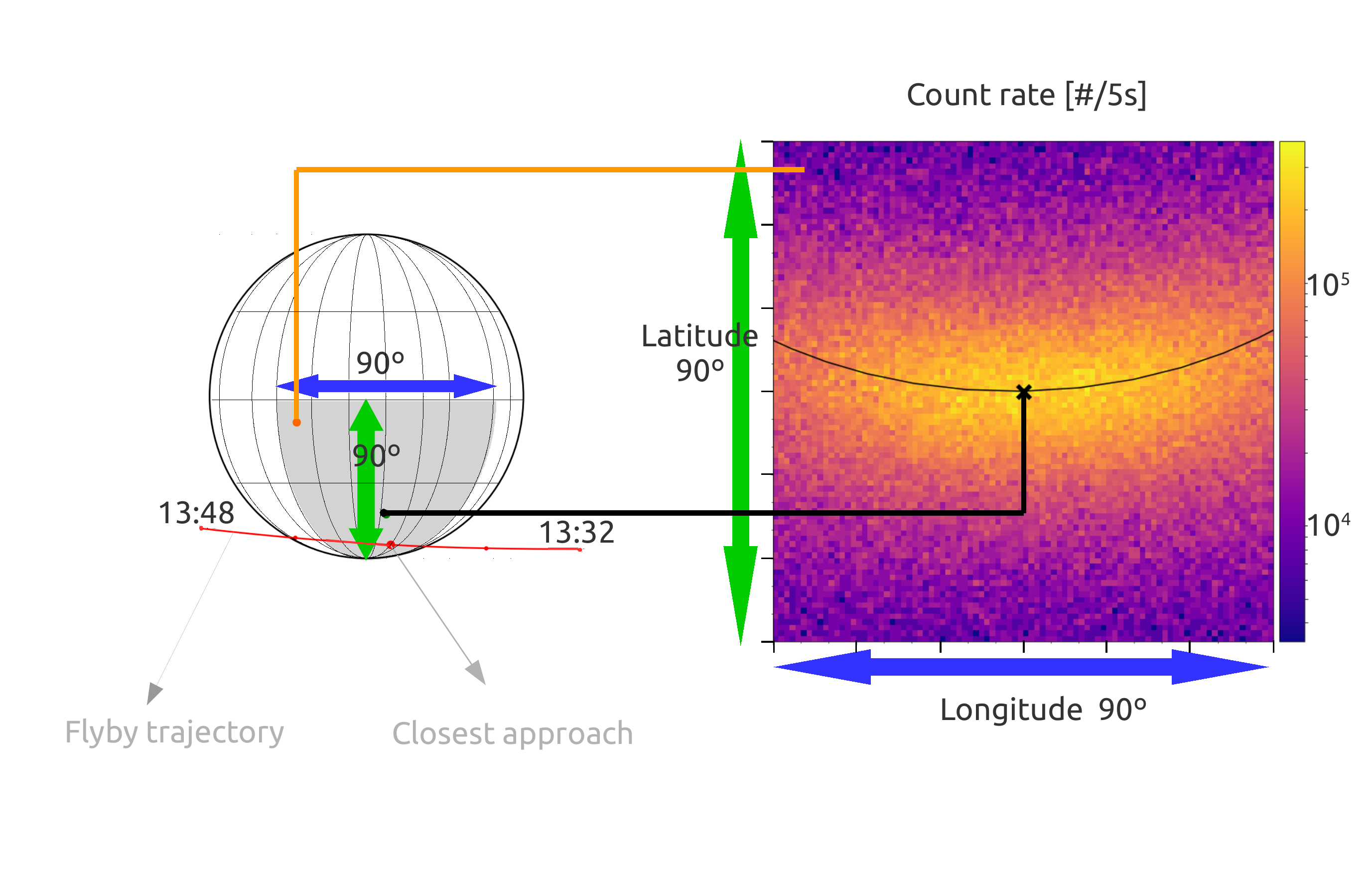}
  \caption{Simulation of the H$_2$O count rate of the NIM neutral particle detector resulting from a plume point source. The right panel shows how the maximum count rate along the flyby varies when the position of the plume point source is varied with respect to the closest approach. A grid of 90 degree latitude and longitude is considered (indicated by the gray area in the left panel, also shown in Figure \ref{img_schematic_orbit}). For each 1 by 1 degree grid cell a simulation of the NIM count rate is done. Indicated in the panel on the right is the the projection of the spacecraft trajectory on the surface. An earlier figure of these simulation results was published in \cite{Huybrighs2017} as Figure 3 (left panel).}
  \label{img_nim_map_solutions}
\end{figure}

In \cite{Huybrighs2015}, it was assumed that the maximum count rate always occurs at the closest approach, no matter the position of the plume. The present work shows that this assumption is not correct. Figure \ref{img_nim_map_solutions} shows how the maximum count rate, measured along the flyby, varies for different positions of the plume on the surface of Europa. The extent of the grid over which different positions are considered is shown in the left panel of Figure \ref{img_nim_map_solutions}. This shows that \cite{Huybrighs2015} underestimated the maximum count rate that would be measured for various positions. The present results show that the count rate along the latitude of the closest approach will in fact stay above 10$^5$ cm$^{-3}$, over the latitude range considered. The signal to noise ratio will also exceed one for every plume position considered for this simulation. Thus, the plume particles should also be detectable if the plume is not located directly below the closest approach.

\begin{figure}[h]
  \centering
  \includegraphics[width=0.75\textwidth]{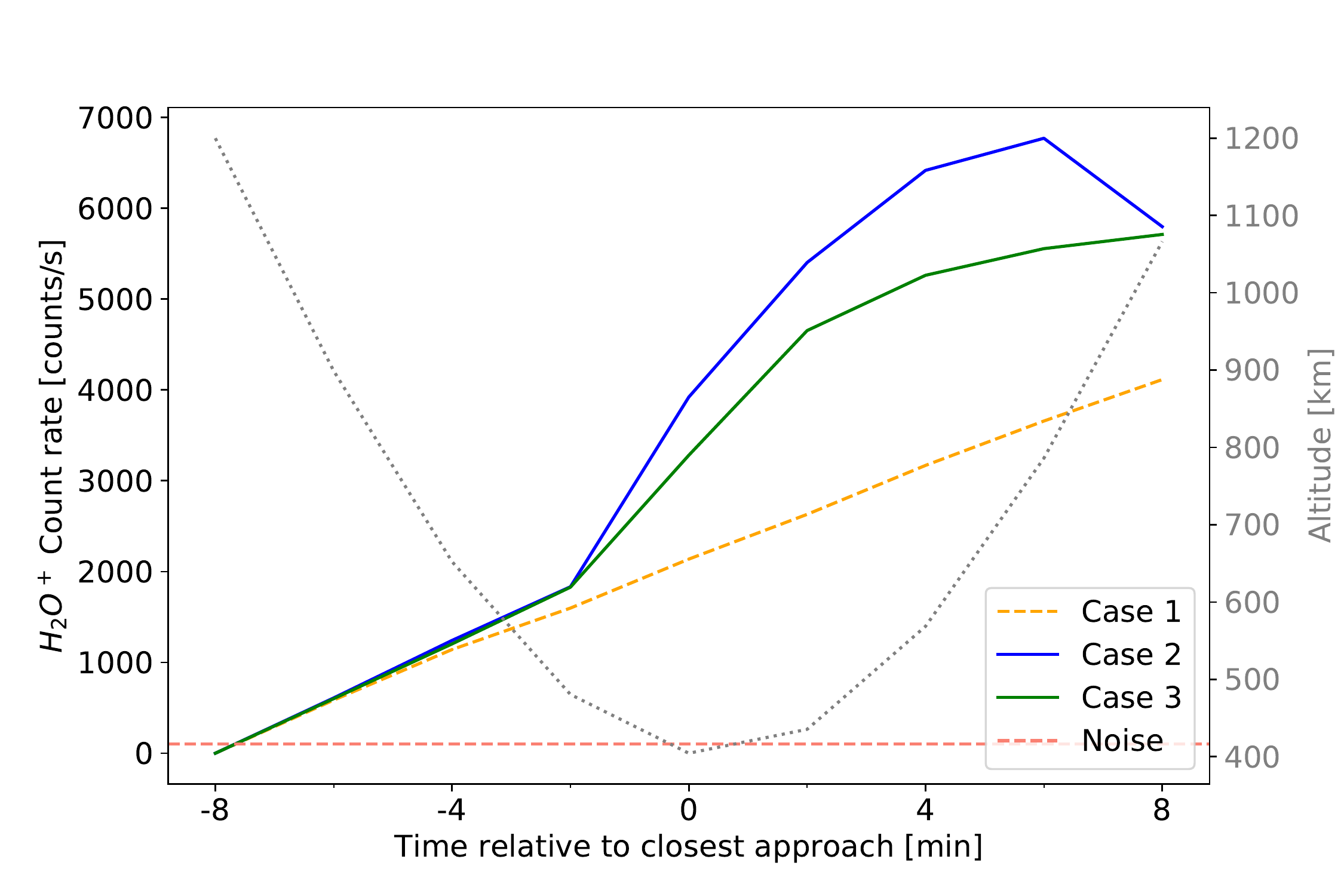}
  \caption{Simulated count rate of H$_2$O$^+$ by the ion detector JDC as a function of time with respect to the closest approach. Results are shown for the optimal looking direction. The results are indicated for Case 1, 2 and 3. The dashed orange line indicates the count rate resulting from the H$_2$O profile from \cite{Shematovich2005}. The red dashed line indicates the noise level due to penetrating radiation. Also indicated by the grey vertical line is the altitude of the spacecraft. An earlier figure of these simulation results was published in \cite{Huybrighs2017} as Figure 7.}
  \label{img_jdc_solutions} 
\end{figure}

Figure \ref{img_jdc_solutions} shows the simulated count rate of H$_2$O$^+$, for Case 1-3. Specifically the results for the optimal looking direction are shown, which is the direction in which the count rate is the highest. The density contribution of the plume in Case 2 and 3 to the environment is like that shown in Figure \ref{img_density}, except that the source of the plume has been moved to directly below the closest approach. Furthermore, the atmospheric profile by \cite{Shematovich2005} has been superposed on it to approximate the global H$_2$O component of Europa's atmosphere. It is clear that the signal of the plume particles exceeds that of the noise (including noise due to penetrating radiation). The signal to noise ratio in the best case is about 33, assuming one second integration time.  These results also show that the introduction of a plume to the environment creates a local increase in the H$_2$O density that can be separated from the atmospheric signal, in Case 2 and 3. Unlike the results for the neutral particles (Figure \ref{img_nim_solutions}), the count rate is not symmetrical with respect to the closest approach. A higher count rate is seen after the spacecraft has passed the closest approach. This is because the spacecraft is moving in the direction of the corotational plasma, which is the direction that the H$_2$O$^+$ ions are being transported in (see Figure \ref{img_schematic_orbit}).

The presence of a plume can be recognized as a local increase in the count rate of the respective instrument in the simulated signal of H$_2$O in both Case 2 and 3  (Figure \ref{img_nim_solutions}). This suggests that small plumes (with a mass flux of 1 kg/s) can indeed by detected by the neutral sensor NIM. Similarly Figure \ref{img_jdc_solutions} shows that the presence of the plume creates a recognizable increase in the  H$_2$O$^+$  count rate that is asymmetric between the upstream and downstream part of the trajectory. Thus a small plume can indeed be detected by the ion sensor JDC.

However, the models of the atmospheric distribution of H$_2$O on which my comparison depends, have not been confirmed with observations. Which is in particular important for detecting those plumes that are not located sufficiently close to the closest approach and for which the observed signature is weaker. Currently only two Europa flybys have been planned for JUICE, thus limiting the opportunities for in situ measurements of Europa's tenuous atmosphere. Complementary observations could be obtained with the remote sensing instruments of JUICE or with measurements of the Europa Clipper mission.

Finally I consider some limitations of the model. Firstly, I neglect the effect of collisions between the neutral particles in the simulations. Previously, \cite{Hurley2015} has stated that a thermal velocity distribution that is a combination of a thermal velocity and a bulk speed, as I have assumed, is an appropriate description outside of the collisional region of the plume. As \cite{Yeoh2015} has shown using a collisional DSMC model, the collisional region is only a few kilometres high in the case of Enceladus' plumes. The mass flux in this model (0.2 kg/s) is comparable to my assumption (1 kg/s), as is the initial velocity of the particles (900 m/s and 450 m/s respectively). This suggests that the size of the collisional region of the 1 kg/s Europa plume I simulated is comparable to the Enceladus case. Therefore, I consider that, at large distances from the collisional region, such as those during the JUICE flyby (> 400 km), the plume model I use is applicable.
Furthermore, \cite{Cai2012} and \cite{Cai2013} suggest that the collision-less flow solutions can still be used to obtain a first order approximation of a collisional supersonic flow. They conclude this by comparing analytical solutions for collision-less flow and collisional DSMC simulations.
They argue that the models agree because the particles travelling downstream with a high bulk speed do not have the time to collide and propagate in the direction perpendicular to the bulk speed.
An important difference between Enceladus and Europa is the lower escape velocity (0.239 km/s and 2.025 km/s, respectively). For a particle with a velocity of about 500 m/s it takes about 20 s to reach 10 km altitude. Now, considering that that Europa's gravitational acceleration is 1.314 m/s$^2$, the particle will be decelerated by about 25 m/s because of gravity within this time. This change in velocity is small compared to the initial velocity and thus gravity will not alter the trajectories much in the first few kilometres above the surface gravity. Therefore, I expect that the difference in escape velocity is unimportant. 

\cite{Berg2016}, which presents a DSMC model of the Europa plume, predicts that for very large mass flux plumes (1000 kg/s) a canopy shock will occur at an altitude of about 400 km in the plume. This shock occurs when there are sufficient particles falling back and interacting with the particles still rising. This shock limits the total altitude of the plume. It does not occur in their models for plumes with a mass flux of 10 kg/s. Furthermore \cite{Berg2016} states that the canopy shock could be a transient feature. My model cannot reproduce this shock feature. This implies that for very high mass flux cases, my simulations could be a best case scenario that represents the case when the (possibly transient) canopy shock is not present.

A second limitation is related to the ions generated by the plume. I have assumed that the ion trajectories are determined only by the ExB drift caused by a uniform electric and magnetic field, which are not influenced by the newly formed ions. In my simulation the density of ions only exceeds 10 cm$^{-3}$ in a region close to the plume source, outside of that the ion density is below 10 cm$^{-3}$. The plasma density near Europa varies from 12 to 170 cm$^{-3}$ \citep{Pappalardo2009_Kivelson}. Assuming the ambient ion density is 100 cm$^{-3}$, the velocity of the ambient ions is 76 km/s, the density of plume originating ions is 10 cm$^{-3}$ and the velocity of the newly formed ions is their ballistic velocity 1 km/s, the average velocity of the particles is still 69 km/s. Thus, in my case the plume ions will not significantly affect the fields. However, if the mass flux of the plume would be much larger, for example the case of \cite{Roth2014a} (1000 kg/s), it can be expected that the newly formed ions will influence the fields. Such a situation cannot be handled by the model that I used. Instead, a different modelling technique is needed to simulate the fields, such as the MHD model in \cite{Jia2018}.

Very high neutral densities will also result in a non-homogeneous ionisation. For an electron temperature of 100 eV, the mean free path of an electron is about 150 km, using electron scattering cross section of $7\times10^{16}$cm$^{-2}$ from \cite{Itikawa2005} and an H$_2$O density of 10$^8$cm$^{-3}$ (the maximum in my simulations). This is larger than the area over which this high density can be found (see Figure \ref{img_density}). Thus this region can be ionized homogeneously. However, if the density were several orders of magnitude larger, electrons would not be able to reach the denser parts of the plumes. Therefore the area closest to the source would be free of newly generated ions.


\section{Main results and recommendations for future work}
I simulated the detection of plume and atmosphere originating H$_2$O and H$_2$O$^+$ by the neutral particles sensor NIM and the ion sensor JDC. I confirm the earlier results of \cite{Huybrighs2015} that the H$_2$O and H$_2$O$^+$ particles originating from small plumes (mass flux 1 kg/s) can be detected with large margins, both in the case when the source of the plume is a point source and a 1000 km long crack. I focussed specifically on small plumes, with a mass flux of 1 kg/s. Small events, which cannot be observed from Earth, could be more frequent than the events that have been measured. Furthermore, to make a convincing case for the detectability of plumes, it is better to have a large margin. The signal to noise ratios, in the optimal case of the plume location with respect to the flyby, are $\sim$ 5700 for H$_2$O and $\sim$33 for H$_2$O$^+$. The noise here is due to penetrating radiation. Neutral species that have an abundance of several orders of magnitude less than H$_2$O would still be above the noise level. Reducing the orbit by 300 km would result in a signal to noise ratio of 286000 in the case of H$_2$O, and would increase the possibility to detect minor species. Furthermore, the results show that the maximum count rate is not necessarily detected at the closest approach. Also, the count rate for plumes, not located directly below the closest approach, still exceeds the noise level.

Here I specifically compared the signature of atmospheric H$_2$O and H$_2$O$^+$  and plume originating H$_2$O and H$_2$O$^+$ particles. The results show that the signal of the plume can be recognized, as a temporal enhancement in the combined H$_2$O count rate and an enhancement in the combined H$_2$O$^+$ count rate in the downstream direction of the trajectory. However, the atmospheric models of H$_2$O have not been confirmed by observations. Since JUICE is only planned to make two flybys of Europa, there is a limited possibility to obtain such measurements from in-situ data. Remote sensing observations with the other JUICE instruments could provide more measurements, or the Europa Clipper mission.

Future simulations of detecting Europa plume models should consider the effect of collisions. In particular for plumes with a large mass flux for which (transient) shocks are predicted that could limit the height of the plume. For those plumes the effect of the plume ions on the fields should be modelled self consistently. The ionization rate cannot be assumed to be homogeneous in those cases.

\chapter{Conclusions and outlook}\label{ch_conclusion}
The question guiding this thesis was:
can the properties (density, scale height) of Europa's tenuous atmosphere or any water vapour plumes be constrained using data from the Galileo in-situ particle detector instruments (PLS, EPD)? The approach chosen to address the main research question was to combine the interpretation of spacecraft observations and simulations.

First, I made an overview of charged particle and magnetic field data collected during the Galileo flybys of Europa. During all the flybys, signatures of Europa's interaction with its magnetospheric environment are visible. I selected the E12 flyby as the most relevant candidate to find signatures of Europa's atmosphere and potential plumes in the charged particle data. 
Signs of interaction of the magnetospheric environment with Europa's atmosphere-derived ionosphere are stronger than during any other flyby. A detailed analysis of the charged particle data collected during this flyby was made, indicating that, near Europa, the plasma flow is disturbed and that energetic protons, oxygen and sulphur ions are depleted.

Next, I simulated the depletion of energetic protons and oxygen ions due to surface impact and charge exchange during Galileo's E12 flyby. 
The depletion region can extend beyond the radius of Europa, because of the size of the gyroradius of the particles involved and the presence of a tenuous atmosphere, which forms an obstacle larger than Europa's solid body. The spatial distribution of the depletion thus depends on the energy, pitch angle and mass of the particles, through the gyroradius. Charge exchange with neutral particles increases the effective impact area of Europa. The simulation shows that the observed depletion strongly depends on the looking direction of the instrument. 
The effect of the flow deflection and Europa's induced dipole on the depletion of energetic ions for upstream flybys is negligible. 
The depletion of energetic protons of energies between 80 and 220 keV in
the data is more than 90\%. This exceeds the expected depletion in my simulation of the atmosphere-free case. I have simulated the depletion in all the energetic proton channels for a range of surface densities and scale heights. I have compared the extent in time and percentage of the resulting depletions to determine which combinations of scale height and density correspond best to the data. The simulations show that the depletions of energetic protons observed in the range 80 to 220 keV can be explained by charge exchange with neutral atmospheric particles, but not by impact on Europa's surface alone.
This suggests that an atmosphere must have been present during the E12 flyby.  The observation of strong magnetic field and flow perturbations upstream of Europa, support this interpretation. My
solutions favour higher surface densities ($10^8$-$10^{10}$ cm$^{-3}$) than previous studies (10$^7$-10$^8$ cm$^{-3}$), but are not sensitive to the scale height. 
I have also simulated the extent of the depletion in the oxygen channels (416-1792 keV). My results suggest that the energetic ions are multiply charged. 
An additional depletion feature is seen in the energetic protons between 540 and 1040 keV, near the closest approach of Europa. This feature is not consistent with a global atmosphere. I demonstrate that a narrow depletion feature in this channel can be explained by a dense, narrow plume. However, the simulations cannot reproduce the depletion signature measured at the same time in the lower energy range 80 and 220 keV, which show a weaker depletion. 
A natural follow-up of this work would be to include an improved description of the electromagnetic fields near Europa, as predicted by MHD or hybrid simulations, in order to capture additional details of the energetic ion motion across localized features (e.g. the observed sharp magnetic field gradients near a plume), which are not included in the current simulations.

Finally, I simulated the detection of plume and atmosphere originating H$_2$O and H$_2$O$^+$ by the neutral particles sensor NIM and the ion sensor JDC on board JUICE. I confirm the earlier results of \cite{Huybrighs2015} that the H$_2$O and H$_2$O$^+$ particles originating from small plumes (mass flux 1 kg/s) can be detected with large margins, both in the case when the source of the plume is a point source and a 1000 km long crack. The signal to noise ratios, in the optimal case of the plume location with respect to the flyby, are $\sim$ 5700 for H$_2$O and $\sim$33 for H$_2$O$^+$. Neutral species that have an abundance of several orders of magnitude less than H$_2$O would still be above the noise level. Reducing the orbit by 300 km would increase the signal to noise ratio to $2.86\times10^5$. 
Furthermore, by comparing the signature of atmospheric H$_2$O and H$_2$O$^+$  and plume originating H$_2$O and H$_2$O$^+$ particles, I show that the signature of the plume can in principle be recognized in the combined atmosphere-plume signal. 
However, the atmospheric models of H$_2$O have not been confirmed by observations. Since JUICE is only planned to make two flybys of Europa, there is a limited possibility to obtain such measurements from in-situ data. Remote sensing observations with the other JUICE instruments or by the Europa Clipper mission could provide more measurements.

\section*{Outlook}
In this work I used a Monte Carlo particle tracing code and applied it to the different scenario's. This method was selected because it allows fast computation and limits the required computer resources. Furthermore, it allows to easily discriminate the contribution of different physical effects.
I have reached the limit of what can be done solely using this simulation method, in particular for the simulations of energetic particles and their interaction with the atmosphere and plumes. The logical next step is to include the result of self consistent models of the fields around Europa. This will allow to identify the effect of the complicated magnetic field topology during the E12 flyby and others. Here I have investigated only the E12 flyby in detail, other flybys show potential signs of interaction with Europa's ionosphere and atmosphere too, in particular E4, E6 and E26. My results show that the depletion in the wake is also affected by the atmosphere, therefore the depletions in the wake flybys E11 and E15 should also be simulated. My simulation method can also be expanded to the flybys of other moons such as Callisto and Ganymede. The tenuous atmospheres of these objects has not been previously analysed using the energetic particle data. The obtained constraints on the properties of these moons will be relevant for the preparation of the measurement campaigns of future missions such as JUICE and Europa Clipper. Furthermore the method I used here to determine atmospheric properties, could be applied to data collected by the high energy particle detectors of JUICE's Particle Environment Package.

Though the most direct measurements of Europa's tenuous atmosphere and plumes will come from optical imaging or the in-stu detection of neutral particles, this works shows that the atmosphere and plumes can influence energetic particle and plasma properties near the moon. The synergy between the optical measurement and the in-situ measurements of neutral and charged particles will allow a detailed characterization of the atmosphere-plume system. Such a capability is available both in the upcoming JUICE and Europa clipper missions, and can be extended to the atmospheres of other moons of Jupiter, such as Ganymede and Callisto.



\clearpage\null\newpage

\appendix
\chapter{Appendix: Galileo data analysis}

Technical details of the data analysis are discussed in this appendix. At the end an overview of the available data is presented. To analyse the EPD data the "Galileo EPD Data Processing Software" is used, which was originally developed by Dr. Andreas Lagg, see for example \cite{Lagg2003}. Products of this software are shown in Figure \ref{img_depletions_xy}, \ref{img_combined_fov_line} and \ref{img_1203_feature}. 
The overview plots in Appendix \ref{app_overview} have been made with the CCATI software, originally developed by Dr. Markus Fr{\"a}nz. For enquiries, contact Dr. Norbert Krupp (krupp@mps.mpg.de) about the EPD software and Dr. Markus Fr{\"a}nz (fraenz@mps.mpg.de) about CCATI \footnote{CCATI website http://www2.mps.mpg.de/projects/mars-express/aspera/ccati/}.

First, the details of the PLS data processing are discussed in Section \ref{a_pls_processing}. Next, I present an overview of the EPD, PLS and MAG data in Section \ref{app_overview}.

\section{Processing of EPD data}
\label{a_epd_tof}
The Energetic Particle Detector EPD is equipped with the CMS time of flight (TOF) mass spectrometer (see Section \ref{ss_epd}). The data from the CMS TOF telescope
is separated in several channels that separate species and energies. Of particular importance for this work are the energetic proton channels TP1, TP2 and TP3 and the energetic oxygen ion channels TO2 and TO3. Here I explain how these channels are defined.

\begin{figure}[h]
  \centering
  \includegraphics[width=0.75\textwidth]{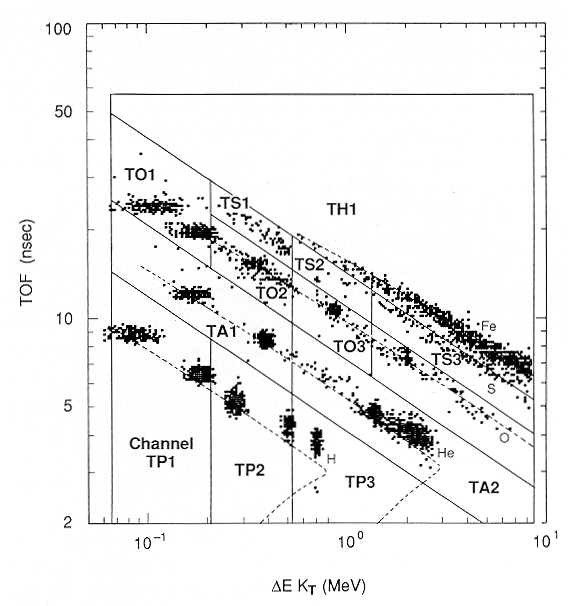}
  \caption{Time Of Flight versus energy measured in detector K$_T$. The solid lines demarcate the different channels (TP1, TO2...), the dashed lines indicate the predicted TOF vs E curves, and the black dots indicate the data from calibrations. [NASA PDS]}
  \label{img_epd_tof}
\end{figure}

Every measurement of the CMS TOF consists of a start signal, a stop signal and the total energy of the particle. The total energy is measured by detector K$_T$ (part of CMS). From the stop and start signal the time of flight can be determined. In Figure \ref{img_epd_tof} it can be seen how the EPD channels are defined as a function of the TOF and the deposited energy. The channels are demarcated by solid lines. The dashed lines indicate the predicted tracks of several elements. The dots indicate calibration data. Note that the TO1 channel was predicted to contain oxygen only, but has later been shown to be contaminated by sulphur too. Therefore only TO2 and TO3 are considered as oxygen channels in this thesis.

\clearpage

\section{Processing of the PLS data}
\label{a_pls_processing}

In this section it is described how the PLS ion data is processed. A basic description of the instrument and its operating principles are given in Section \ref{ss_galileo}. As described previously a major challenge for processing and analysing the PLS data is the poor spatial and time coverage, resulting from the malfunction of the main antenna. An additional challenge is the noise introduced in the data by penetrating radiation. This penetrating radiation consist of highly energetic particles (such as electrons) that penetrate the shielding of the instrument and directly hit the detector, without passing through the electrostatic analyser.

For the processing of the data a modified version of the CCATI software, originally developed by Dr. Markus Fr{\"a}nz, was utilized. Prior to this thesis already several fundamental steps of PLS data processing were implemented in this software. Development of the software continued as part of this thesis project. 

The raw PLS data is obtained from NASA's Planetary Data System (PDS). In this thesis I am only using the HIGH{\_}RES product. The raw data is available in the unit of counts per second. Besides the measurement, for every data point a time tag, the number of the respective detector, the energy bin number and the sector number are available. The sector number represents the spin angle at which the measurement was made.

\subsection{Energy spectrum}
\begin{figure}[h]
  \centering
  \includegraphics[width=1.0\textwidth]{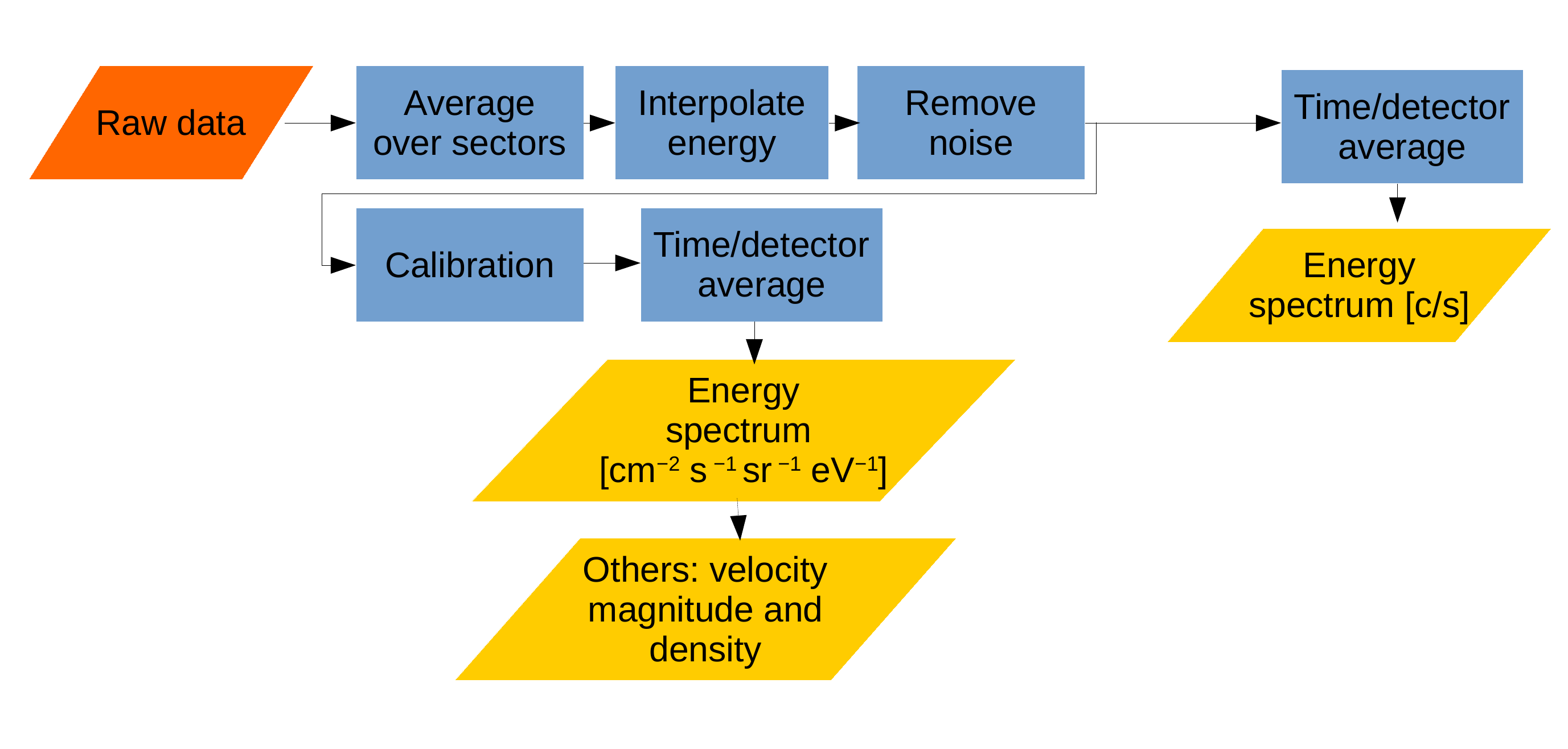}
  \caption{Visual overview of the PLS data processing}
  \label{img_pls_processing}
\end{figure}
The PLS data is processed in several steps to compensate for the poor coverage in time and energy, and remove the effect of the background noise. Here it is described how the raw data in count per second is converted into an omnidirectional energy spectrum, and converted into differential flux with unit of cm$^{-2}$ s$^{-1}$ sr$^{-1}$ eV$^{-1}$ (sometimes referred to as intensity). A visual overview of the different steps is provided in Figure \ref{img_pls_processing}.

Per detector, for the same timetag, multiple data points with different sector numbers can be available in the raw data. To deal with this, the data is averaged over the sectors for each detector separately. This works as follows. Every single data value from a certain energy bin is averaged with the data from the previous and next time step, corresponding to the same energy bin, but having different sector numbers.

The next issue to deal with is the sparse coverage of energy bins. After the sector averaging, a cubic spline interpolation is applied to fill in the empty energy bins. 

Subsequently the noise due to penetrating radiation is removed. First the contribution of the noise is estimated based on the count rate in the four highest energy bins. The average and standard deviation are calculated from the count rate data in these four bins. Empty bins are neglected for this procedure. The sum of the average and the standard deviation multiplied by 0.2  is then considered as the noise, and the obtained noise is subtracted from the count rate data of the other energy channels. This procedure of noise removal is equivalent to previous PLS studies, such as \cite{Paterson1999} and \cite{Frank1999b}, when comparing to the published results.

\begin{figure}[h]
  \centering
  \includegraphics[width=1.0\textwidth]{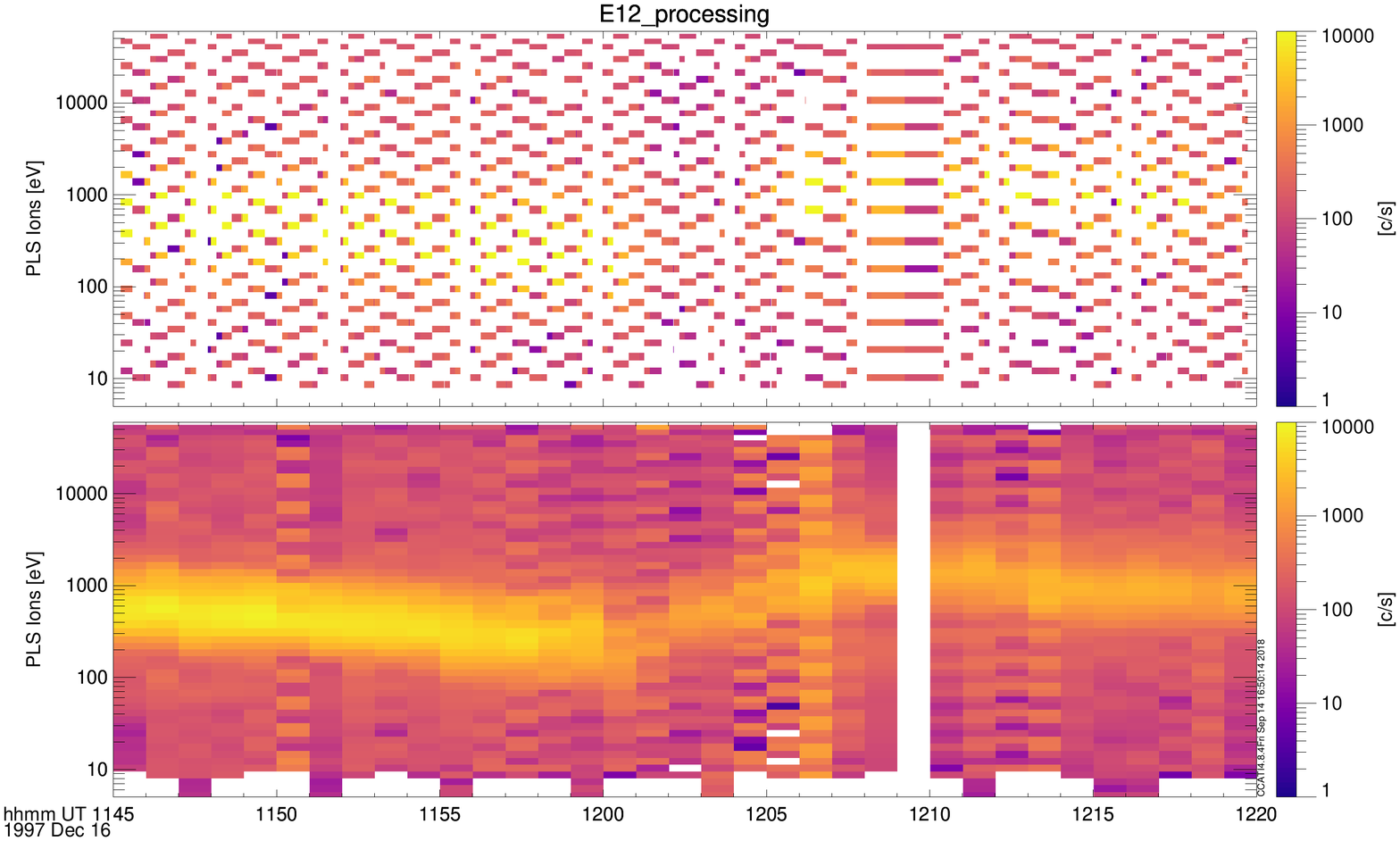}
  \caption{Top panel: unprocessed data, bottom panel: processed data. This example is for the E12 flyby.}
  \label{img_pls_processing_compare}
\end{figure}

Next, averaging in time is performed together with averaging over the seven detectors. The 'box car' averaging method is used for this. This is implemented as follows. First the time frame is divided in equally sized bins and then all the values in each energy-time bin (taken from all the detectors) are summed up. Finally this sum is divided by the number of values in the energy-time bin. Here I average over 60 seconds. Figure \ref{img_pls_processing_compare} shows the data before and after processing in units of count/s. This data product can be seen for the other flybys in the figures in Section \ref{app_overview}.

\newpage

The the data (prior to box car averaging) can be converted from counts/s to differential flux (cm$^{-2}$ s$^{-1}$ sr$^{-1}$ eV$^{-1}$). This is done by using the calibration data provided in the PDS. Calibration data are provided as a factor that has to be multiplied with the data. This factor is referred to as 'chi-factor'. The chi-factor was obtained by the PLS team from calibration experiments in laboratory. Box car averaging is performed in the same way as described previously. The resulting data product in units of cm$^{-2}$ s$^{-1}$ sr$^{-1}$ eV$^{-1}$ can be seen in the figures in Section \ref{app_overview}. Two higher level data products can be derived from the differential flux: the plasma velocity magnitude and the density. These are discussed in the next section. 

The PLS data collected during the close Europa flybys has generally been obtained with two different scanning modes: one low-energy mode that only scans below 15 eV and one that scans above 15 eV. Because of uncertainties in the processing of the low energy scans, I only consider data above 15 eV to be reliable.

\subsection{Other data products}
In Section \ref{app_overview} the following additional data products are also shown:
\begin{itemize}
\item \textbf{Density: }plasma density is calculated from the differential flux, by integrating the differential flux in every energy bin and taking the sum. This is done using Equation \ref{eq_pls_dens}, which is based on \cite{Fraenz2007}.
\begin{equation}
\rho = 4\pi \sum J_i \Delta E_i \frac{\sqrt{m_{factor}}}{\sqrt{E_i}}1.686e^{-8}
\label{eq_pls_dens}
\end{equation}
In this Equation $J_i$ is the differential flux of the bin in cm$^{-2}$ s$^{-1}$ sr$^{-1}$ eV$^{-1}$, $\Delta E_i$ the energy width of the bin in eV, ${E_i}$ the energy of the bin in eV and $m_{factor}$ expresses the mass of the particles in electron mass, density $\rho$ is then in $cm^{-3}$. I use a mass factor of 18.5, based on Table \ref{tab_properties_magnetosphere}.

\item \textbf{Velocity magnitude: }The velocity magnitude of the corotational plasma is determined by first multiplying every energy bin with the corresponding energy bin width and converting the energy of the bin with the highest value into a velocity. The multiplication of differential flux and energy bin width scales with the count rate. This method thus looks for the bin with the highest probability of particle detections. I consider that this a reliable way to estimate the velocity of the corotational plasma. Since the corotational plasma is the dominant source of plasma, the bin with the most detected particles should correspond to the corotational plasma.
\end{itemize}

\subsection{PLS instrument pointing}
Determining the pointing of the PLS instrument and linking observed features to a direction in physical space has been traditionally difficult. This is due to an absence of SPICE kernels for the PLS instrument and a lack of clear and consistent documentation. \cite{Bagenal2016} and supplementary material therein have documented several key points needed to understand instrument pointing and provided corrections for the documentation provided in \cite{Frank1992}. Following the suggestions of \cite{Bagenal2016} I implemented my own code to understand the looking direction of PLS. I do this by providing figures of PLS data in a non-spinning spacecraft frame, which has a constant orientation with respect to physical directions such as the corotational plasma.  This frame shares its z-axis with the spin axis of the spacecraft. I plot the data as a function of the polar angle by separating it in polar bins and azimuth bins, both in the non-spinning spacecraft frame. I determine the polar and azimuth angles as follows, using information from \cite{Bagenal2016}. For the polar angle:
\begin{itemize}
\item I define the polar angle of the data as the polar angle with respect to the z-axis (spin-axis) in the spacecraft frame. The polar angles of the PLS detectors in the spinning spacecraft frame are known. The number of polar bins is seven. There are seven detectors. Each detector has a different polar angle.
\item A direction of interest (such as that of the corotational plasma) can be transformed to the spacecraft frame (available in SPICE as SC{\_}BUS = -77000), in which the polar angle of the direction can be calculated. It is then clear which detectors are best covering this polar angle.
\end{itemize}
For the azimuth angle:
\begin{itemize}
\item I define the azimuth angle as the azimuth angle in a non-rotating spacecraft frame.
\item Since Galileo is spinning, the spin angle of each detector is continuously changing. To account for the spin, every data point has a sector number. This sector number expresses the spin angle corresponding to the measurement. The key difficulty is finding which direction sector 0 corresponds to.
\item Sector number 0 occurs when PLS detector 1 and 7 are in the plane containing the north pointing vector ([0,0,1]) in the ECL50 reference frame. ECL50 is an abbreviation of the Earth Mean Ecliptic equinox 1950 frame.
\item No SPICE kernel is available for the ECL50 frame, but ECL50 north is (approximately) the same as +Z in the ECLIPB1950 reference frame, which is available in SPICE.
\item By knowing the orientation of directions of interest (such as that of the corotational plasma) in the ECLIPB1950 frame it can be determined which sector numbers cover which directions. I divide the azimuth range in 8 parts and separate the data for each part, so that a plot can be made which shows the data in each part of the full azimuth range.
\end{itemize}
I have converted the following physically interesting directions in the non-spinning spacecraft frame:
\begin{itemize}
\item The direction of Jupiter. I use the IAU Europa x-axis
\item The direction of the corotating plasma. I use the y-axis of the IAU Europa frame
\item The rotation axis of Europa, which is the z-axis of the IAU Europa frame
\item The ram direction of the spacecraft, which is the velocity of the spacecraft in the IAU frame, namely the velocity relative to Europa.
\end{itemize}

\section{Overview of the data}
\label{app_overview}

\begin{figure}[h]
  \centering
  \includegraphics[width=1.0\textwidth]{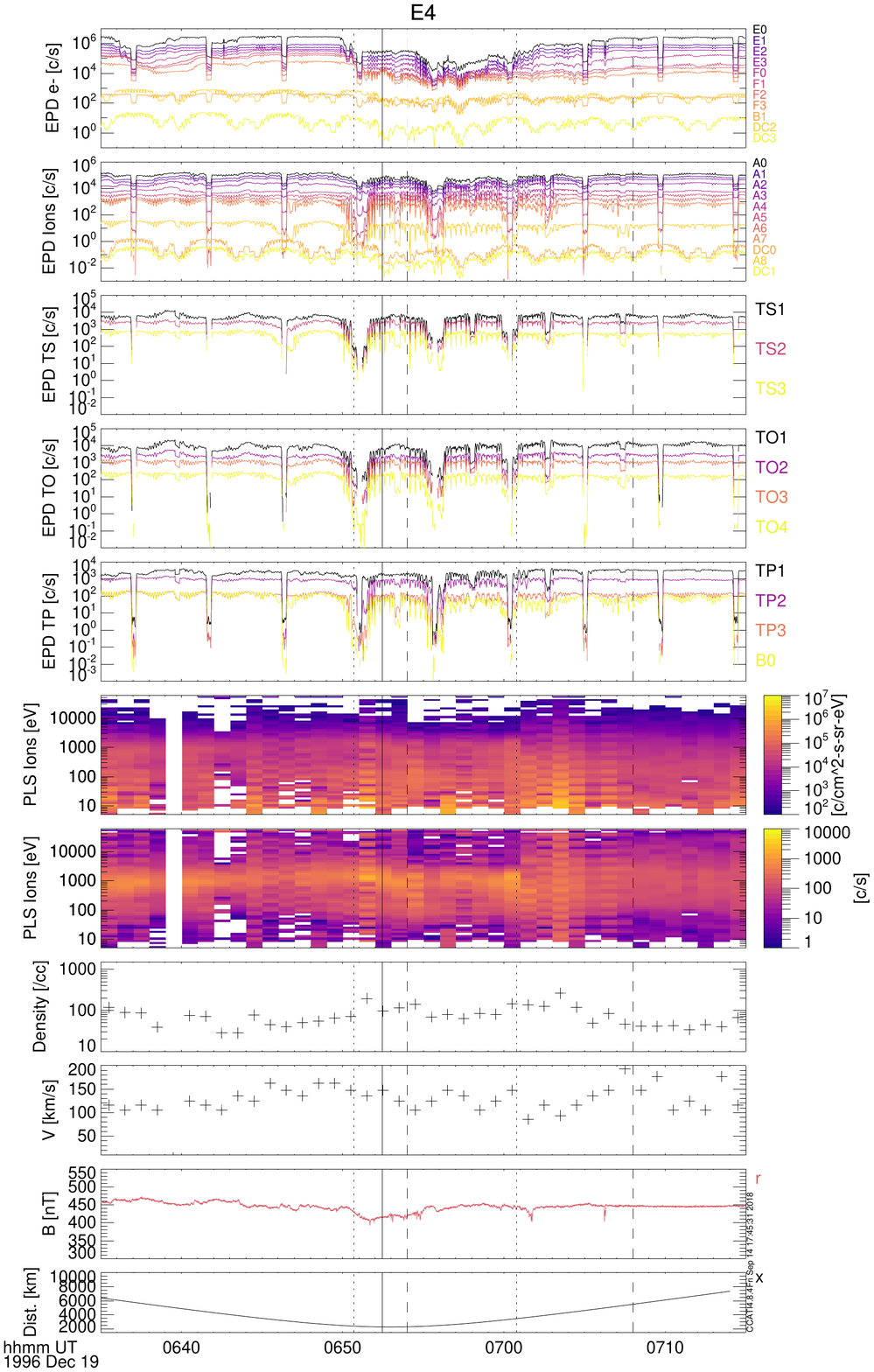}
  \caption{E4 flyby data. From the top: EPD data (panel 1-5), PLS data (panel 6-9), magnetic field data (panel 10) and altitude (panel 11). The dashed vertical lines indicate the geometrical wake. The dotted lines indicate the region in which \cite{Paterson1999} reports deflection of the corotational plasma based on PLS data.}
  \label{img_E4_all}
\end{figure}

\begin{figure}[h]
  \centering
    \includegraphics[width=1.0\textwidth]{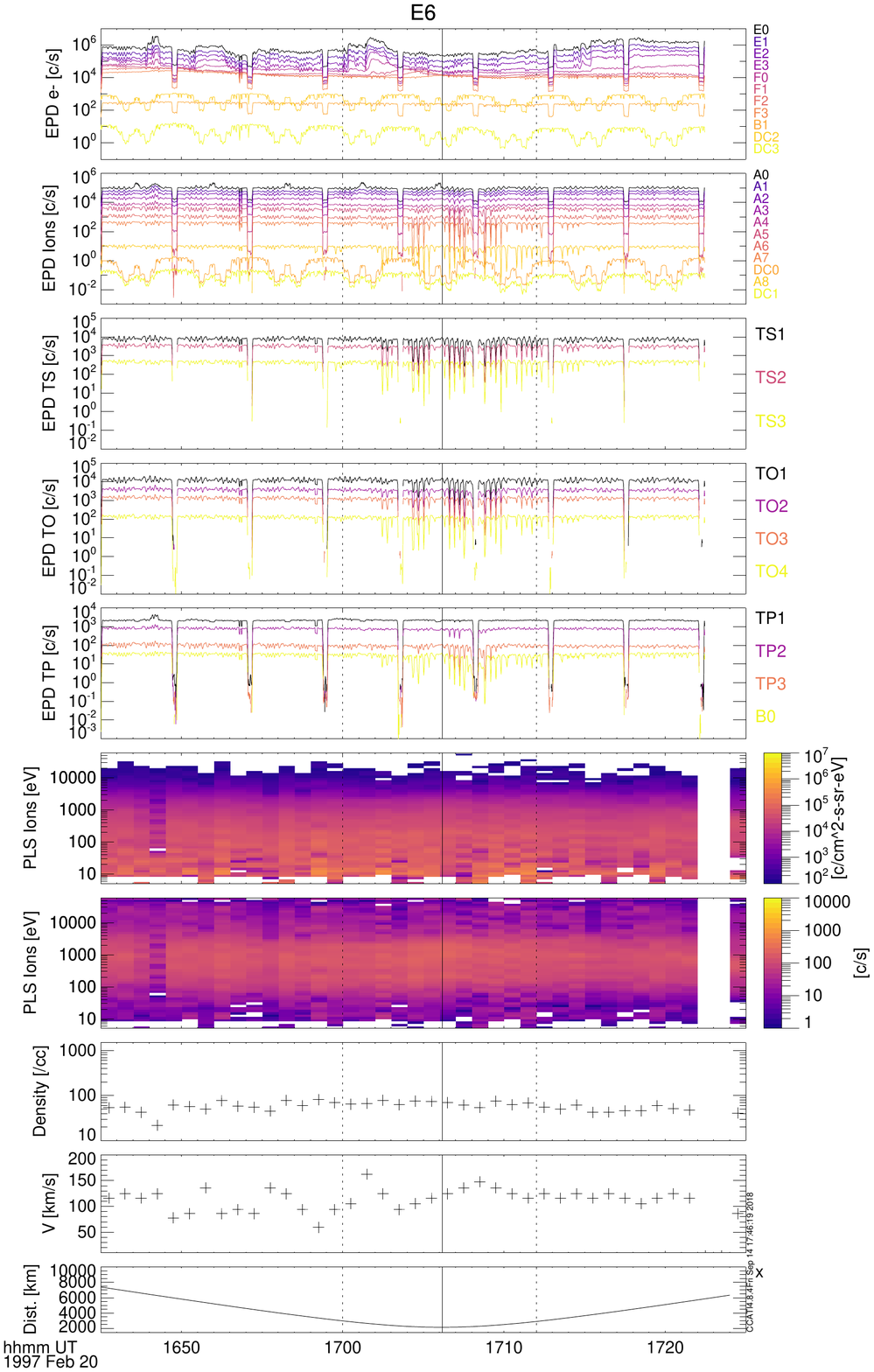}
  \caption{E6 flyby data in the same format as Figure \ref{img_E4_all}. The dotted lines indicate the region in which \cite{Paterson1999} reports deflection of the corotational plasma based on PLS data. Note that no magnetic field data is available for this flyby.}
  \label{img_E6_all}
\end{figure}

\begin{figure}[h]
  \centering
  \includegraphics[width=1.0\textwidth]{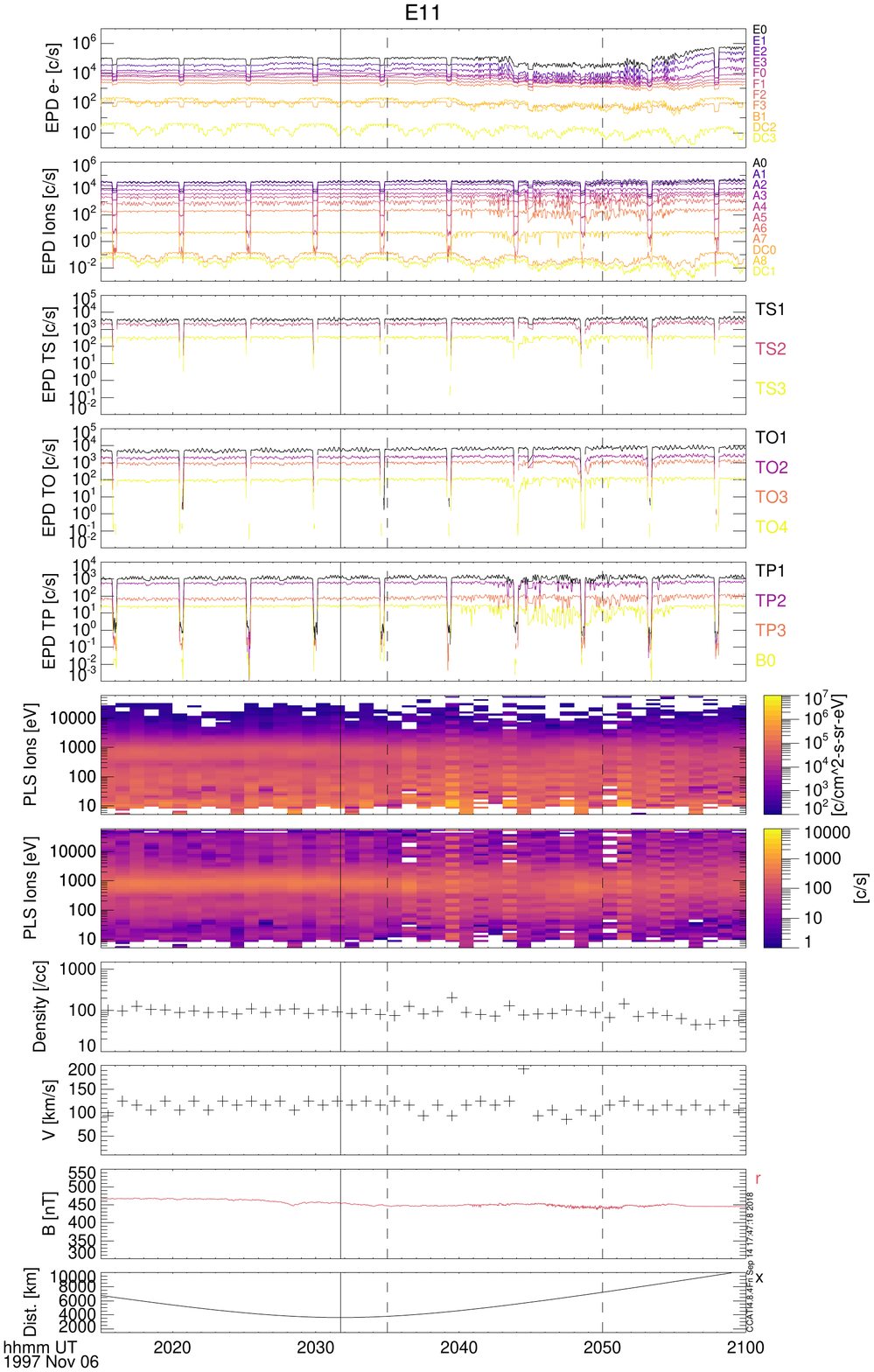}
  \caption{E11 flyby data in the same format as Figure \ref{img_E4_all}. The dashed vertical lines indicate the geometrical wake.}
  \label{img_E11_all}
\end{figure}

\begin{figure}[h]
  \centering
  \includegraphics[width=1.0\textwidth]{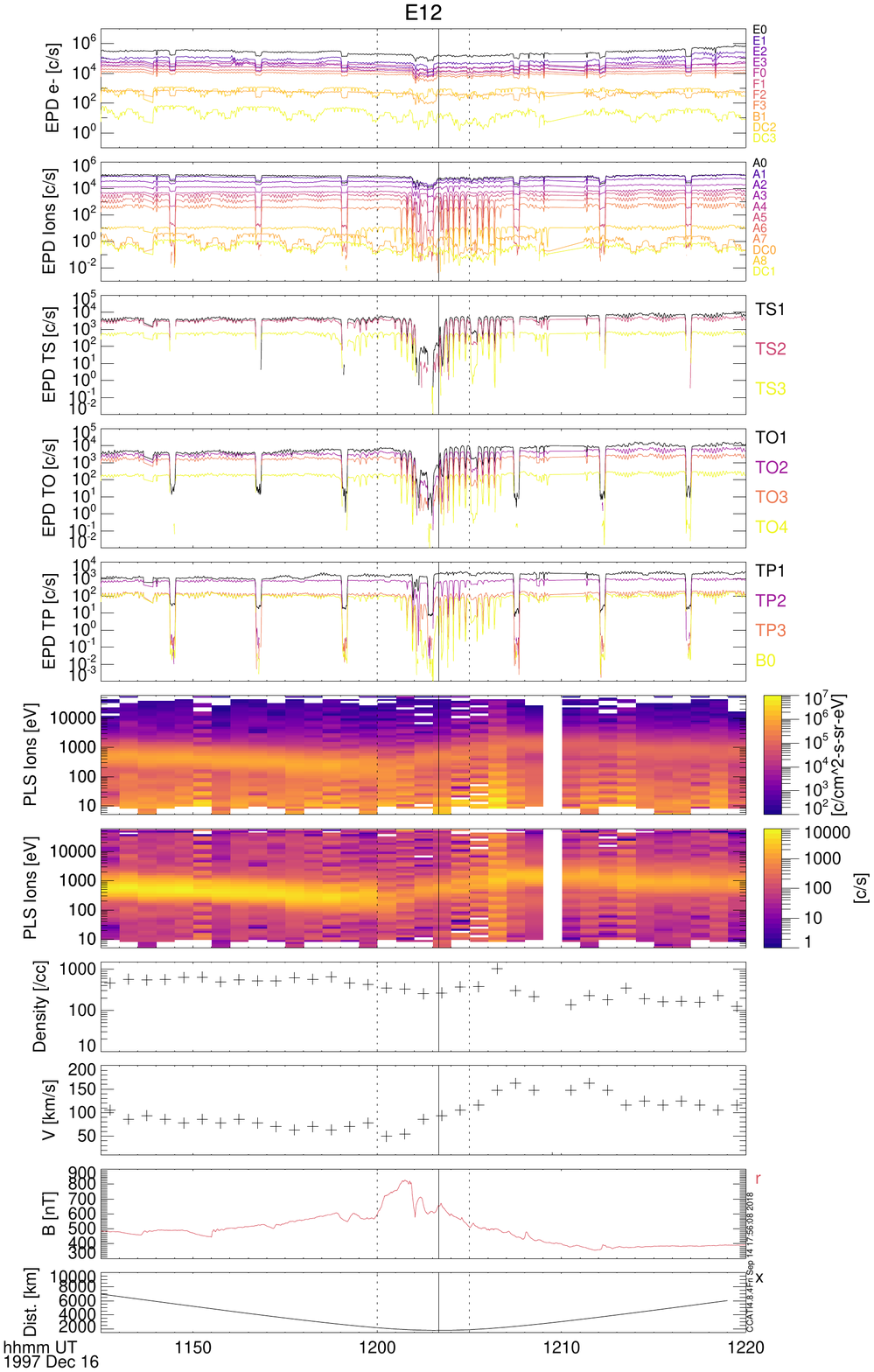}
  \caption{E12 flyby data in the same format as Figure \ref{img_E4_all}. The dotted lines indicate the region during which the direction of the corotational plasma changes according to the analysis of the PLS data presented in this thesis, see Chapter \ref{ch_comparison}. \cite{Jia2018} reports a change in magnetic field direction between the first dotted line and the closest approach (solid line), possibly related to a plume.}
  \label{img_E12_all}
\end{figure}

\begin{figure}[h]
  \centering
  \includegraphics[width=1.0\textwidth]{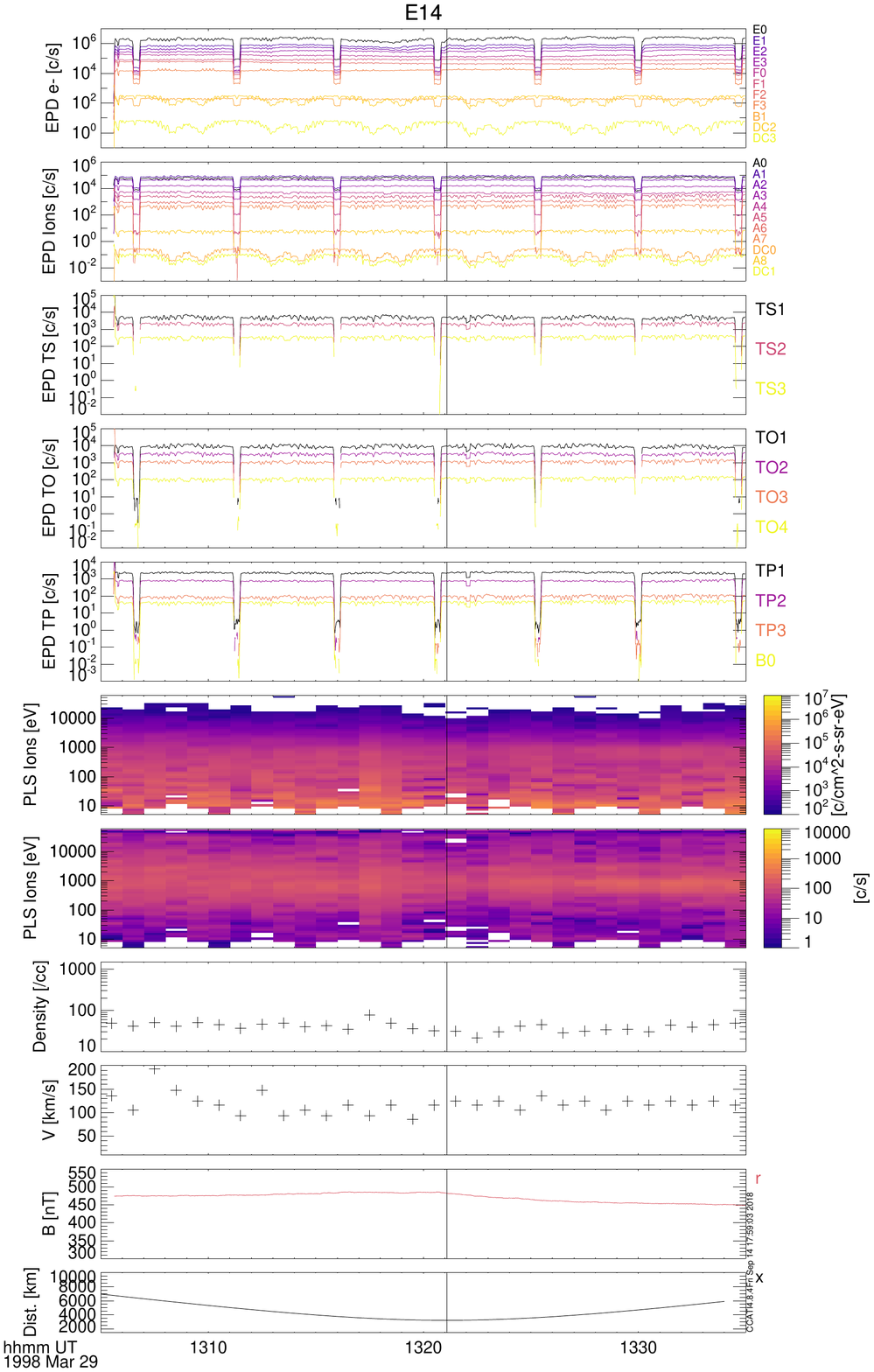}
  \caption{E14 flyby data in the same format as Figure \ref{img_E4_all}.}
  \label{img_E14_all}
\end{figure}

\begin{figure}[h]
  \centering
   \includegraphics[width=1.0\textwidth]{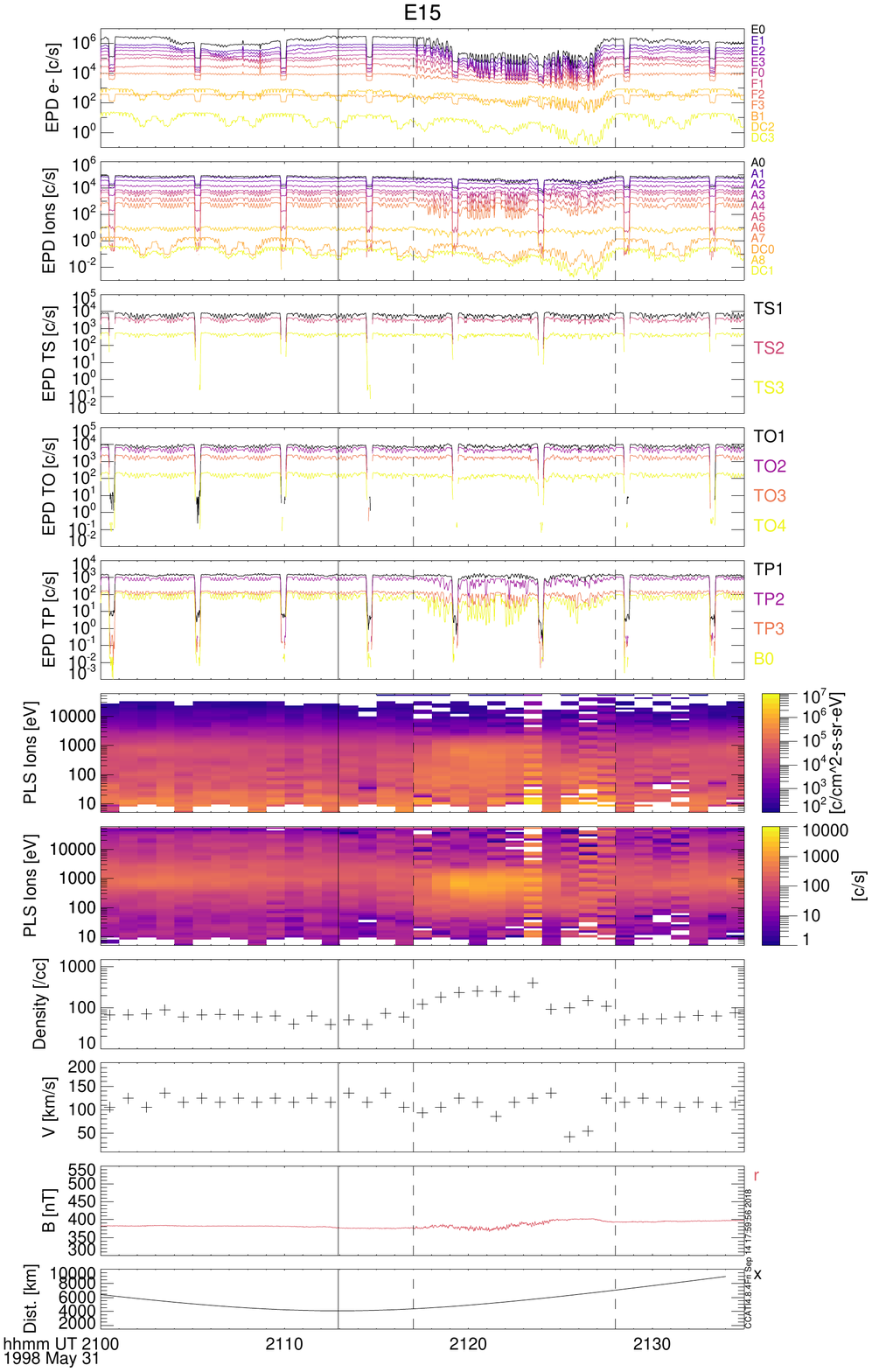}
  \caption{E15 flyby data in the same format as Figure \ref{img_E4_all}. The dashed vertical lines indicate the geometrical wake.}
  \label{img_E15_all}
\end{figure}

\begin{figure}[h]
  \centering
  \includegraphics[width=1.0\textwidth]{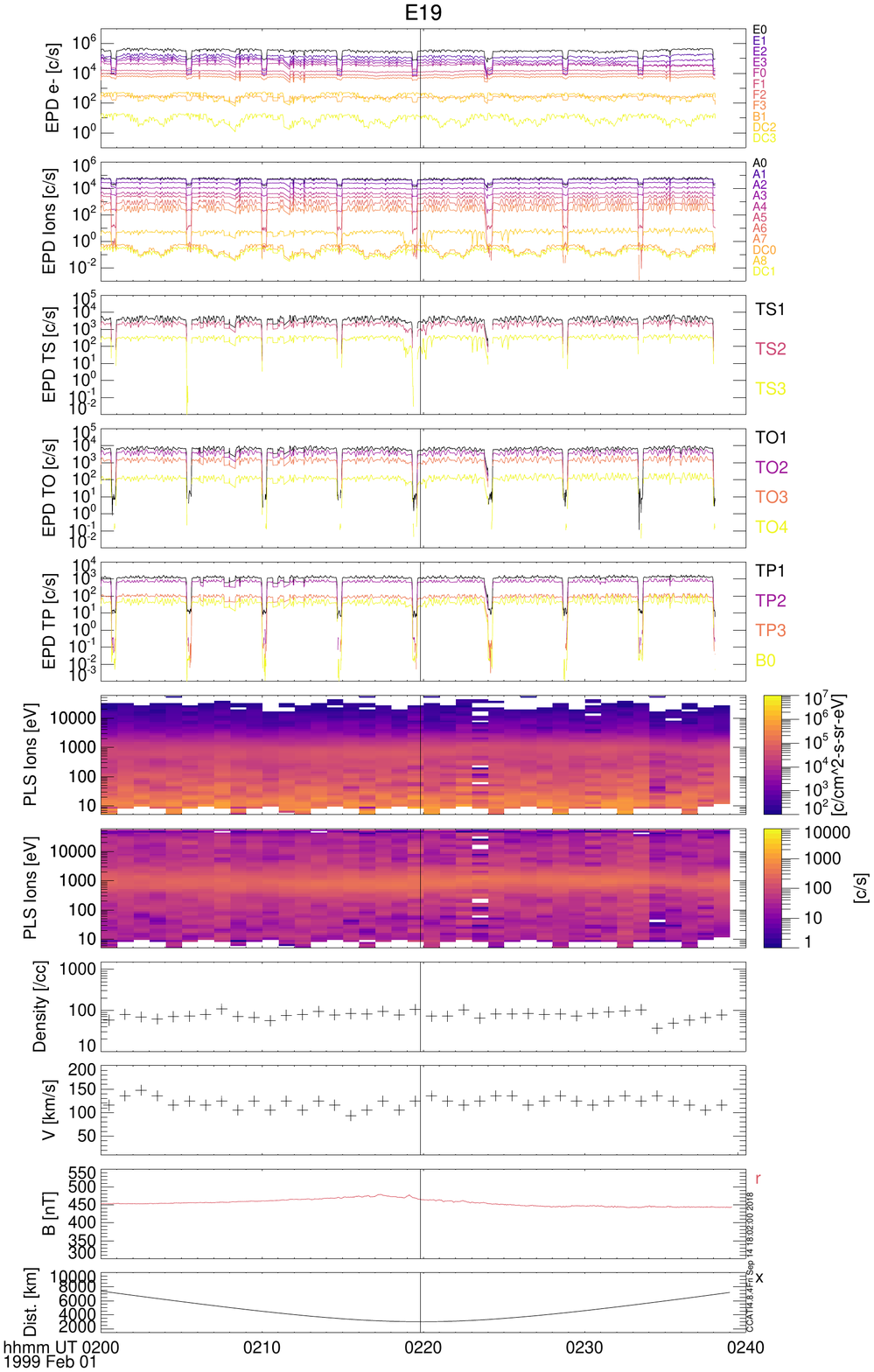}   \caption{E19 flyby data in the same format as Figure \ref{img_E4_all}.}
  \label{img_E19_all}
\end{figure}

\begin{figure}[h]
  \centering
  \includegraphics[width=1.0\textwidth]{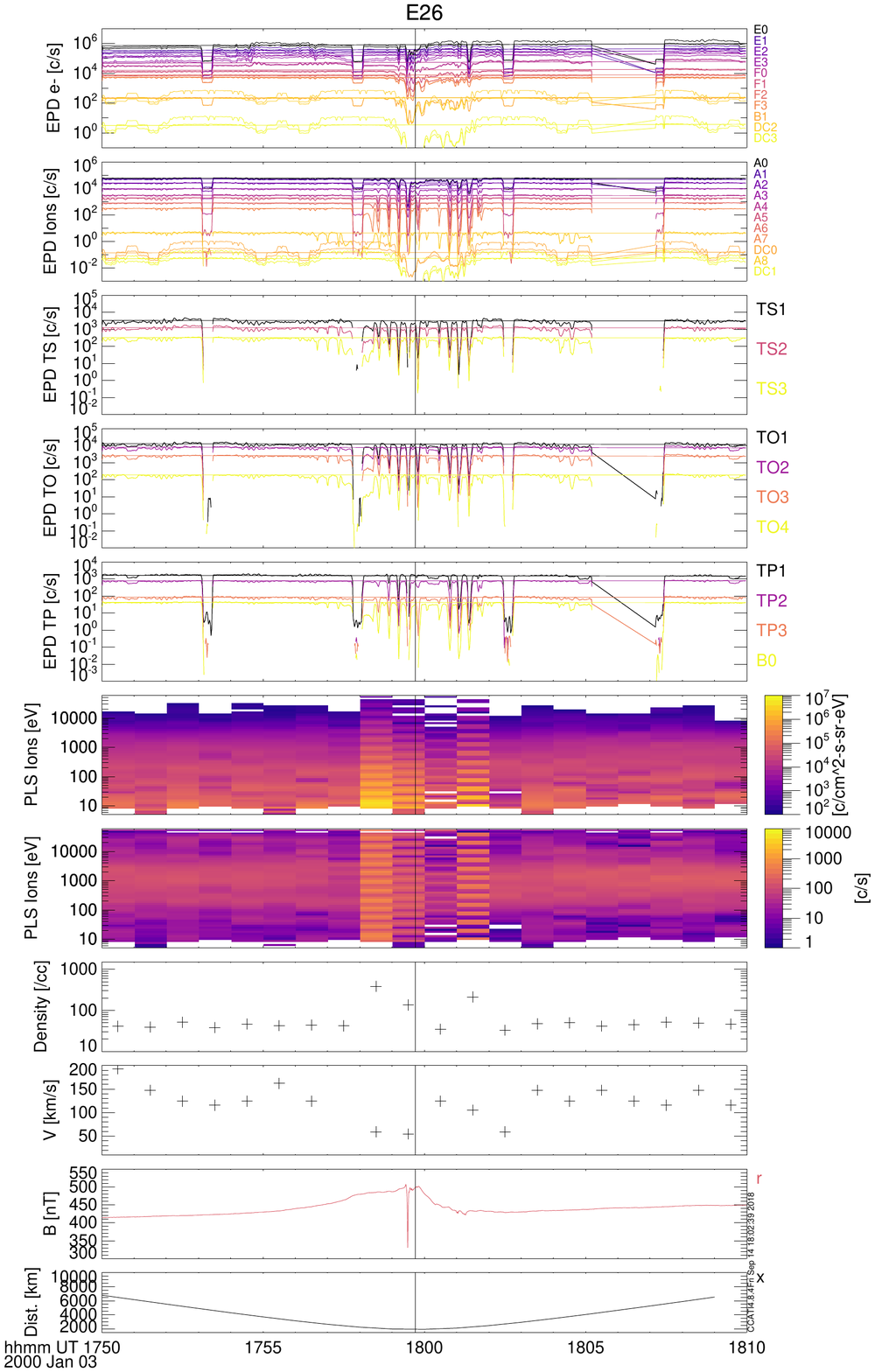}
  \caption{E26 flyby data in the same format as Figure \ref{img_E4_all}.}
  \label{img_E26_all}
\end{figure}

In this Appendix an overview of the EPD, PLS and MAG data collected during the Europa flybys of Galileo is provided. Data products that are presented are the count rates in the EPD E, A, TS, TO and TP channels (panel 1-5 from the top). For PLS both the differential flux, the count rate, the plasma density and magnitude of the velocity are presented (panel 6-9 from the top). For MAG the magnitude of the magnetic field is shown (panel 10). The flybys data are shown in sequential order in Figures \ref{img_E4_all} to \ref{img_E26_all}. In every figure the solid vertical line indicates the time of the closest approach.

\clearpage\null\newpage

\bibliographystyle{thesis} 
%
%

\bibliography{database} 

\clearpage\null\newpage

\chapter*{Acknowledgements}
\addcontentsline{toc}{chapter}{Acknowledgements}
\label{ch_acknowledgements}
I worked on this project at three institutes, the Swedish Institute of Space Physics in Kiruna, the Max Planck Institute for Solar System Research in G\"ottingen and the technical University of Braunschweig, there are people at each of these locations that I would like to thank for helping me during the past few years.

First I would like to thank Prof. Dr. Stas Barabash, Dr. Norbert Krupp, Dr. Sonja Schuh, Prof. Dr. Karl-Heinz Glassmeier, the International Max Planck Research School for Solar System Research at the Max Planck Institute for Solar System Research, the Swedish Institute of Space Physics and the Braunschweig University of Technology who have made it possible for me to conduct this PhD project on Europa. I couldn't have imagined a more interesting project. 

Secondly I would like to thank my supervisors. Starting with Dr. Yoshifumi Futaana, who was my daily supervisor back in Kiruna. Futaana, thank you for getting me started with particle simulations and the many, many teachings by the drawing board in your office.
Next, Dr. Elias Roussos and Dr. Norbert Krupp my two supervisors in G\"ottingen. Elias, I remember the first talk by you that I listened to, it inspired me by showing how interesting magnetospheric physics really can be. Thank you for always being ready to help and sharing your endless knowledge and ideas. Norbert, I remember our first meeting at the PEP team meeting in Kiruna, thanks for taking  me as a PhD student and for your advice, support and ideas. Thanks also go to Prof. Dr. Karl-Heinz Glassmeier for supporting the project and giving advice. Finally thanks go to Prof. Dr. Stas Barabash. Stas, I remember our first meeting at the airport in Schiphol, thank you for getting me started and giving me a chance to do the research I'm most interested in.

Beyond my direct supervisors I would like to acknowledge a few more people. Thanks go to Dr. Markus Fr\"anz for the countless times you have patiently helped me understand CCATI and deal with the PLS data. Also, thanks go to Dr. Kostas Dialynas for sharing your vast database of charge exchange cross sections, to Boris Semenov at NAIF for answering my questions about the Galileo SPICE files, to Dr. Andreas Lagg for helping me with the EPD software and to Dr. Jesper Lindkvist for helping me implement the induced dipole.

Thanks go to all my friends and colleagues at IRF and MPS for advice and support, I started out with listing you all, but I decided not to in the end, thank you all! Finally, I would also like to thank my family and friends back home, for having to miss me during my stay in the North and East.

\vspace{5mm}
\noindent
\textit{Europa is the new Mars.}\\
\textit{I'm a believer, I believe in the plume. - Pontus C. Brandt}\\
\textit{Hans is looking for monsters on Europa. - Stas Barbash}\\
\textit{You got the Americans pretty scared. - Magnus Emanuelsson}\\

\chapter*{Curriculum Vitae}
\addcontentsline{toc}{chapter}{Curriculum Vitae}
\label{ch_cv}

\tolerance=600
\begin{cv}{}
  \begin{cvlist}{Personal information\\}
  \item Name: Hans Leo Frans Huybrighs
  \item Date of birth: 8 June 1990
  \item Place of birth: Etterbeek (Belgium)
  \end{cvlist}
  \begin{cvlist}{Education}
  \item[06/2015--12/2018] Braunschweig University of Technology (Braunschweig, Germany) \\
    PhD thesis at the Max Planck Institute for Solar System Research, International Max Planck Research School, IMPRS (G\"ottingen, Germany). \\
  \item[09/2012--05/2015] Delft University of Technology (Delft, the Netherlands) \\
    Master of Science: Aerospace Engineering: Space Flight \\
    \textit{Thesis: The feasibility of in-situ observations of Europa's water
vapour plumes}. Thesis work and internship conducted at the Swedish Institute of Space Physics, IRF (Kiruna, Sweden).
  \item[09/2008--08/2012] Delft University of Technology (Delft, the Netherlands) \\
    Bachelor: Aerospace Engineering
  \end{cvlist}
  
\end{cv}




\end{document}